\documentclass[useAMS,usenatbib]{mnras}
\usepackage[T1]{fontenc}
\usepackage{ae,aecompl}
\usepackage{soul}
\usepackage{array}
\usepackage{times}
\usepackage{graphicx}
\usepackage{mathptmx}
\usepackage{mathtools}
\usepackage{amsmath}
\usepackage{amssymb}
\usepackage[usenames]{color}
\usepackage{pdflscape}
\usepackage{hyperref}
\usepackage[all]{hypcap}
\usepackage{caption}
\def\mnras{MNRAS}
\def\apjl{ApJL }
\def\aj{AJ }
\def\apj{ApJ }
\def\pasj{PASJ}
\def\pasp{PASP }

\def\apjs{ApJS }
\def\araa{ARAA }
\def\aap{A\&A }
\def\nat{Nature }
\hypersetup{colorlinks=true,citecolor=blue,linkcolor=purple,filecolor=black,runcolor=black,breaklinks=true}
\usepackage{etoolbox}
\makeatletter
\patchcmd\@combinedblfloats{\box\@outputbox}{\unvbox\@outputbox}{}{%
   \errmessage{\noexpand\@combinedblfloats could not be patched}%
}%
 \makeatother
\begin{document}
\newcommand{\comment}[1]{}
\definecolor{purple}{RGB}{160,32,240}
\newcommand{\peter}[1]{\textcolor{purple}{(\bf #1)}}
\newcommand{\macc}{M_\mathrm{acc}}
\newcommand{\mpeak}{M_\mathrm{peak}}
\newcommand{\mnow}{M_\mathrm{now}}
\newcommand{\vacc}{v_\mathrm{acc}}
\newcommand{\vpeak}{v_\mathrm{peak}} 
\newcommand{\vnow}{v^\mathrm{now}_\mathrm{max}}

\newcommand{\onesigdist}{$16^\mathrm{th}-84^\mathrm{th}$ percentile range}
\newcommand{\Mnfw}{M_\mathrm{NFW}}
\newcommand{\Msun}{\;\mathrm{M}_{\odot}}
\newcommand{\mvir}{M_\mathrm{vir}}
\newcommand{\rvir}{R_\mathrm{vir}}
\newcommand{\vmax}{v_\mathrm{max}}
\newcommand{\vmac}{v_\mathrm{max}^\mathrm{acc}}
\newcommand{\mvac}{M_\mathrm{vir}^\mathrm{acc}}
\newcommand{\sfr}{\mathrm{SFR}}
\newcommand{\plotgrace}[1]{\includegraphics[width=\columnwidth,type=pdf,ext=.pdf,read=.pdf]{#1}}
\newcommand{\plotssgrace}[1]{\includegraphics[width=0.95\columnwidth,type=pdf,ext=.pdf,read=.pdf]{#1}}
\newcommand{\plotgraceflip}[1]{\includegraphics[width=\columnwidth,type=pdf,ext=.pdf,read=.pdf]{#1}}
\newcommand{\plotlargegrace}[1]{\includegraphics[width=2\columnwidth,type=pdf,ext=.pdf,read=.pdf]{#1}}
\newcommand{\plotlargegraceflip}[1]{\includegraphics[width=2\columnwidth,type=pdf,ext=.pdf,read=.pdf]{#1}}
\newcommand{\plotminigrace}[1]{\includegraphics[width=0.5\columnwidth,type=pdf,ext=.pdf,read=.pdf]{#1}}
\newcommand{\plotmicrograce}[1]{\includegraphics[width=0.25\columnwidth,type=pdf,ext=.pdf,read=.pdf]{#1}}
\newcommand{\plotsmallgrace}[1]{\includegraphics[width=0.66\columnwidth,type=pdf,ext=.pdf,read=.pdf]{#1}}
\newcommand{\plotappsmallgrace}[1]{\includegraphics[width=0.33\columnwidth,type=pdf,ext=.pdf,read=.pdf]{#1}}
\newcommand{\plotmodestgrace}[1]{\includegraphics[width=0.86\columnwidth,type=pdf,ext=.pdf,read=.pdf]{#1}}

\newcommand{\vmp}{v_\mathrm{Mpeak}}
\newcommand{\fq}{f_\mathrm{Q}}
\newcommand{\sfrq}{SFR_\mathrm{Q}}
\newcommand{\sfrsf}{SFR_\mathrm{SF}}
\newcommand{\sigq}{\sigma_\mathrm{Q}}
\newcommand{\sigsf}{\sigma_\mathrm{SF}}
\newcommand{\fit}{fitting}

\newcommand{\hinv}{h^{-1}}
\newcommand{\mpc}{\rm{Mpc}}
\newcommand{\hmpc}{$\hinv\mpc$}

\newcommand{\mstar}{M_\ast}
\newcommand{\risa}[1]{\textcolor{red}{\textbf{RW: {#1}}}}
\newcommand{\aph}[1]{\textcolor{purple}{\textbf{APH: {#1}}}}

\title[The Galaxy -- Halo Assembly Correlation]{\textsc{UniverseMachine}: The Correlation between Galaxy Growth and Dark Matter Halo Assembly from $z=0-10$}

\author[Behroozi, Wechsler, Hearin, Conroy]{Peter~Behroozi$^{1}$,\thanks{E-mail:     behroozi@email.arizona.edu} Risa H.~Wechsler$^{2,3}$, Andrew P.~Hearin$^{4}$, Charlie Conroy$^{5}$
\\
\\
$^{1}$ Department of Astronomy and Steward Observatory, University of Arizona, Tucson, AZ 85721, USA\\
$^{2}$ Kavli Institute for Particle Astrophysics and Cosmology and Department of Physics, Stanford University, Stanford, CA 94305, USA\\
$^{3}$ Department of Particle Physics and Astrophysics,  SLAC National  Accelerator Laboratory, Stanford, CA 94305, USA\\
$^{4}$ High-Energy Physics Division, Argonne National Laboratory, Argonne, IL 60439, USA\\
$^{5}$ Department of Astronomy, Harvard University, Cambridge, MA 02138, USA\\
}
\date{Released \today}
\pubyear{2019}

\maketitle
 
\begin{abstract}
We present a method to flexibly and self-consistently determine individual galaxies' star formation rates (SFRs) from their host haloes' potential well depths, assembly histories, and redshifts.  The method is constrained by galaxies' observed stellar mass functions, SFRs (specific and cosmic), quenched fractions, UV luminosity functions, UV--stellar mass relations, IRX--UV relations, auto- and cross-correlation functions (including quenched and star-forming subsamples), and quenching dependence on environment; each observable is reproduced over the full redshift range available, up to $0<z<10$.  Key findings include: galaxy assembly correlates strongly with halo assembly; quenching correlates strongly with halo mass; quenched fractions at fixed halo mass decrease with increasing redshift; massive quenched galaxies reside in higher-mass haloes than star-forming galaxies at fixed galaxy mass; star-forming and quenched galaxies' star formation histories at fixed mass differ most at $z<0.5$; satellites have large scatter in quenching timescales after infall, and have modestly higher quenched fractions than central galaxies; \textit{Planck} cosmologies result in up to $0.3$ dex lower stellar---halo mass ratios at early times; and, nonetheless, stellar mass--halo mass ratios rise at $z>5$.  Also presented are revised stellar mass---halo mass relations for all, quenched, star-forming, central, and satellite galaxies; the dependence of star formation histories on halo mass, stellar mass, and galaxy SSFR; quenched fractions and quenching timescale distributions for satellites; and predictions for higher-redshift galaxy correlation functions and weak lensing surface densities.  The public data release (DR1) includes the massively parallel ($>10^5$ cores) implementation (the \textsc{UniverseMachine}), the newly compiled and remeasured observational data, derived galaxy formation constraints, and mock catalogues including lightcones.
 \end{abstract}

\section{Introduction}

\label{s:introduction}

In $\Lambda$CDM cosmologies, galaxies form at the centres of gravitationally self-bound, virialized dark matter structures (known as \textit{haloes}).  Haloes form hierarchically, and the largest collapsed structure in a given overdensity (i.e., a \textit{central halo}) can contain many smaller self-bound structures (\textit{satellite haloes}).  While the broad contours of galaxy formation physics are known \citep[see][for reviews]{Silk12,Somerville15}, a fully predictive framework from first principles does not yet exist \citep[see][for a review]{Naab17}.

Traditional theoretical methods include hydrodynamical and semi-analytic models, which use known physics as a strong prior on how galaxies may form.  For example, current implementations attempt to simulate the effects of supernovae, radiation pressure, multiphase gas, black hole accretion, photo- and collisional ionization, and chemistry \citep[see][for reviews]{Somerville15,Naab17}.  All such methods approximate physics below their respective resolution scales (galaxies for semi-analytic models; particles and/or grid elements for hydrodynamical simulations), and different reasonable approximations lead to different resulting galaxy properties \citep{YuLu14,Kim16}.  

These methods are complemented by empirical modeling, wherein the priors are significantly weakened and the physical constraints come almost entirely from observations.  Current empirical models constrain physics averaged over galaxy scales, similar to semi-analytical models.  Indeed, as empirical modeling has grown in complexity and self-consistency, as well as in the number of galaxy \citep[e.g.,][]{Moster17,RP17,Somerville18}, gas \citep[e.g.,][]{Popping15}, metallicity \citep[e.g.,][]{RP16b}, and dust \citep{Imara18} observables generated, the mechanics of semi-analytic and empirical models have become increasingly similar.  For example, the techniques of post-processing merger trees from N-body simulations, comparing to galaxy correlation functions, and using orphan galaxies were commonplace in semi-analytical models well before they were used in empirical ones \citep{Roukema97,Kauffmann99}.

Nonetheless, the presence or absence of strong physical priors remains a key difference between empirical and semi-analytic models.  While semi-analytic models can therefore obtain tighter parameter constraints for the same data (or lack thereof), empirical models can reveal physics that was not previously expected to exist \citep[e.g.,][]{Behroozi13,BehrooziHighZ}.  In cases where traditional methods have strong disagreements (e.g., on the mechanism for galaxy quenching), this latter quality can be very powerful, and is hence a strong motivation for using empirical modeling here.

Most current empirical models relate galaxy properties to properties of their host dark matter haloes.  Larger haloes host larger galaxies, with relatively tight scatter in the stellar mass---halo mass relation \citep{more-09,yang-09,Leauthaud12,Reddick12,Watson13,Tinker13,Gu16}.  Hence, it has become common to investigate average galaxy growth via a connection to the average growth of haloes (\citealt{Zheng07,White07,cw-08,Firmani10,Leitner11,Bethermin13,Wang12,Moster12,BWC13,BehrooziND,Mutch13,Birrer14,Marchesini14,Lu14,Lu15,Papovich15,Li16}; see \citealt{Wechsler18} for a review).  These studies have found that the stellar mass---halo mass relation is relatively constant with redshift from $0<z<4$ \citep{Behroozi13}, but may evolve significantly at $z>4$ \citep{BehrooziHighZ,Finkelstein15b,Sun16}.

If galaxy mass is tightly correlated with halo mass on average, it is natural to expect that individual galaxy assembly could be correlated with halo assembly.  This assembly correlation for individual galaxies has strong observational support. For example, satellite galaxies in clusters have redder colours (implying lower star formation rates; SFRs)  and more elliptical morphologies than similar-mass galaxies in the field \citep[and references thereto]{Hubble31}.  At the same time, the satellite haloes hosting these satellite galaxies have undergone significant stripping due to cluster tidal forces \citep{Tormen98b,Kravtsov04b,Knebe06,Hahn09,Wu13,BehrooziMergers}.  Thus, there is a \textit{correlation} between the assembly histories of satellite galaxies and satellite haloes, regardless of whether there is a direct \textit{causation}.

For central galaxies (i.e., the main galaxies in central haloes), several studies (\citealt{Tinker12}, \citealt{Berti16}, \citealt{Wang18}, and this study) have also found correlations between these galaxies' quenched fractions (i.e., the fraction not forming stars) and the surrounding environment.  At the same time, environmental density strongly correlates with halo accretion rates \citep{Hahn09,BehrooziMergers,Lee16}.  This would again suggest a correlation (and again not necessarily causation) between central galaxies' star formation rates and their host halo matter accretion rates.  

Empirical models that correlate galaxy star formation rates or colours with halo concentrations (correlated with halo formation time; \citealt{Wechsler02}) have shown success in matching galaxy autocorrelation functions, weak lensing, and radial profiles of quenched galaxy fractions around clusters \citep{Hearin13,Hearin13b,Watson15}.  Models that relate galaxy SFRs linearly to halo mass accretion rates \citep[albeit non-linearly to halo mass;][]{Popp15,Becker15,RP16,Sun16,Mitra16,Cohn16,Moster17} have also shown success in this regard.  To date, all such models have made a strong assumption that galaxy formation is perfectly correlated to a chosen proxy for halo assembly.

In our approach, we do not impose an \textit{a priori} correlation between galaxy assembly and halo assembly.  Instead, given that galaxy clustering depends strongly on this correlation, we can directly measure it.   Our method first involves making a guess for how galaxy SFRs depend on host halo potential well depth, assembly history, and redshift.  This ansatz is then self-consistently applied to halo merger trees from a dark matter simulation, resulting in a mock universe; this mock universe is compared directly with real observations to compute a Bayesian likelihood.  A Markov Chain Monte Carlo algorithm then makes a new guess for the galaxy SFR function, and the process is repeated until the range of SFR functions that are compatible with observations is fully sampled.  

Observational constraints used here include stellar mass functions, UV luminosity functions, the UV--stellar mass relation, specific and cosmic SFRs, galaxy quenched fractions, galaxy autocorrelation functions, and the quenched fraction of central galaxies as a function of environmental density.  We also compare to galaxy-galaxy weak lensing.  High-redshift constraints have improved dramatically in the past five years due to the CANDELS, 3D-HST, ULTRAVISTA, and ZFOURGE surveys \citep{Grogin11,Brammer12,McCracken12}.  At the same time, pipeline and fitting improvements have made significant changes to the inferred stellar masses of massive low-redshift galaxies \citep{Bernardi13}.  As with past analyses \citep{Behroozi10,BWC13}, we marginalize over many systematic uncertainties, including those from stellar population synthesis, dust, and star formation history models.  

We present the simulations and the new compilation of observational data in \S \ref{s:data}, followed by the methodology in \S \ref{s:methodology}.  The main results, discussion, and conclusions are presented in \S \ref{s:results}, \S \ref{s:discussion}, and \S \ref{s:conclusions}, respectively.  Appendices discuss alternate parametrizations for halo assembly history (\ref{a:alternatives}), the need for orphan satellites (\ref{a:orphans}), the compilation of and uncertainties in the observational data (\ref{a:data}), revised fits to UV--stellar mass relations (\ref{a:uvsm}), the functional forms used (\ref{a:fforms}), code parallelization and performance (\ref{a:code}), results for non-universal stellar initial mass functions (\ref{a:imf}), the best-\fit{} model and 68\% parameter confidence intervals (\ref{a:bestfit}), parameter correlations (\ref{a:correlations}), and fits to stellar mass--halo mass relations (\ref{a:smhm_fits}).

For conversions from luminosities to stellar masses, we assume the \cite{Chabrier03} stellar initial mass function, the \cite{bc-03} stellar population synthesis model, and the \cite{calzetti-00} dust law.  We adopt a flat, $\Lambda$CDM cosmology with parameters ($\Omega_m=0.307$, $\Omega_\Lambda = 0.693$, $h=0.678$, $\sigma_8 = 0.823$, $n_s = 0.96$) consistent with \textit{Planck} results \citep{Planck15}.  Halo masses follow the \cite{mvir_conv} spherical overdensity definition and refer to peak historical halo masses extracted from the merger tree ($\mpeak$) except where otherwise specified.

\section{Simulations \& Observations}

\label{s:data}

\begin{table*}
\caption{Summary of Observational Constraints}
\begin{tabular}{lcccc}
\hline
Type & Redshifts & Primarily Constrains & Details \& References\\
\hline
Stellar mass functions$^\ast$ & $0-4$ & SFR$-\vmp$ relation & Appendix \ref{a:smf}\\
Cosmic star formation rates$^\ast$ & $0-10$ & SFR$-\vmp$ relation & Appendix \ref{a:csfr_ssfr}\\
Specific star formation rates$^\ast$  & $0-8$ & SFR$-\vmp$ relation &  Appendices \ref{a:csfr_ssfr},  \ref{a:tension}\\
UV luminosity functions & $4-10$ & SFR$-\vmp$ relation &  Appendix \ref{a:uvlfs}\\
Quenched fractions$^\ast$ & $0-4$ & Quenching$-\vmp$ relation & Appendix \ref{a:qf}\\
Autocorrelation functions for quenched/SF/all galaxies from SDSS$^\dag$ & $\sim 0$ & Quenching/assembly history correlation& Appendix \ref{a:cf}\\
Cross-correlation functions for galaxies from SDSS$^\dag$ & $\sim 0$ & Satellite disruption & Appendix \ref{a:cf}\\
Autocorrelation functions for quenched/SF galaxies from PRIMUS$^\ast$ & $\sim 0.5$ & Quenching/assembly history correlation& Appendix \ref{a:cf}\\
Quenched fraction of primary galaxies as a function of neighbour density$\dag$ & $\sim 0$ & Quenching/assembly history correlation & Appendix \ref{a:ecq}\\
Median UV--stellar mass relations$\dag$ & $4-8$ & Systematic Stellar Mass Biases &  Appendix \ref{a:uvsm}\\
IRX--UV relations & $4-7$ & Dust &  Appendix \ref{a:uvsm}\\
\hline
\end{tabular}
\parbox{2.0\columnwidth}{\textbf{Notes.} SDSS: the Sloan Digital Sky Survey.  PRIMUS: the PRIsm MUlti-object Survey.  $\vmp$: $v_\mathrm{max}$ at the time of peak historical halo mass.\\ $^\ast$: renormalized/converted in this study to more uniform modeling assumptions.  $\dag$: newly measured or reanalyzed in this study.}
\label{t:obs_summary}
\end{table*}

\subsection{Simulations}

\label{s:simulations}
We use the \textit{Bolshoi-Planck} dark matter simulation \citep{Klypin14,RP16b} for halo properties and assembly histories.  \textit{Bolshoi-Planck} follows a periodic, comoving volume 250 $h^{-1}$ Mpc on a side with 2048$^3$ particles ($\sim 8\times 10^9$), and was run with the \textsc{art} code \citep{kravtsov_etal:97,kravtsov_klypin:99}.  The simulation has high mass ($1.6 \times 10^{8} h^{-1}\Msun$), force (1 $h^{-1}$ kpc), and time output (180 snapshots spaced equally in $\log(a)$) resolution.  The adopted cosmology (flat $\Lambda$CDM; $h=0.678$, $\Omega_m = 0.307$, $\sigma_8=0.823$, $n_s = 0.96$) is compatible with \textit{Planck}15 results \citep{Planck15}.  We also use the \textit{MDPL2} dark matter simulation \citep{Klypin14,RP16b} to calculate covariance matrices for auto- and cross-correlation functions.  \textit{MDPL2} adopts an identical cosmology to \textit{Bolshoi-Planck}, except for assuming $\sigma_8 = 0.829$, and follows a 1 $h^{-3}$ Gpc$^3$ region with 3840$^3$ particles ($\sim 57 \times 10^{9}$).  The  mass ($2.2 \times 10^9\Msun$) and force (5 $h^{-1}$ kpc) resolution are coarser than for \textit{Bolshoi-Planck}.  For both simulations, halo finding and merger tree construction used the \textsc{rockstar} \citep{Rockstar} and \textsc{Consistent Trees} \citep{BehrooziTree} codes, respectively.

\subsection{Observations}

\label{s:obs}

As summarized in Table \ref{t:obs_summary}, we combine recent constraints from stellar mass functions (SMFs; Table \ref{t:smf}), cosmic star formation rates (CSFRs; Table \ref{t:csfr}), specific star formation rates (SSFRs; Table \ref{t:ssfr}), quenched fractions (QFs), UV luminosity functions (UVLFs), UV--stellar mass relations (UVSM relations), and infrared excess--UV relations (IRX--UV relations) with measurements of galaxy auto- and cross-correlation functions (CFs) and the environmental dependence of central galaxy quenching.  Full details are presented in Appendices \ref{a:data} and \ref{a:uvsm}.

Briefly, stellar mass function (SMF) constraints include data from the Sloan Digital Sky Survey (SDSS), the PRIsm MUlti-object Survey (PRIMUS), UltraVISTA, the Cosmic Assembly Near-infrared Deep Extragalactic Legacy Survey (CANDELS), and the FourStar Galaxy Evolution Survey (ZFOURGE).  These constraints cover $0 < z < 4$ and were renormalized as necessary to ensure consistent modeling assumptions (Table \ref{t:smf_assumptions}) and photometry for massive galaxies (Fig.\ \ref{f:mous_corr}).  As noted in \cite{Kravtsov14}, improved photometry for massive galaxies significantly increases their stellar mass to halo mass ratios as compared to \cite{BWC13}.  In addition, based on null findings in \cite{Williams16}, there was no need to perform surface brightness corrections for low-mass galaxies as in \cite{BWC13}.

Specific SFRs and cosmic SFRs cover $0 < z < 10.5$ and were only renormalized to a \cite{Chabrier03} initial mass function, as matching other modeling assumptions does not increase self-consistency between SFRs and the growth of SMFs \citep{Madau14,Leja15,Tomczak15}.  These data are taken from a wide range of surveys (including SDSS, GAMA, ULTRAVISTA, CANDELS, ZFOURGE) and techniques (including UV, $24\mu$m, radio, H$\alpha$, and SED fitting).

Quenched fractions as a function of stellar mass, from \cite{Bauer13}, \cite{Moustakas13}, and \cite{Muzzin13}, cover the range $0 < z < 3.5$.  As discussed in Appendix \ref{a:qf}, these papers use different definitions for ``quenched'' (cuts in SSFR and UVJ luminosities, respectively), which we self-consistently model when comparing to each paper's results.  Stellar masses were renormalized as for SMFs.

Autocorrelation functions for all, quenched, and star-forming galaxies are newly measured from the SDSS (Appendix \ref{a:cf}), with covariance matrices measured from identical sky masks in mock catalogues of significantly greater volume.  Cross-correlation functions of massive galaxies ($M_\ast > 10^{11}\Msun$) with Milky-Way mass galaxies ($M_\ast \sim 10^{10.4}\Msun$) are also newly measured from the SDSS to help constrain satellite disruption.  At $z\sim0.5$, we use the correlation functions for quenched and star-forming galaxies from PRIMUS \citep{Coil17}.  As with the SDSS, covariance matrices are measured from mock catalogues; redshift errors are remeasured from a cross-comparison between the G10/COSMOS redshift catalogue \citep{Davies15} and the PRIMUS DR1 catalogue \citep{Coil11}.

Correlation functions are primarily sensitive to \textit{satellite} quenching, so constraining \textit{central} galaxy quenching requires a different measurement.  Here, we use the quenched fraction for primary galaxies (i.e., those that are the largest in a given surrounding volume) as a function of the number counts of lower-mass neighbours \citep[Appendix \ref{a:ecq}; see also][]{Berti16}.  This signal is significantly stronger and more robustly measurable than two-halo galactic conformity, and is a plausible cause thereof \citep{Hearin15}.  In addition, \cite{Lee16} has shown that halo mass accretion rates correlate strongly with environmental density for central haloes, so this statistic helps constrain the correlation between halo mass and galaxy assembly for central haloes.

Finally, existing stellar mass functions at $z>4$ often depend on uncertain UV--stellar mass conversions, which results in significant interpublication scatter even on the same underlying data sets \citep[e.g., Fig. 6\ in][]{Moster17}.  We explore the underlying reason for these uncertainties in Appendix \ref{a:uvsm}, finding that uncertainties in the SED for star formation history and dust can be reduced by combining additional observables.  As a result, we develop a new SED-fitting tool (\textsc{SEDition}; Appendix \ref{a:uvsm}) and use it to remeasure median UV--stellar mass relations from the \cite{Song15} SED stacks for $z=4-8$ galaxies, combined with star formation history constraints from UV luminosity functions' evolution and dust constraints from ALMA \citep{Bouwens16}.  The resulting UV--stellar mass relations, combined with UV luminosity functions from \cite{Finkelstein15} and \cite{Bouwens15b}, replace constraints on SMFs at $z>4$.

\section{Methodology}

\label{s:methodology}

We summarize our approach in \S \ref{s:overview}, followed by details for the SFR parameterization (\S \ref{s:fform}), galaxy mergers (\S \ref{s:mergers}), stellar masses and luminosities (\S \ref{s:sm_uv}), and observational systematics including dust (\S \ref{s:systematics}).

\begin{figure*}
\includegraphics[width=1.5\columnwidth]{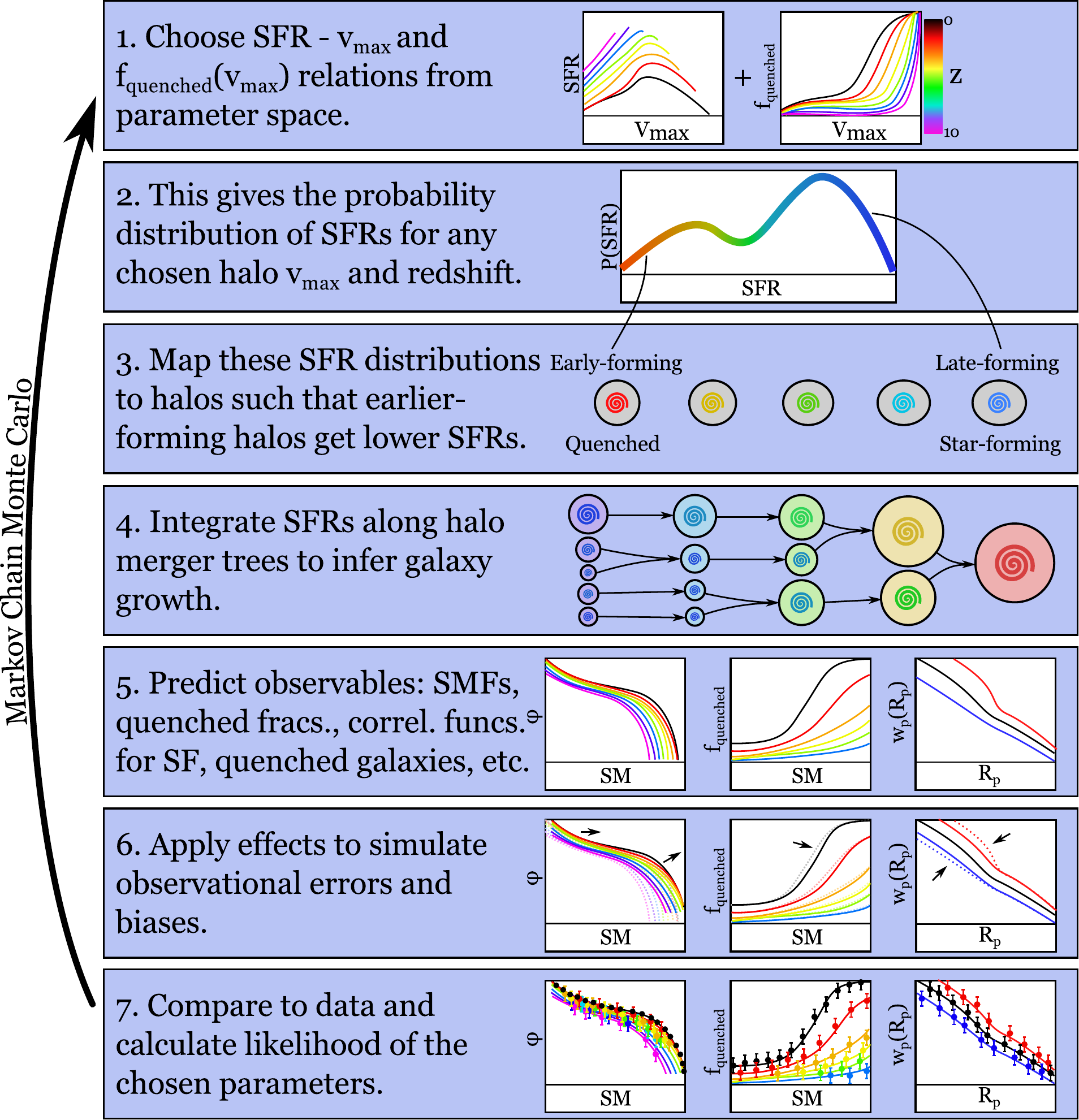}
\caption{Visual summary of the method for linking galaxy growth to halo growth (\S \ref{s:methodology}).}
\label{f:summary}
\end{figure*}

\begin{table*}
\caption{Table of parameters.}
\begin{tabular}{lcccc}
\hline
Symbol & Description & Equation & Parameters & Section\\
\hline
$\sigsf(z)$ & Scatter in SFR for star-forming galaxies & \ref{e:sig_sf_a} & 2 & \ref{s:fform}\\
$V(z)$ & Characteristic $\vmp$ in SFR -- $\vmp$ relation & \ref{e:v} & 4 & \ref{s:fform}\\
$\epsilon(z)$ & Characteristic SFR in SFR -- $\vmp$ relation & \ref{e:sfr} & 4 & \ref{s:fform}\\
$\alpha(z)$ & Faint-end slope of SFR -- $\vmp$ relation & \ref{e:alpha} & 4 & \ref{s:fform}\\
$\beta(z)$ & Massive-end slope of SFR -- $\vmp$ relation & \ref{e:beta} & 3 & \ref{s:fform}\\
$\gamma(z)$ & Strength of Gaussian efficiency boost in SFR -- $\vmp$ relation & \ref{e:gamma} &  3 & \ref{s:fform}\\
$\delta$ & Width of Gaussian efficiency boost in SFR -- $\vmp$ relation & \ref{e:delta} &  1 & \ref{s:fform}\\
\hline
$Q_\mathrm{min}(z)$ & Minimum quenched fraction & \ref{e:fqmin} & 2 & \ref{s:fform}\\
$V_Q(z)$ & Characteristic $\vmp$ for quenching & \ref{e:vq} & 3 & \ref{s:fform}\\
$\sigma_{VQ}(z)$ & Characteristic $\vmp$ width over which quenching happens & \ref{e:sig_q} & 3 & \ref{s:fform}\\
\hline
$r_c(z)$ & Rank correlation between halo assembly history ($\Delta \vmax$) and SFR & \ref{e:rc} & 4 & \ref{s:fform}\\
$\tau_R(z)$ & Correlation time for long-timescale random contributions to SFR rank & $-$ & 0 & \ref{s:fform}\\
$f_\mathrm{short}$ & Fraction of short-timescale random contributions to SFR rank & \ref{e:fshort} & 1 & \ref{s:fform}\\
\hline
$T_\mathrm{merge}$ & Threshold for $\vmax / \vmp$ at which disrupted haloes are no longer tracked & $-$ & 2 & \ref{s:mergers}\\  
$f_\mathrm{merge}$ & Fraction of host halo's radius below which disrupted satellites merge into the central galaxy & $-$ & 1 & \ref{s:mergers}\\
\hline
$\alpha_\mathrm{dust}$ & Characteristic rate at which dust increases with UV luminosity & \ref{e:dust} & 1 & \ref{s:sm_uv}\\
$M_\mathrm{dust}(z)$ & Characteristic UV luminosity for dust to become important & \ref{e:dust2} & 2 & \ref{s:sm_uv}\\
\hline
$\mu(z)$ & Systematic offset in both observed stellar masses and SFRs & \ref{e:mu} & 2 & \ref{s:systematics}\\
$\kappa(z)$ & Additional systematic offset in observed SFRs & \ref{e:kappa} & 1 &  \ref{s:systematics}\\
$\sigma_\mathrm{SM,obs}(z)$ & Random error in recovering stellar masses & \ref{e:smobs} & 1 & \ref{s:systematics}\\
$\sigma_\mathrm{SFR,obs}(z)$ & Random error in recovering SFRs & \ref{e:sfrobs} & 0 & \ref{s:systematics}\\
\hline
\end{tabular}
\parbox{2\columnwidth}{\textbf{Notes.} $\vmp$: $v_\mathrm{max}$ at the time of peak historical halo mass.  $\Delta \vmax$ is described by Eq.\ \ref{e:dvmax} in \S \ref{s:overview}.  Symbols followed by ``$(z)$'' depend on redshift and are described by multiple parameters (see equation references above).  $\tau_R$ is fixed to the halo dynamical time, $\left(\frac{4}{3}\pi G\rho_\mathrm{vir}\right)^{-1/2}$.  The total number of free model parameters is 44.}
\label{t:params}
\end{table*}

\begin{table*}
\caption{Table of priors.}
\begin{tabular}{lccc}
\hline
Symbol & Description & Equation & Prior\\
\hline
$T_\mathrm{orphan,300}$ & Threshold for $\vmax / \vmp$ at which disrupted haloes are no longer tracked around 300 km s$^{-1}$ hosts & $-$ & $U(0.2,1)$\\  
$T_\mathrm{orphan,1000}$ & Threshold for $\vmax / \vmp$ at which disrupted haloes are no longer tracked around 1000 km s$^{-1}$ hosts & $-$ & $U(0.2,1)$\\  
$f_\mathrm{merge}$ & Fraction of host halo's virial radius below which disrupted satellites are merged with the central galaxy & $-$ & $U(0,2)$\\
$\mu_0$ & Value of $\mu$ at $z=0$, in dex & \ref{e:mu} & $G(0,0.14)$\\
$\mu_a$ & Redshift scaling of $\mu$, in dex & \ref{e:mu} & $G(0,0.24)$\\
$\kappa$ & Additional offset in observed vs.\ true SFR, in dex & \ref{e:kappa} & $G(0,0.24)$\\
$\sigma_\mathrm{SM,z}$ & Redshift scaling of $\sigma_\mathrm{SM}$ & \ref{e:smobs} & G(0.05, 0.015)\\
\hline
\end{tabular}
\parbox{2.15\columnwidth}{\textbf{Notes.} $G(x,y)$ is a Gaussian distribution with center $x$ and width $y$.  $U(x,y)$ is a uniform distribution over $[x,y]$.  Remaining parameters do not have explicit priors; $V$, $V_Q$, and $M_D$ are explored in logarithmic space, whereas the remainder are explored in linear space.}
\label{t:prev}
\end{table*}

\subsection{Design Overview}

\label{s:overview}

Our approach (Fig.\ \ref{f:summary}) parametrizes galaxy SFRs as a function of halo potential well depth, redshift, and assembly history.  For potential well depth, past works used peak historical halo mass \citep[e.g.,][]{Moster12,BWC13} or peak historical $\vmax$ \citep[e.g.,][]{Reddick12}, where $v_\mathrm{max}$ is the maximum circular velocity of the halo ($\equiv\max(\sqrt{GM(<R)/R})$).  Peak $\vmax$ better matches galaxy clustering \citep{Reddick12} and avoids pseudo-evolution issues \citep{Diemer13}.  Yet, strong, transient $\vmax$ peaks occur following major halo mergers \citep{BehrooziMergers}.  Hence, we use the value of $\vmax$ at the redshift where the halo reached its peak mass ($\vmp \equiv \vmax(z_\mathrm{M_{peak}})$) so that transient peaks after mergers do not affect long-term SFRs.

Previous studies have varied SFRs with halo mass accretion rates \citep[e.g.,][]{Becker15,RP16,Moster17} and concentrations \citep{Hearin13,Watson15}.  Satellites are problematic for both approaches, as neither satellite mass accretion nor concentration are robustly measured by halo finders \citep{Onions12,BehrooziNotts}, and most solutions (e.g., using the time since accretion; \citealt{Moster17}) cannot capture orbit-dependent effects.  

Here, we use the $\vmax$ accretion history, which is robustly measurable for satellites \citep{Onions12} and yields more clearly orbit-- and profile--dependent satellite SFRs.  A rapid increase in $\vmax$ means both a large influx of gas and a better ability to retain existing gas, both of which would suggest higher SFRs.  A rapid decrease implies either strong tidal stripping (as for satellites) or that the halo's $\vmax$ peaked during a major merger and is now dropping rapidly (as for post-starburst galaxies), both suggesting lower SFRs.  However, recent changes in $\vmax$ likely matter less for extremely stripped satellites---which we would expect to remain quenched.  Hence, we adopt the following $\vmax$ history parameter:
\begin{equation}
\Delta \vmax \equiv \frac{\vmax(z_\mathrm{now})}{\vmax(\max(z_\mathrm{dyn}, z_\mathrm{M_{peak}}))},
\label{e:dvmax}
\end{equation}
where $z_\mathrm{dyn}(t)$ is the redshift a dynamical time ($\equiv(\frac{4}{3}\pi G\rho_\mathrm{vir})^{-\frac{1}{2}}$) ago and $\rho_\mathrm{vir}(t)$ is the virial overdensity according to \cite{mvir_conv}.  For most haloes, $\Delta \vmax$ corresponds to the relative change in $\vmax$ over the past dynamical time.  However, for extremely stripped satellites, $\Delta \vmax$ corresponds to the relative change in $\vmax$ since they last accreted mass, preventing them from resuming star formation.

$\Delta\vmax$ is not the only option for parameterizing halo assembly (see Appendix \ref{a:alternatives} for alternatives).  However, as discussed above, there are physical reasons for it to correlate well with galaxy SFRs.  As long as there is some correlation between $\Delta \vmax$ and galaxy SFRs (regardless of the underlying \textit{causation}), our approach remains valid to determine the correlation strength.

Any choice for the function $SFR(\vmp,z,\Delta\vmax)$ (see \S \ref{s:fform} and Table \ref{t:params} for our parametrization; see Table \ref{t:prev} for priors) fully determines galaxy SFRs in every halo at every redshift in a dark matter simulation (Fig.\ \ref{f:summary}).  For each halo, the galaxy stellar mass and UV luminosity are self-consistently calculated from star formation histories along the halo's assembly and merger history.  This results in a mock observable universe (including galaxy positions, redshifts, stellar masses, SFRs, and luminosities) that can be directly compared to the real Universe.  We compute the likelihood for a given point in parameter space using galaxy stellar mass functions, specific SFRs, cosmic SFRs, quenched fractions, UV luminosity functions, UV--stellar mass relations, IRX--UV relations, auto-correlation functions (for all, star-forming, and quiescent galaxies), cross-correlation functions, and measurements of central galaxy quenching with environment (\S \ref{s:obs}), using covariance matrices where available.  The likelihood function ($\exp(-\chi^2/2)$) is processed through an MCMC algorithm (a hybrid of adaptive Metropolis and stretch-step MCMC algorithms; \citealt{Haario01,Goodman10}), resulting in empirical constraints on how galaxy growth correlates with halo growth.

\subsection{SFR Distribution}

\label{s:fform}

We parametrize SFRs in haloes as a function of $\vmp$ ($\vmax$ at the redshift of peak halo mass), $z$, and $\Delta\vmax$ (logarithmic growth in $\vmax$ over the past dynamical time), summarized in Table \ref{t:params}.  At fixed $\vmp$ and $z$, we assume that the SFR distribution is the sum of two log-normal distributions, corresponding to a quenched population and a star-forming population:
\begin{equation}
P(SFR|\vmp,z) = \fq G(\sfrq, \sigq) + (1-\fq) G(\sfrsf, \sigsf),
\end{equation}
where $G(\mu, \sigma)$ is a log-normal distribution with median $\mu$ and scatter $\sigma$; $\fq$ is the fraction of quenched galaxies, $\sfrsf$ and $\sfrq$ are the median SFRs for star-forming and quenched galaxies, respectively, and $\sigsf$ and $\sigq$ are the corresponding scatters.

All of these parameters ($\fq,\sfrq,\sfrsf,\sigq,\sigsf$) could vary with $\vmp$ and $z$.  However, scatter in SFRs is not observed to vary with either stellar mass or redshift \citep{Speagle14}, and the tight connection between stellar mass and halo mass (or $\vmax$) required to match galaxy clustering and weak lensing \citep{Leauthaud12,Tinker13,Reddick12} suggests that $\sigsf$ need not vary much with $\vmp$ or $z$, either.  However, we allow redshift flexibility to test this assumption, setting a maximum of 0.3 dex:
\begin{equation}
\sigsf = \min(\sigma_\mathrm{SF,0} + (1-a)\sigma_\mathrm{SF,1}, 0.3) \;\mathrm{dex},\label{e:sig_sf_a}
\end{equation}
where $a$ is the scale factor.  SFRs for quenched galaxies have large systematic uncertainties \citep{Brinchmann04,Salim07,Wetzel11,Hayward14}.  As long as $\sfrsf \gg \sfrq$, the exact value of $\sfrq$ does not impact galaxy stellar masses or colours.  Hence, $\sfrq$ is set such that median specific SFRs for quenched galaxies are $10^{-11.8}$ yr$^{-1}$ and $\sigq$ is fixed at $0.36$ dex, matching SDSS values for $L^*$ galaxies \citep{BehrooziMM}.

The functional form for $\sfrsf(\vmp,z)$ is based on the best-\fit{} $\langle SFR(\mpeak,z) \rangle$ constraints from \cite{BWC13}.  We determined $\mpeak(\vmp,z)$ (i.e., the median halo mass as a function of $\vmp$ and $z$) from the \textit{Bolshoi-Planck} simulation (see \S \ref{s:simulations}), and find in Appendix \ref{a:sfr_sf} that a double power law plus a Gaussian is a good fit to $\langle \sfrsf(\mpeak(\vmp, z),z) \rangle$.  We hence adopt:
\begin{eqnarray}
\sfrsf & = & \epsilon\left[\left(v^\alpha + v^\beta\right)^{-1} + \gamma \exp\left(-\frac{\log_{10}(v)^2}{2\delta^2}\right)\right] \label{e:sfrsf}\\
v & = & \frac{\vmp}{V\cdot\mathrm{km}\;\mathrm{s}^{-1}}\\
\log_{10}(V) & = & V_0 + V_a (1-a) + V_{la} \ln(1+z) + V_z z  \label{e:v}\\
\log_{10}(\epsilon) & = & \epsilon_0 + \epsilon_a (1-a) + \epsilon_{la} \ln(1+z) + \epsilon_z z \label{e:sfr}\\
\alpha & = & \alpha_0 + \alpha_a (1-a) +  \alpha_{la} \ln(1+z) + \alpha_z z \label{e:alpha}\\
\beta & = & \beta_0 + \beta_a (1-a) + \beta_z z \label{e:beta}\\
\log_{10}(\gamma) & = & \gamma_0 + \gamma_a(1-a)+\gamma_z z \label{e:gamma}\\
\delta & = & \delta_0 .\label{e:delta}
\end{eqnarray}
For the parameter redshift scaling (Eqs.\ \ref{e:v}--\ref{e:delta}), we generally follow \cite{BWC13} in having variables to control the parameter value at $z=0$, the scaling to intermediate redshift ($z\sim1-2$), and the scaling to high redshift ($z>3$); we add one more parameter to decouple the moderately high redshift scaling ($z=3-7$) from the very high-redshift scaling ($z>7$).  For $\beta$, we do not include this extra parameter, as the massive-end slope is ill-constrained at such high redshifts.  The width of the Gaussian part of $\sfrsf$ seems not to change significantly in fits to \cite{BWC13} constraints, so we keep $\delta$ fixed over the entire redshift range. 

For the functional form for $\fq(\vmp,z)$, we adopt:
\begin{eqnarray}
\fq & = & Q_\mathrm{min} + (1.0-Q_\mathrm{min})\times \nonumber\\
& & \left[0.5 + 0.5 \mathrm{erf}\left(\frac{\log_{10}(\frac{\vmp}{V_Q\cdot \mathrm{km}\;\mathrm{s}^{-1}})}{\sqrt{2} \cdot \sigma_{VQ}}\right)\right], \label{e:fq_1}
\end{eqnarray}
where \textit{erf} is the error function.  This function smoothly rises from $Q_\mathrm{min}$ to 1 over a characteristic width $\sigma_{VQ}$, with the halfway point at the velocity $V_Q$.  The adopted redshift scaling is:
\begin{eqnarray}
Q_\mathrm{min} & = & \max(0,Q_\mathrm{min,0} + Q_\mathrm{min,a} (1-a)) \label{e:fqmin}\\
\log_{10}(V_Q) & = & V_{Q,0} + V_{Q,a} (1-a) + V_{Q,z} z \label{e:vq} \label{e:fq_2}\\
\sigma_{VQ} & = & \sigma_{VQ,0} + \sigma_{VQ,a} (1-a) + \sigma_{VQ,la} \ln(1+z). \label{e:sig_q} \label{e:fq_3}
\end{eqnarray}
We verify that this functional form is sufficiently flexible to match observed quenched fractions in Appendix \ref{a:fq}.

For haloes at a given $\vmp$ and $z$, the above parametrization determines the SFR distribution.  In our approach, we assign higher SFRs to haloes with higher values of $\Delta \vmax$ (similar to the conditional abundance matching approach in \citealt{Watson15}), allowing for random scatter in this assignment.  Because satellites on average have much lower $\Delta \vmax$ values than centrals, zero scatter in the assignment would result in the largest quenched fractions for satellites.  Increasing the scatter decreases the quenched fraction of satellites while increasing the quenched fraction of centrals.  Observationally, this is strongly constrained by the ratio of autocorrelation strengths for quenched vs.\  star-forming galaxies.

We let $r_c$ be the correlation coefficient between haloes' rank orders in $\Delta \vmax$ and their rank orders in SFR (both at fixed $\vmp$ and $z$), and allow this correlation to depend on $\vmp$ and $z$:
\begin{eqnarray}
r_c(\vmp, z) &= & r_\mathrm{min} + (1.0 - r_\mathrm{min})\times\nonumber\\
&& \left[0.5 - 0.5 \mathrm{erf}\left(\frac{\log_{10}\left(\frac{\vmp}{V_R \cdot \mathrm{km}\;\mathrm{s}^{-1}}\right)}{\sqrt{2}\cdot r_\mathrm{width}}\right)\right]\label{e:rc}\\
\log_{10}(V_R) & = & V_{R,0} + V_{R,a}(1-a).
\end{eqnarray}
Similar to the functional form for $f_Q$, this function declines smoothly from $1$ to $r_\mathrm{min}$, with a characteristic width $r_\mathrm{width}$ and the halfway point at the velocity $V_R$.  We allow $r_\mathrm{width}$ to be negative, which would result in $r_c$ increasing (instead of declining) from $r_\mathrm{min}$ to $1$ with increasing $\vmp$.  
In principle, $r_\mathrm{min}$ and $r_\mathrm{width}$ could vary with redshift, but as we do not have enough $z>0$ autocorrelation data to constrain the redshift dependence of these parameters, we leave them fixed.  A halo's resulting (cumulative) percentile rank in the SFR distribution ($\equiv C_\mathrm{SF}$) is given by:
\begin{equation}
C_\mathrm{SF} = C\left(r_c \cdot C^{-1}(C_{\Delta \vmax}) + \sqrt{1-r_c^2} \cdot R\right),
\end{equation}
where $C(x)$ is the cumulative distribution for a Gaussian with unit variance (i.e., $C(x) \equiv 0.5 + 0.5\mathrm{erf}(x/\sqrt{2})$), $C_{\Delta \vmax}$ is the halo's (cumulative) percentile rank in $\Delta\vmax$, and $R$ is a random normal with unit variance.

Along with the fraction of random variations in galaxy SFRs, the random variations' timescales also need parameterization.    Longer-timescale random variations can arise from galactic feedback interacting with the circumgalactic medium and larger-scale environment.  At the same time, short-timescale ($\sim10-100$ Myr) variations occur due to internal processes affecting local galactic cold gas.   We find that the random component $R$ must have some correlation on longer timescales; otherwise, it becomes difficult to produce galaxies quenched according to their UVJ colours.  However, simulations suggest that short-timescale variations are nonetheless common \citep[e.g.,][]{Sparre17}.  We thus generate a time-varying standard normal variable for each halo, composed of a sum of a short-timescale random variable ($S_\mathrm{short}(t)$, which is uncorrelated across simulation timesteps) and a long-timescale random variable:
\begin{equation}
R(t) = f_\mathrm{short}S_\mathrm{short}(t) + \sqrt{1-f_\mathrm{short}^2} S_\mathrm{long}(t) \label{e:fshort},
\end{equation}
where $f_\mathrm{short}$ is the relative contribution of short-timescale variations.  We take $S_\mathrm{long}(t)$ to be a random unit Gaussian time series with correlation time parameterized by $\tau_R$; i.e., the correlation coefficient between $S_\mathrm{long}(t)$ and $S_\mathrm{long}(t+\Delta t)$ is $\exp\left[-\Delta t / (\tau_R)\right]$.  We find that our present observational data do not robustly constrain $\tau_R$, so we fix $\tau_R$ to the halo dynamical time, $t_\mathrm{dyn}= (\frac{4}{3}\pi G\rho_\mathrm{vir})^{-\frac{1}{2}}$.

\subsection{Galaxy Mergers}

\label{s:mergers}

We assume that satellite galaxies survive until their host subhaloes reach a threshold $T_\mathrm{merge}$ for $\frac{\vmax}{\vmp}$, after which the satellite is considered disrupted.  If a subhalo stops being detectable in the N-body simulation before that threshold is reached, it is tracked via a simple gravitational evolution algorithm and the mass-- and $\vmax$--loss prescriptions from \cite{Jiang16}; full details are in Appendix \ref{a:orphans}.  We allow for different survival thresholds around low- and high-mass host haloes via the parameterization:
\begin{eqnarray}
T_\mathrm{merge} & =  & T_\mathrm{merge,300} + (T_\mathrm{merge,1000}-T_\mathrm{merge,300}) \times\nonumber\\
&& \left[0.5 + 0.5 \mathrm{erf}\left(\frac{\log_{10}\left(\frac{v_\mathrm{Mpeak,host}}{\mathrm{km}\;\mathrm{s}^{-1}}\right) - 2.75}{0.25\sqrt{2}}\right)\right]
\end{eqnarray}
This threshold rises smoothly from $T_\mathrm{merge,300}$ (for host haloes with $\vmp\sim$300 km s$^{-1}$) to $T_\mathrm{merge,1000}$  (for host haloes with $\vmp\sim$1000 km s$^{-1}$).

When a satellite is disrupted, we use the distance between the subhalo's last position and the host halo's center to decide the fate of the disrupted material.  As galaxy sizes scale approximately with the virial halo radius \citep{vanderWel14,Shibuya15}, we set the maximum threshold distance for merging with the host halo's galaxy at $f_\mathrm{merge} \cdot R_\mathrm{vir,host}$, with $f_\mathrm{merge}$ a free parameter.  If galaxies disrupt outside of this distance, we instead add their stars to the intrahalo light (IHL) of the host halo.

\subsection{Stellar Masses and Luminosities}

\label{s:sm_uv}

For every halo, full star formation histories (SFHs) are recorded separately for stars in the central galaxy and in the intrahalo light (IHL).  During merger events, the SFH of the merging halo is added either to the central galaxy or IHL for the host halo, as determined in \S \ref{s:mergers}.  Given a SFH, the stellar mass remaining is:
\begin{equation}
\mstar(t_\mathrm{now}) = \int_0^{t_\mathrm{now}} SFH(t) (1-f_\mathrm{loss}(t_\mathrm{now} - t)) dt,
\end{equation}
where $f_\mathrm{loss}(t)$ is computed using the \textsc{FSPS} package \citep{Conroy09,Conroy10} for a \cite{Chabrier03} IMF, and is fit in \cite{BWC13}:
\begin{equation}
f_\mathrm{loss}(t) = 0.05 \ln\left(1+\frac{t}{1.4\; \mathrm{Myr}}\right).
\end{equation}

Johnson U-, Johnson V-, and 2MASS J-band luminosities are calculated as in \cite{Behroozi14}.   Briefly, we use FSPS v3.0 \citep{Conroy09,Conroy10,Byler17} to tabulate the simple stellar population (SSP) luminosity per unit stellar mass as a function of age, metallicity, and dust ($L(t,Z,D)$), assuming a \cite{Chabrier03} IMF and the \cite{calzetti-00} dust model.   We adopt the median metallicity relation of \cite{Maiolino08} and extrapolate the relation to higher redshifts (Eqs.\ \ref{e:maiolino1}-\ref{e:maiolino3}), setting a lower metallicity floor of $\log_{10}(Z/Z_\odot)=-1.5$ to avoid unphysically low metallicities at high redshifts and low stellar masses.  For comparison to UV luminosity functions, we generate M$_\textrm{1500,UV}$ in the same manner.  As shown in Appendix \ref{a:qf}, the UVJ quenching diagnostic used in \cite{Muzzin13} is relatively robust to uncertainties in dust and metallicity except for very metal-poor populations ($\log_{10}(Z/Z_\odot)\sim-2$) and dust-free metal-poor ($\log_{10}(Z/Z_\odot)\sim-1$) rising star formation histories (SFHs).  To avoid issues with dust-free metal-poor populations, we calculate UVJ luminosities assuming a dust optical depth of $\tau=0.3$.  A more detailed dust model is required for UV luminosities; we parameterize the net attenuation as:
\begin{eqnarray}
A_\mathrm{1500,UV} & = & 2.5\log_\mathrm{10}(1+10^{0.4\alpha_\mathrm{dust}( M_\mathrm{dust} - M_\mathrm{1500,UV,intrinsic})}) \label{e:dust}\\
M_\mathrm{dust} & = & M_\mathrm{dust,4} +  M_\mathrm{dust,z}(\max(z,4)-4),\label{e:dust2}
\end{eqnarray}
where $M_\mathrm{1500,UV,obs} = M_\mathrm{1500,UV,intrinsic} + A_\mathrm{1500,UV}$ and where $\alpha_\mathrm{dust}$, $M_\mathrm{dust,4}$, and  $M_\mathrm{dust,z}$ are free parameters.  Since we do not constrain the model with any UV data at $z<4$, the UV luminosities generated in this way are only expected to be realistic at $z>4$.

\begin{figure*}
\vspace{-5ex}
\plotgrace{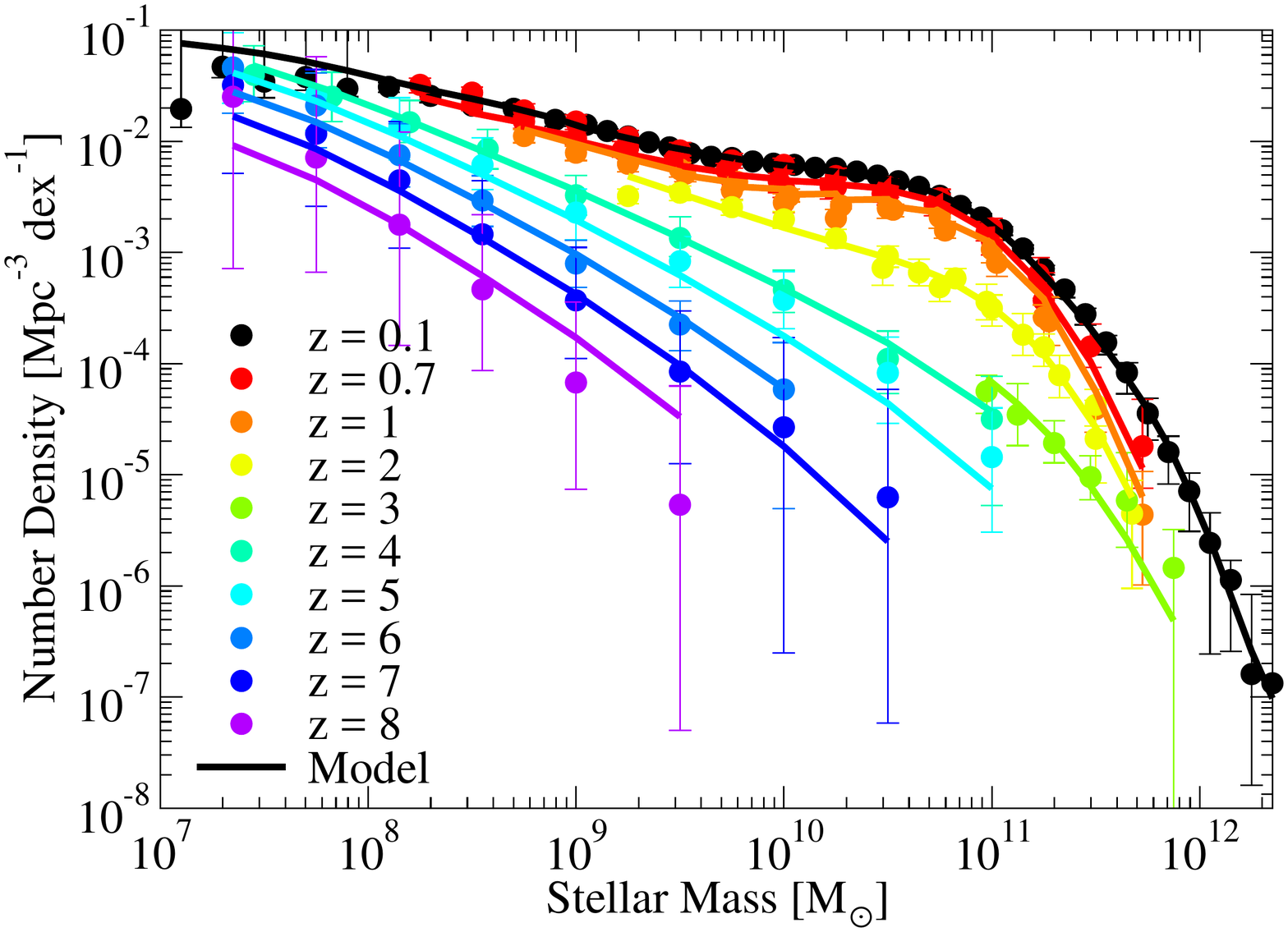}\plotgrace{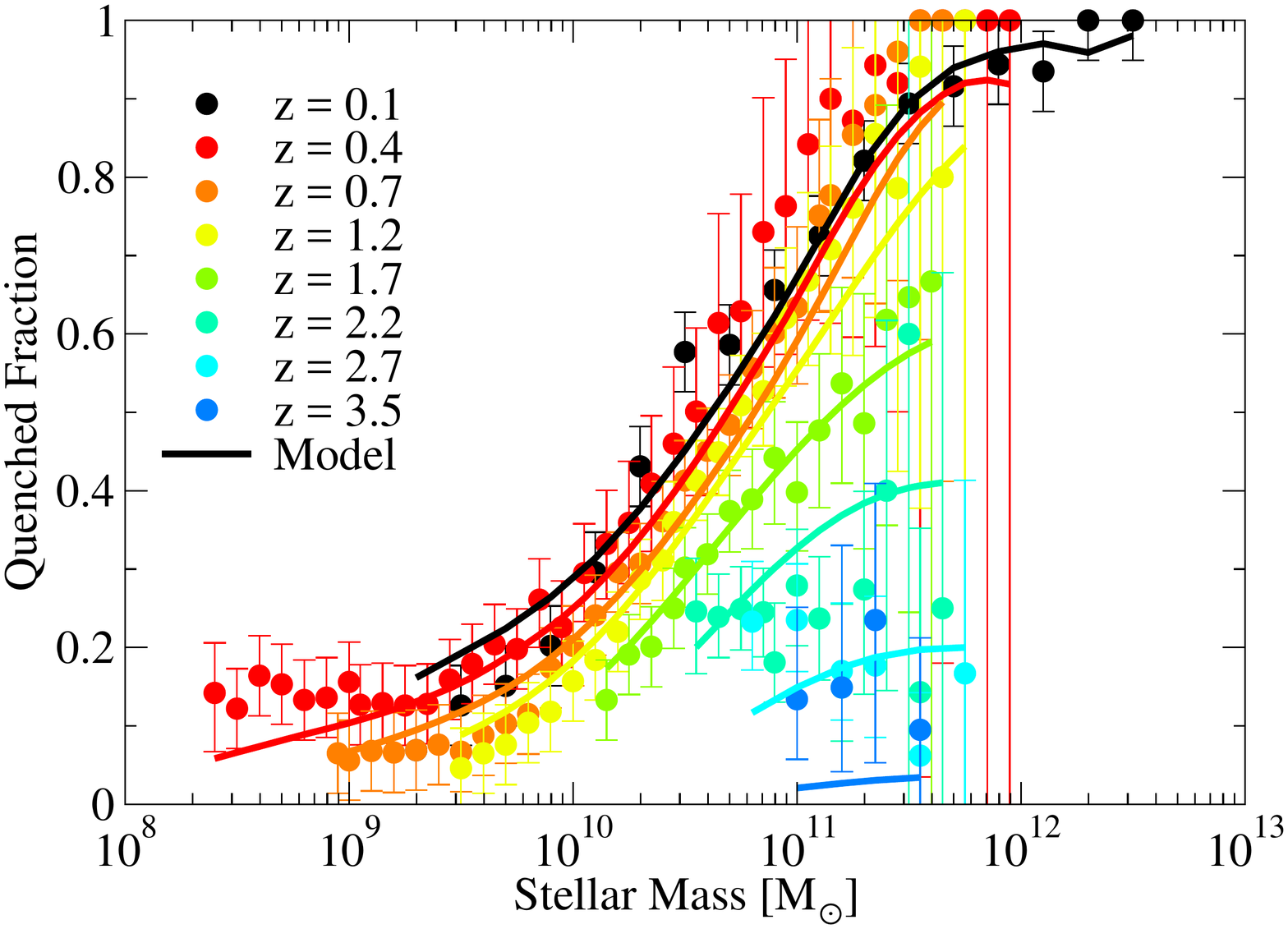}\\[-5ex]
\caption{\textbf{Left} panel: Comparison between observed stellar mass functions (Appendix \ref{a:smf}) and the best-\fit{} model.  References for observations are in Table \ref{t:smf}. \textbf{Right} panel: Comparison between observed quenched fractions (Appendix \ref{a:qf}) and the best-\fit{} model.  Observed quenched fractions are adapted from \protect\cite{Bauer13}, \protect\cite{Moustakas13}, and \protect\cite{Muzzin13}.  \textbf{Notes:} almost all data from both panels were used to constrain the best-\fit{} model.  The exceptions are the $z=4-8$ SMFs from \protect\cite{Song15}, which are shown for comparison only; these were not used in the fitting as the same underlying data is already represented in the $z=4-8$ UVLFs and the UV--SM relations (Fig.\ \ref{f:uv_comp}). The \textit{Bolshoi-Planck} simulation used is incomplete for low-mass haloes, contributing to an underestimation of the SMF below $10^{7}\Msun$ at $z=0$, a limit which rises smoothly to $10^{8}\Msun$ by $z\sim 8$.}
\label{f:smf_comp}
 \label{f:qf_comp}
\vspace{-5ex}
\plotgrace{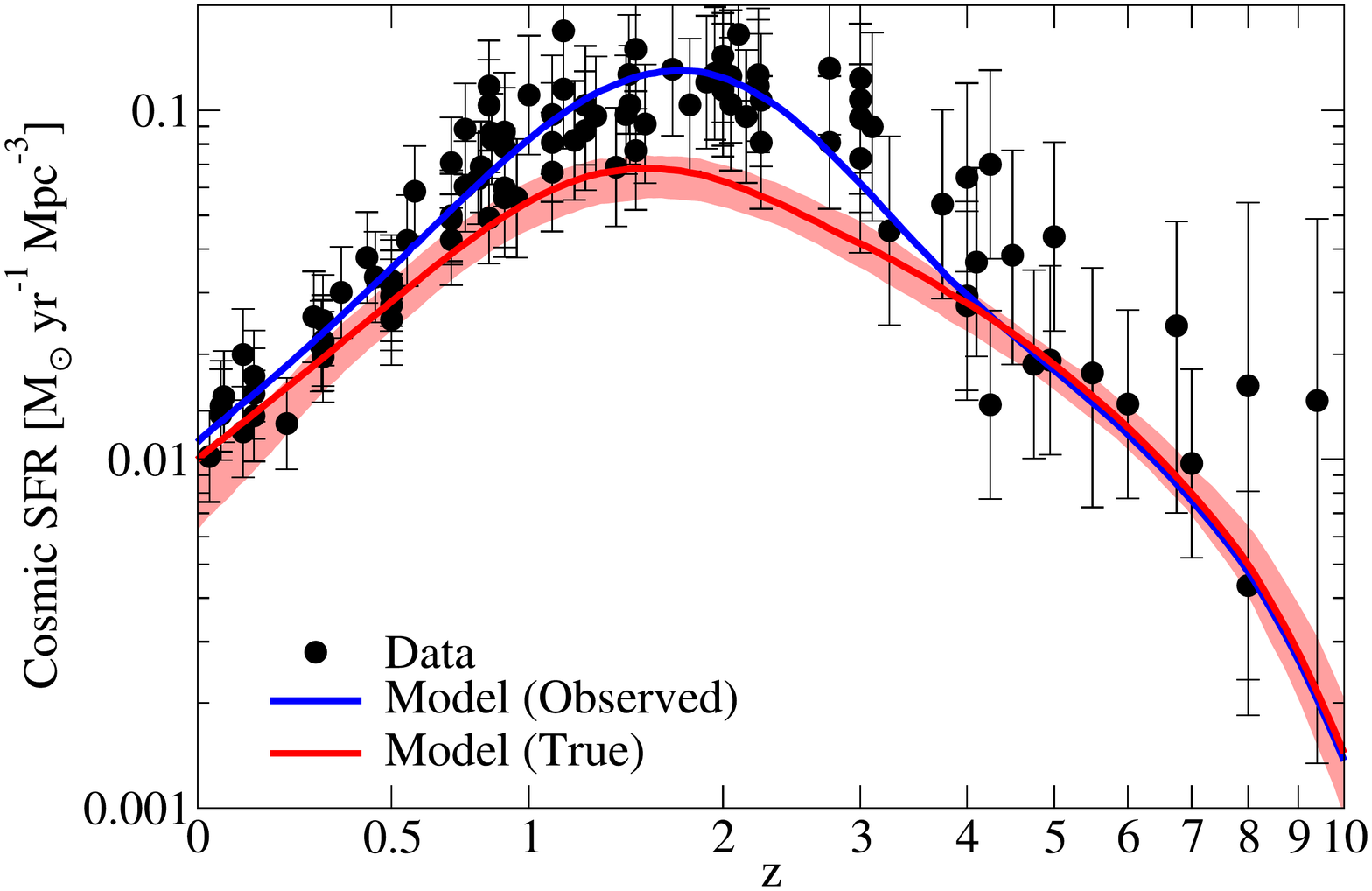}\plotgrace{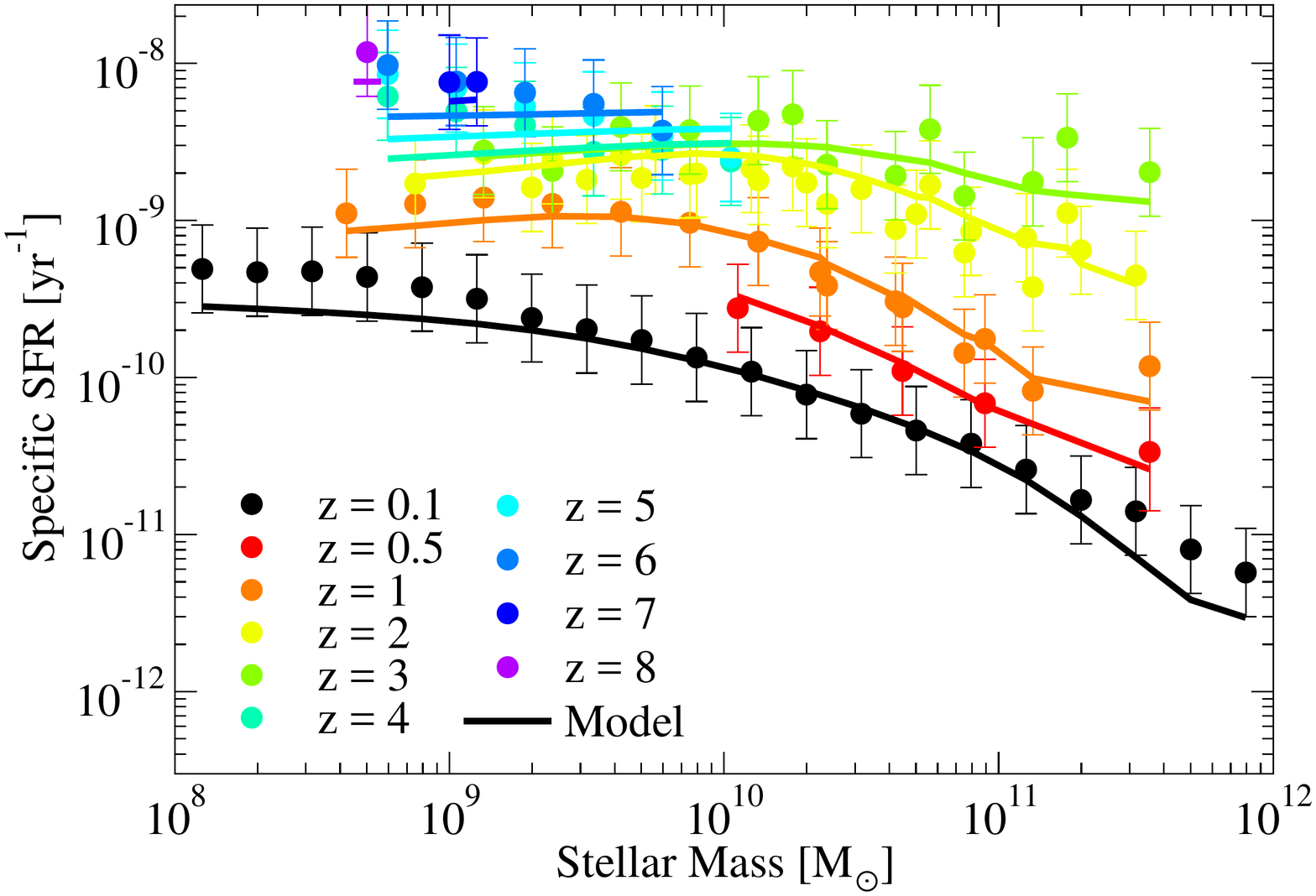}\\[-5ex]
\caption{\textbf{Left} panel: Comparison between observed cosmic star formation rates (CSFRs; Appendix \ref{a:csfr_ssfr}) and the best-\fit{} model; references are in Table \ref{t:csfr}.  The \textit{red line} shows the inferred true cosmic star formation rate, and the \textit{red shaded region} shows the \onesigdist{} from the posterior distribution.  The \textit{blue line} shows the best-\fit{} model after accounting for redshift-dependent observational systematic offsets. \textbf{Right} panel: Comparison between observed specific star formation rates (Appendix \ref{a:csfr_ssfr}) and the best-\fit{} model; references are in Table \ref{t:ssfr}.  \textbf{Notes:} all data from both panels were used to constrain the best-\fit{} model.  For CSFRs at $z>4$, data from both magnitude-limited ($M_\mathrm{1500}<-17$) and total CSFRs (from long GRBs) are shown.  In \textit{Bolshoi-Planck}, resolution limits mean that the total CSFR for all modeled galaxies is nearly identical to the CSFR for galaxies with $M_\mathrm{1500}<-17$.  See Appendix \ref{a:csfr_ssfr} for further discussion.}
 \label{f:ssfr_comp}
 \label{f:csfr_comp}
\vspace{-5ex}
\plotgrace{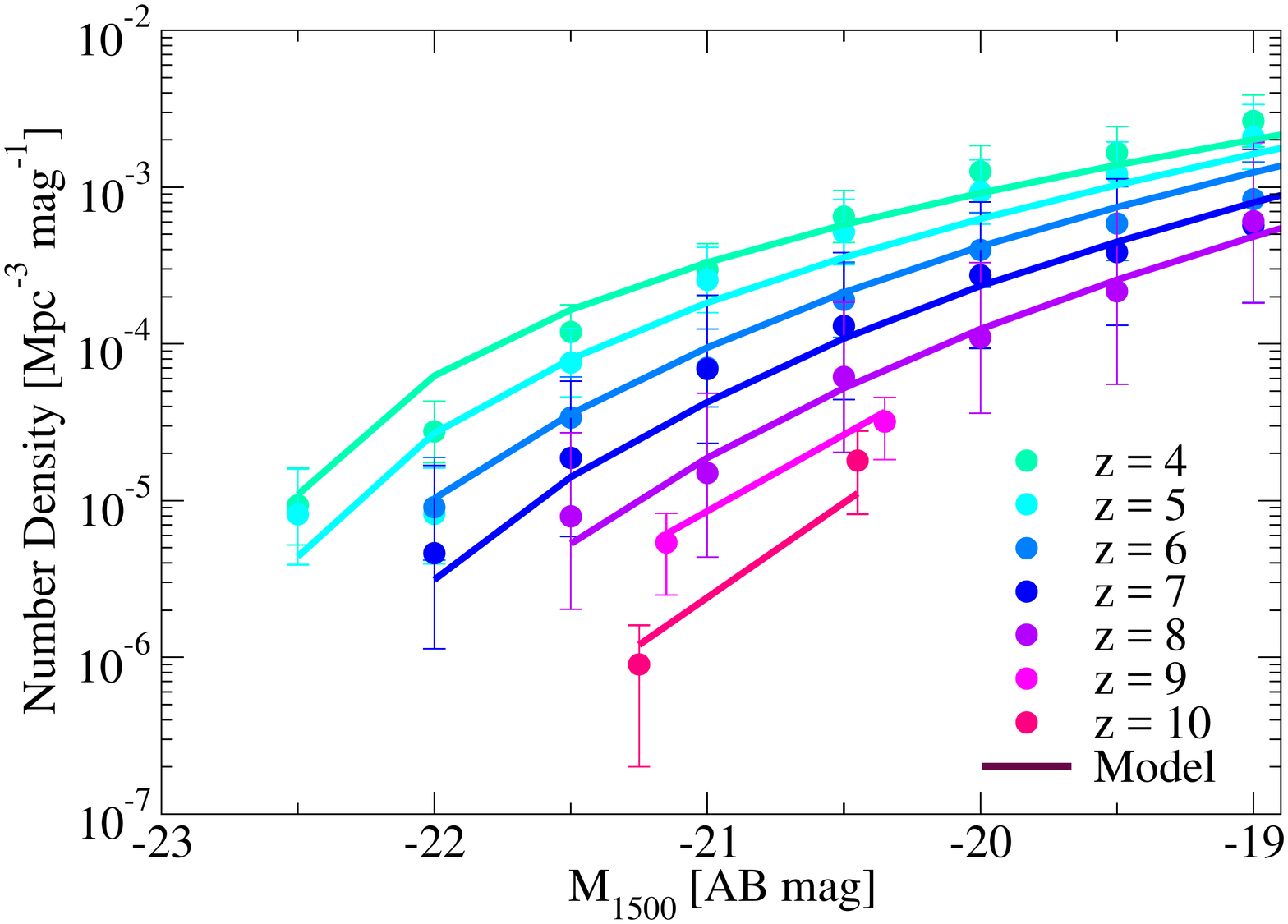}\plotgrace{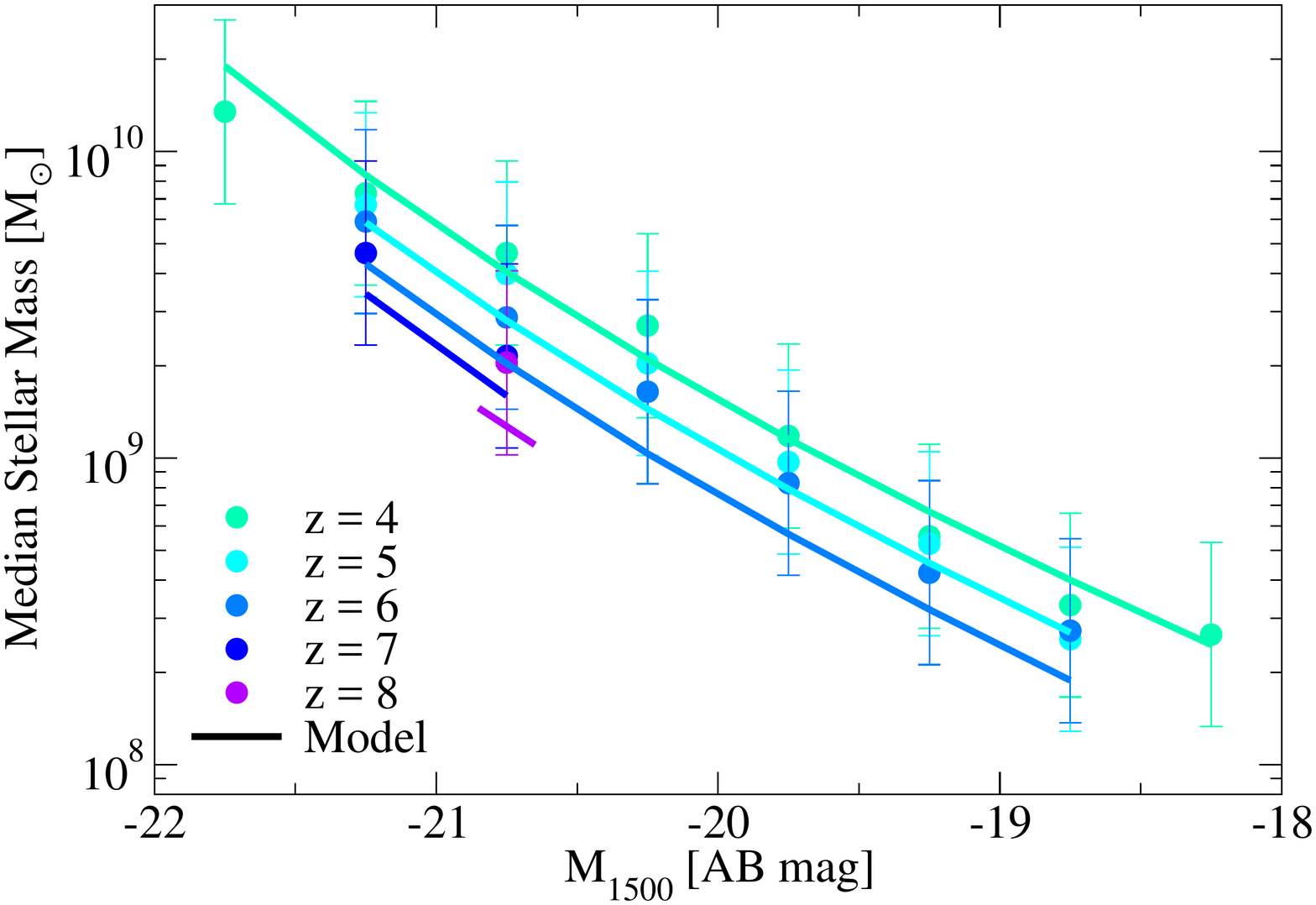}\\[-5ex]
\caption{\textbf{Left} panel: Comparison between observed UV luminosity functions (Appendix \ref{a:uvlfs})  from \protect\cite{Finkelstein15} and \protect\cite{Bouwens15b} and the best-\fit{} model. \textbf{Right} panel: comparison between median UV--stellar mass relationships (Appendix \ref{a:uvsm}) for the best-\fit{} model and the observed results rederived in this paper from SED stacks in \protect\cite{Song15}.  \textbf{Notes:} all data from both panels were used to constrain the best-\fit{} model. The \textit{Bolshoi-Planck} simulation used is incomplete for low-mass haloes, so we do not fit to UV luminosity functions for $M_{1500}>-19$.}
\label{f:uv_comp}
\end{figure*}

\begin{figure*}
\vspace{-5ex}
\plotgrace{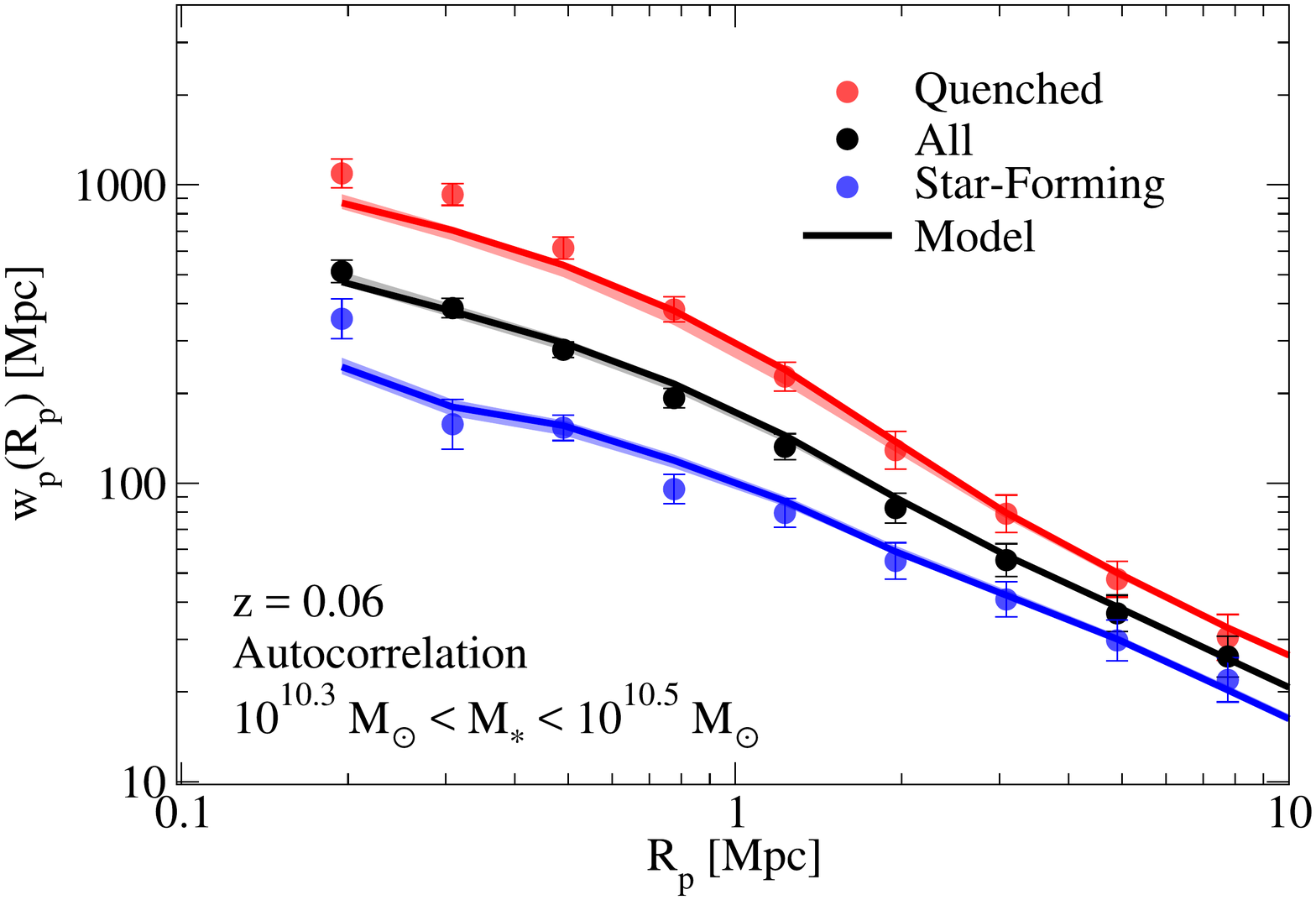}\plotgrace{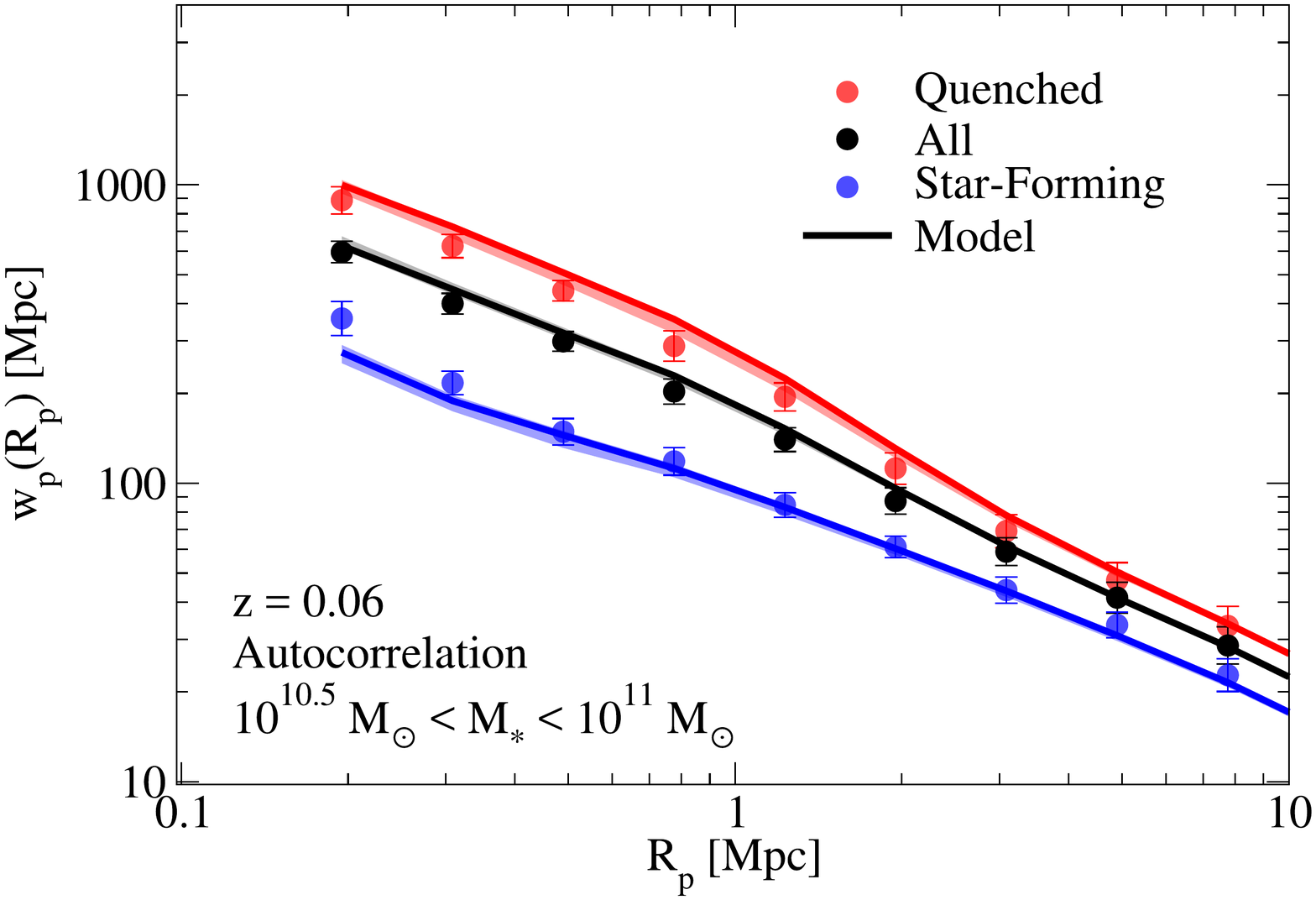}\\[-5ex]
\plotgrace{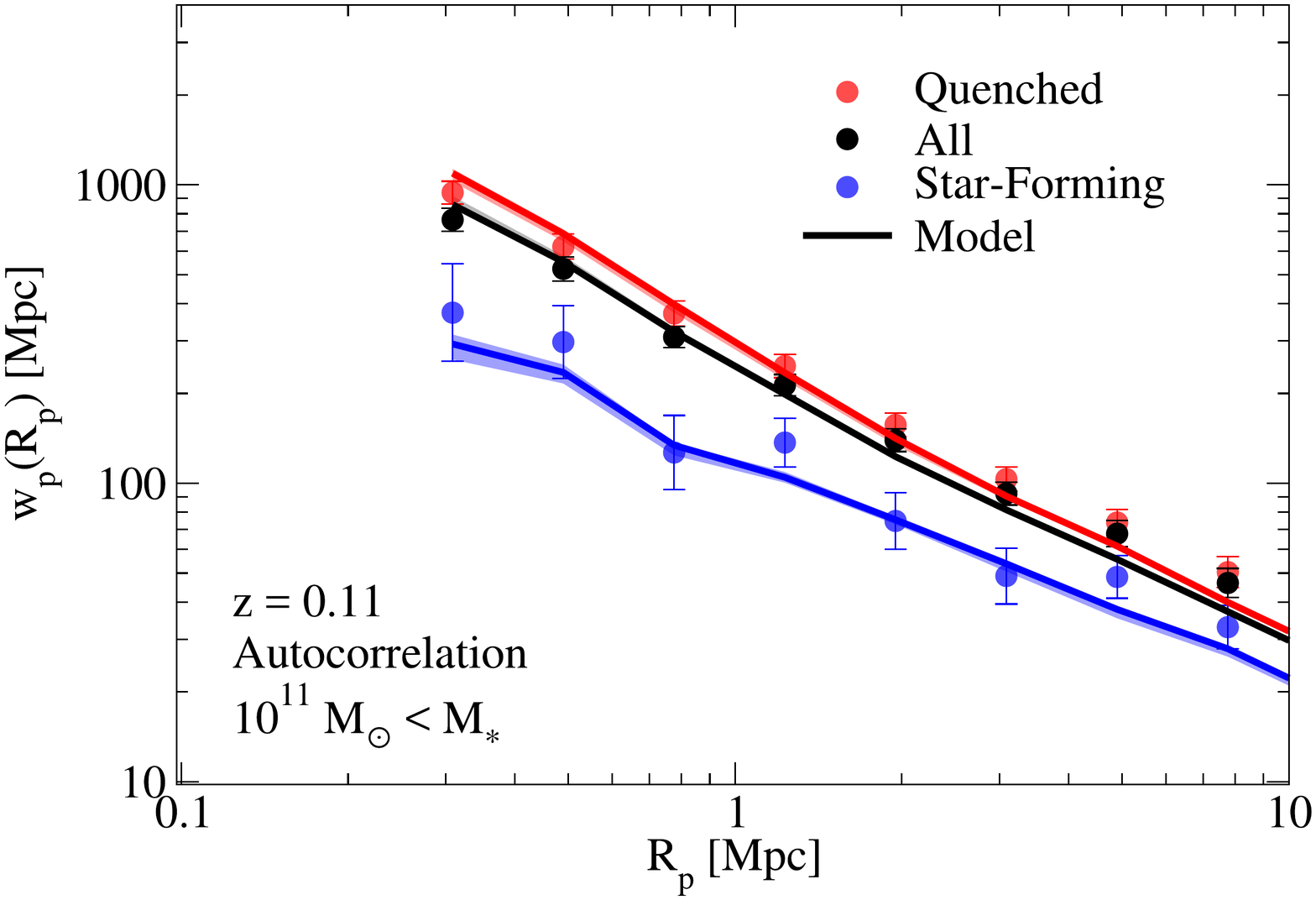}\plotgrace{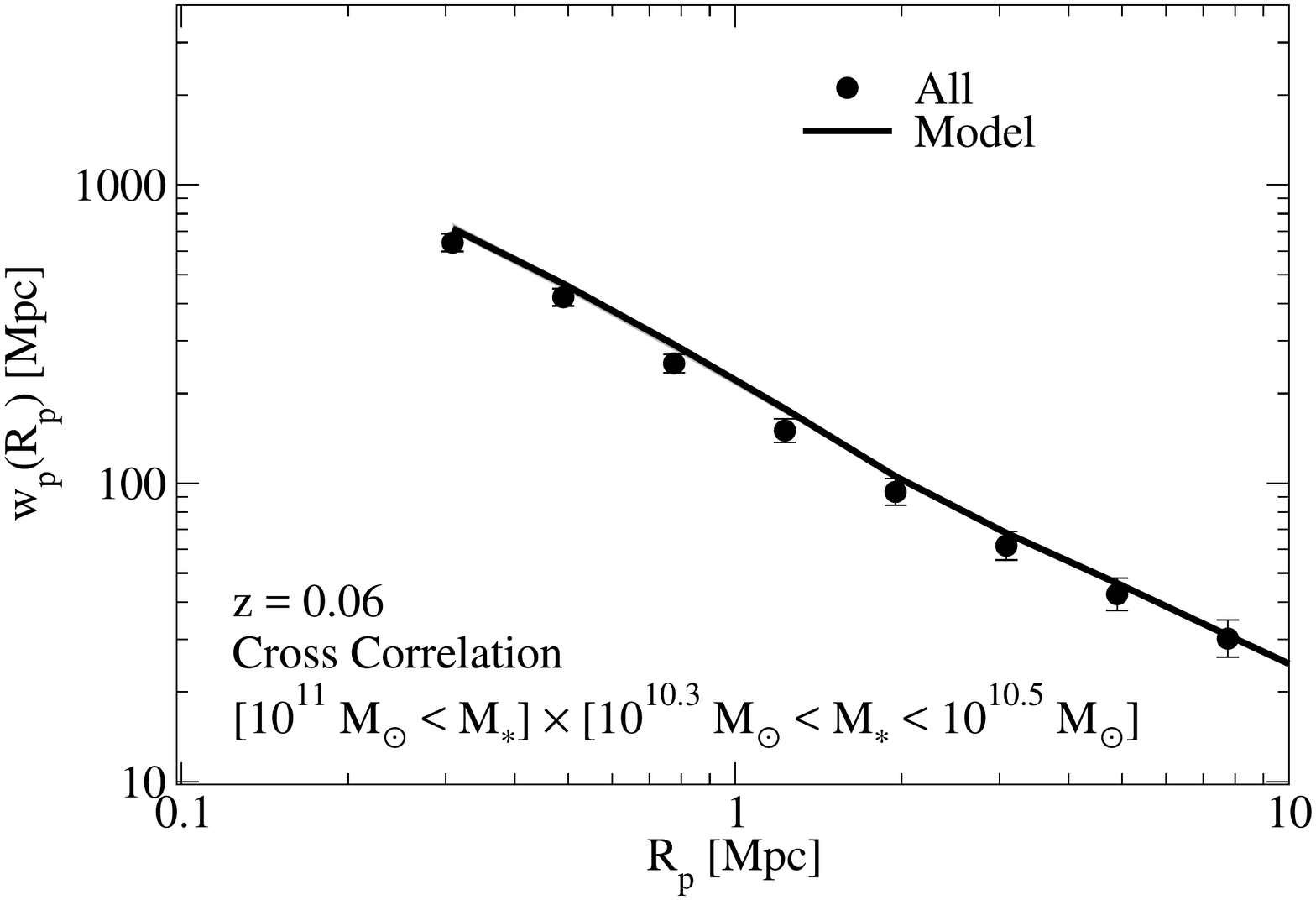}\\[-5ex]
\caption{Comparison between galaxy autocorrelation and cross correlation functions at $z\sim 0$ for the best-\fit{} model and the observed results rederived from the SDSS in Appendix \ref{a:cf}.  \textbf{Notes:} all data from all panels were used to constrain the best-\fit{} model.  Observational errors shown are jackknife estimates from the observational sample.  Actual fitting used covariance matrices as detailed in Appendix \ref{a:cf}.}
\label{f:cf_comp}
\end{figure*}

\begin{figure}
\vspace{-5ex}
\plotgrace{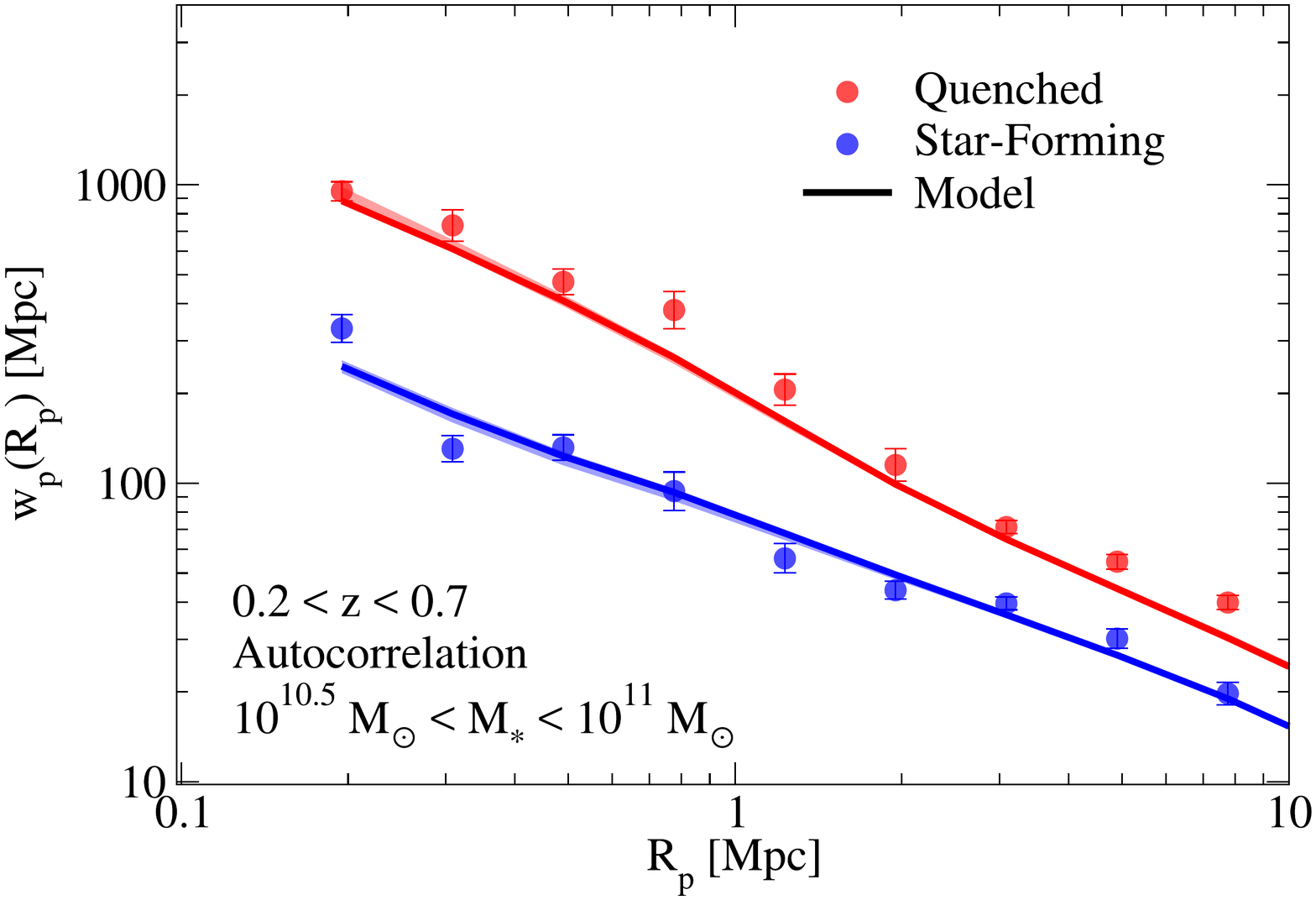}\\[-5ex]
\caption{Comparison between observed galaxy autocorrelation functions at $z\sim 0.5$ \protect\citep[i.e., PRIMUS]{Coil17} and the best-\fit{} model.  \textbf{Notes:} all data from this panel were used to constrain the best-\fit{} model.  Redshift errors for PRIMUS were rederived according to Appendix \ref{a:cf}.}
\label{f:cf_comp_primus}
\vspace{-5ex}
\plotgrace{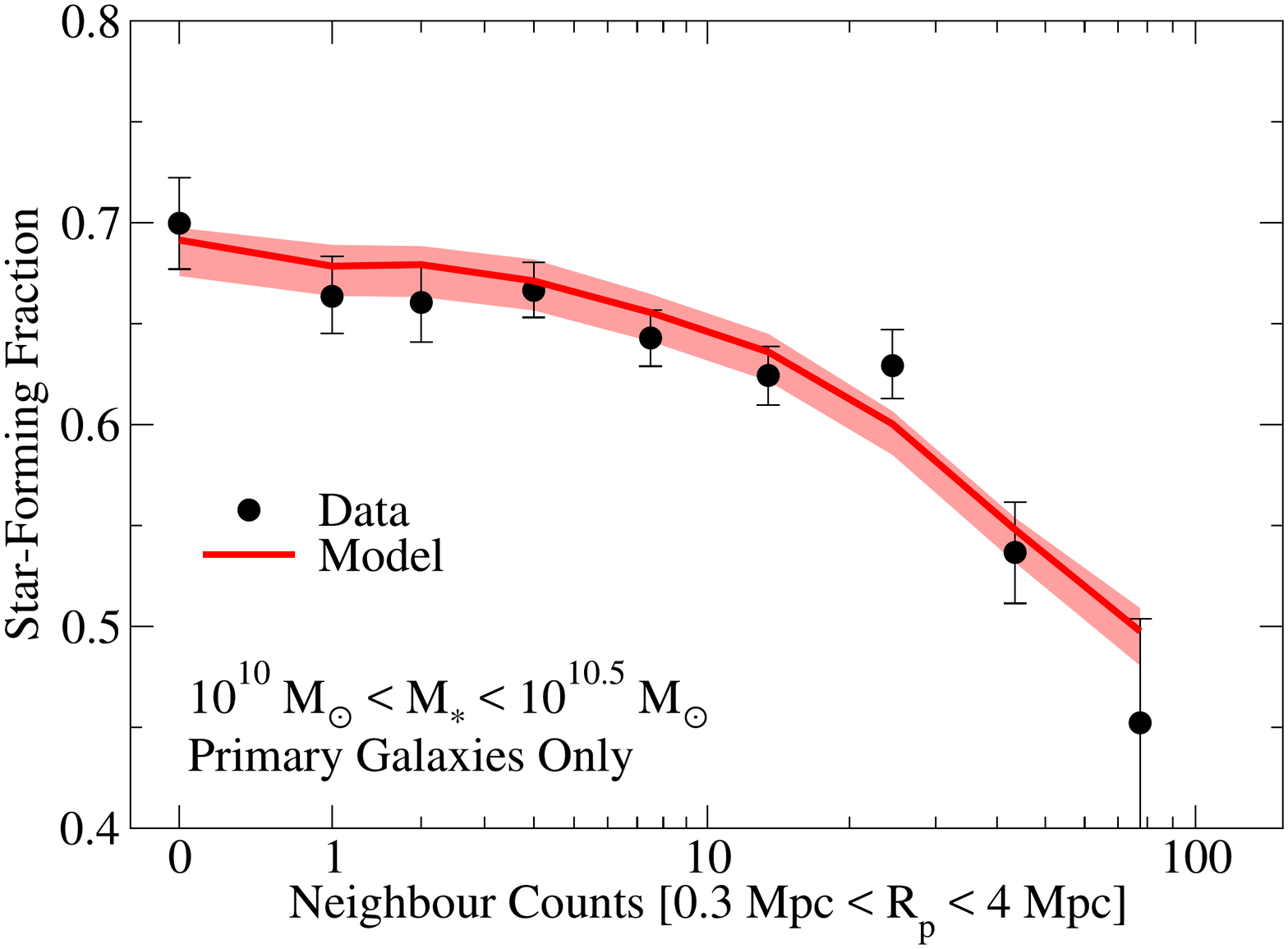}\\[-5ex]
\caption{Comparison between primary galaxy quenched fractions as a function of neighbour density in the SDSS (derived in Appendix \ref{a:ecq}) and the best-\fit{} model (\textit{red line}; \onesigdist{} shown as \textit{red shaded region}).  As discussed in \S \ref{s:obs}, this provides an approximate probe of the correlation between central halo and galaxy assembly.   Primary galaxies are defined as being the largest galaxy within a projected distance of 500 kpc and a redshift distance of 1000 \mbox{km s$^{-1}$}.  Neighbours are defined as galaxies with masses within a factor $0.3-1$ of the primary galaxy.  \textbf{Notes:} all data from this panel were used to constrain the best-\fit{} model.  }
\label{f:ecq_comp}
\vspace{-5ex}
\plotgrace{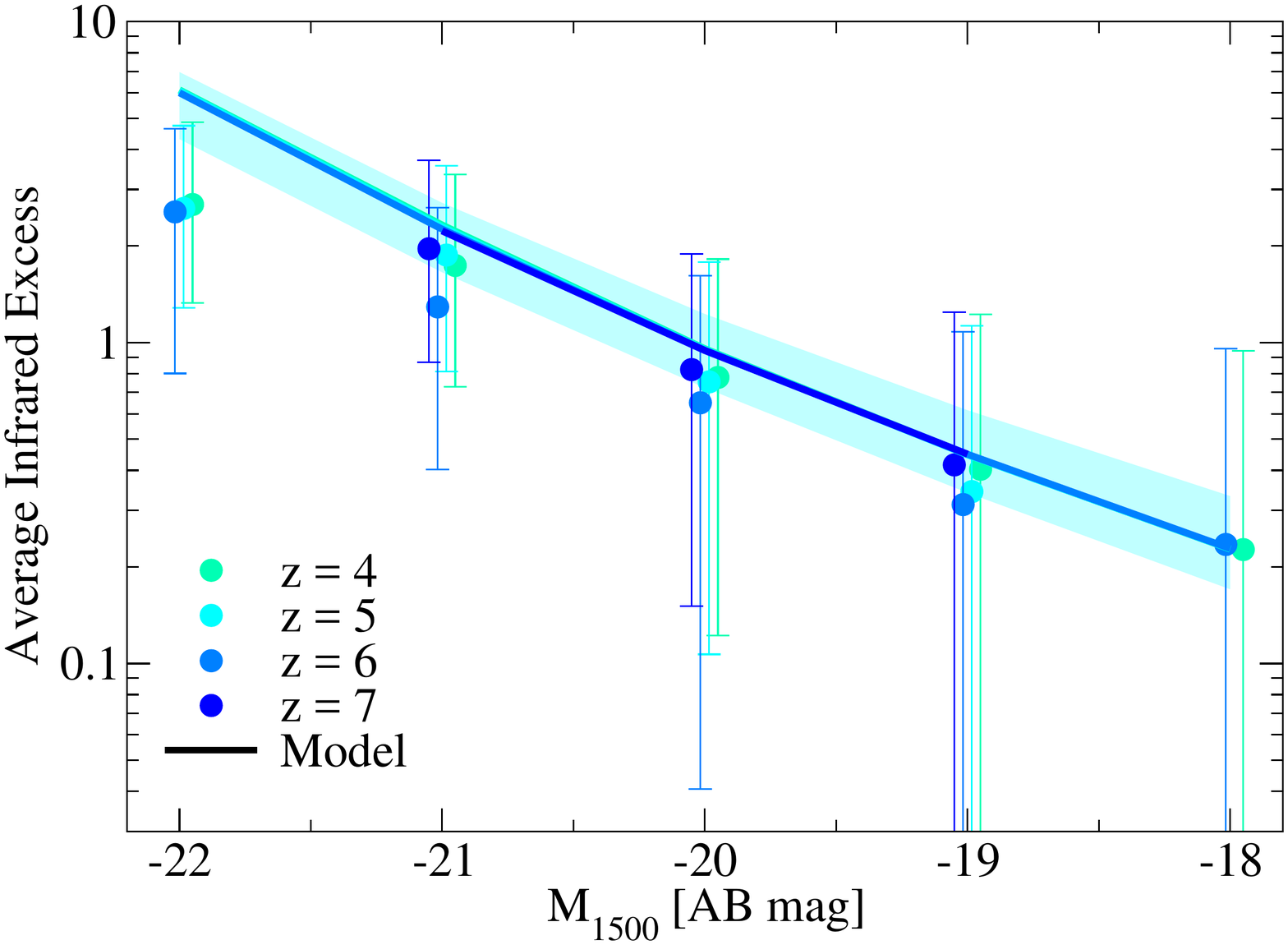}\\[-5ex]
\caption{Average infrared excess (IRX) as a function of observed UV luminosity from ALMA data at $z=4-7$ \citep{Bouwens16} and the best-\fit{} model (\textit{lines}; \onesigdist{} at $z=4$ shown as \textit{cyan shaded region}).  \textbf{Notes:} all data from this panel were used to constrain the best-\fit{} model.  Observational data are offset slightly in UV luminosity to increase clarity between different redshifts.  The best-\fit{} model evolves very little with redshift, so model lines are not distinguishable.}
\label{f:irx_comp}
\end{figure}

\subsection{Observational Systematics}

\label{s:systematics}

Systematic uncertainties in stellar masses and SFRs arise from modeling assumptions for stellar population synthesis (SPS), dust, metallicity, and star formation history \citep{Conroy09,Conroy13b,Behroozi10,BWC13}.  The dominant effect is a redshift-dependent offset between true and observed values for both stellar masses and SFRs, parametrized here as
\begin{equation}
\mu \equiv SM_\mathrm{obs} - SM_\mathrm{true}  = \mu_0 + \mu_a (1-a). \label{e:mu}
\end{equation}
This is constrained by tension between CSFRs and evolution in SMFs \citep[e.g.,][]{Wilkins08,Yu16}, as well as tension between SSFRs and observed UVLFs.  Following \cite{BWC13}, we set the prior width on $\mu_0$ and $\mu_a$ to 0.14 and 0.24 dex, respectively.

We also include a redshift-dependent offset that affects only SFRs, motivated by strong tensions between observed radio and UV+IR SSFRs and evolution in SMFs that peaks at $z=2$ \citep[][see also Appendix \ref{a:tension}]{Leja15}.  This coincides with existing tensions between SSFRs from observations and SSFRs from modern hydrodynamical simulations \citep[e.g.,][]{Sparre15,Dave16}, as well as tensions between IR and SED-fit SFR indicators \citep{Fang17}:
\begin{equation}
SFR_\mathrm{obs} - SFR_\mathrm{true} = \mu + \kappa \exp\left(-\frac{(z-2)^2}{2}\right). \label{e:kappa}
\end{equation}
The prior width on $\kappa$ is set to 0.24 dex (Table \ref{t:prev}), matching the prior on $\mu_a$.

Offsets could also be mass-dependent (see \citealt{li-2009} and Appendix \ref{a:data}) and SSFR-dependent \citep{BWC13}.  Tension between SSFRs and SMFs could also constrain a mass-dependent offset; however, the fact that errors on the SMF are much tighter than errors on the SSFR would result in our MCMC algorithm recruiting the parameter so as to better fit the shape of the SMF (i.e., overfitting).  SSFR-dependent offsets could be constrained by the amplitude of the autocorrelation function---however, this is extremely degenerate with sample variance.  Hence, we use the simple Eqs.\ \ref{e:mu} and \ref{e:kappa} in this work to parametrize systematic offsets.  

Random errors in recovering stellar masses can also cause Eddington bias in the massive-end shape of the SMF \citep{Behroozi10,BWC13,Grazian15}.  Since the errors grow with redshift, they impact the inferred mass growth of massive galaxies.  Here, we use the same redshift dependence as \cite{BWC13}, but limit the maximum scatter at high redshift:
\begin{equation}
\sigma_\mathrm{SM,obs} = \min(\sigma_{SM,0} + \sigma_{SM,z} z, 0.3) \,\mathrm{dex}.
\label{e:smobs}
\end{equation}
For example, \cite{Grazian15} find a scatter of $\sim 0.2$ dex at $z=6$ in the distribution of SMs for $\sim 10^{11}\Msun$ galaxies; this expands to 0.3 dex after accounting for additional scatter from photometric redshifts, photometry, code choices (including finite age/metallicity grids), and other sources (see \citealt{Mobasher15} for a review).

Similarly, random log-normal errors in observed SFRs can broaden the observed SFR distribution and lead to enhanced average CSFRs, as the average (in linear space) of a log-normal distribution is higher than the median.  We adopt
\begin{equation}
\sigma_\mathrm{SFR,obs} = \sqrt{0.3^2 - \sigsf^2} \mathrm{dex}, \label{e:sfrobs}
\end{equation}
so that the combined intrinsic plus observed main-sequence scatter is $0.3$ dex, consistent with \cite{Speagle14}.

All observables are subject to volume-weighting effects; some are also subject to binning effects.  When modeling observables, we use identical binning, and we also use volume-weighting across the reported redshift range.  For a given observable $X$ reported for $z_1 < z < z_2$, we thus simulate the observation as 
\begin{equation}
X_\mathrm{model} = \frac{\int_{z_1}^{z_2} X(z) dV(z)}{V(z_2) - V(z_1)},
\end{equation}
where $V(z)$ is the enclosed volume out to redshift $z$.  For correlation functions and higher-order statistics that depend on spectroscopic redshifts, we include redshift-space distortions from halo peculiar velocities.  We also include 30 km s$^{-1}$ of combined galaxy--halo velocity bias and redshift-fitting errors, assumed to be normally distributed \citep{Guo15}.  For the grism-based redshifts in PRIMUS, we model the redshift errors as $\sigma_z/(1+z) = 0.0033$, as discussed in Appendix \ref{a:cf}.

The initial mass function (IMF) is known to vary with galaxy velocity dispersion \citep{Conroy12,Conroy13,Geha13,MartinNavarro15,LaBarbera15,vanDokkum17}.  However, broadband photometric luminosities depend largely on the mass in $>1 \Msun$ stars, resulting in a constant overall mass offset for different IMF assumptions.  As a result, the main body of this paper adopts the same assumption as for all the observational results with which we compare---namely, a universal \cite{Chabrier03} IMF.  Appendix \ref{a:imf} shows how derived stellar mass---halo mass relationships would change for a halo mass-dependent IMF.

\section{Results}

\label{s:results}
We discuss best-\fit{} parameters and the comparison to observables in \S \ref{s:obs_comparison}, the stellar mass--halo mass relation in \S \ref{s:smhm}, average SFRs and quenched fractions in dark matter haloes in \S \ref{s:average_sfrs}, average star formation histories in \S \ref{s:avg_sfr}, individual stochasticity in SFRs in \S \ref{s:stochasticity}, correlations between galaxy and halo assembly in \S \ref{s:corr}, satellite quenching in \S \ref{s:satellites}, \textit{in-situ} vs.\ \textit{ex-situ} star formation in \S \ref{s:in_situ}, predictions for future observations in \S \ref{s:corr_pred}, systematic uncertainties in \S \ref{s:syst_uncertainties}, and additional online data in \S \ref{s:online_data}.

\subsection{Best-\fit{} Parameters and Comparison to Observables}

\label{s:obs_comparison}

We explored model posterior space with 100 simultaneous MCMC walkers, totaling $\sim$500k MCMC steps and $400$k CPU hours.  Convergence was approached by running the chains for $10$ autocorrelation times.  The best-\fit{} model was found by starting from the average of all walker positions during the final 400 steps and then using a gradient descent algorithm to converge on the model with lowest $\chi^2$.  

The best-\fit{} model is able to match all data in \S \ref{s:obs}, including stellar mass functions (SMFs; Fig.\ \ref{f:smf_comp}, left panel), quenched fractions (QFs; Fig.\ \ref{f:qf_comp}, right panel), cosmic star formation rates (CSFRs; Fig.\ \ref{f:csfr_comp}, left panel), specific star formation rates (SSFRs; Fig.\ \ref{f:ssfr_comp}, right panel), high-redshift UV luminosity functions (UVLFs; Fig.\ \ref{f:uv_comp}, left panel), high-redshift UV--stellar mass relations (UVSMs; Fig.\ \ref{f:uv_comp}, right panel), correlation functions (CFs; Figs.\ \ref{f:cf_comp} and \ref{f:cf_comp_primus}), the dependence of the quenched fraction of central galaxies as a function of environment (Fig.\ \ref{f:ecq_comp}), and the average infrared excess as a function of UV luminosity (Fig.\ \ref{f:irx_comp}).  Calculating the true number of degrees of freedom for the observational data  is difficult; for example, covariance matrices are unavailable for most SMFs, QFs, UVLFs, etc.\ in the literature.  Yet, for 1069 observed data points and 44 parameters, the naive reduced $\chi^2$ of the best-\fit{} model is 0.36, suggesting a reasonable fit.

The best-\fit{} model and 68\% confidence intervals for parameters are presented in Appendix \ref{a:bestfit}, and parameter correlations are discussed in Appendix \ref{a:correlations}.   Key physical aspects of the parameterization include the star formation rate for star-forming galaxies and the quenched fraction as a function of $\vmp$ (shown in Fig.\ \ref{f:sfr_vmax}), as well as the correlation between galaxy and halo growth (shown in Fig.\ \ref{f:rank_correl}).  Posterior distributions of many other quantities (e.g., the stellar mass--halo mass relation, cross-correlation functions, satellite and quenching statistics, etc.) are described in the following sections and are available \href{https://www.peterbehroozi.com/data}{\textbf{online}}.

\subsection{The Stellar Mass -- Halo Mass Relation for \textit{z}=0 to \textit{z}=10}

\label{s:smhm}

\subsubsection{Stellar Mass -- Halo Mass Ratios}

\begin{figure*}
\vspace{-7ex}
\plotgrace{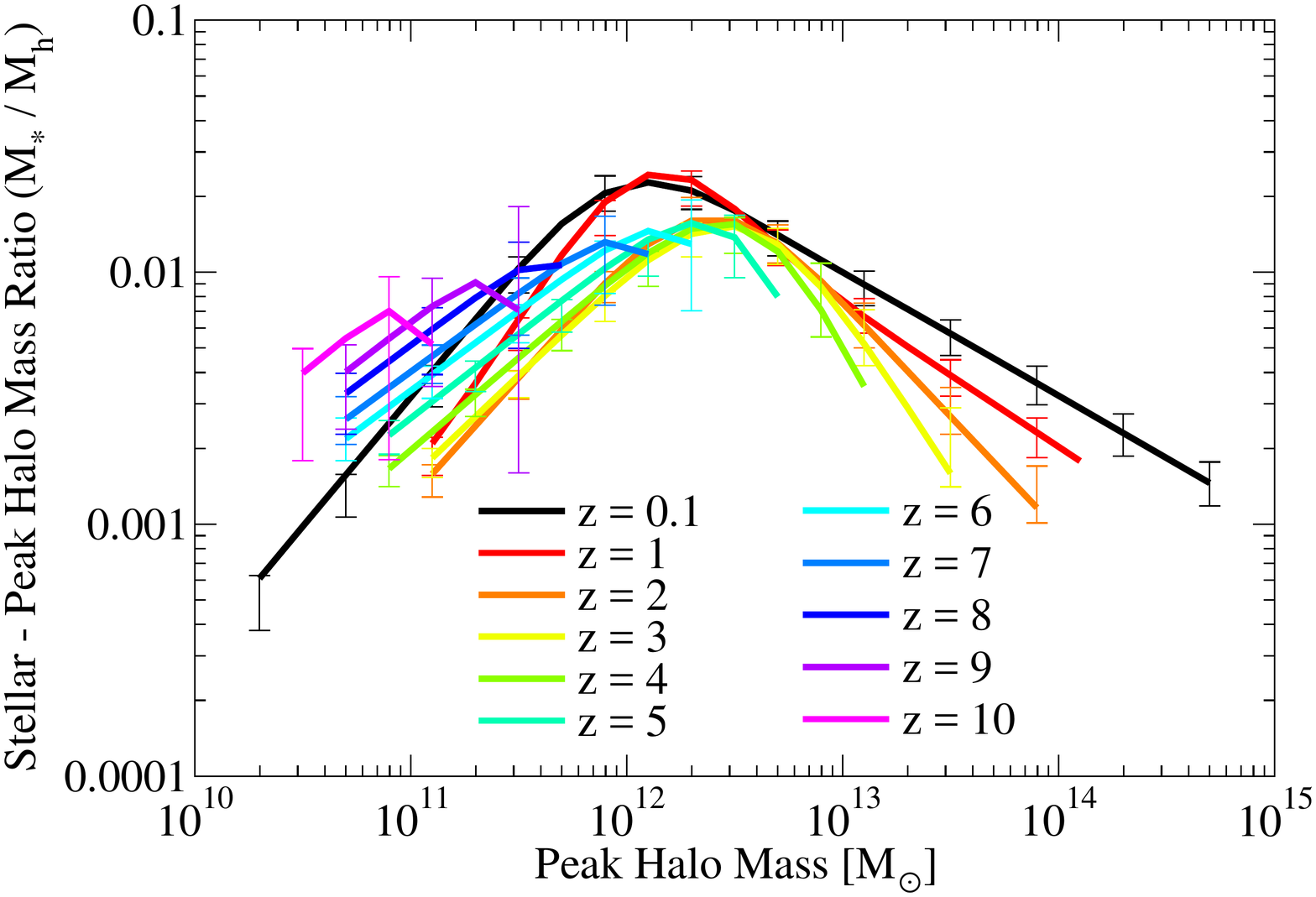}\plotgrace{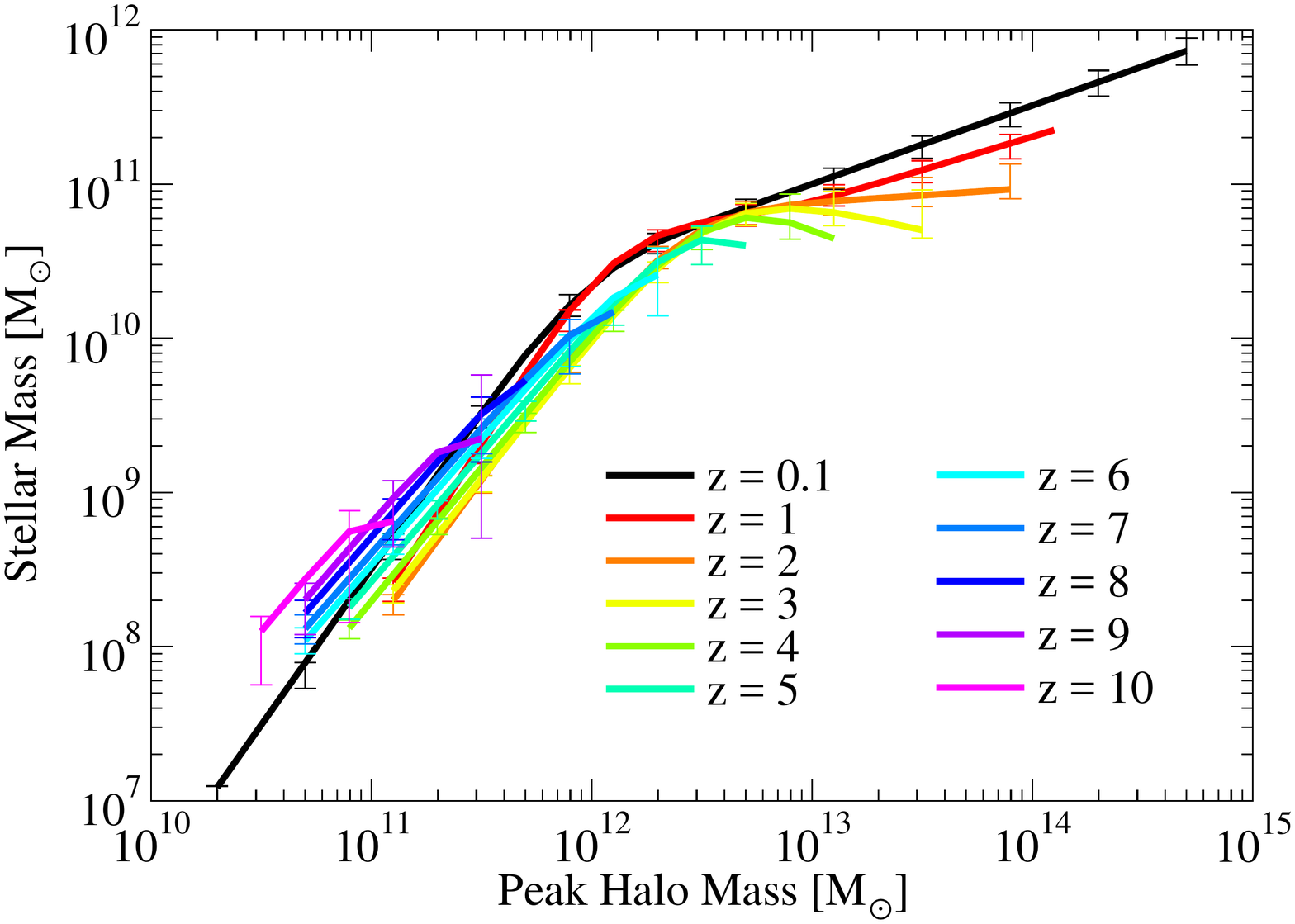}\\[-4.5ex]
\caption{\textbf{Left} panel: best-\fit{} median ratio of stellar mass to peak halo mass ($\mpeak$) as a function of $\mpeak$ and $z$.  \textbf{Right} panel: best-\fit{} median stellar mass as a function of $\mpeak$ and $z$.  Error bars in both panels show the 68\% confidence interval for the model posterior distribution.  \textbf{Notes:} see Figs.\ \ref{f:comp_z0}-\ref{f:comp_z4} for a comparison with past results.  \textbf{See Appendix \ref{a:smhm_fits} for fitting formulae.}}
\label{f:smhm}
\vspace{-5ex}
\plotgrace{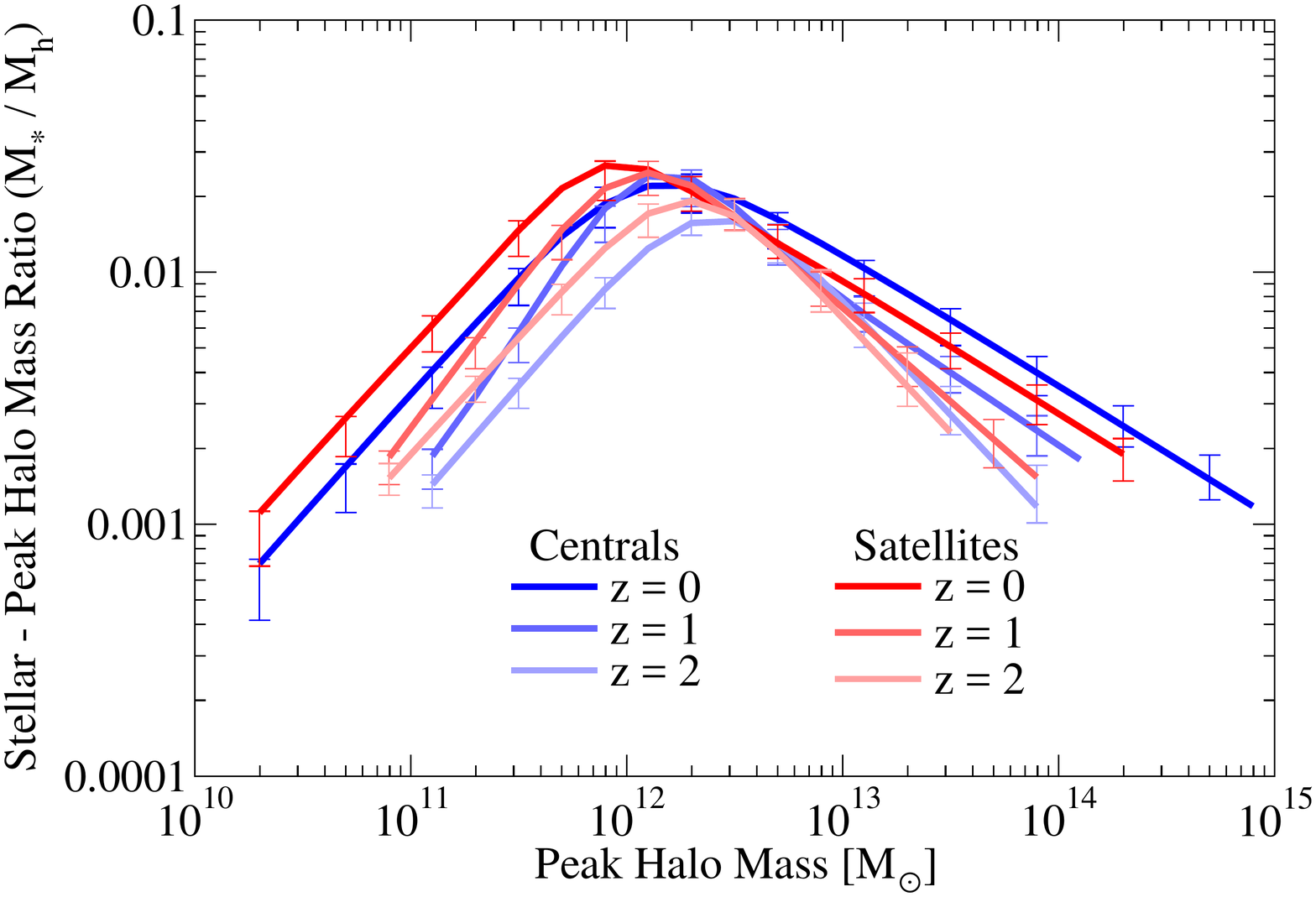}\plotgrace{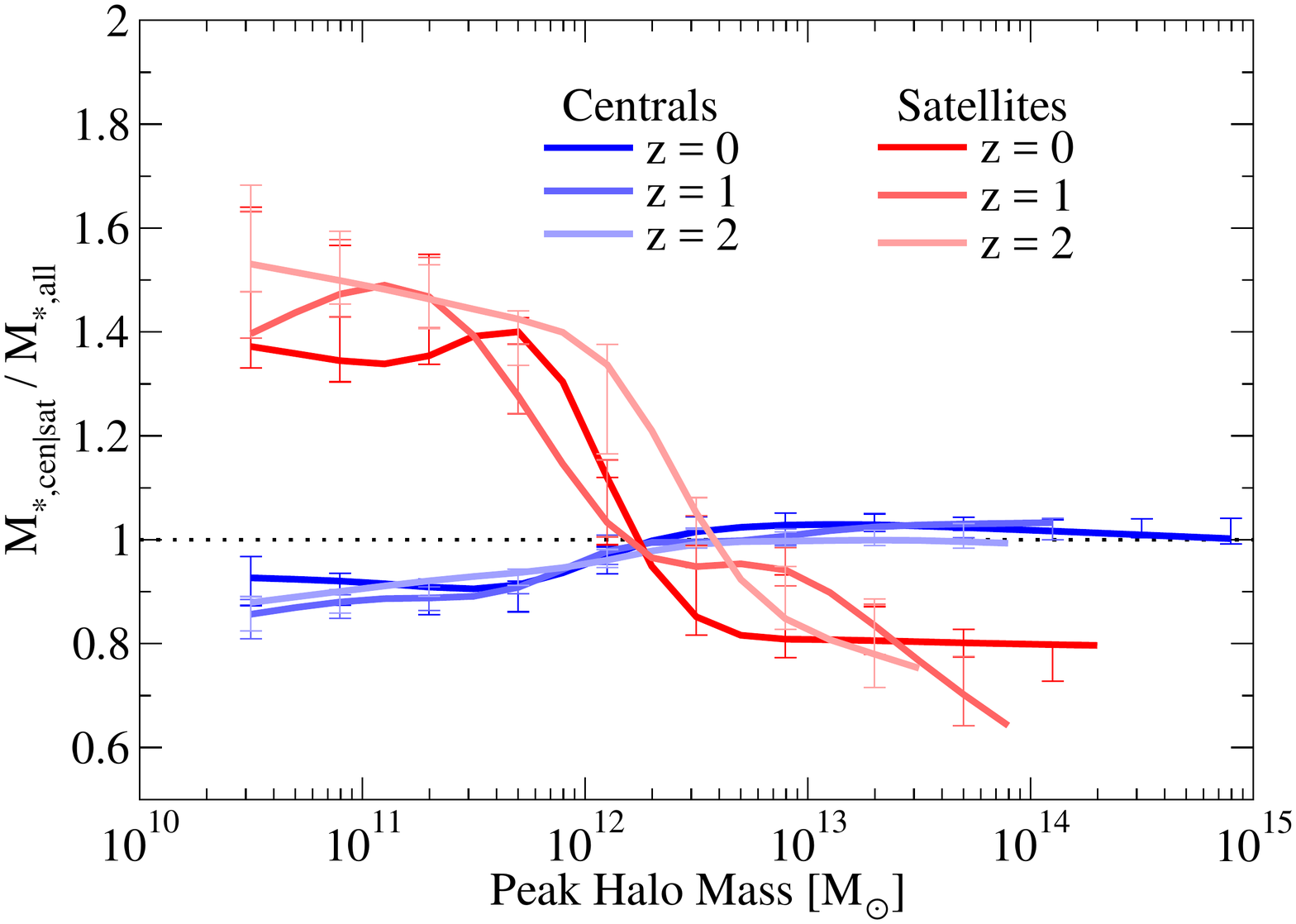}\\[-4.5ex]
\caption{The best-\fit{} median ratio of stellar mass to peak halo mass ($\mpeak$) for central and satellite galaxies (\textbf{left} panel) compared to the ratio for all galaxies (\textbf{right} panel).  Error bars in both panels show the 68\% confidence interval for the model posterior distribution. \textbf{See Appendix \ref{a:smhm_fits} for fitting formulae.}}
\label{f:smhm_cen_sat}
\vspace{-5ex}
\plotgrace{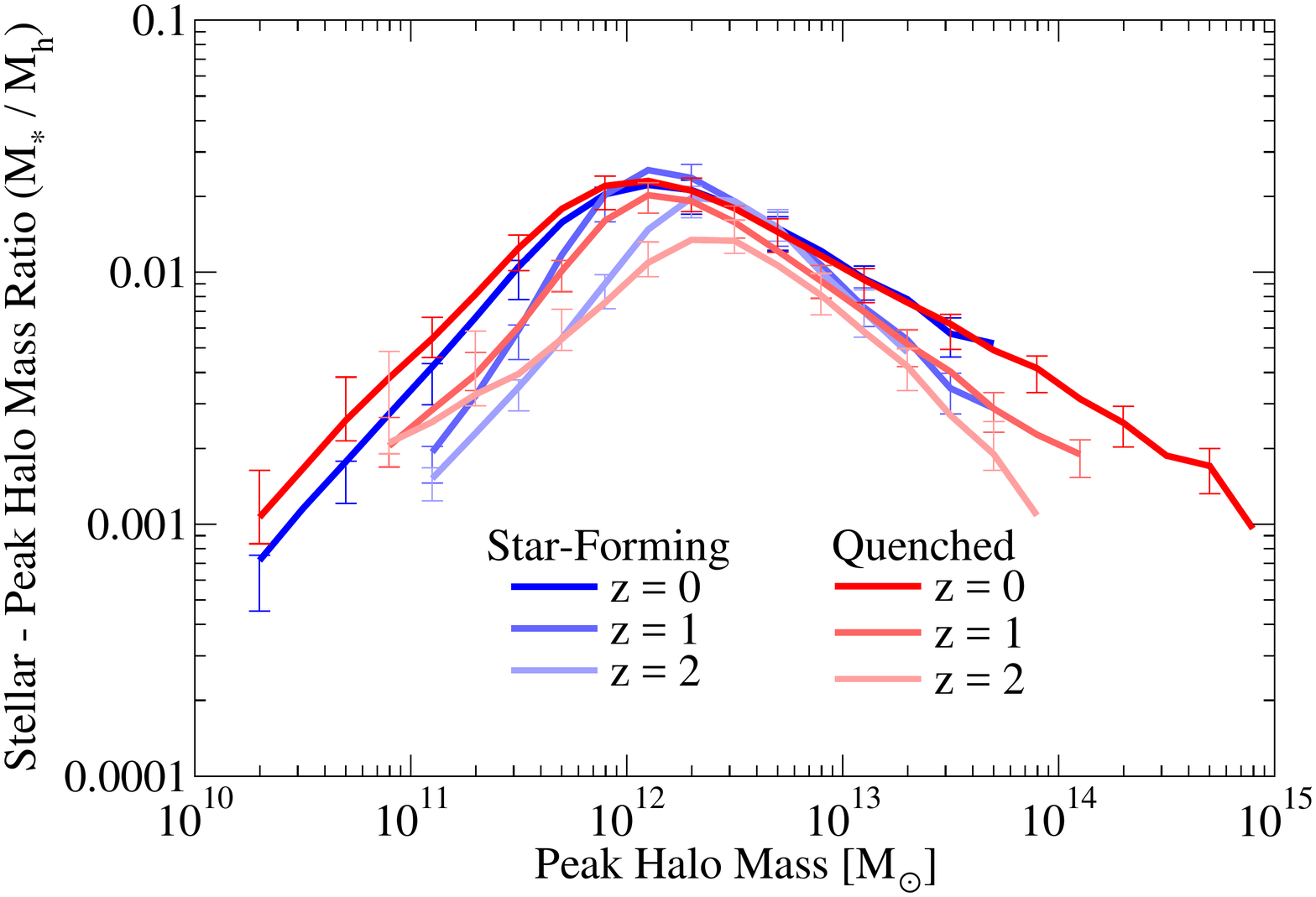}\plotgrace{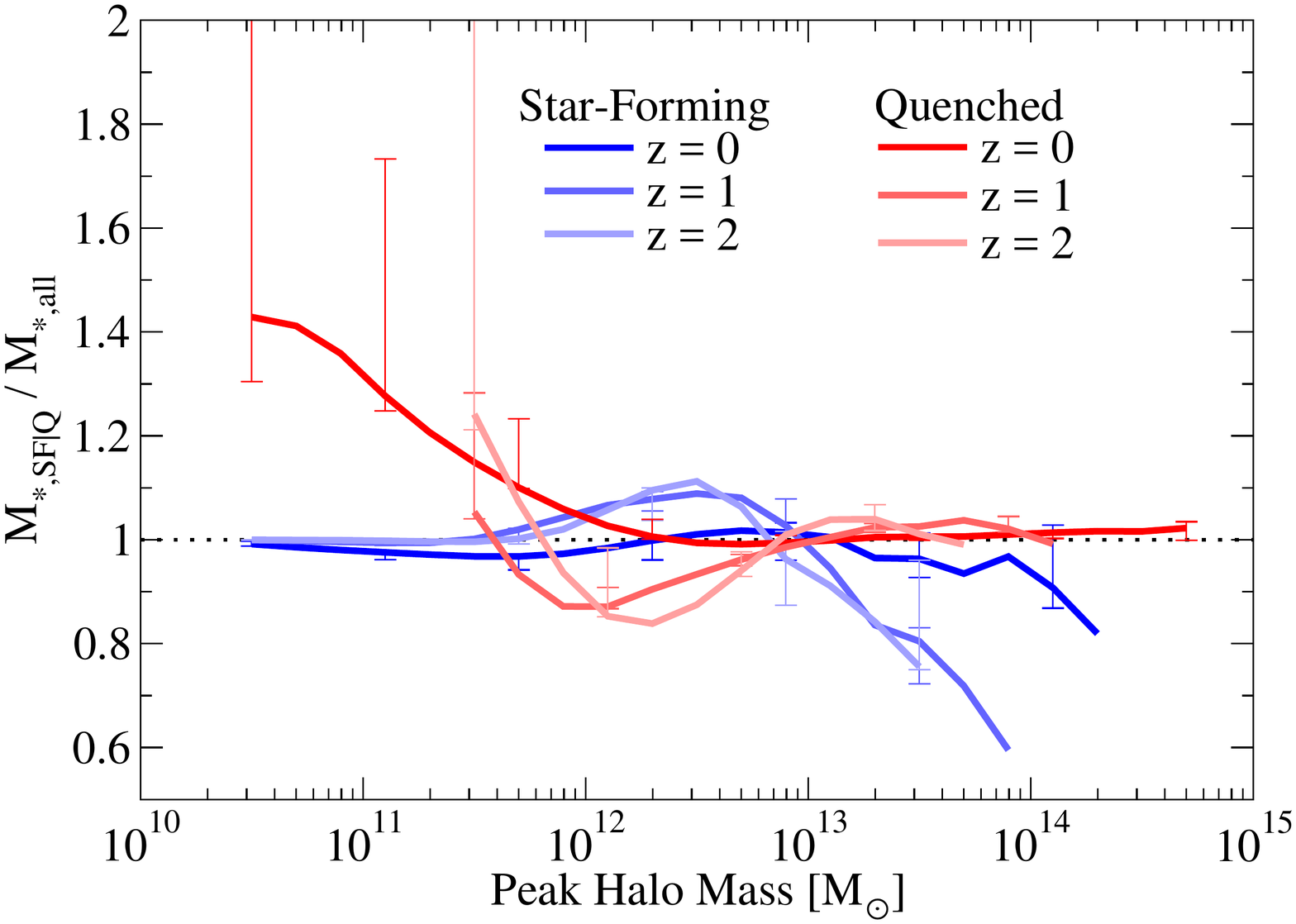}\\[-4.5ex]
\caption{The best-\fit{} median ratio of stellar mass to peak halo mass ($\mpeak$) for star-forming and quenched galaxies  (\textbf{left} panel) compared to the ratio for all galaxies  (\textbf{left} panel).  Error bars in both panels show the 68\% confidence interval for the model posterior distribution.  \textbf{Notes:} quiescent galaxies in low-mass ($<10^{12}\Msun$) haloes are very rare at $z\ge 1$, contributing to substantial uncertainties in their stellar mass--halo mass relations.  Due to scatter in stellar mass at fixed halo mass, a higher stellar mass for star-forming galaxies at fixed halo mass does not necessarily imply a lower halo mass for star-forming galaxies at fixed galaxy mass; see \S \ref{s:smhm_comp} for discussion. \textbf{See Appendix \ref{a:smhm_fits} for fitting formulae.}}
\label{f:smhm_sf_q}
\end{figure*}

\begin{figure*}
\vspace{-5ex}
\plotgrace{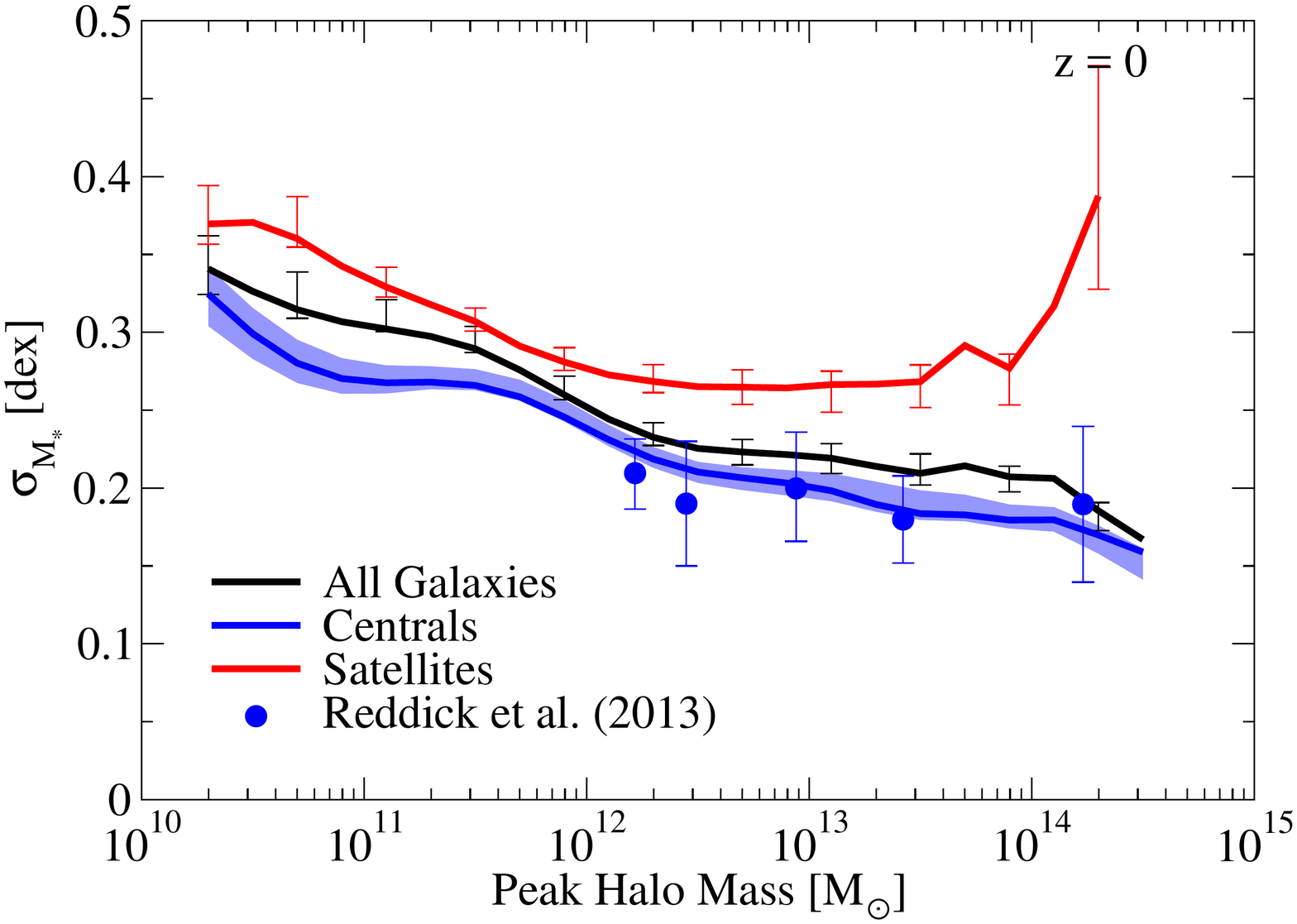}\plotgrace{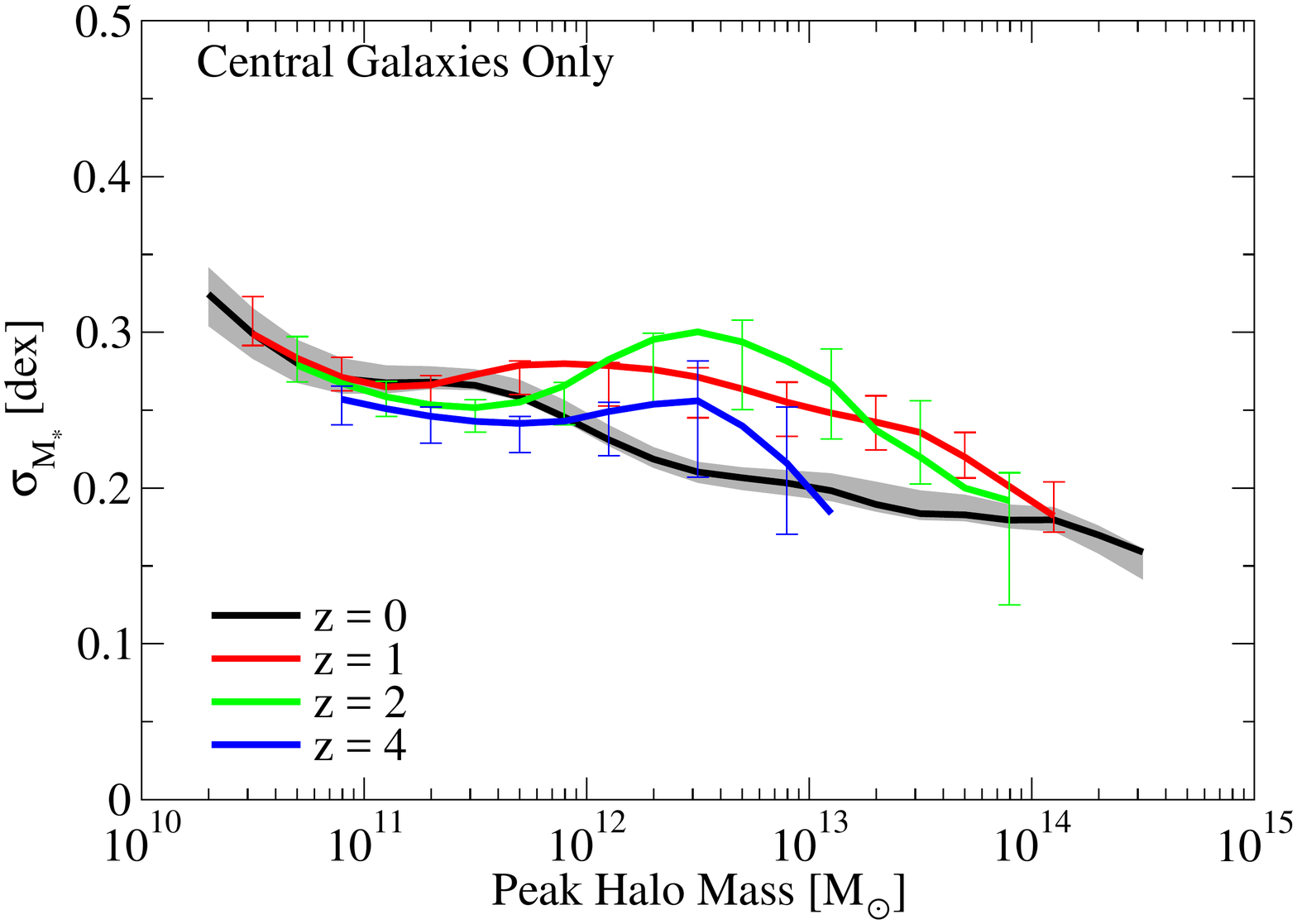}\\[-4.5ex]
\caption{\textbf{Left} panel: best-\fit{} scatter in stellar mass at fixed $\mpeak$, split for central and satellite galaxies at $z=0$, compared with the results for central galaxies in \protect\cite{Reddick12}.  \textbf{Right} panel: best-\fit{} scatter in stellar mass at fixed peak halo mass ($\mpeak$) for central galaxies as a function of $\mpeak$ and $z$.  Error bars and shaded regions in both panels show the 68\% confidence interval for the model posterior distribution.}
\label{f:smhm_scatter}
\vspace{-2ex}
\plotgrace{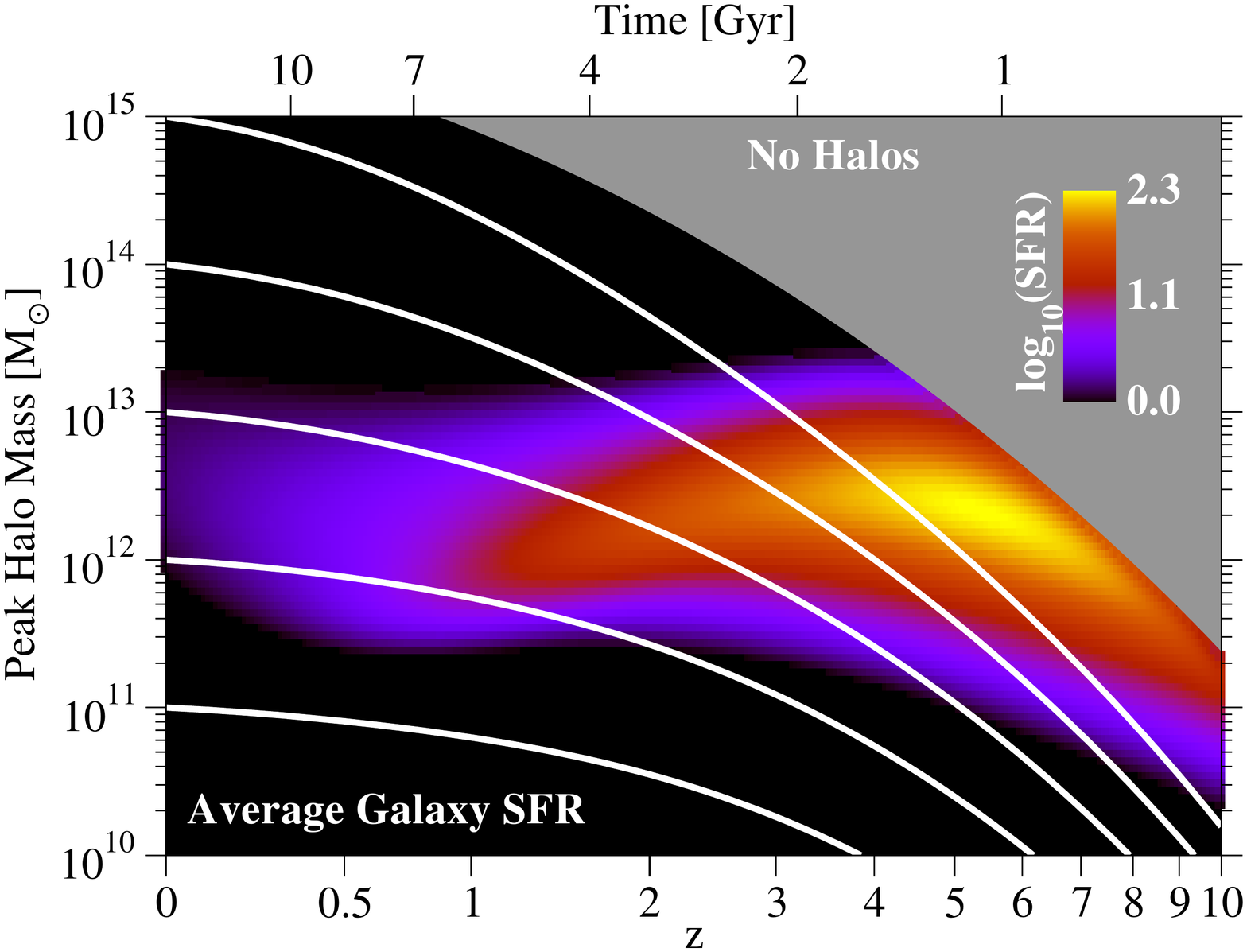}\plotgrace{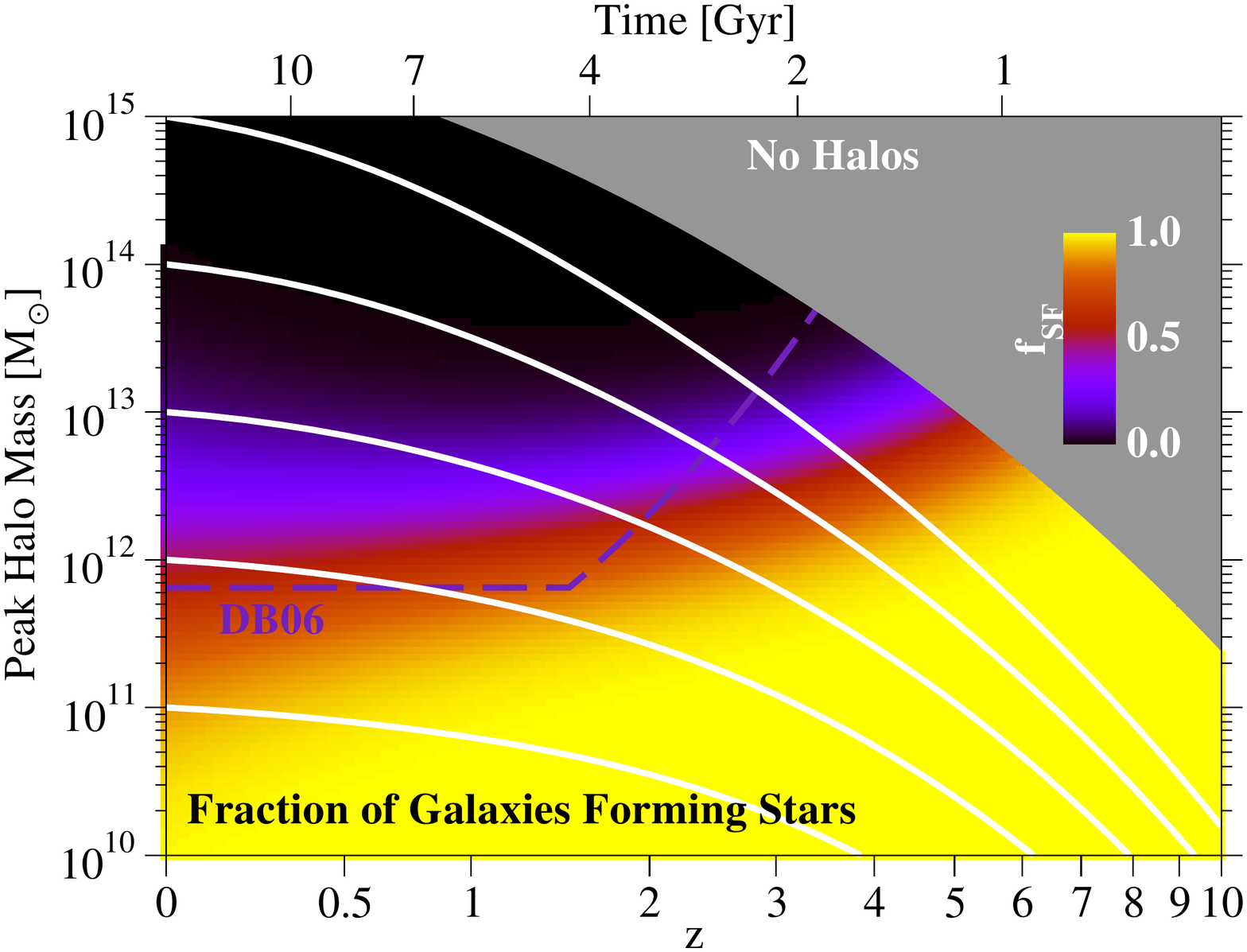}\\[-4.5ex]
\caption{\textbf{Left} panel: Average star formation rates in galaxies as a function of halo mass and redshift. \textbf{Right} panel: Average star-forming fractions as a function of halo mass and redshift.  The \textit{purple line} marks the predicted transition in \protect\cite{Dekel06} between cold flows reaching the central galaxy (below the line) and not reaching it due to shock heating (above the line).  In both panels, \textit{white lines} mark median halo growth trajectories, and the \textit{gray region} marks where no haloes are expected to exist in the observable Universe.  A robust upturn in the star-forming fraction to higher redshifts is visible (\S \ref{s:average_sfrs}).}
\label{f:avg_sfr}
\vspace{-2ex}
\plotgrace{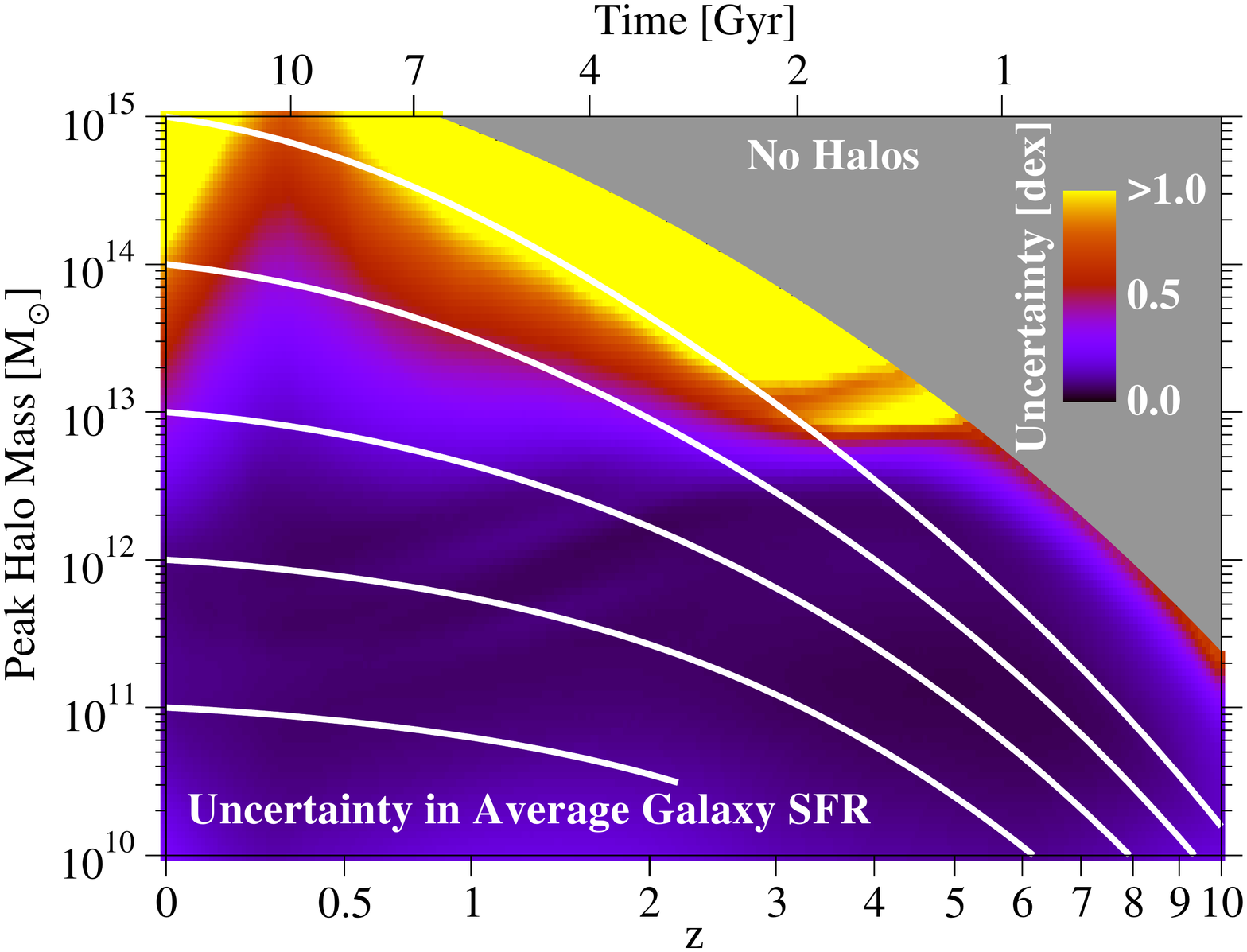}\plotgrace{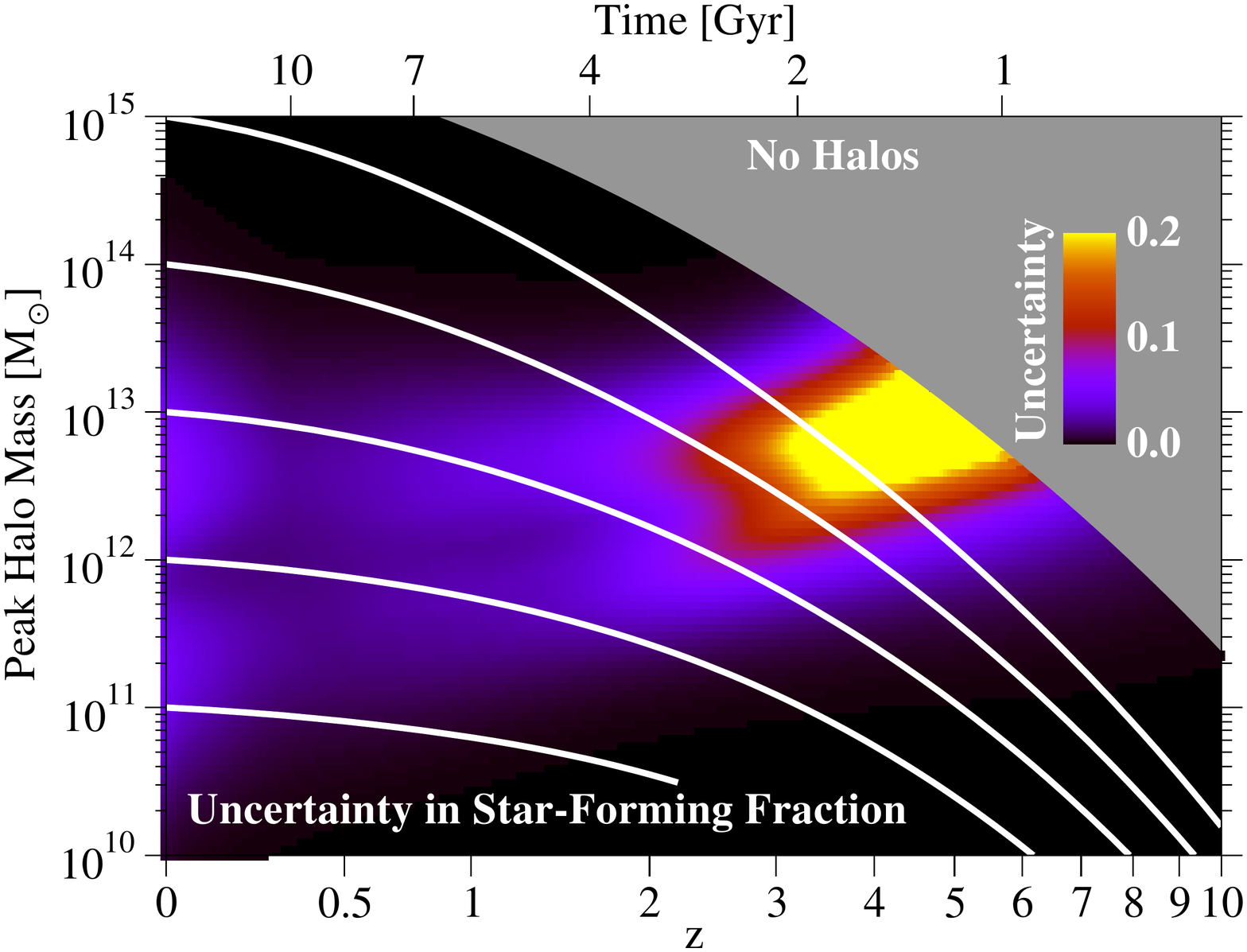}\\[-4.5ex]
\caption{\textbf{Left} panel: formal model uncertainty (half of the \onesigdist{}) in average galaxy SFRs (Fig.\ \ref{f:avg_sfr}, left panel).  \textbf{Right} panel: formal model uncertainty (half of the \onesigdist{}) in average galaxy star-forming fractions (Fig.\ \ref{f:avg_sfr}, right panel).}
\label{f:sfr_errors}
\end{figure*}
\begin{figure*}
\plotgrace{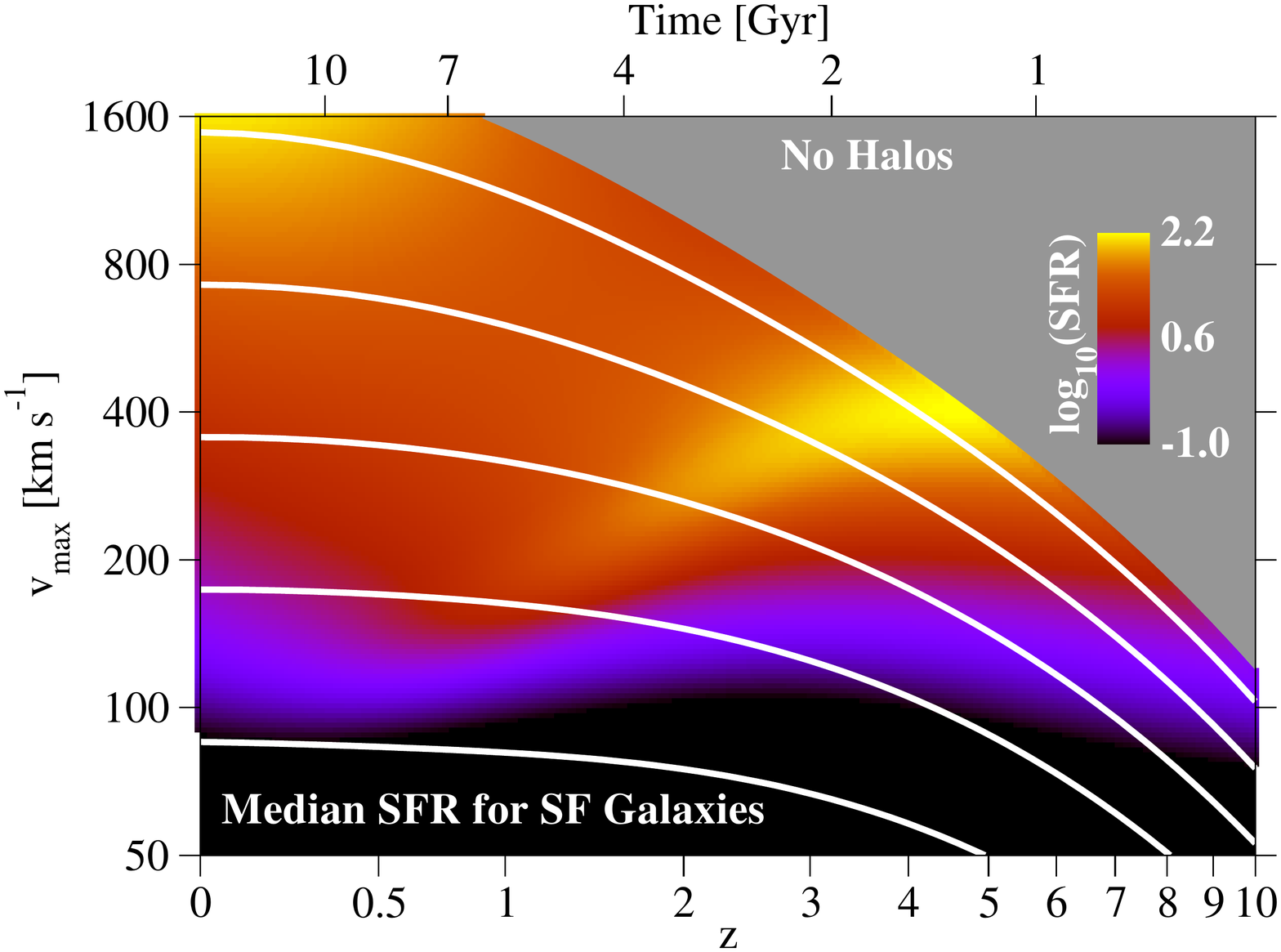}\plotgrace{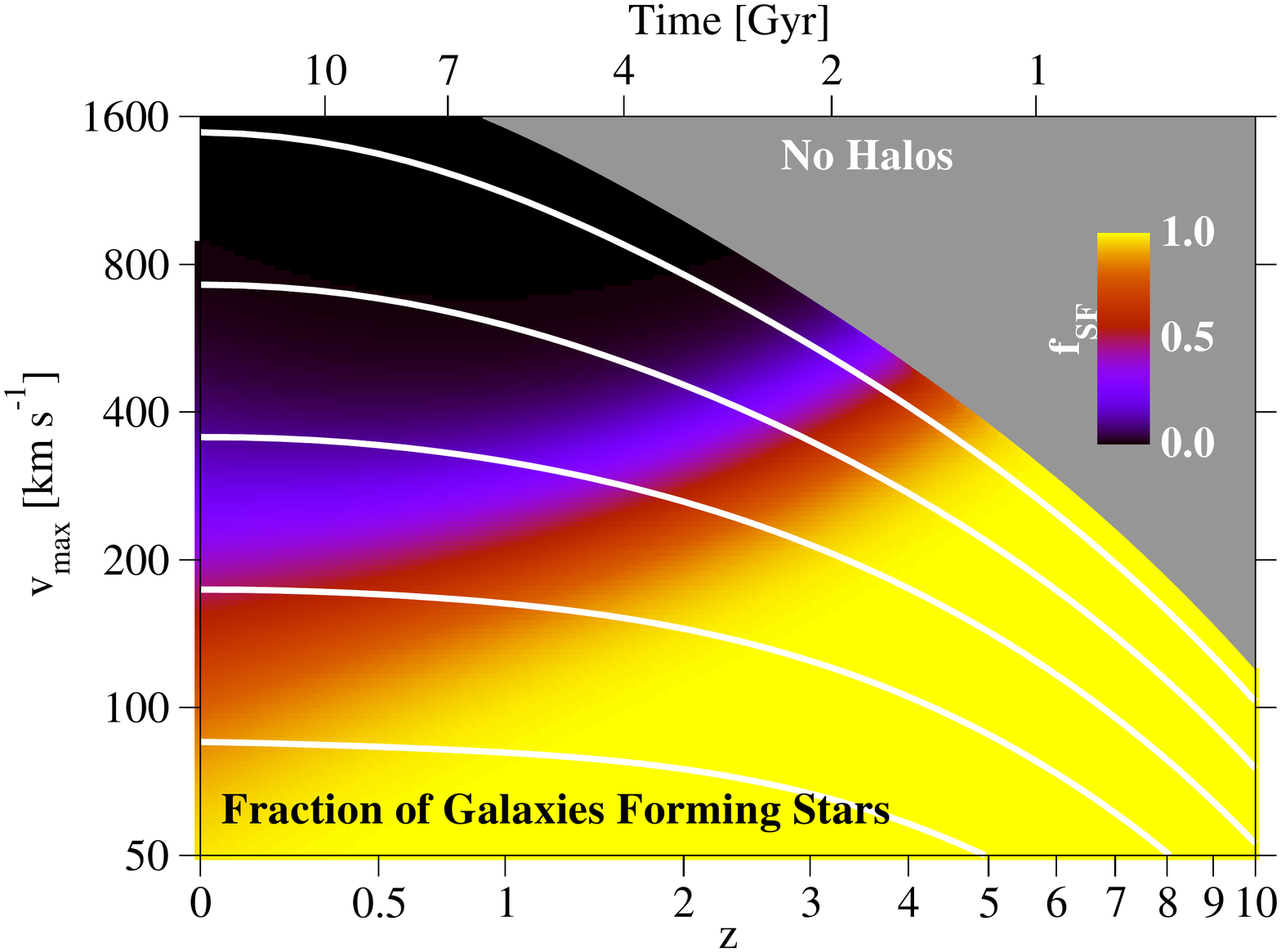}\\[-5ex]
\caption{\textbf{Left} panel: Median star formation rate for star-forming haloes as a function of $\vmp$ and $z$ ($\sfrsf$; Eqs.\ \ref{e:sfrsf}$-$\ref{e:delta}).  \textbf{Right} panel: star-forming fraction ($1-\fq$) as a function of  $\vmp$ and $z$ (Eqs.\ \ref{e:fq_1}$-$\ref{e:fq_3}).}
\label{f:sfr_vmax}
\end{figure*}

 We show the median stellar mass --- peak halo mass ratio (SMHM ratio) for all galaxies in Fig.\ \ref{f:smhm}, which agrees with past measurements (\S \ref{s:smhm_comp}).  Although the SMHM ratio has little net change from $z=0$ to $z\sim 5$, this study supports significant evolution at $z>5$ (see also \S \ref{s:smhm_evolution}).  \textbf{Fitting formulae for median SMHM ratios are presented in Appendix \ref{a:smhm_fits}}.

As shown in Fig.\ \ref{f:smhm_cen_sat}, we find that central and satellite haloes have significantly different SMHM ratios.  At low halo masses, satellite quenching timescales are long (\S \ref{s:satellites}), so they grow in stellar mass while $\mpeak$ remains fixed, leading to higher SMHM ratios.  At high halo masses, the dominant growth channel is via mergers (\S \ref{s:in_situ}), which are reduced for satellites due to high relative velocities; hence, they have lower SMHM ratios than centrals.  

We also find that star-forming and quenched galaxies have significantly different SMHM ratios (Fig.\ \ref{f:smhm_sf_q}) except at $z\sim0$.  At low masses ($\mpeak \ll 10^{12}\Msun$) and redshifts $z>0$, most quenched galaxies stopped forming stars only recently, leading to relatively small differences.  At high masses ($\mpeak \gg 10^{12}\Msun$), the only star-forming galaxies are those whose haloes have formed very recently, resulting in less time for satellites to merge and contribute stellar mass.

For intermediate masses ($\mpeak \sim 10^{12}\Msun$), the picture is more complex.  These haloes quench and rejuvenate (\S \ref{s:stochasticity}) while mass accretion continues.  Hence, galaxies that are star-forming tend to have higher SMHM ratios (galaxies growing faster relative to their haloes), whereas those that are quenched have lower SMHM ratios (no galaxy growth but continued halo growth).  These differences are more evident at $z=2$, where the ratio of galaxy SSFRs to halo specific accretion rates is higher \citep[e.g.,][]{BehrooziHighZ}.  At $z=0$, galaxy growth is less rapid, and so galaxies have less time to grow significantly between periods of quenching and rejuvenation driven by halo mass accretion.  The difference between SMHM ratios for quiescent and star-forming galaxies is thus sensitive to the amount of quenching and rejuvenation, but in practice, the observed difference is just as sensitive to systematic errors in the stellar masses used (Appendix \ref{a:smf}).  

\subsubsection{Scatter in the Stellar Mass -- Halo Mass Relation}

We also show constraints on scatter in the SMHM relation in Fig.\ \ref{f:smhm_scatter}.  Our primary observational constraint on scatter comes from correlation functions (Figs.\ \ref{f:cf_comp} and \ref{f:cf_comp_primus}), but this is somewhat degenerate with the orphan fraction (see Appendix \ref{a:orphans}).  Without orphans, our model cannot match autocorrelation functions for low-mass galaxies, which are largely unaffected by scatter.  As a result, autocorrelation functions for larger galaxies (which are more sensitive to scatter) can be reproduced with somewhat larger scatters than previous works that did not include orphans \citep[e.g.,][]{Reddick12}.  It is possible that a more complicated orphan model could reduce the need for additional scatter; constraining such a model would require additional observational data beyond what is used here (see \S \ref{s:orphans}).

\begin{figure*}
\vspace{-5ex}
\plotgrace{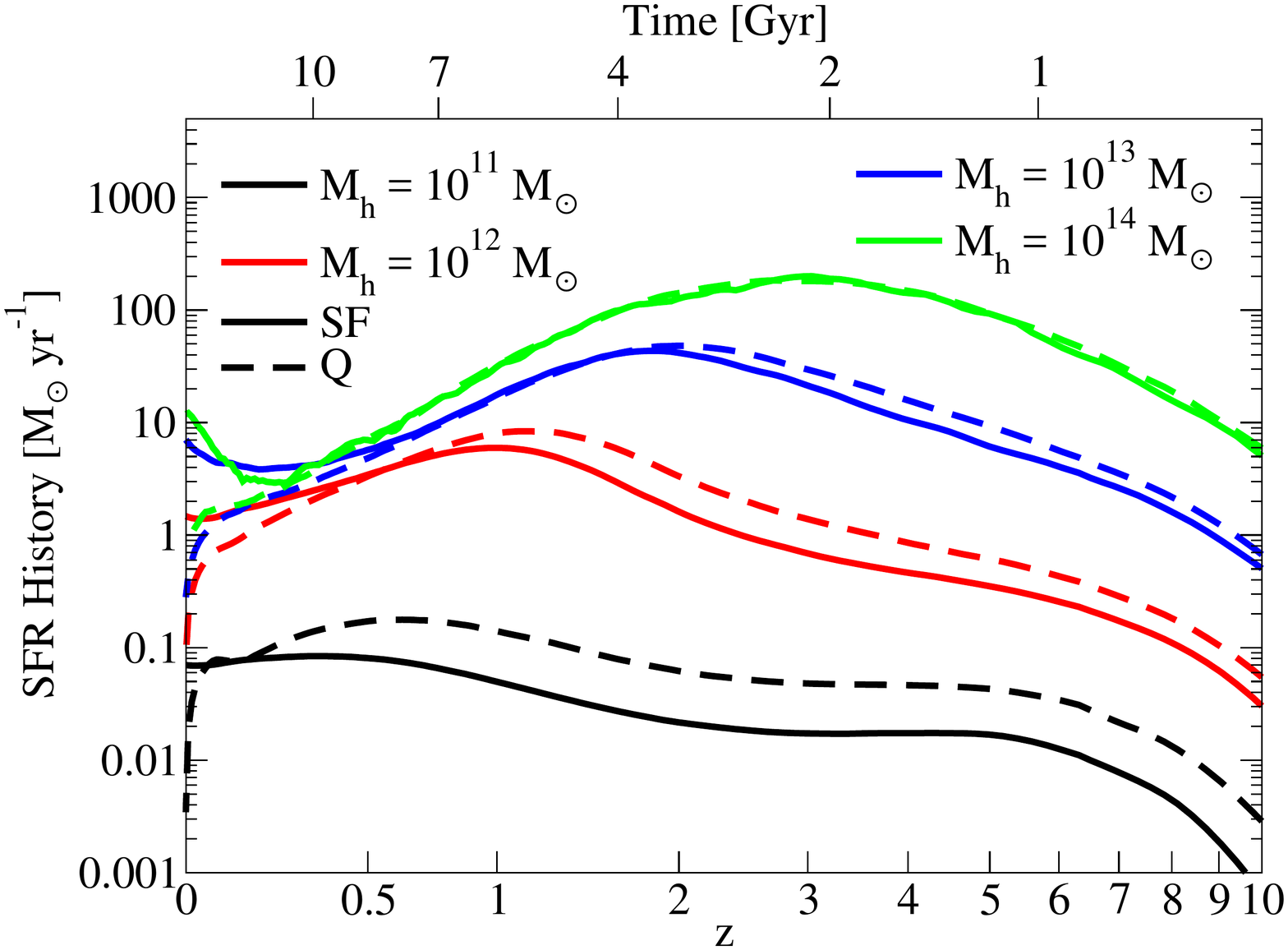}\plotgrace{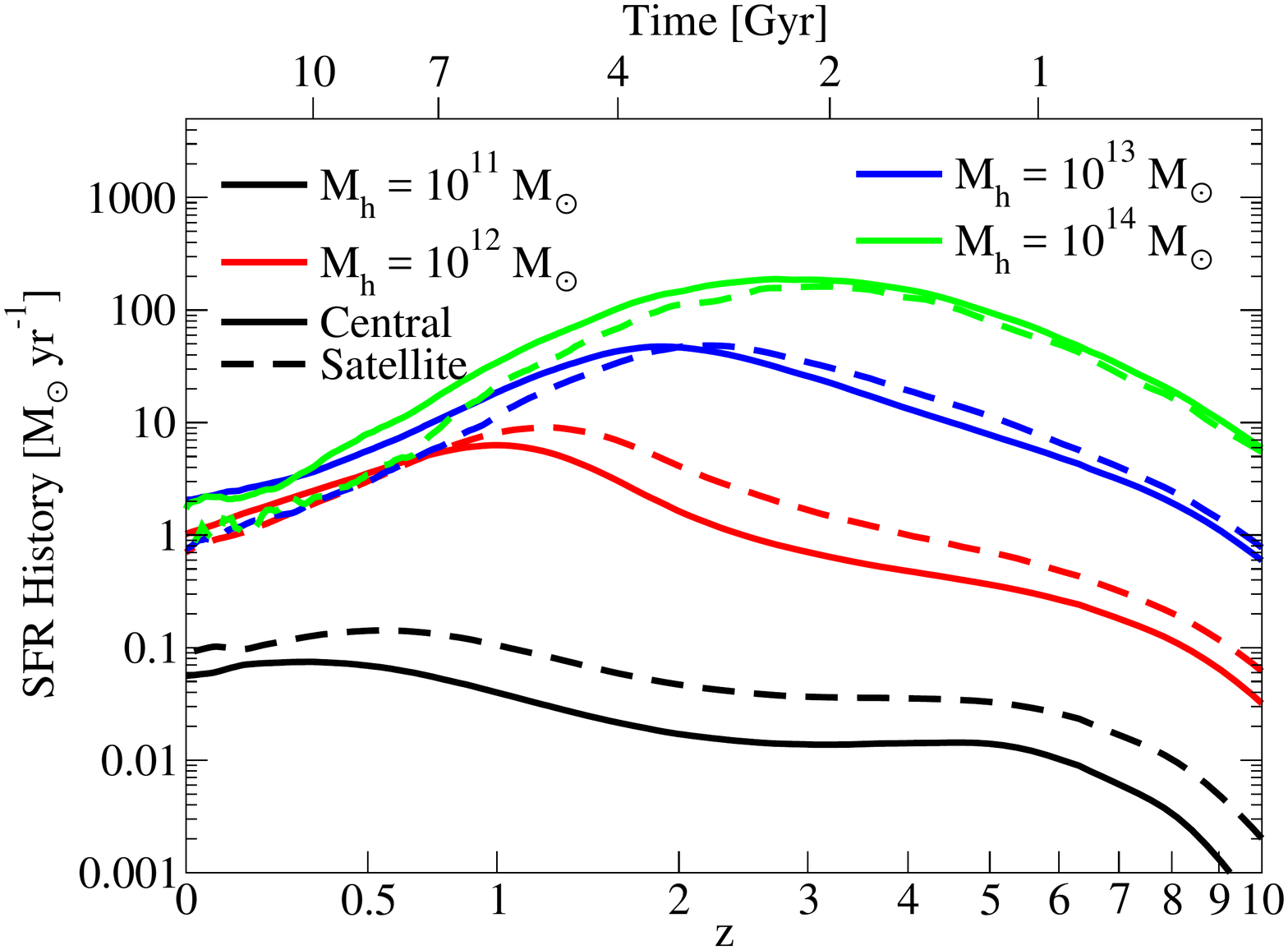}\\[-5ex]
\caption{\textbf{Left} panel: average star formation histories for star-forming and quiescent galaxies, as a function of peak $z=0$ halo mass ($\mpeak$) for our best-\fit{} model.  \textbf{Right} panel: same, except split for central and satellite galaxies.  Notably, satellites are binned according to the peak historical mass of the \textit{satellite} halo, as opposed to their host halo.  For both panels, bin widths are $\pm 0.25$ dex; e.g.,  the label $M_h = 10^{11}\Msun$ corresponds to $10.75 < \log_{10}(\mpeak/\Msun) < 11.25$.}
\label{f:sfh_m}
\plotgrace{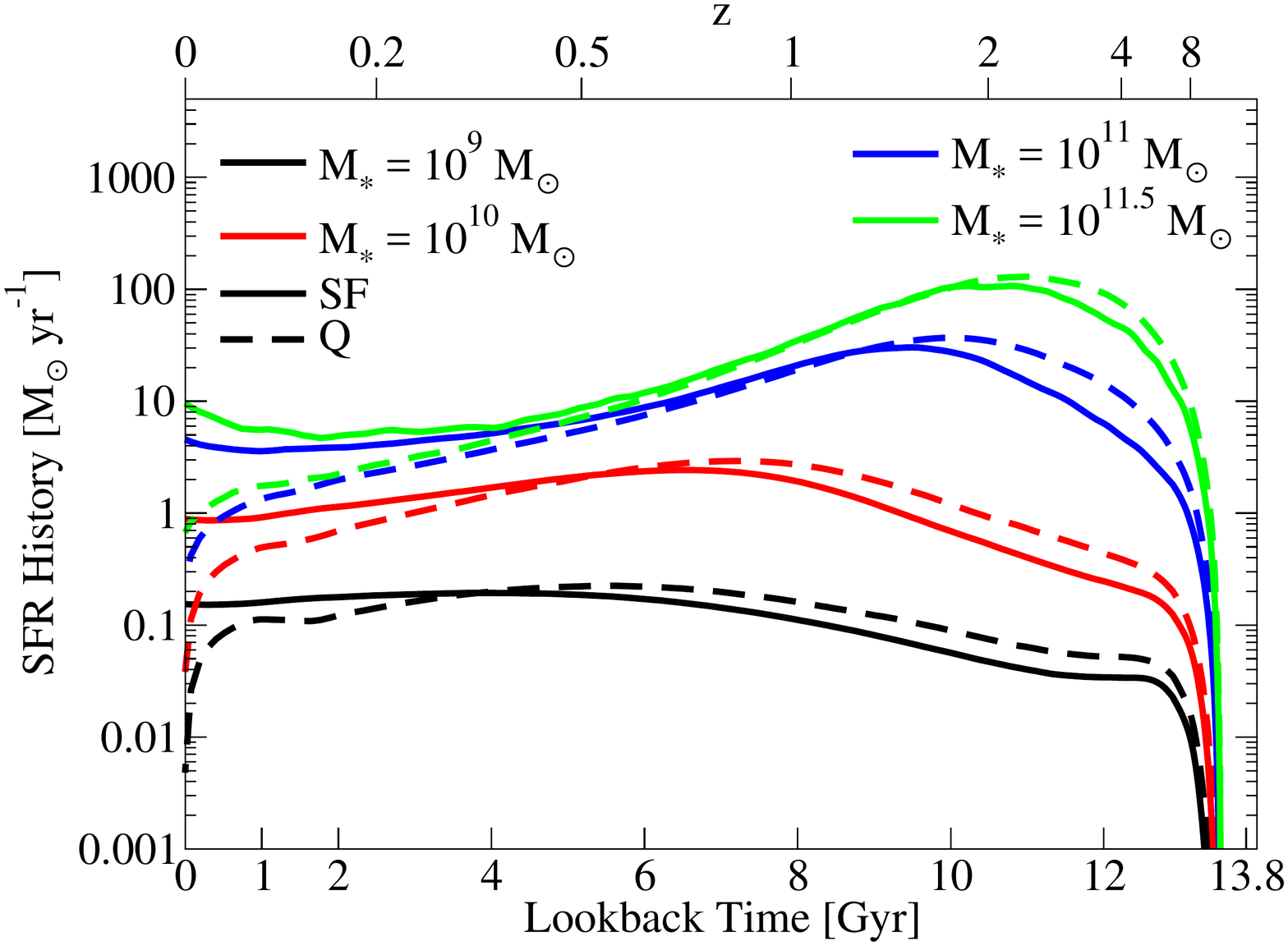}\plotgrace{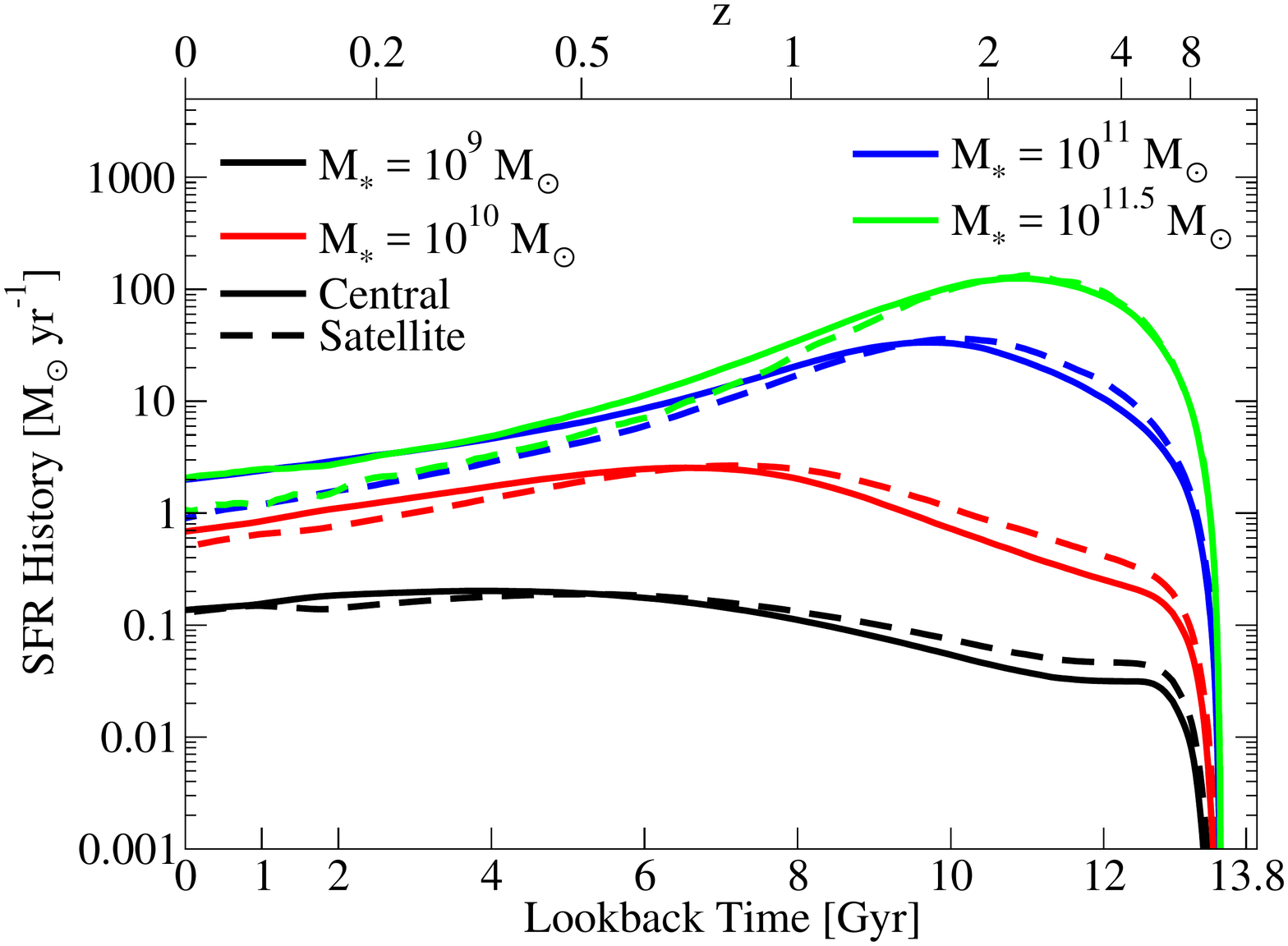}\\[-5ex]
\caption{\textbf{Left} panel: average star formation histories for star-forming and quiescent galaxies, as a function of $M_\ast$ at $z=0$ for our best-\fit{} model.  These are shown as a function of lookback time to emphasize the recent differences that have greater observable effects. \textbf{Right} panel: same, except split for central and satellite galaxies.  For both panels, bin widths are $\pm 0.25$ dex; e.g.,  the label $M_\ast = 10^{11}\Msun$ corresponds to $10.75 < \log_{10}(M_\ast/\Msun) < 11.25$.}
\label{f:sfh_sm}
\end{figure*}

In the model, satellites have much larger scatter than central galaxies (Fig.\ \ref{f:smhm_scatter}, left panel), due to the orbit-dependence of continued star formation after infall.  Similarly, quenched galaxies have larger scatter than star-forming galaxies, as quenched populations have larger satellite fractions.  We also find lower scatter in stellar mass towards higher halo masses.  This results from an increasing fraction of mass growth via mergers \citep[Fig.\ \ref{f:merger_stats}; see also ][]{Moster12,BWC13,Gu16}; other empirical models \citep[e.g.,][]{Moster17} show similar trends.

Our current constraints are consistent with either no redshift evolution in the scatter or a slight increase toward higher redshifts (Fig.\ \ref{f:smhm_scatter}, right panel).  Increased scatter is most prominent for haloes near $10^{12}\Msun$.  Haloes near this mass grow primarily by star formation, and so are dramatically affected by quenching (see also Fig.\ \ref{f:random_sfhs}, right panel) if it is not perfectly correlated with mass growth.  Galaxies in lower-mass haloes are mostly star-forming and hence have smaller variations in star formation histories (see also Fig.\ \ref{f:random_sfhs}, left panel).  Galaxies in larger haloes grow primarily by merging, which is more correlated with halo mass growth.  Indeed, had our model correlated quenching directly with halo mass growth instead of change in $\vmax$, the overall scatter would be lower and the feature near $10^{12}\Msun$ would not exist \citep[see][]{Moster17}.

\subsection{Average SFRs and Star-Forming Fractions}

\label{s:average_sfrs}

Average SFRs and star-forming fractions for the best-\fit{} model are shown as a function of $\mpeak$ and $z$ for all galaxies in Fig.\ \ref{f:avg_sfr}.  Similar to past results \citep[e.g.,][]{BWC13}, high mass haloes exhibit a short period of very intense star formation and then quench, whereas lower-mass haloes have much more extended star formation histories.  The most notable difference from previous modeling is an improved treatment of quenching in massive haloes (\S \ref{s:fform}), which reduces their expected star formation rates.  We caution that star formation rates for central galaxies in massive haloes are nonetheless very hard to measure observationally, so the values in Fig.\ \ref{f:avg_sfr} for massive quenched haloes should be treated as upper limits (see also the formal uncertainties in Fig.\ \ref{f:sfr_errors}, left panel).

At $z>1$, Fig. \ref{f:avg_sfr} shows a strong correlation between halo mass and quenching; a difference of $\lesssim$1.5 dex in host halo mass separates populations that are nearly $100\%$ star-forming from those that are nearly 100\% quenched.  For $z<1$, satellite quenching becomes more important, and so quenched galaxies appear over a broader range in halo mass.  As discussed in later sections, haloes with moderate quenched fractions (30-70\%) are more susceptible to quenching via differences in assembly rates.

At fixed halo mass, average quenched fractions for galaxies decrease significantly with increasing redshift.  The SMHM relation evolves relatively little from $z=0-4$ (Fig.\ \ref{f:smhm}), whereas $\fq(M_\ast)$ evolves significantly (Fig.\ \ref{f:qf_comp}), requiring $\fq(M_h)$ to evolve significantly with redshift as well.  This is qualitatively (but not quantitatively) in agreement with \cite{Dekel06}, as discussed in \S \ref{s:central_quenching}.   Formal uncertainties are shown in Fig.\ \ref{f:sfr_errors} (right panel), and are under 10\% for almost all redshifts and halo masses.

We show the underlying constraints on $\sfrsf(\vmp, z)$ and $\fq(\vmp,z)$ in Fig.\ \ref{f:sfr_vmax}.  The average SFR in Fig.\ \ref{f:avg_sfr} is the product of the left and right panels of Fig.\ \ref{f:sfr_vmax}, with a small correction for scatter.  The left panel suggests that when galaxies in massive haloes are able to form stars, they do so extremely rapidly---qualitatively consistent with observations of the Phoenix cluster \citep{McDonald13} and precipitation theory \citep{Voit15}.  However, the highest star-formers on average are in lower-mass haloes at higher redshifts, where the quenched fractions are much lower.

\begin{figure*}
\vspace{-5ex}
\plotgrace{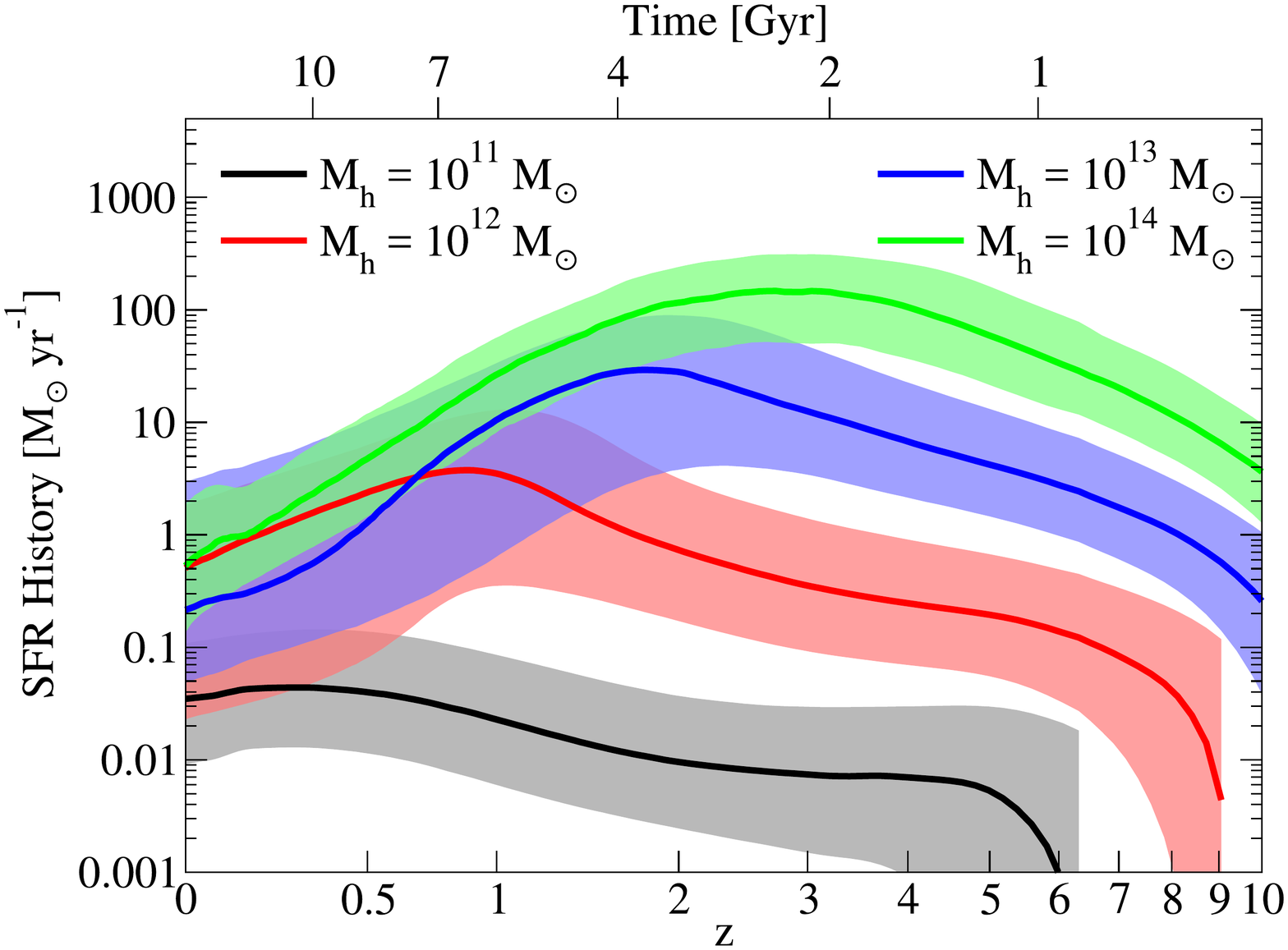}\plotgrace{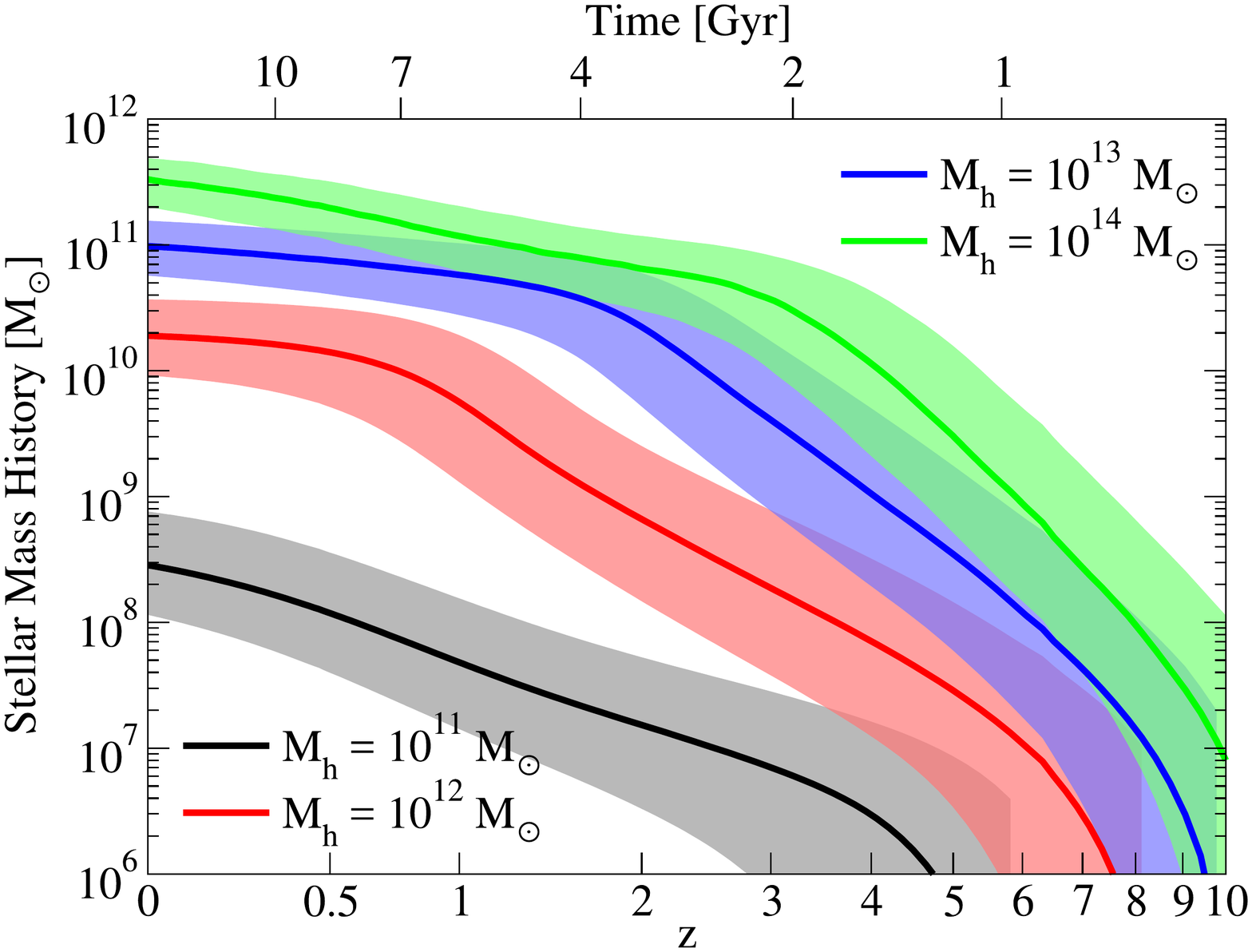}\\[-5ex]
\plotgrace{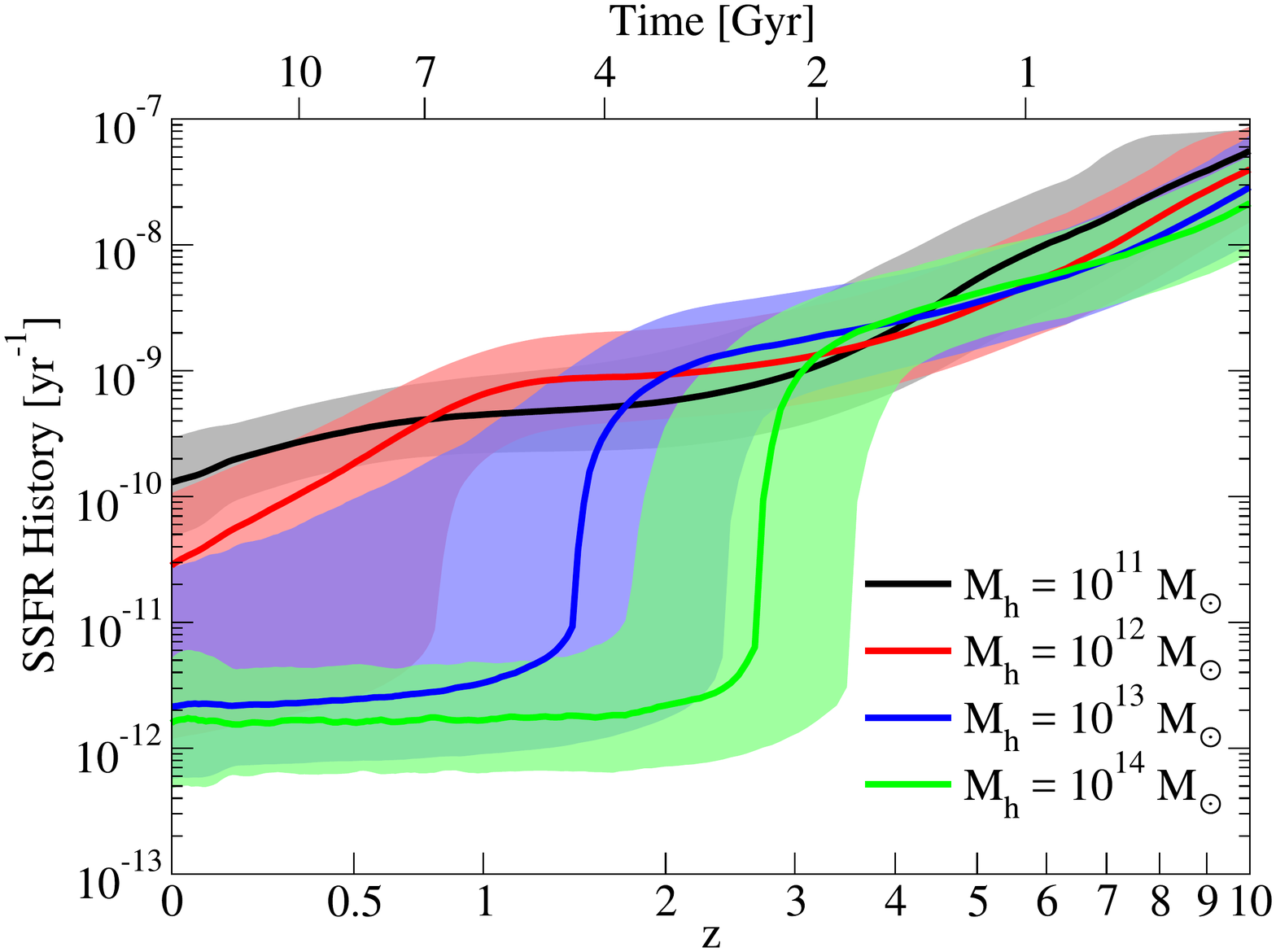}\plotgrace{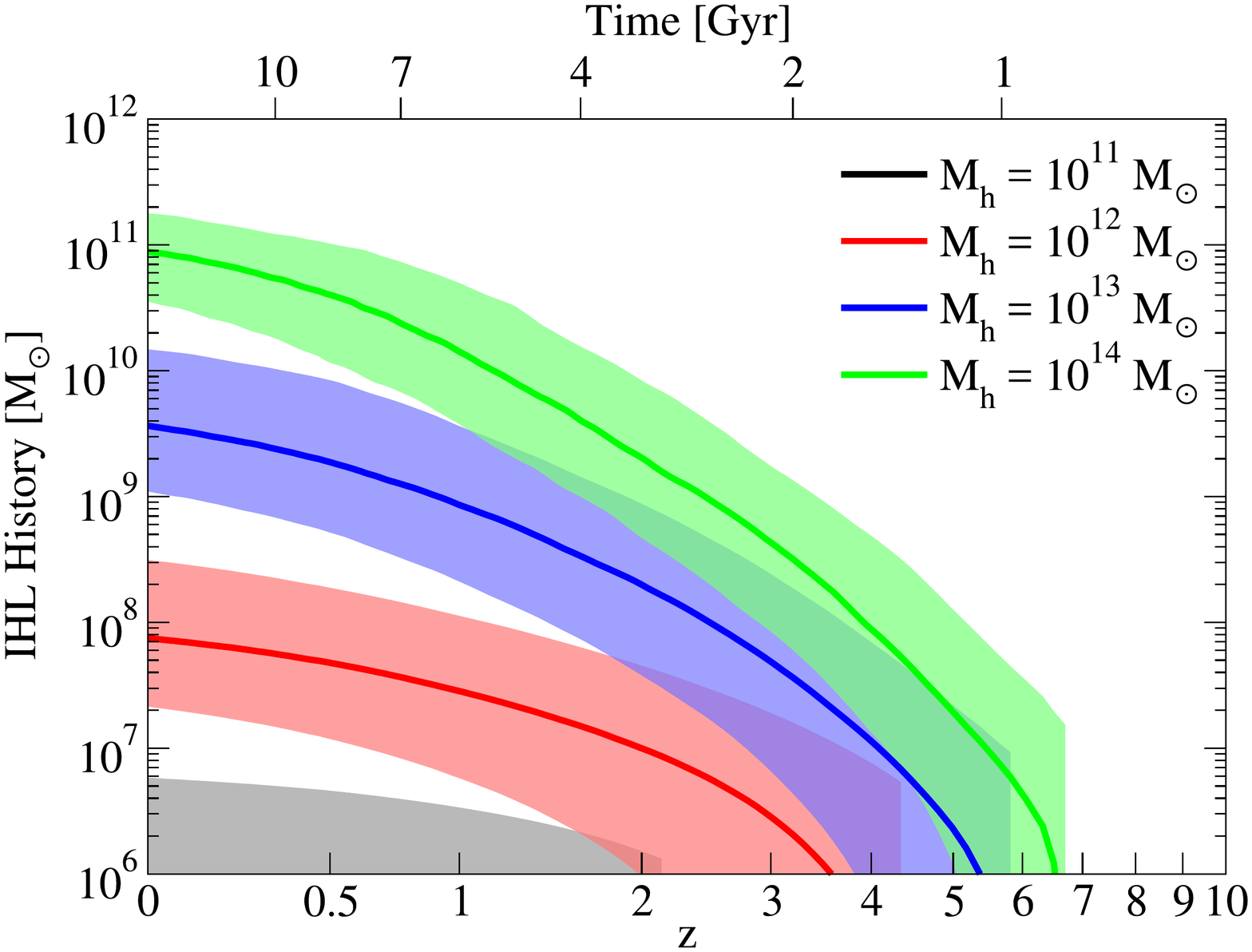}\\[-5ex]
\plotgrace{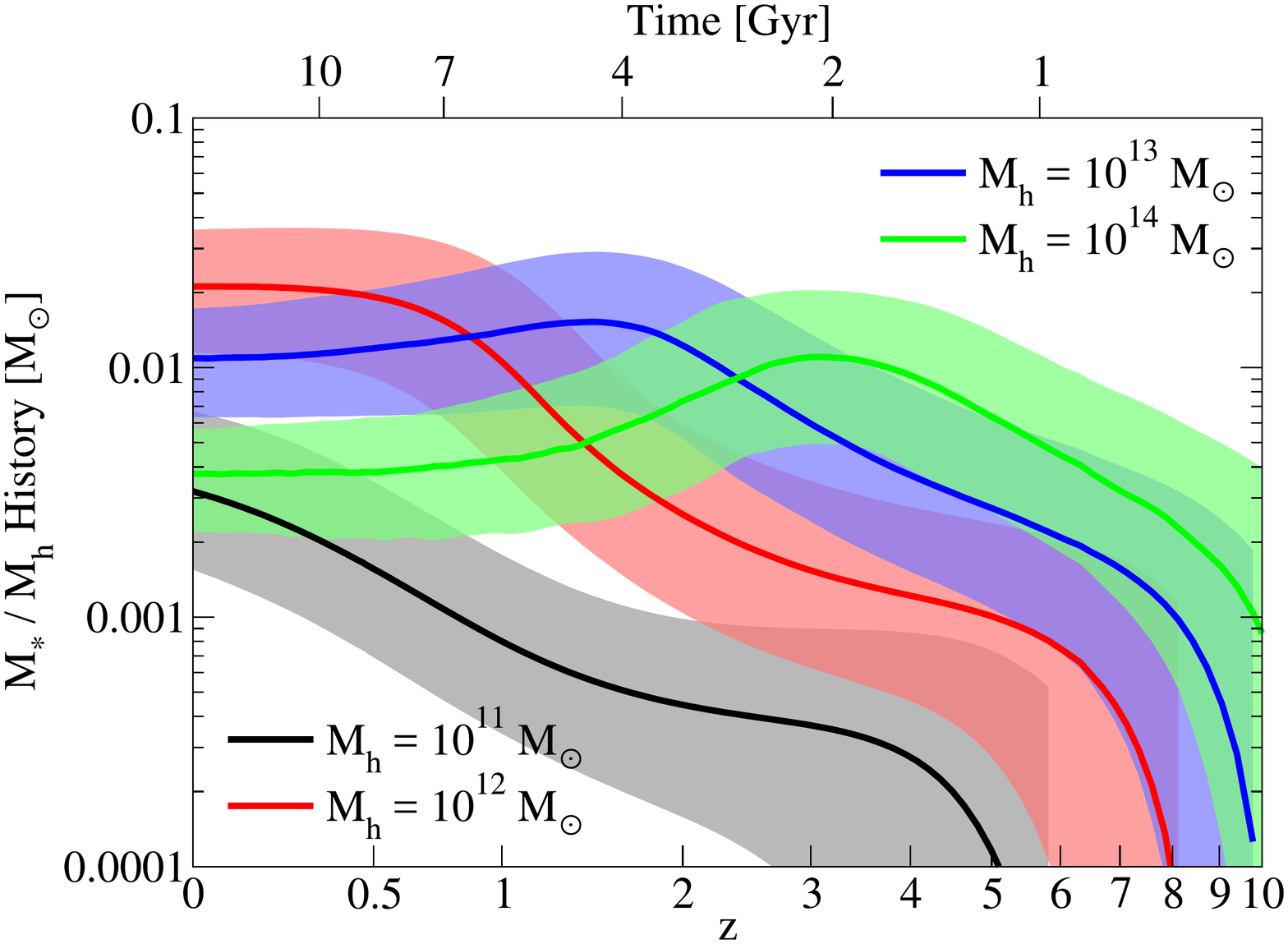}\plotgrace{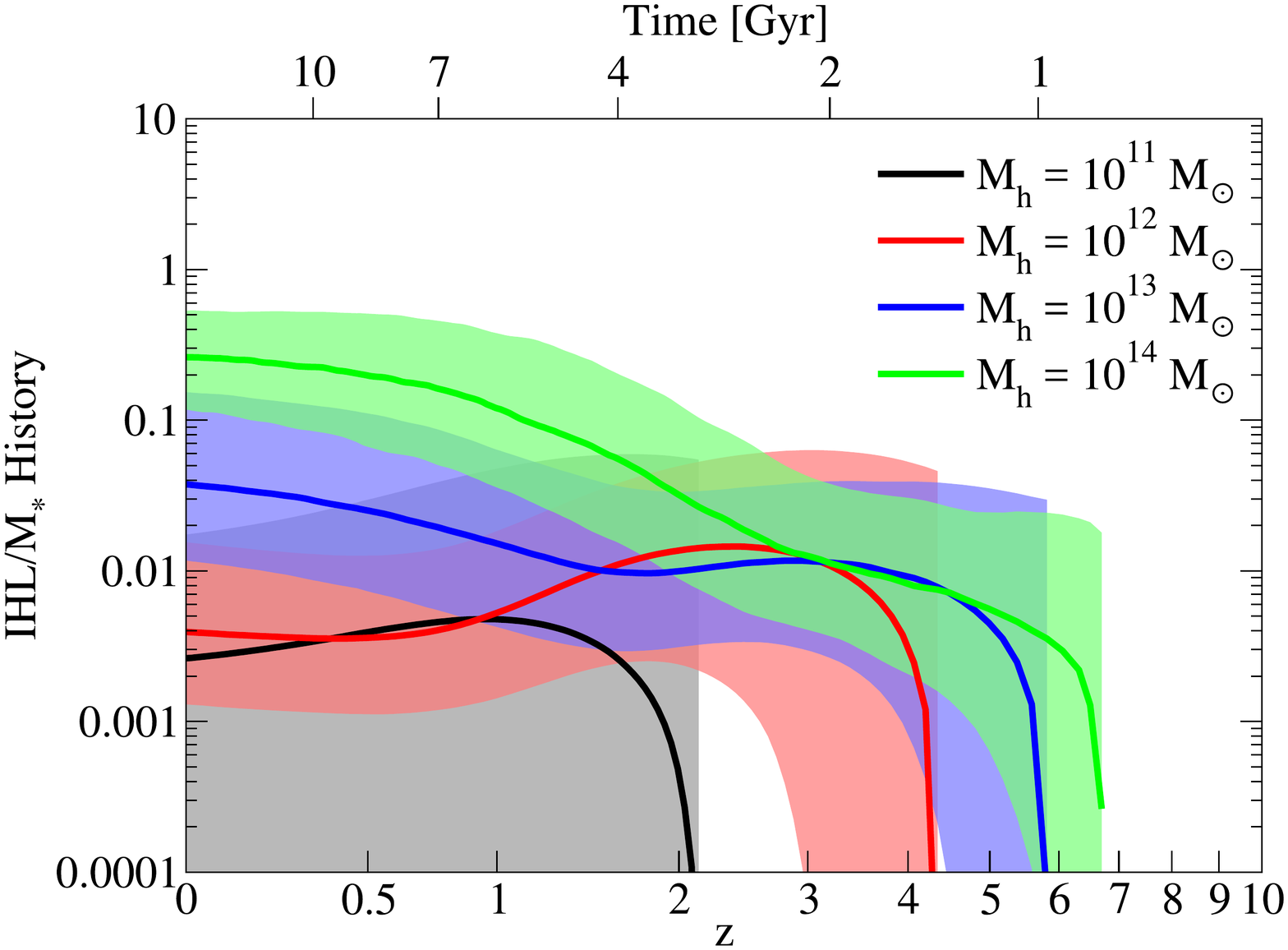}\\[-5ex]
\caption{\textbf{Top left}: total (including mergers) star formation histories for haloes in bins of $\mpeak$ for our best-\fit{} model; \textit{coloured regions} indicate the \onesigdist{} among different haloes in the best-\fit{} model.  \textbf{Top right}: same, with main progenitor stellar mass histories.  \textbf{Middle left}: same, with main progenitor SSFR histories.  \textbf{Middle right}: same, with main progenitor intrahalo light (IHL) histories.  \textbf{Bottom left}: same, with main progenitor $M_\ast / \mpeak$ ratio histories.  \textbf{Bottom right}: same, with main progenitor IHL / $M_\ast$ ratio histories.  For all panels, bin widths are $\pm 0.25$ dex; e.g.,  the label $M_h = 10^{11}\Msun$ corresponds to $10.75 < \log_{10}(\mpeak/\Msun) < 11.25$.  \textbf{Notes:} Quiescent SSFRs in our models are fixed to $10^{-11.8}$ yr$^{-1}$ (\S \ref{s:fform}), explaining why massive haloes' SSFRs are close to this value at low redshifts.}
\label{f:indiv_sfr}
\end{figure*}

\begin{figure*}
\vspace{-5ex}
\hspace{-10ex}\plotmodestgrace{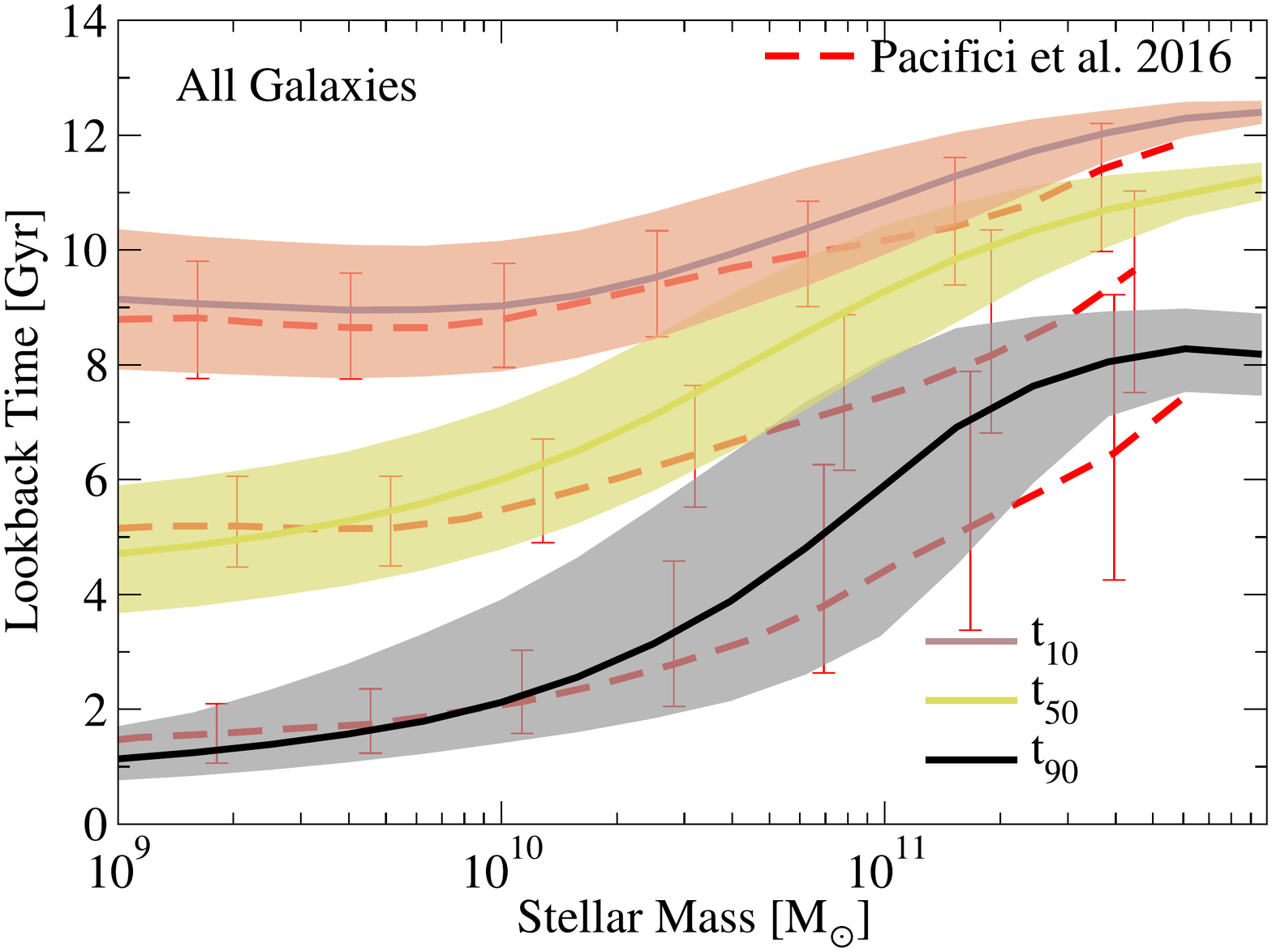}\hspace{-13ex}\plotmodestgrace{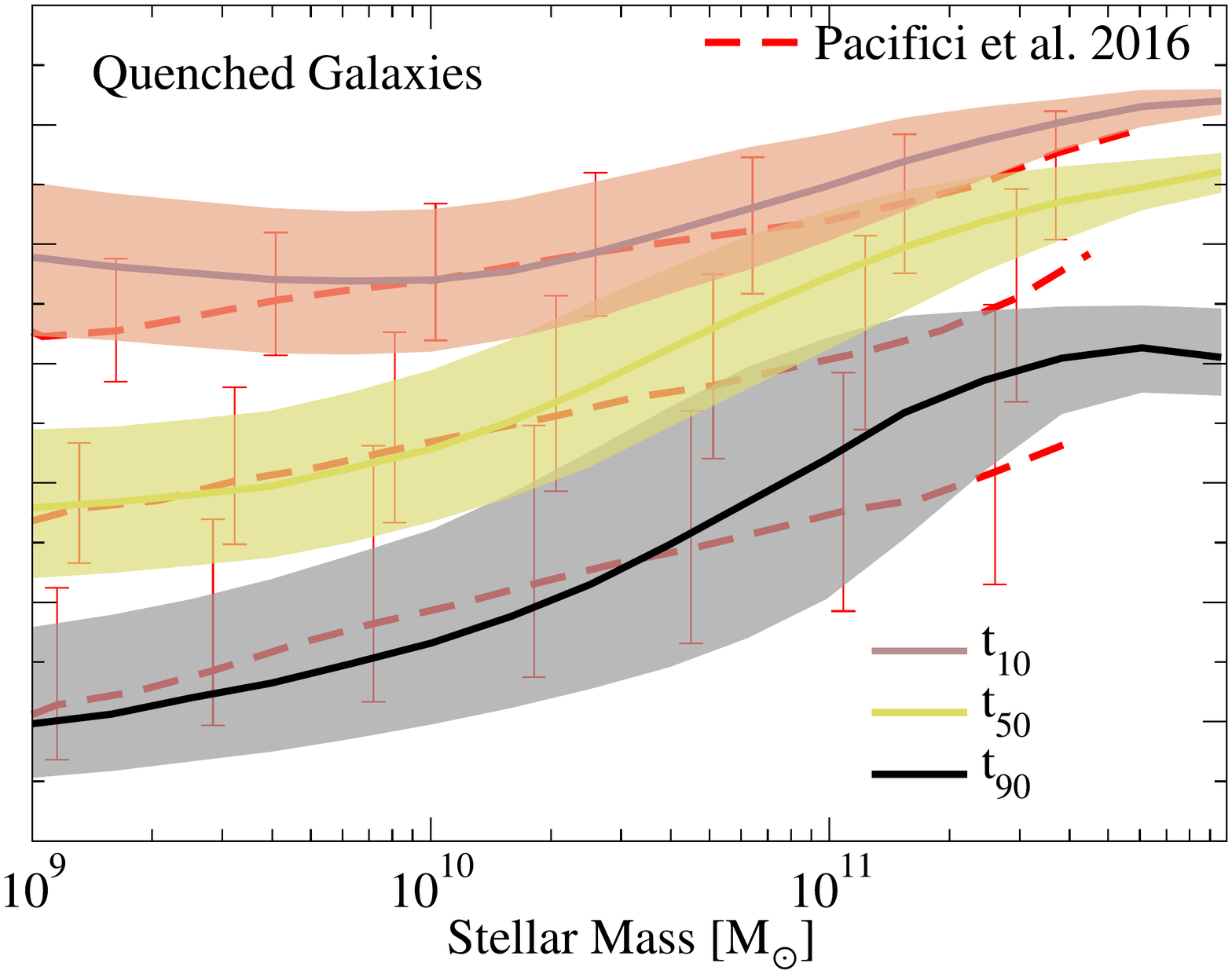}\hspace{-13ex}\plotmodestgrace{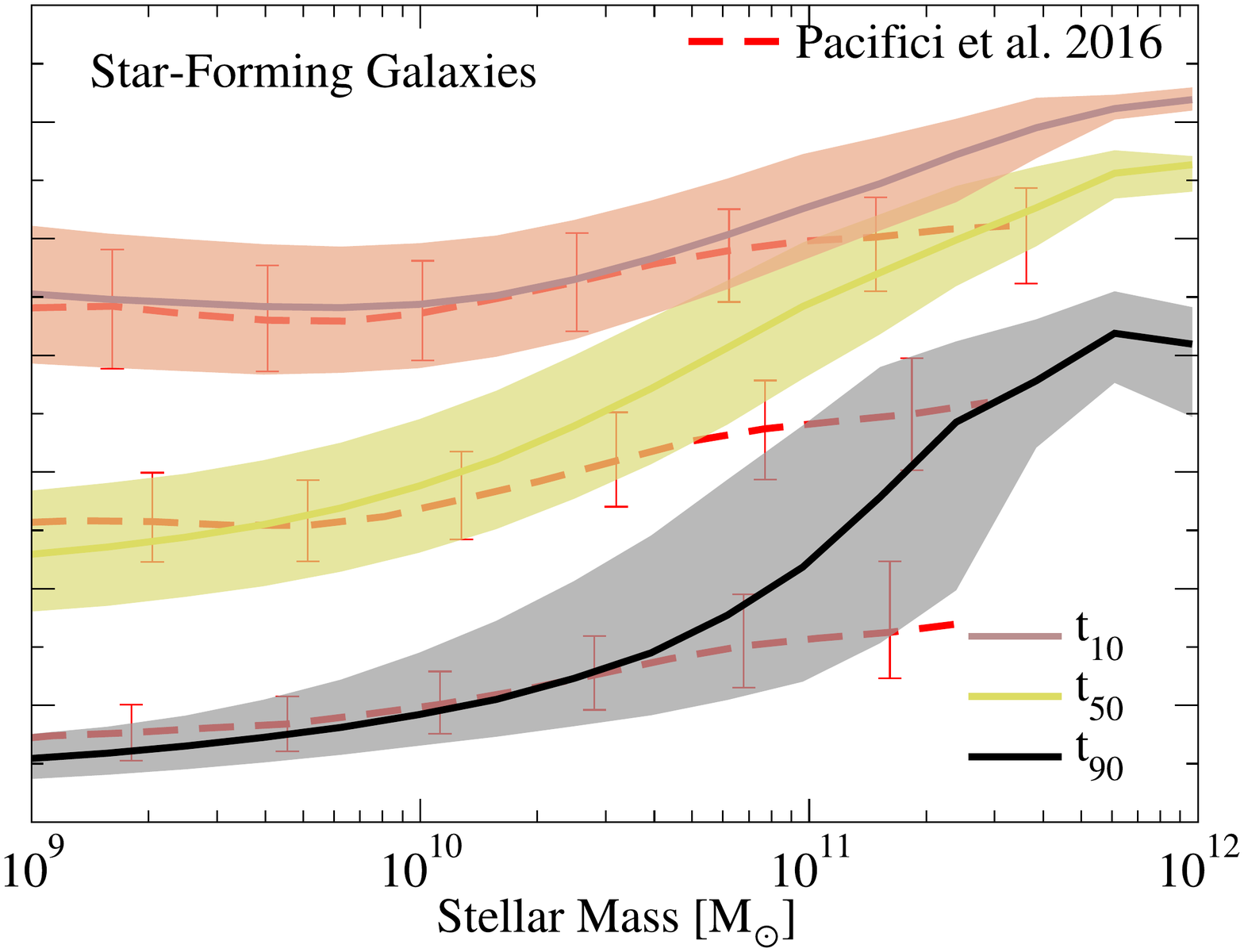}\\[-3ex]
\caption{Lookback times for stellar mass assembly thresholds.  \textbf{Left} panel: median lookback times for our best-\fit{} model at which progenitors of $z=0$ galaxies reached $10\%$ (\textit{brown line}), 50\% (\textit{sandy line}) and 90\% (\textit{black line}) of their $z=0$ stellar mass.  Shaded regions show the 16$^\mathrm{th}-84^\mathrm{th}$ percentile ranges of lookback times across different $z=0$ galaxies.  \textit{Red dashed lines} show the comparison with \protect\cite{Pacifici16}; error bars show the equivalent \onesigdist{}.  \textbf{Middle} panel: same, for quenched galaxies.  \textbf{Right} panel: same, for star-forming galaxies.}
\label{f:sf_formation_times}
\end{figure*}

\begin{figure}
\vspace{-5ex}
\plotgrace{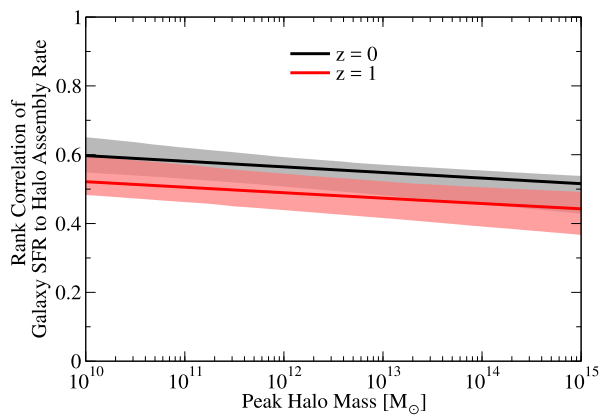}\\[-3ex]
\caption{Constraints on the rank correlation coefficient  ($r_c$; Eq.\ \ref{e:rc}) between halo growth ($\Delta\vmax$; Eq.\ \ref{e:dvmax}) and galaxy SFR, as a function of host halo mass at $z=0$.  \textit{Shaded regions} show the \onesigdist{} across the model posterior space.}
\label{f:rank_correl}
\end{figure}

\begin{figure*}
\vspace{-5ex}
\plotgrace{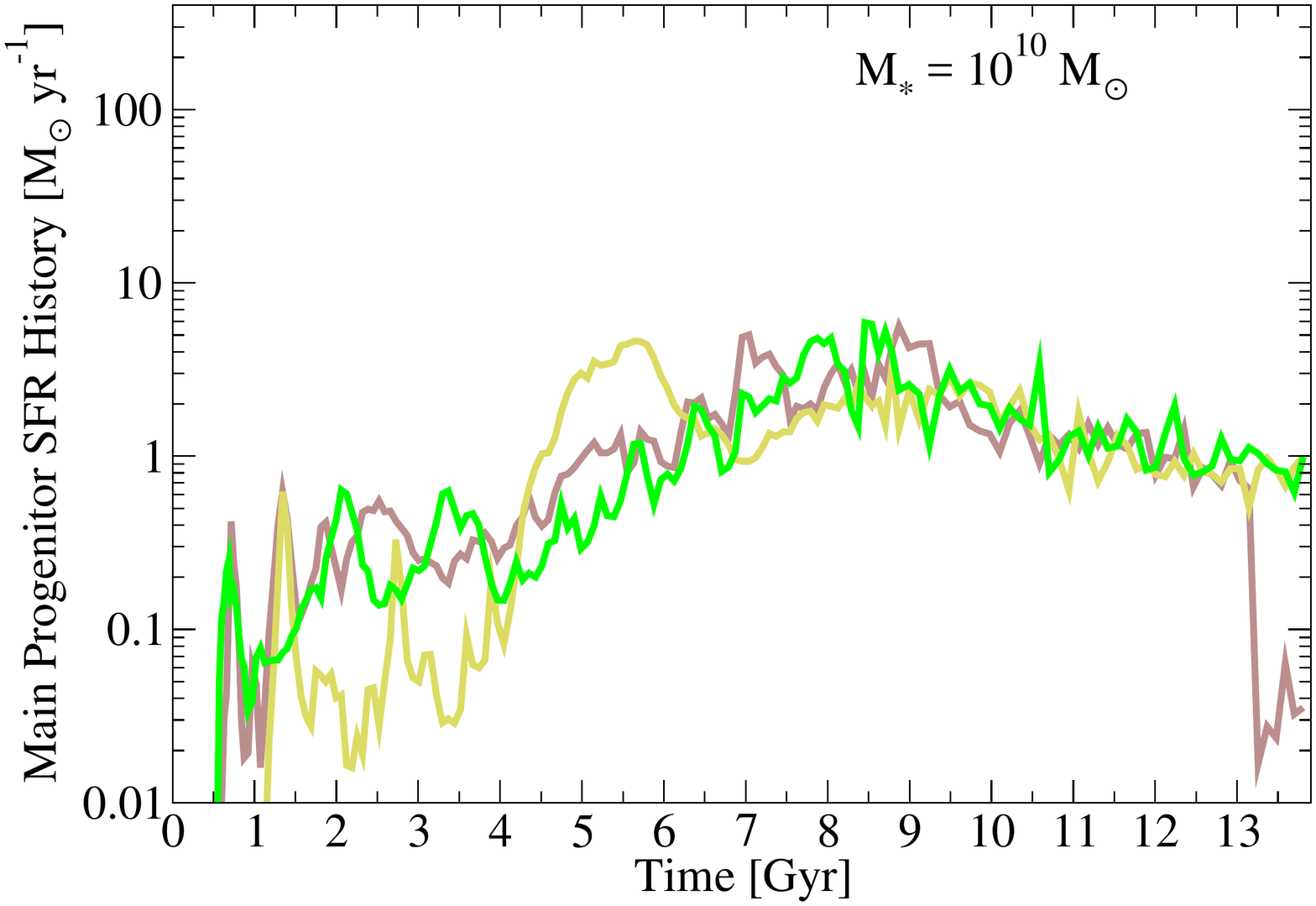}\plotgrace{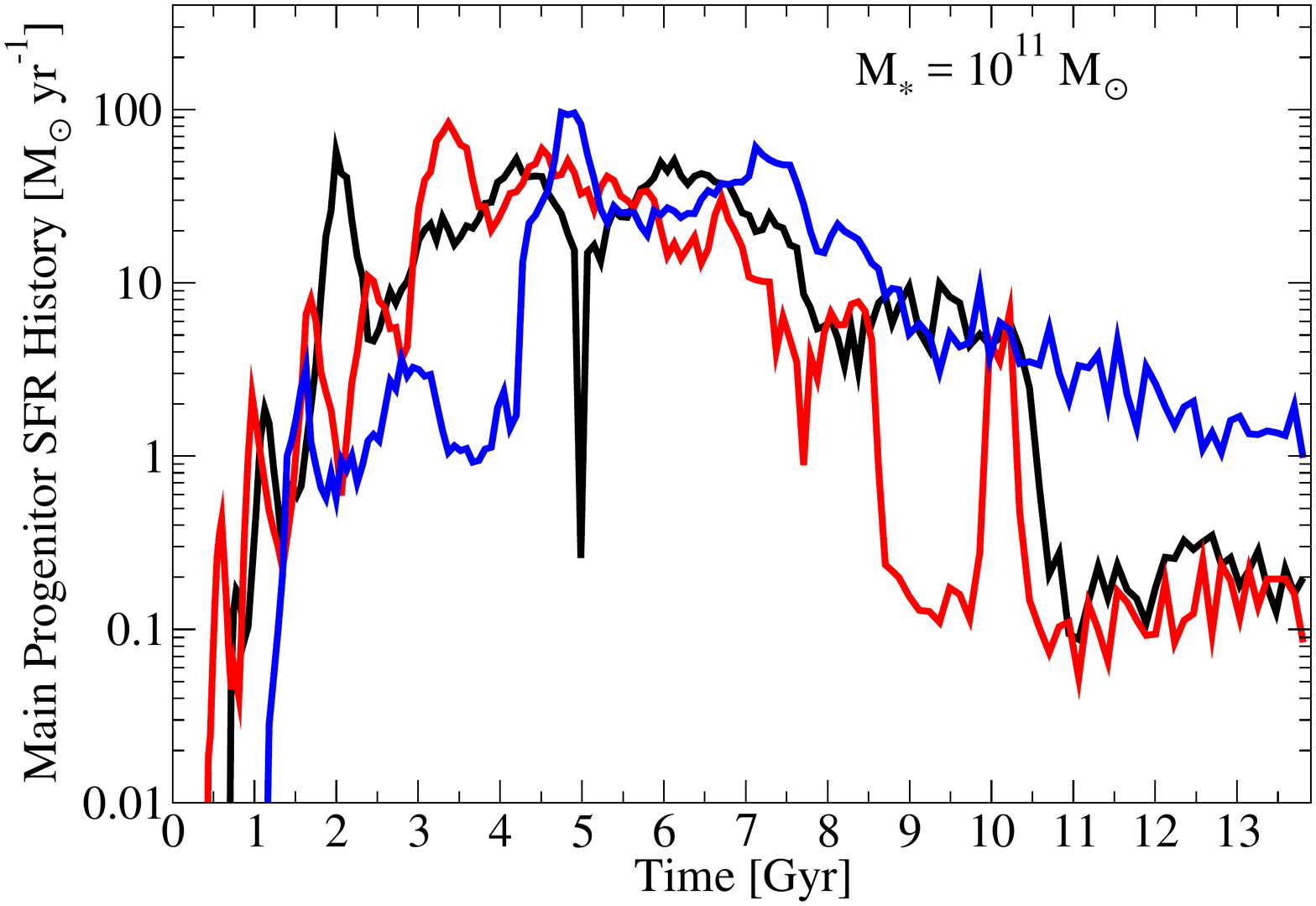}\\[-3ex]
\caption{\textbf{Left} panel: main progenitor star formation histories for several $M_\ast = 10^{10}\Msun$ galaxies at $z=0$ from the best-\fit{} model.  \textbf{Right} panel: progenitor star formation histories for several $M_\ast = 10^{11}\Msun$ galaxies at $z=0$ from the best-\fit{} model.  These show significantly more variation than their lower-mass counterparts, due to periods of quenching and rejuvenation.}
\label{f:random_sfhs}
\vspace{-5ex}
\plotgrace{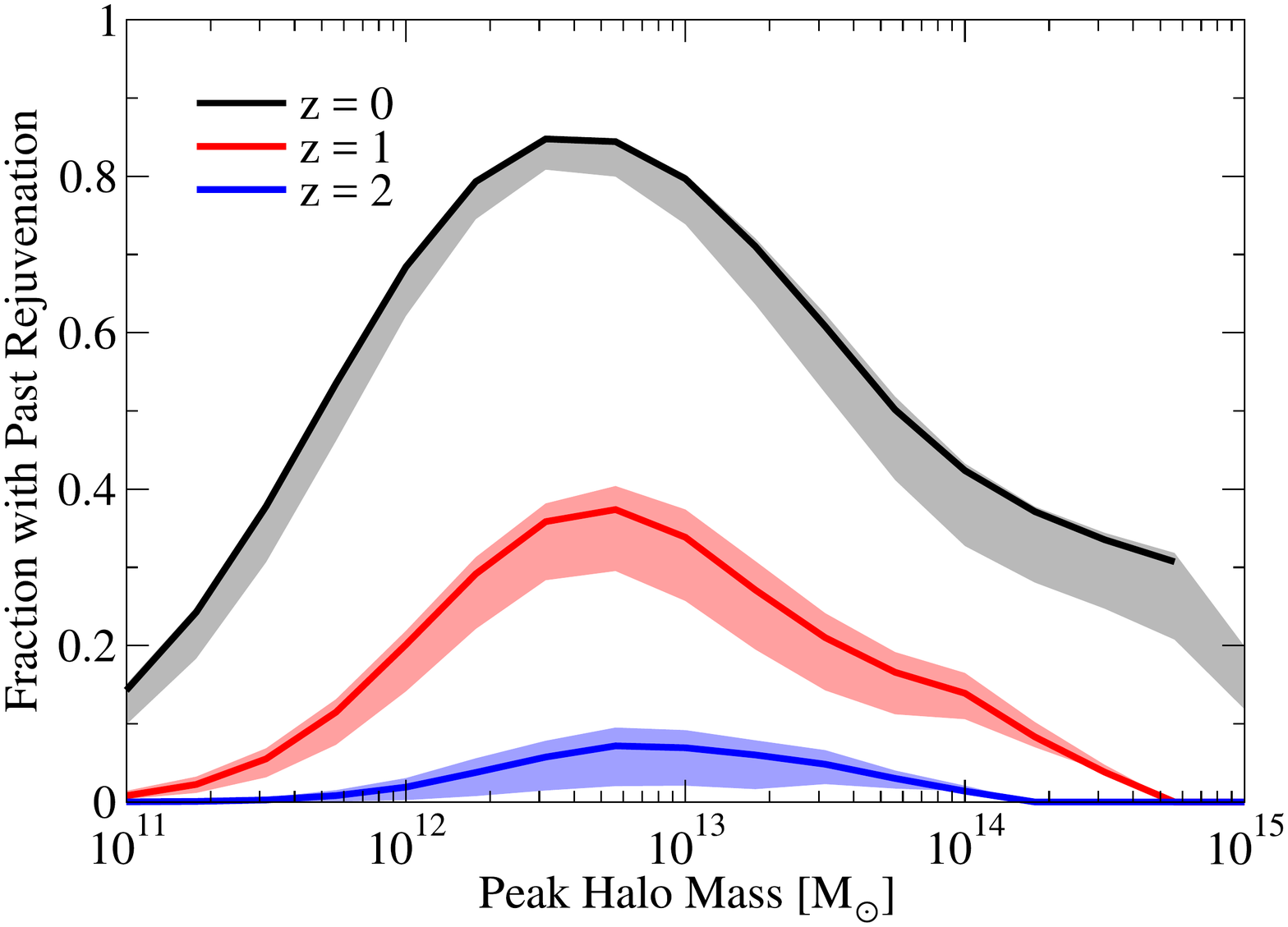}\plotgrace{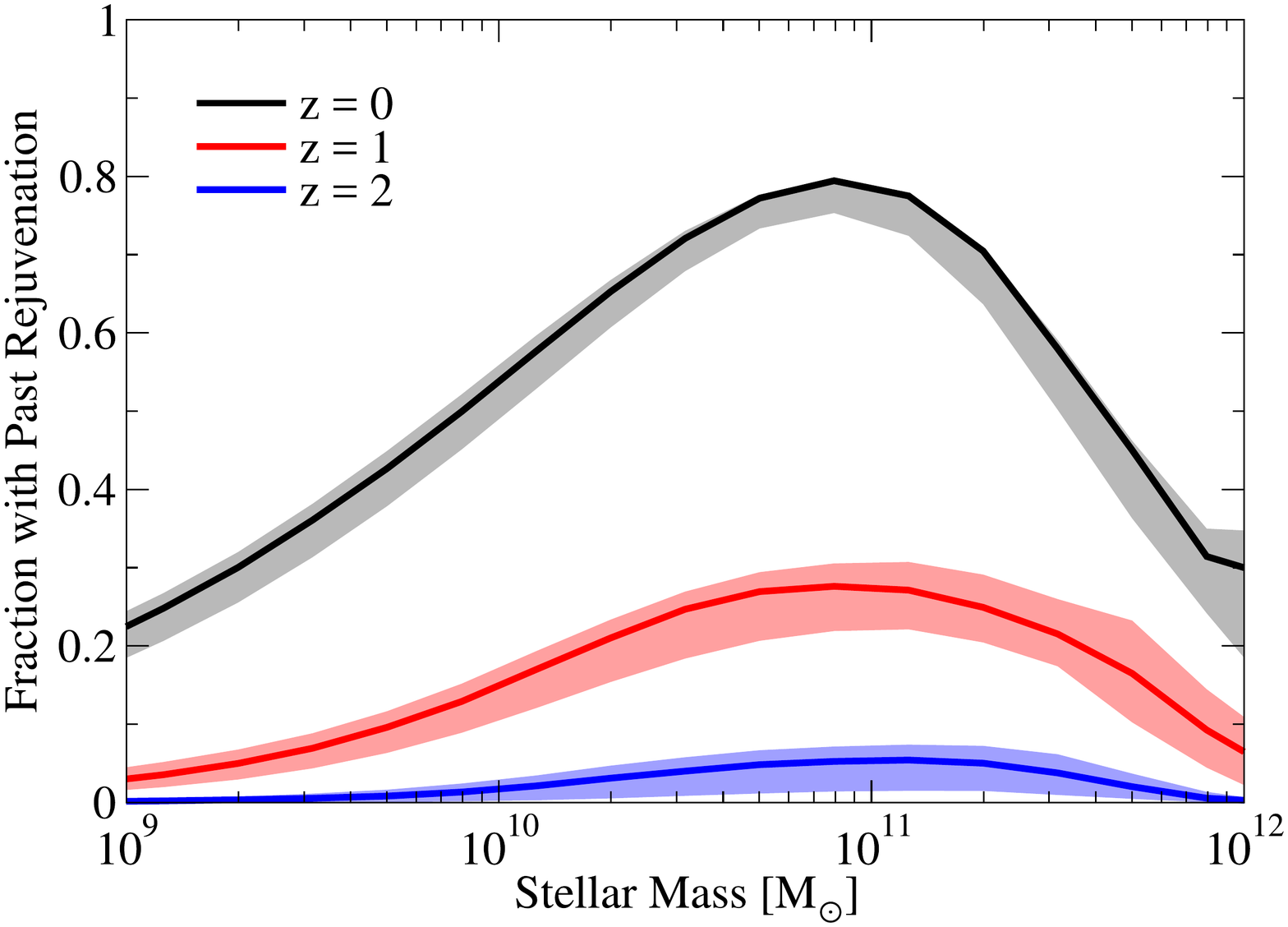}\\[-3ex]
\caption{\textbf{Left} panel: fraction of $z=0$ haloes whose galaxies quenched and then rejuvenated in the past; rejuvenation is defined as at least $300$ Myr of quiescence, followed by at least 300 Myr of star formation. \textbf{Right} panel: same, as a function of $z=0$ galaxy mass.  In both panels, \textit{shaded regions} show the \onesigdist{} across the model posterior space.}
\label{f:rejuv}
\vspace{-5ex}
\plotgrace{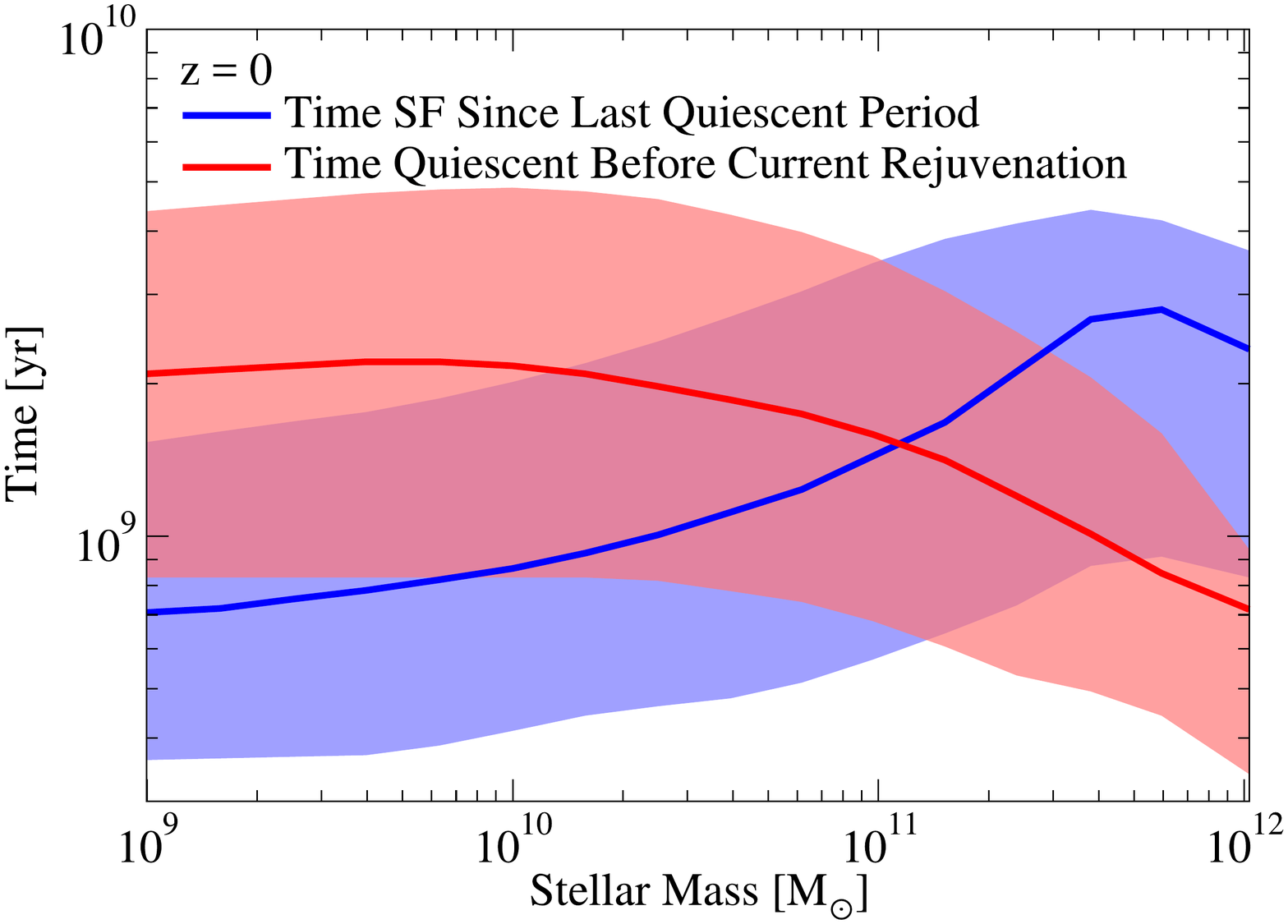}\plotgrace{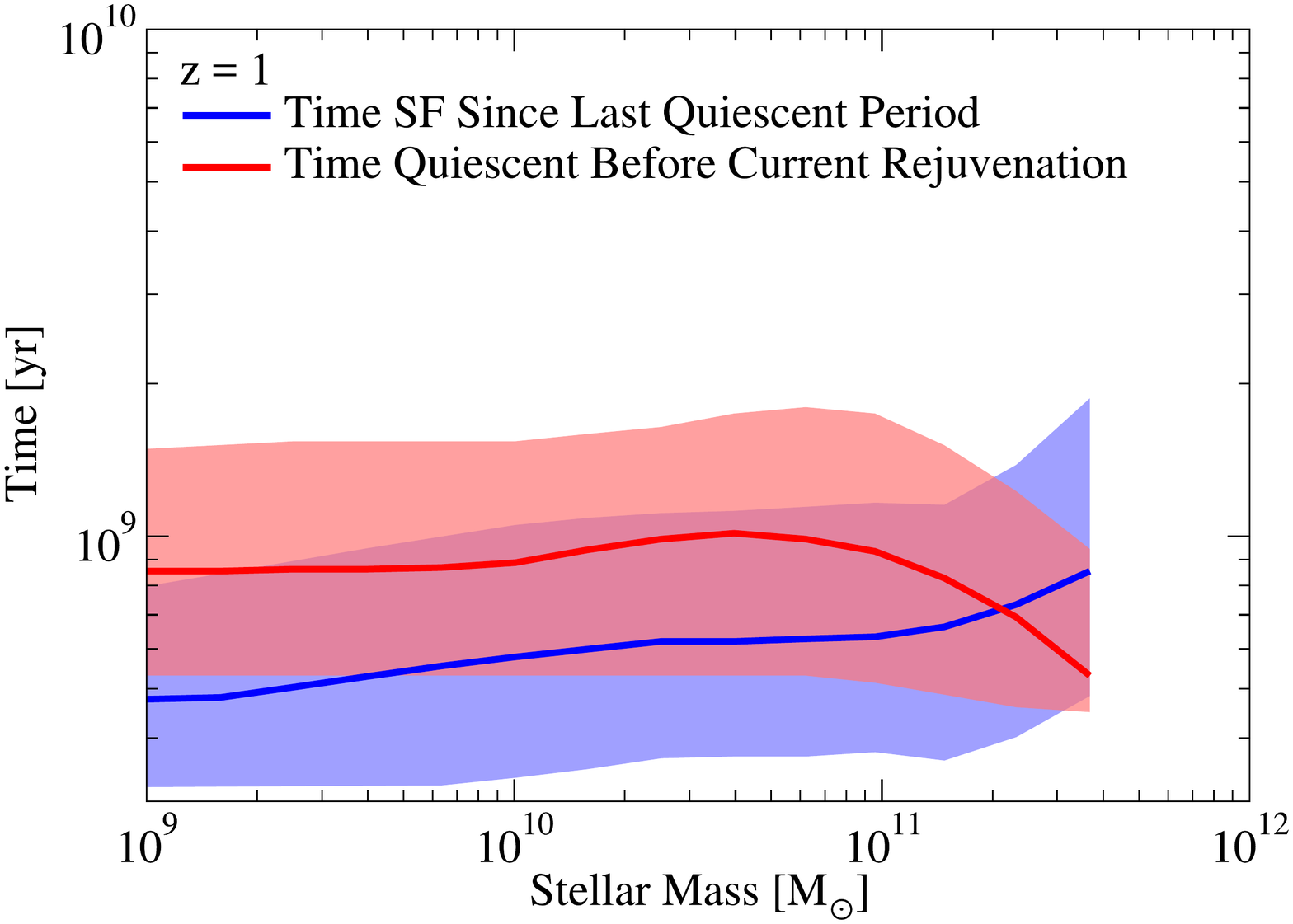}\\[-3ex]
\caption{\textbf{Left} panel: For rejuvenated galaxies at $z=0$ (see text), the median times spent in the rejuvenated star-forming phase (\textit{blue} line) and in the quiescent phase prior to rejuvenation (\textit{red} line).  Shaded regions show the \onesigdist{} for individual galaxies in the best-\fit{} model.  \textbf{Right} panel: same, at $z=1$.}
\label{f:rejuv_times}
\vspace{10ex}
\end{figure*}

\begin{figure*}
\vspace{-9ex}
\plotgrace{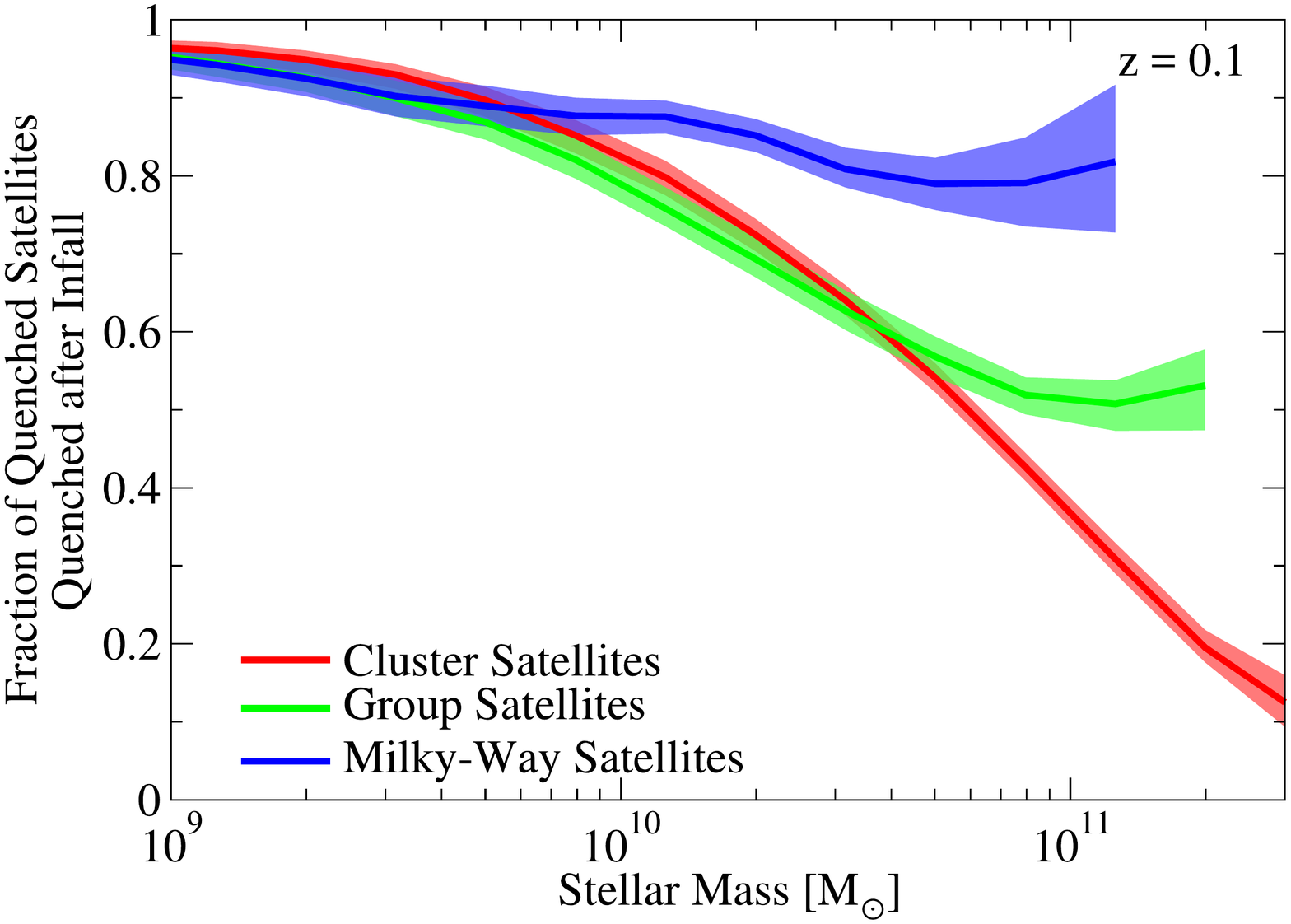}\plotgrace{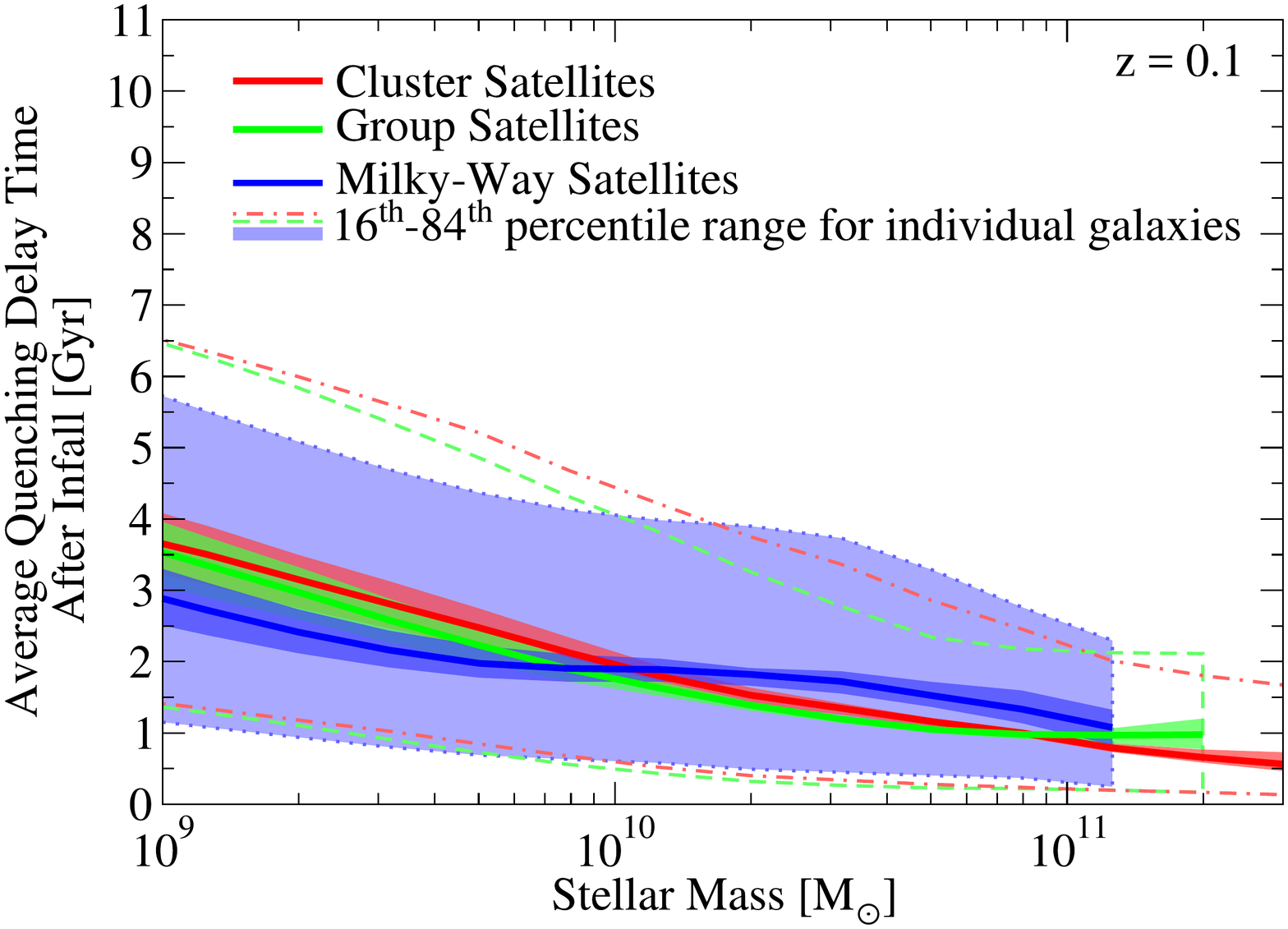}\\[-4.5ex]
\caption{\textbf{Left} panel: the fraction of quenched $z=0$ satellite galaxies that became quenched post-infall.  \textit{Solid lines} show the median relation; \textit{dark shaded regions} show the \onesigdist{} across the model posterior space.  \textbf{Right} panel: the average delay times after first infall for satellite quenching (considering quenched satellites only).  \textit{Solid lines} show the median of the average delay time for the model posterior space; \textit{dark shaded regions} show the \onesigdist{} of the average delay time across the model posterior space.  The \textit{light shaded region} shows the \onesigdist{} of quenching times for individual satellites of Milky-Way mass haloes; \textit{dashed} and \textit{dash-dotted lines} show the same for individual satellites of group and cluster-mass haloes.  \textbf{Notes:} For all panels, Milky-Way host haloes are defined to have $11.75 < \log_{10}(M_h/\Msun) < 12.25$, group-mass host haloes to have $12.75 < \log_{10}(M_h/\Msun) < 13.25$, and cluster-mass host haloes to have $13.75 < \log_{10}(M_h/\Msun) < 14.25$.}
\label{f:sat_q}
\vspace{-4ex}
\plotgrace{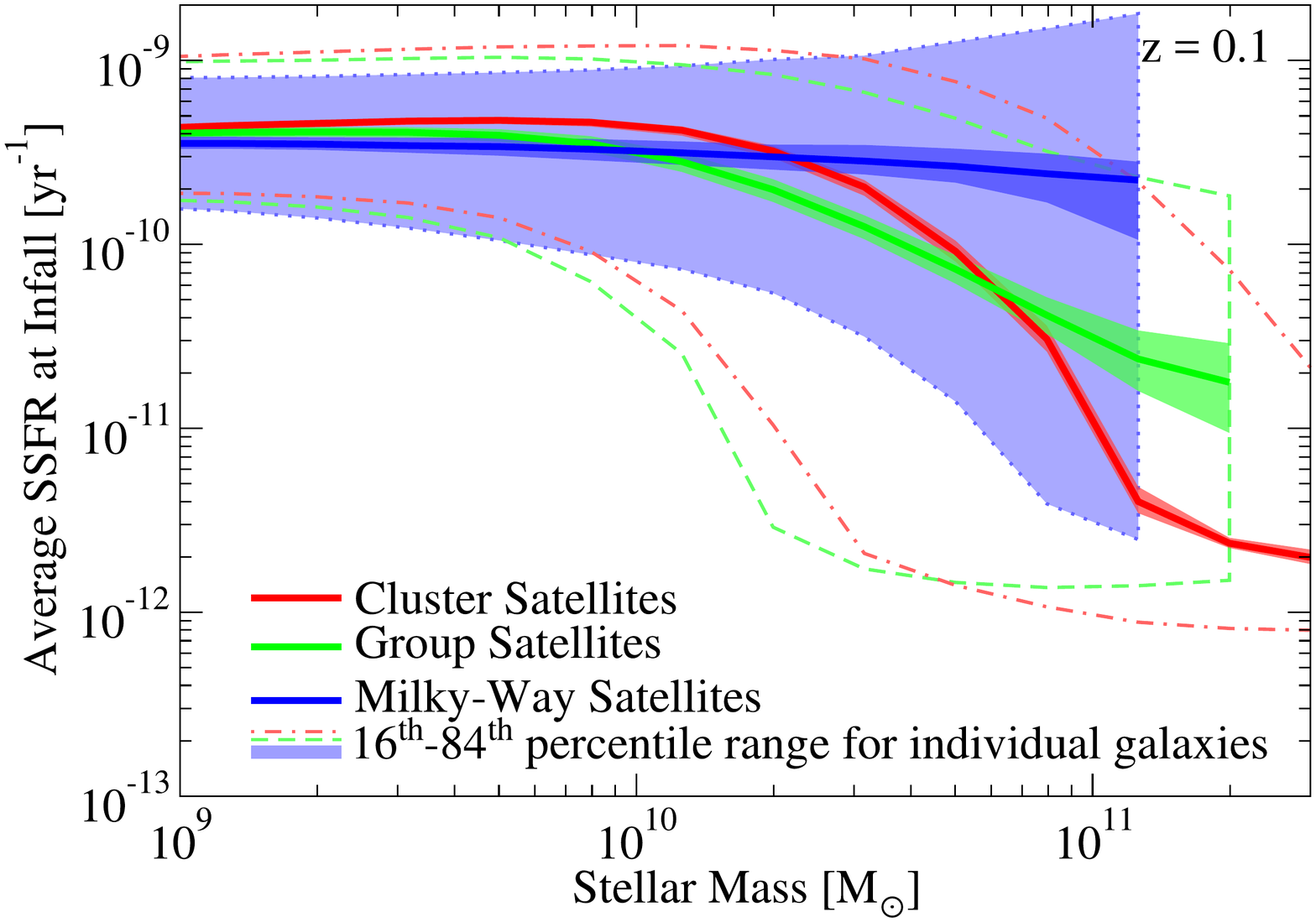}\plotgrace{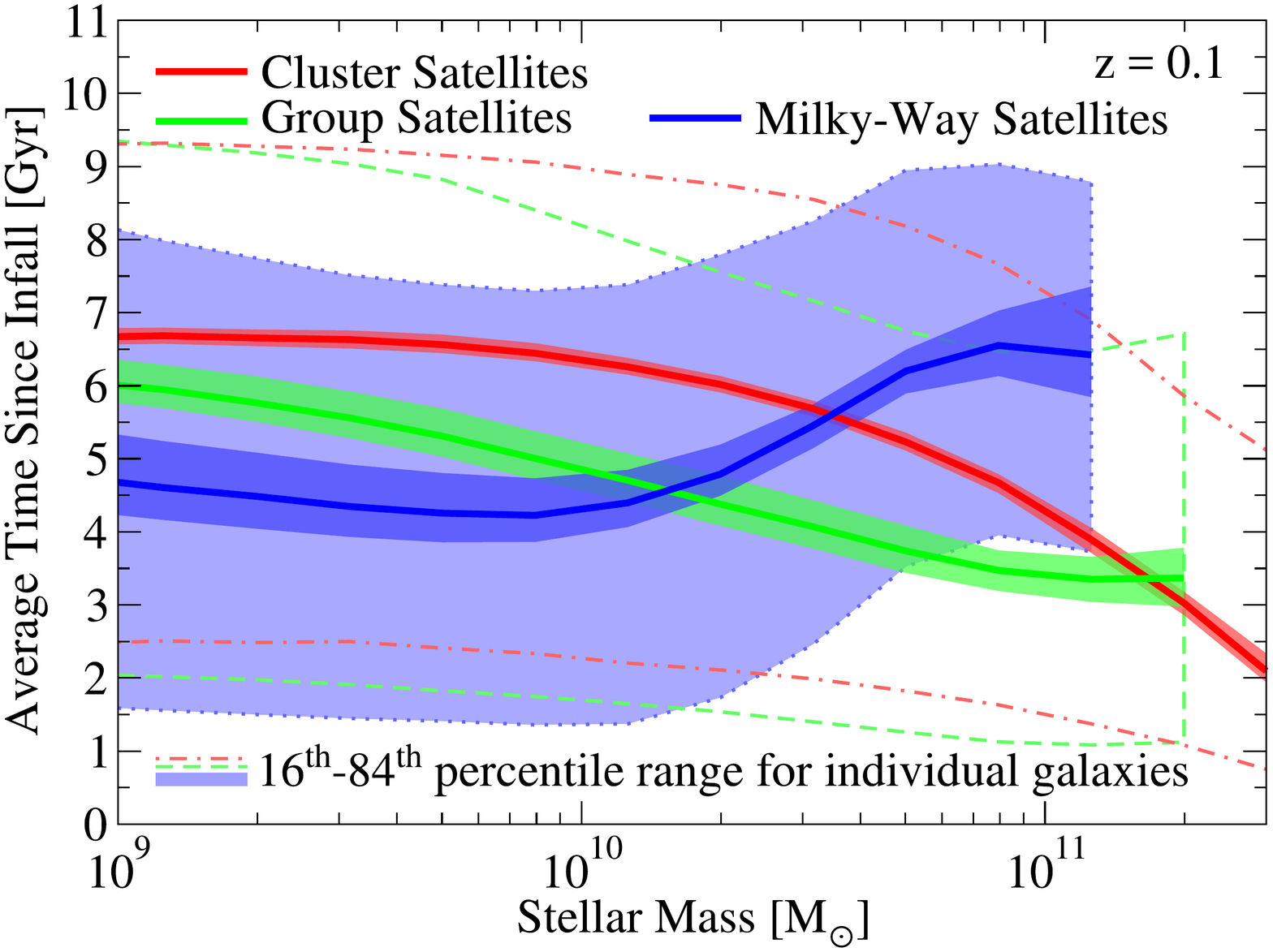}\\[-4.5ex]
\caption{\textbf{Left} panel: average SSFRs at first infall for all $z=0$ satellite galaxies.  \textbf{Right} panel: average time since first infall for $z=0$ satellite galaxies.  Quenched Milky-way satellites have shorter average quenching delay times because there is, on average, less time since first infall.  In both panels,  \textit{solid lines} show the median of the average value in the model posterior space and \textit{dark shaded regions} show the \onesigdist{} of the average value across the model posterior space.  The \textit{light shaded region} shows the \onesigdist{} for individual satellites of Milky-Way mass haloes; \textit{dashed} and \textit{dash-dotted lines} show the same for individual satellites of group and cluster-mass haloes.  As in Fig.\ \ref{f:sat_q}, Milky-Way host haloes have $11.75 < \log_{10}(M_h/\Msun) < 12.25$, group-mass host haloes have $12.75 < \log_{10}(M_h/\Msun) < 13.25$, and cluster-mass host haloes have $13.75 < \log_{10}(M_h/\Msun) < 14.25$.}
\label{f:sat_q_ssfr}
\vspace{-4ex}
\plotgrace{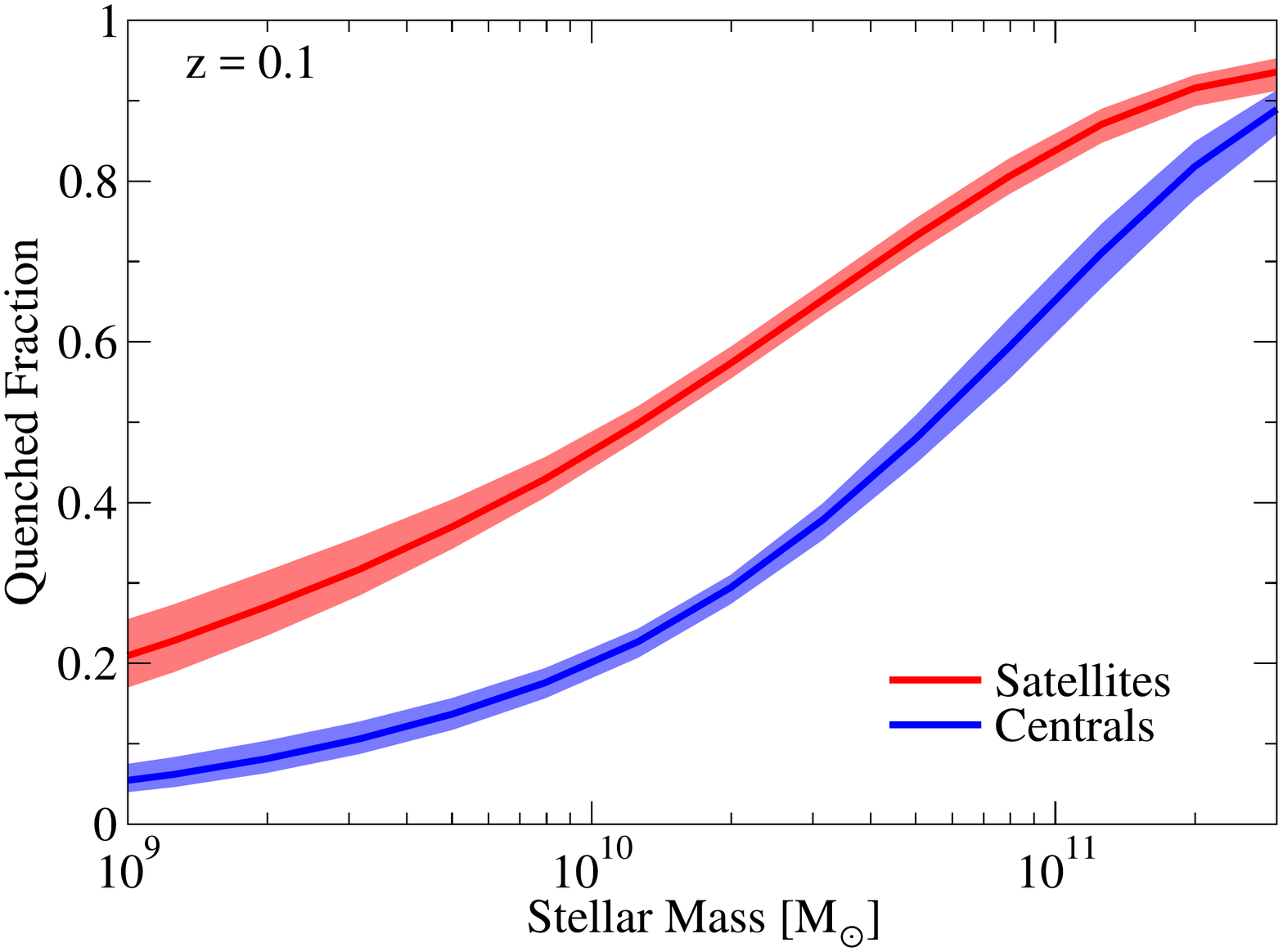}\plotgrace{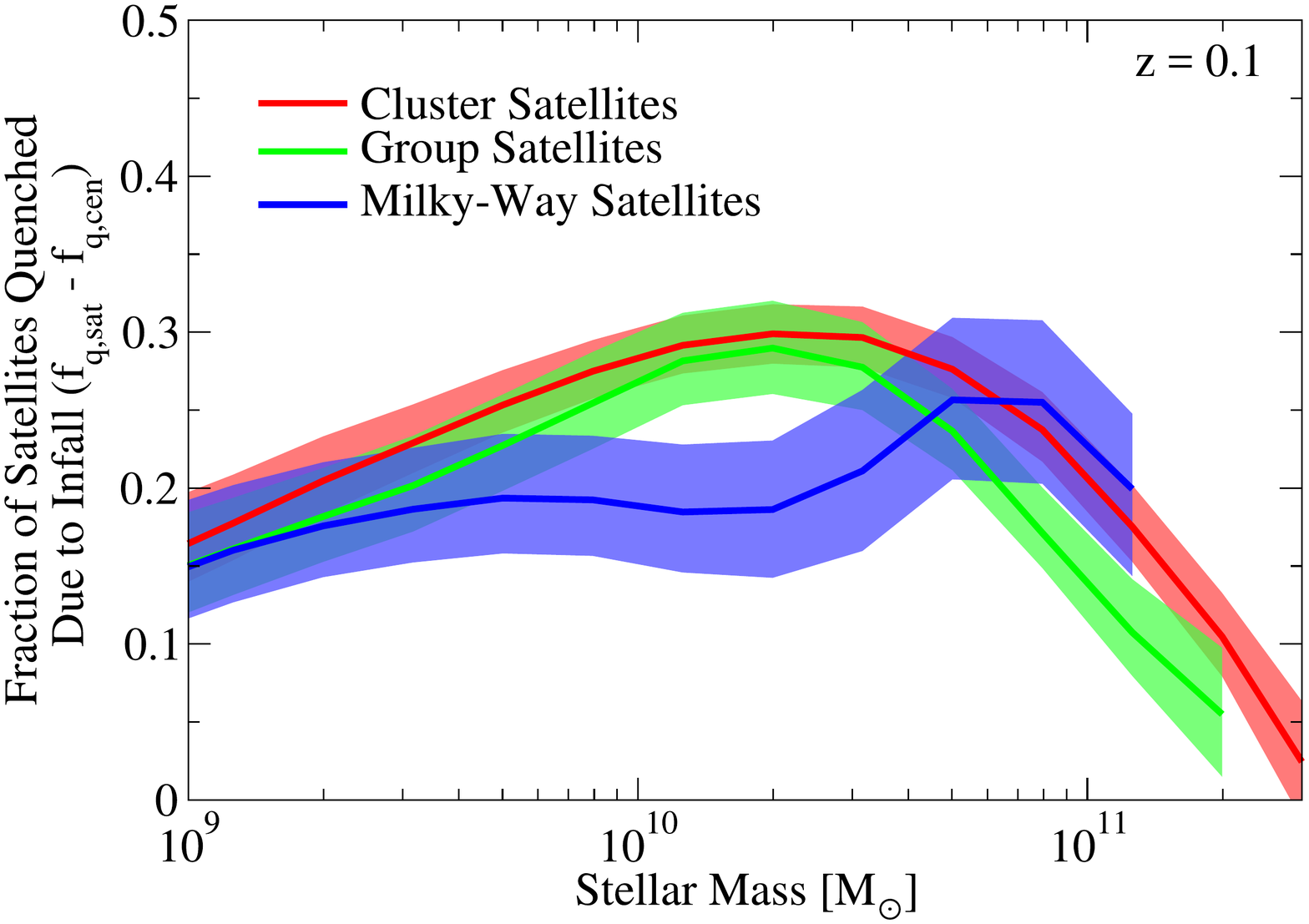}\\[-4.5ex]
\caption{\textbf{Left} panel: Fraction of quenched central and satellite galaxies, as a function of stellar mass.  \textbf{Right} panel: Fraction of satellites quenched due to infall, defined as $(f_\mathrm{q,sat} - f_\mathrm{q,cen})$.  In both panels, \textit{solid lines} show the median relation and \textit{dark shaded regions} show the \onesigdist{} across the model posterior space.  As in Fig.\ \ref{f:sat_q}, Milky-Way host haloes have $11.75 < \log_{10}(M_h/\Msun) < 12.25$, group-mass host haloes have $12.75 < \log_{10}(M_h/\Msun) < 13.25$, and cluster-mass host haloes have $13.75 < \log_{10}(M_h/\Msun) < 14.25$.}
\label{f:sat_cen_q}
\end{figure*}

\subsection{Average Star Formation Histories}

\label{s:avg_sfr}

Average SFHs are shown in Fig.\ \ref{f:sfh_m} as a function of halo mass and redshift.  Quenched galaxies have lower recent SFHs and higher early SFHs, which is expected for any model that correlates quenching with assembly history.  That is, lower present-day SFRs imply an earlier halo formation history, which then gives higher SFRs at early times.  Central galaxies' SFRs are more similar to satellites' SFRs as compared to what might have been naively expected.  As discussed in \S \ref{s:satellites}, star-forming satellites have similar SSFRs as star-forming central galaxies.  Hence, the ratio between their average late-time SFHs is approximately the ratio of the star-forming central fraction to the star-forming satellite fraction.  This ratio is never large: small haloes are mostly star-forming regardless of being centrals or satellites, and large haloes' star-formation rates do not depend as much on assembly history (\S \ref{s:corr}).  The fact that satellite SFHs are higher on average than central SFHs is related to the fact that satellites' peak halo masses do not grow after infall; as a result, they have more stellar mass at a given $\mpeak$ (see \S \ref{s:smhm}).

Average SFHs for galaxies are shown in Fig.\ \ref{f:sfh_sm} as a function of lookback time.  Stellar populations older than $1-2$ Gyr have very similar colours \citep{Conroy09}, so differences beyond that time are very difficult to observe.  Galaxies broadly follow the same trends as haloes, with quenched galaxies and satellite galaxies having earlier formation histories than star-forming galaxies and central galaxies; the most significant differences occur within $\sim 3$ Gyr of $z=0$.  

\subsection{Distribution of Individual Galaxies' Star Formation Histories}
\label{s:stochasticity}

Turning to individual halo histories reveals tremendous diversity, as shown in Fig.\ \ref{f:indiv_sfr}.  The significantly overlapping total star formation histories for $M_h > 10^{12}\Msun$ suggest that the $z=0$ halo mass alone gives limited information on the galaxy's recent star formation history ($z<1$).  The halo mass is instead a better predictor of when the galaxy's star formation rates peaked, as well as their early star formation history---i.e., at times when the progenitors had masses less than $10^{12}\Msun$.  This is partially because it is observationally difficult to constrain SFRs in quenched galaxies, and partially because significant fractions of galaxy growth in $M_h > 10^{12}\Msun$ haloes are from mergers (\S \ref{s:in_situ}), so that contrast between their histories is diminished.  

For SSFR histories (Fig.\ \ref{f:indiv_sfr}, middle-left panel), the $z=0$ halo mass strongly influences the range of redshifts over which quenching takes place.  As noted in \S \ref{s:average_sfrs}, $10^{13}\Msun$ haloes experience quenching over a very extended period of time, leading to more opportunities for rejuvenation.  Quenching in $10^{12}\Msun$ and smaller haloes only began recently ($z<0.5$) for the majority of galaxies.

Halo mass is a much better predictor of intrahalo light (IHL) histories as compared to stellar mass histories (Fig.\ \ref{f:indiv_sfr}, middle-right and top-right panels).  IHL depends only on mergers, which are significantly more scale-free than the process of galaxy formation \citep{Fakhouri10,Behroozi13}.  That said, since central galaxy stellar masses increase with dark matter halo mass, one dex increase in halo mass results in more than one dex increase in IHL; this is especially evident for low-mass haloes where the stellar mass --- halo mass (SMHM) relation's slope is greatest (\S \ref{s:smhm}).

We find broad scatter in the stellar haloes of low-mass haloes, with $\sigma_{IHL/M_\ast} \sim 0.6$ dex for $10^{12}\Msun$ and $\sim 1$ dex for $10^{11}\Msun$ haloes (Fig.\ \ref{f:indiv_sfr}, bottom panel).  These are somewhat higher than predictions in \cite{Gu16} of $\sigma_{IHL/M_\ast} \sim 0.38$ dex (combining 0.2 dex scatter in $M_\ast$ with 0.32 dex scatter in $M_\mathrm{IHL}$).  Our inclusion of orphan galaxies (Appendix \ref{a:orphans}) explains part of this difference; this choice reduces galaxy merger rates by a factor $\sim 2$, thus increasing Poisson scatter.  In addition, our use of the \cite{Bernardi13} corrections to low-redshift SMFs results in more light being associated with the central galaxy instead of the IHL.  This in turn decreases the fraction of mergers that disrupt into the IHL instead of the central galaxy, explaining the rest of the increase in scatter.  For massive haloes, the IHL becomes $>10\%$ of $M_\ast$ at $z\sim 1$.  However, given that photometric surveys have only recently started capturing most of $M_\ast$ (Appendix \ref{a:smf}) at these redshifts, and given the difficulty of removing satellites (including unresolved sources) from galaxy light profiles, this is subject to the methodology employed.

Despite the diversity of stellar mass histories, galaxy progenitors adhere to well-defined SMHM relations (Fig.\ \ref{f:indiv_sfr}, left panel).  The broadening of the scatter at early times is mostly due to halo progenitors no longer being resolved in the simulation.  As in \cite{BWC13}, haloes reach peak SMHM ratios when their halo masses reach $10^{12}\Msun$.  Haloes that have not reached this mass at $z=0$ show increasing SMHM ratio histories, whereas those that have passed $10^{12}\Msun$ at $z=0$ show decreasing histories.  The tightness of the scatter in SMHM histories depends on how well halo assembly correlates with galaxy assembly (\S \ref{s:corr}), which is in turn constrained via star-forming vs.\ quenched galaxies' correlation functions.  Hence, future measurements of SSFR-split correlation functions at $z>1$ will be an important test of this model (\citealt{BehrooziDecadal}; see also \S \ref{s:corr_pred}).

Finally, we show derived constraints on fractional assembly times for $z=0$ galaxies in Fig.\ \ref{f:sf_formation_times}.  We find the same general trends as \cite{Pacifici16}---e.g., that more massive galaxies form more of their stars at early times over a shorter time period, and that star-forming galaxies have more recent assembly histories.  That said, more massive galaxies in the model do have earlier formation times.  This is especially apparent for massive star-forming galaxies, in which recent star formation can make it very difficult to distinguish between a very old underlying population of stars and an only moderately old population.  As a result, SED-fitting techniques will sample the prior space evenly, resulting in lower average stellar ages.

\subsection{Correlations between Galaxy and Halo Assembly and the Permanence of Quenching}

\label{s:corr}

The rank correlation between galaxy SFR and halo assembly rate ($\equiv\Delta\vmax$, Eq.\ \ref{e:dvmax}) is significant and unequivocally detected (Fig.\ \ref{f:rank_correl}).  There may also be a weak trend with halo mass.  Small haloes' SFRs are more correlated with their assembly history ($r_c \sim 0.6$), whereas large haloes' SFRs may be more independent ($r_c \sim 0.5$).  Observationally, the clearest effect of a strong halo---galaxy assembly correlation is that satellites are quenched much more often than centrals; this also causes large separations in quenched vs.\ star-forming galaxies' correlation functions.  Almost all quenched dwarf galaxies ($M_\ast < 10^{9}\Msun$) are satellites \citep{Geha12}, implying a strong assembly correlation.  Yet, the relative difference in quenched fractions for centrals and satellites becomes less with increasing mass \citep{Wetzel11}, as does the relative difference in clustering strength between quenched and star-forming galaxies (Fig.\ \ref{f:cf_comp}).  Massive haloes hence plausibly have weaker correlations between galaxy SFRs and halo assembly.  At the same time, galaxies in massive haloes grow mainly via mergers, so that this lower correlation does not cause increased scatter in the SMHM relation (Fig.\ \ref{f:smhm_scatter}).

In models where galaxy assembly correlates with halo assembly, galaxy rejuvenation (i.e., the resumption of star formation following a period of quiescence) is a generic feature (e.g., Fig.\ \ref{f:random_sfhs}, right panel).   Almost by definition, proxies of halo assembly change significantly over a dynamical time (e.g., from mergers, accretion, or infall into another larger halo); these changes will in turn affect galaxy SFRs.  If a galaxy population's quenched fraction changes slowly compared to halo dynamical times, changes in halo assembly rates will have the most opportunity to switch galaxies from being star-forming to quenched and vice versa.  This is especially the case for galaxies in $10^{12}-10^{13}\Msun$ haloes, which quench at a rate of $<10\%$ per dynamical time (Fig.\ \ref{f:sfh_m}, right panel; see also Fig.\ \ref{f:indiv_sfr}, middle-left panel).  More massive haloes become fully quenched too rapidly, and less-massive haloes never have large enough quenched fractions for rejuvenation to be as common.

We find exactly this behaviour arising in the models (Fig.\ \ref{f:rejuv}).  Here, we define rejuvenation as at least $300$ Myr of quiescence, followed by at least 300 Myr of star formation; this prevents brief spikes of quiescence (as in the black curve at $t=5$Gyr in Fig.\ \ref{f:random_sfhs}, right panel) or star formation from counting as a rejuvenation event.  The majority of $10^{12}-10^{13}\Msun$ haloes at $z=0$ experienced at least one rejuvenation event in their past.  This fraction falls significantly for both higher and lower halo masses, as well as at $z\ge 1$ when galaxy quenched fractions were significantly lower.  Unfortunately, this behaviour is difficult to observe in integrated colours or spectra.  Rejuvenated galaxies at $z=0$ typically spent $\sim 1$ Gyr forming stars since their last quiescent period (Fig.\ \ref{f:rejuv_times}, left panel), which typically lasted $\sim 2$ Gyr.  The brightness of young stars and the similarity in colours of $2$ Gyr vs. $4$ Gyr stellar populations thus make this very difficult to detect.  Timescales at $z=1$ are somewhat more amenable to observations (Fig.\ \ref{f:rejuv_times}, right panel), perhaps with LEGA-C \citep{vdWel16}, although a much smaller fraction of galaxies at that cosmic time had been rejuvenated (Fig.\ \ref{f:rejuv}).

\begin{figure*}
\vspace{-5ex}
\plotgrace{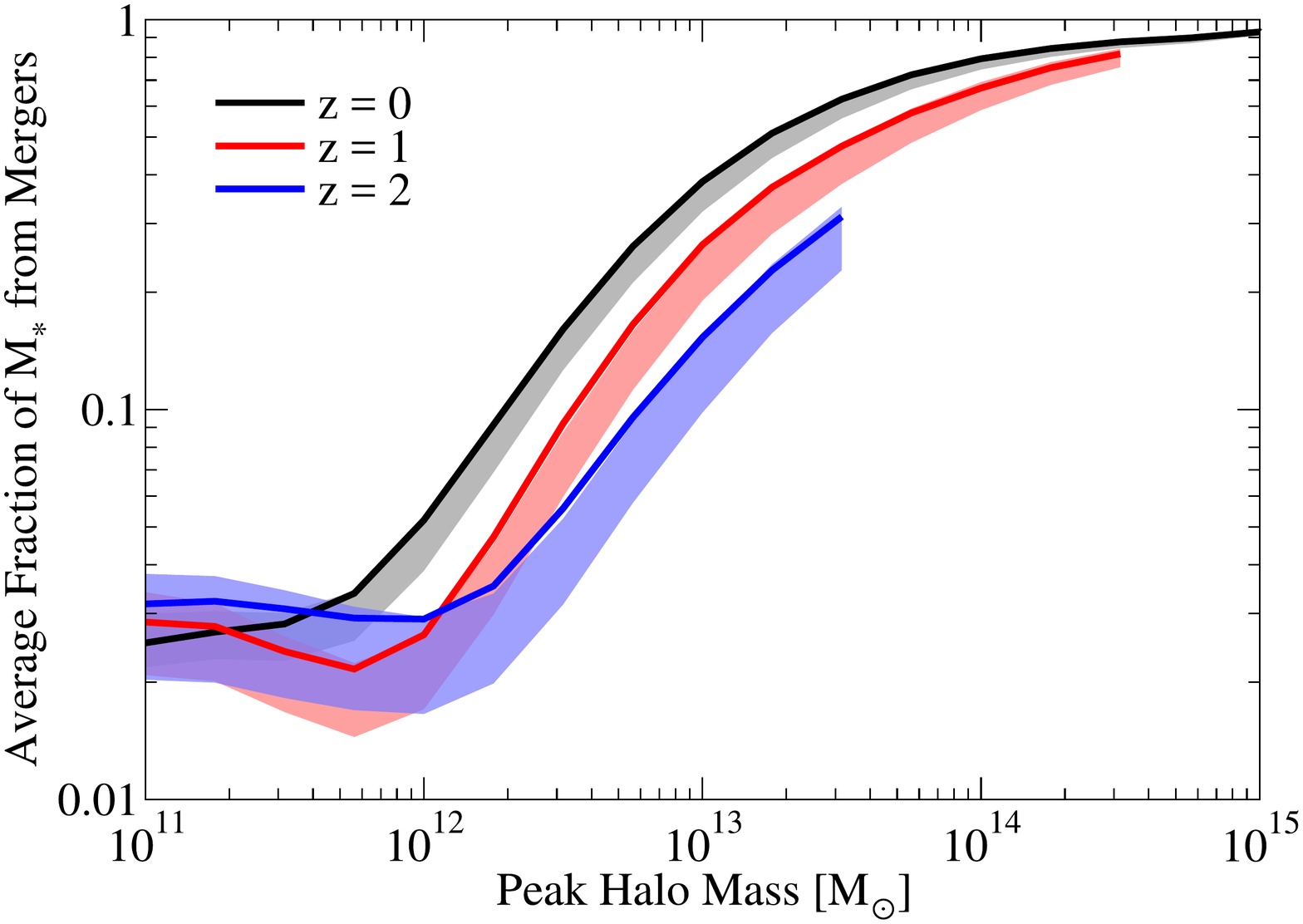}\plotgrace{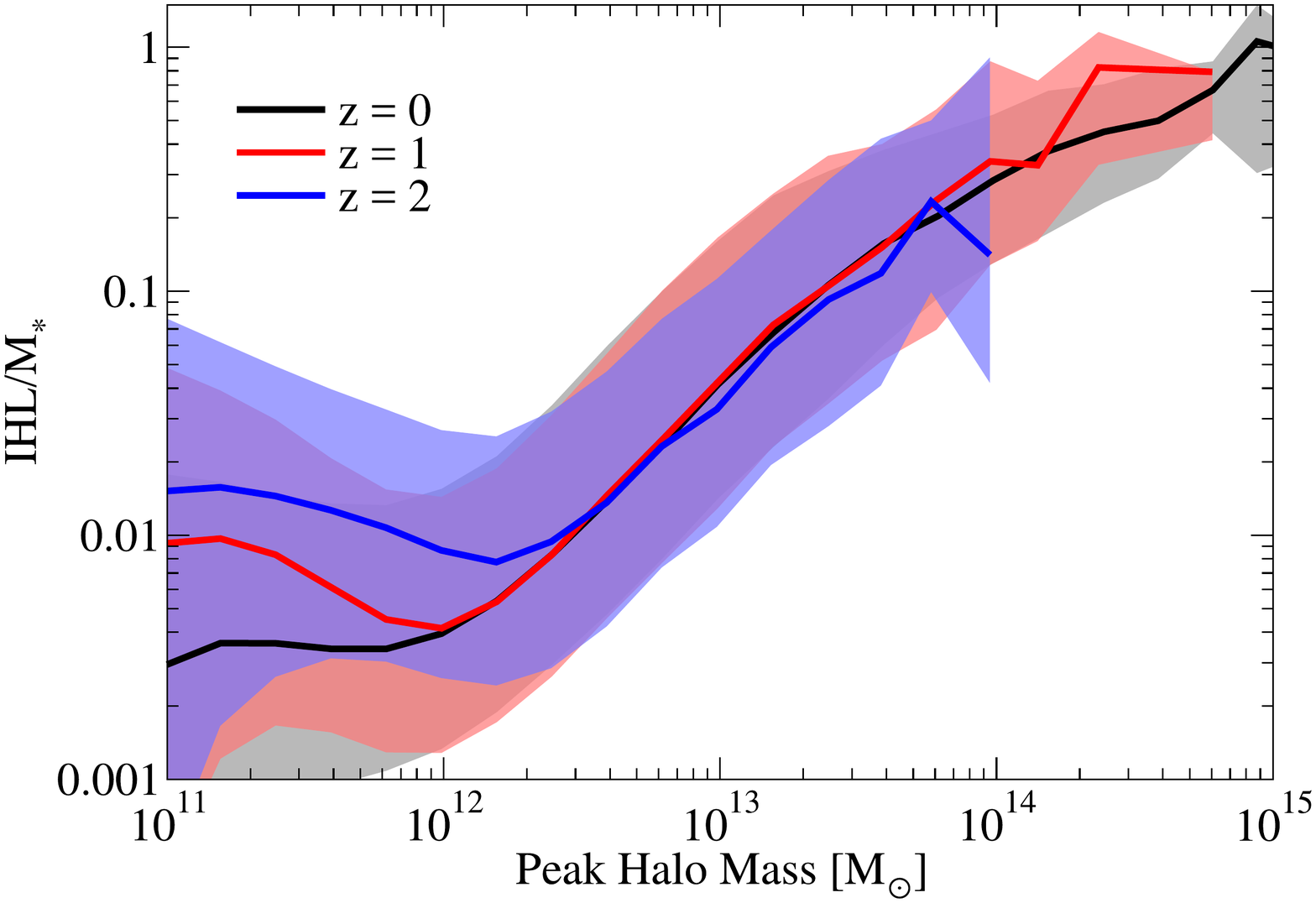}\\[-5ex]
\caption{\textbf{Left} panel: average fraction of galaxy mass that was formed \textit{ex-situ}---i.e., that came in via mergers.  \textit{Shaded regions} show the 68\% confidence interval from the model posterior distribution.  \textbf{Right} panel: median ratio of stellar mass in the intrahalo light (IHL) to $M_\ast$.  \textit{Shaded regions} show the \onesigdist{} for individual galaxies in the best-\fit{} model.  \textbf{Notes:} the \textit{Bolshoi-Planck} simulation resolves haloes down to $10^{10}\Msun$, with the result that mergers are well-resolved for haloes down to $10^{11}\Msun$.  Including mergers from haloes with $M_h < 10^{10}\Msun$ would contribute negligibly to \textit{ex-situ} and IHL fractions due to the steepness of the SMHM relation (Fig.\ \ref{f:smhm}).}
\label{f:merger_stats}
\end{figure*}

\subsection{The Fate of Satellite Galaxies Post-Infall}

\label{s:satellites}

Except for massive galaxies, most quenched satellites at $z=0$ became quenched after infall into a larger halo (Fig.\ \ref{f:sat_q}, left panel).  Past investigations of satellite quenching \citep[e.g.,][]{Wetzel13b,Wetzel15b,Wheeler14,Oman16} assumed that all satellites quench after the same delay time following infall.  We find the same basic trends as these previous works (Fig.\ \ref{f:sat_q}, right panel); low-mass galaxies quench on average much longer after infall than high-mass galaxies, and there is little dependence on delay timescales with host halo mass for $M_\mathrm{host} > 10^{13}\Msun$.

A key assumption of the uniform time delay models is that satellites quench in order of infall time.  This is not true for the model and presumably the real Universe as well, as satellites arrive at their host haloes with a wide variety of SSFRs (Fig.\ \ref{f:sat_q_ssfr}, left panel).  Some are on the verge of quenching, and so quench rapidly after infall, whereas some remain on the star-forming main sequence until $z=0$.  In addition, satellites have a wide variety of post-infall trajectories: some satellites on very radial orbits are stripped and quenched very quickly, whereas others remain on more circular orbits and experience much less disruption.  This results in the broad distribution of quenching time delays we find (Fig.\ \ref{f:sat_q}, right panel), and also introduces a correlation between the average quenching time delay and the average infall time.  For example, as satellites in Milky Way-like hosts ($M_\mathrm{host} \sim 10^{12}\Msun$) had later average infall times (Fig.\ \ref{f:sat_q_ssfr}, right panel), a smaller fraction of the satellites that will eventually quench had time to do so by $z=0$.  As a result, average delay times for quenched galaxies in these haloes are systematically lower than those inferred by uniform delay time models.

Delay time models often implicitly assume that post-infall quenching is entirely due to interactions with the host halo.  Being a satellite certainly results in a higher probability of being quenched (Fig.\ \ref{f:sat_cen_q}, left panel).  However, central galaxies are also quenching at the same time.  Just as a ``quenching delay time'' is meaningless for a central galaxy, it is meaningless for the large fraction of satellite galaxies that would have quenched even if they had been in the field.  Indeed, for galaxies with $M_\ast > 10^{10.5}\Msun$, the majority of quenched $z=0$ satellites would have quenched without any host interactions (Fig.\ \ref{f:sat_cen_q}, left panel).  Considering the excess quenched fraction of satellites in absolute terms ($f_\mathrm{q,sat}(M_\ast) - f_\mathrm{q,cen}(M_\ast)$), we find that this never exceeds 30\% (Fig.\ \ref{f:sat_cen_q}, right panel), similar to past results \citep{Wetzel11,Wang18}. Regardless of how one then defines ``quenching due to infall,'' this result suggests that it does not happen to most satellites.  As discussed in \S \ref{s:sat_quenching}, cluster images typically show only the most visually interesting inner regions (where most satellites are quenched) instead of the outskirts (where many satellites are still star-forming), leading to the common misperception that most satellites are quenched.

\begin{figure*}
\vspace{-5ex}
\plotgrace{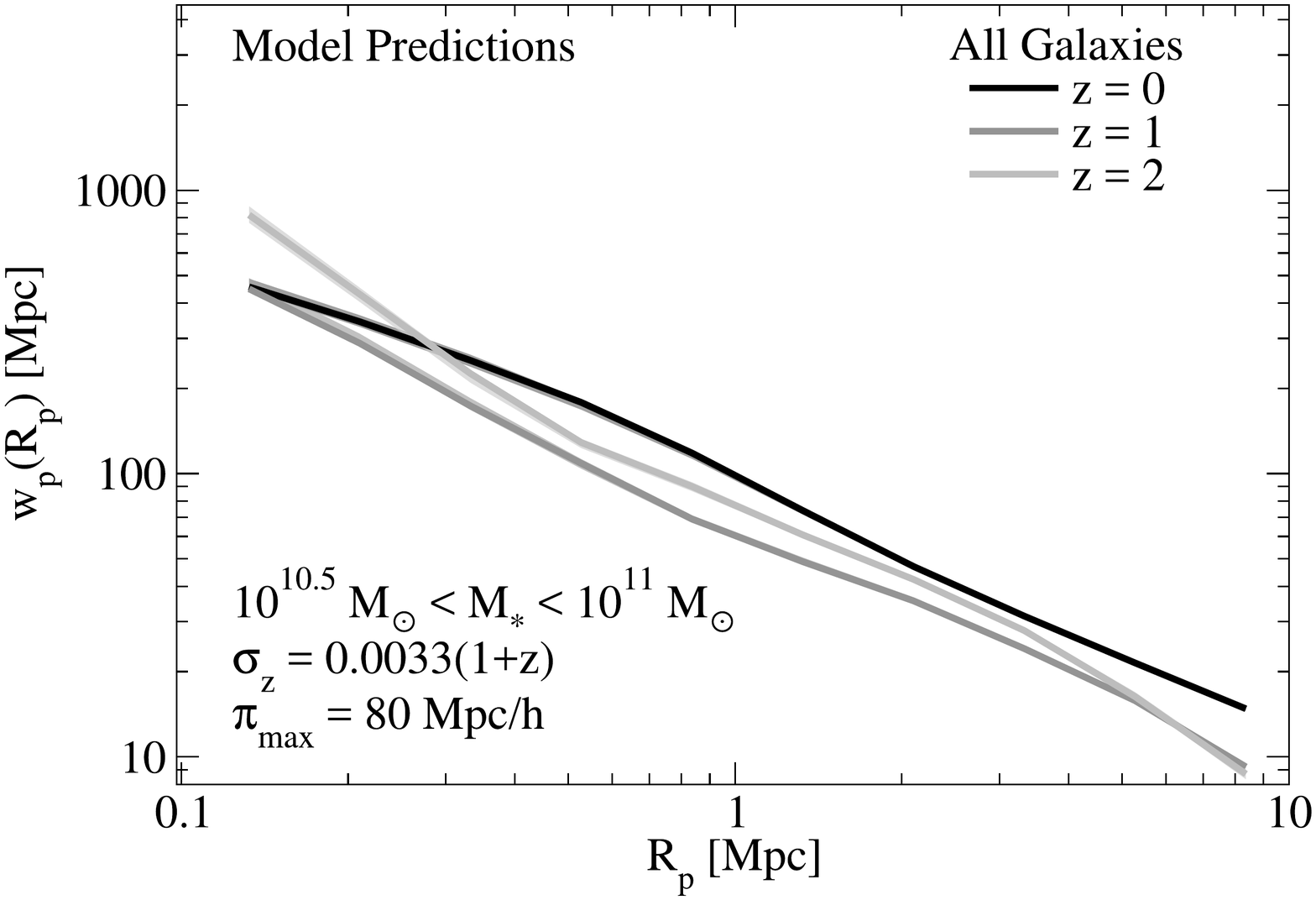}\plotgrace{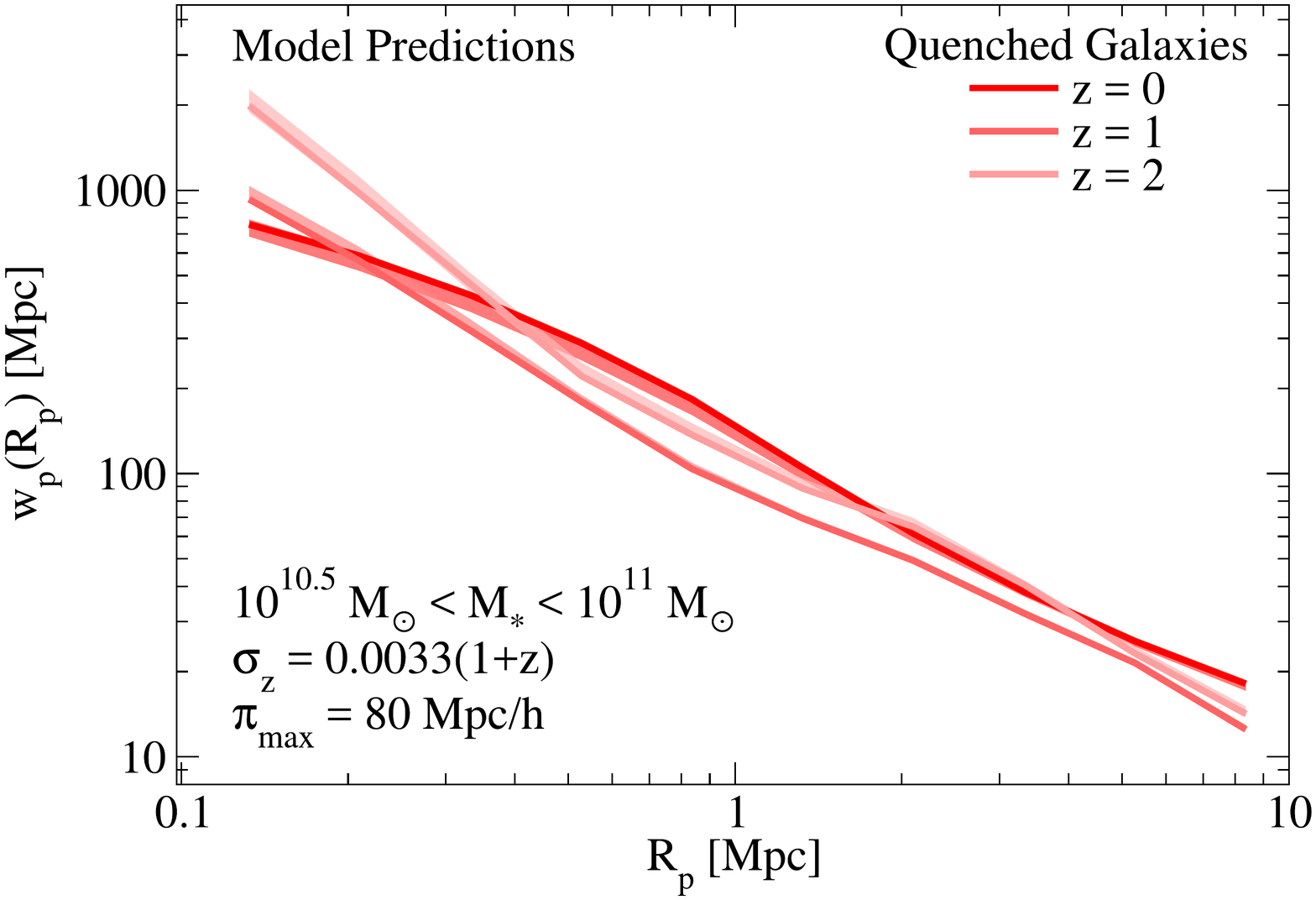}\\[-5ex]
\plotgrace{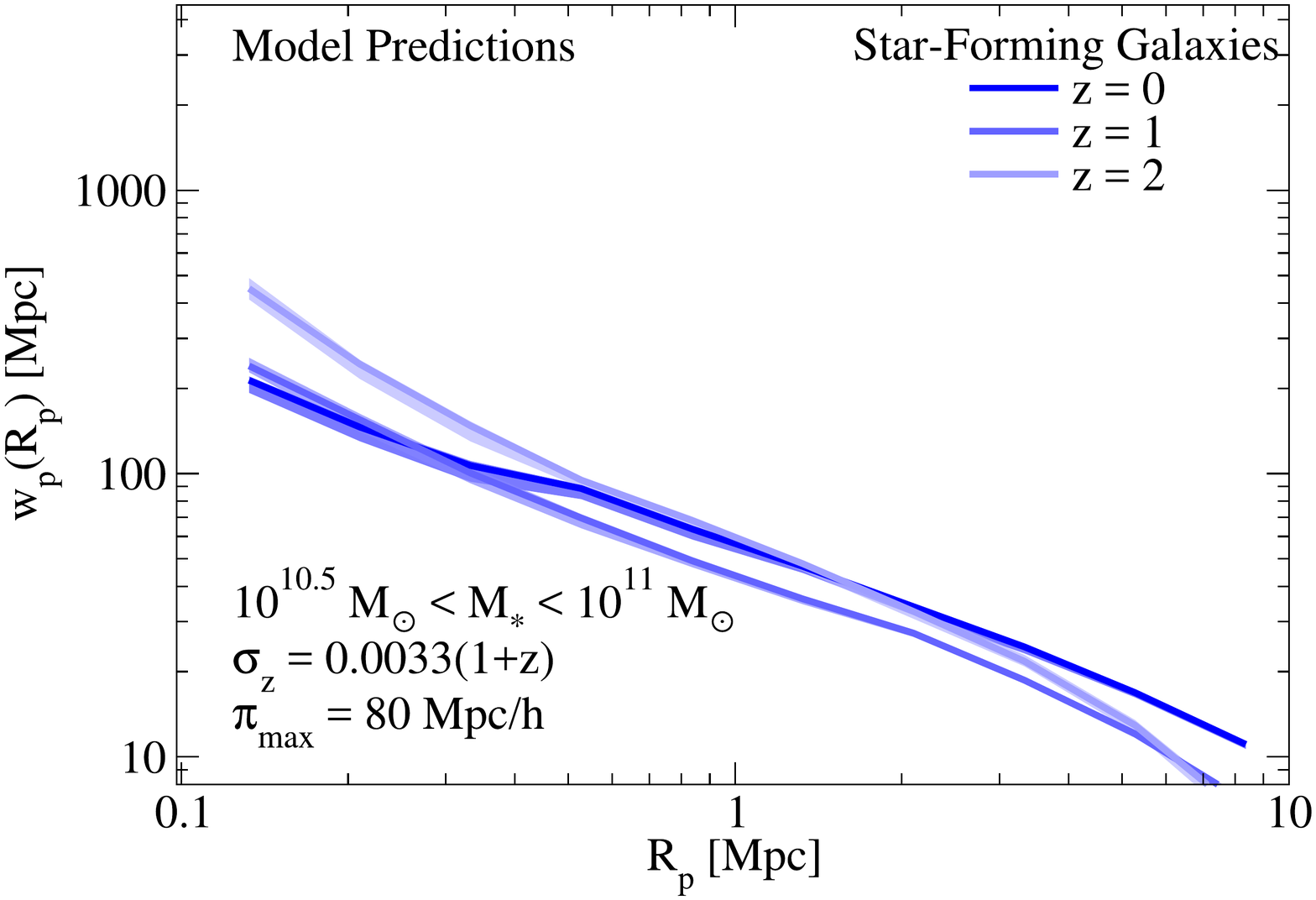}\plotgrace{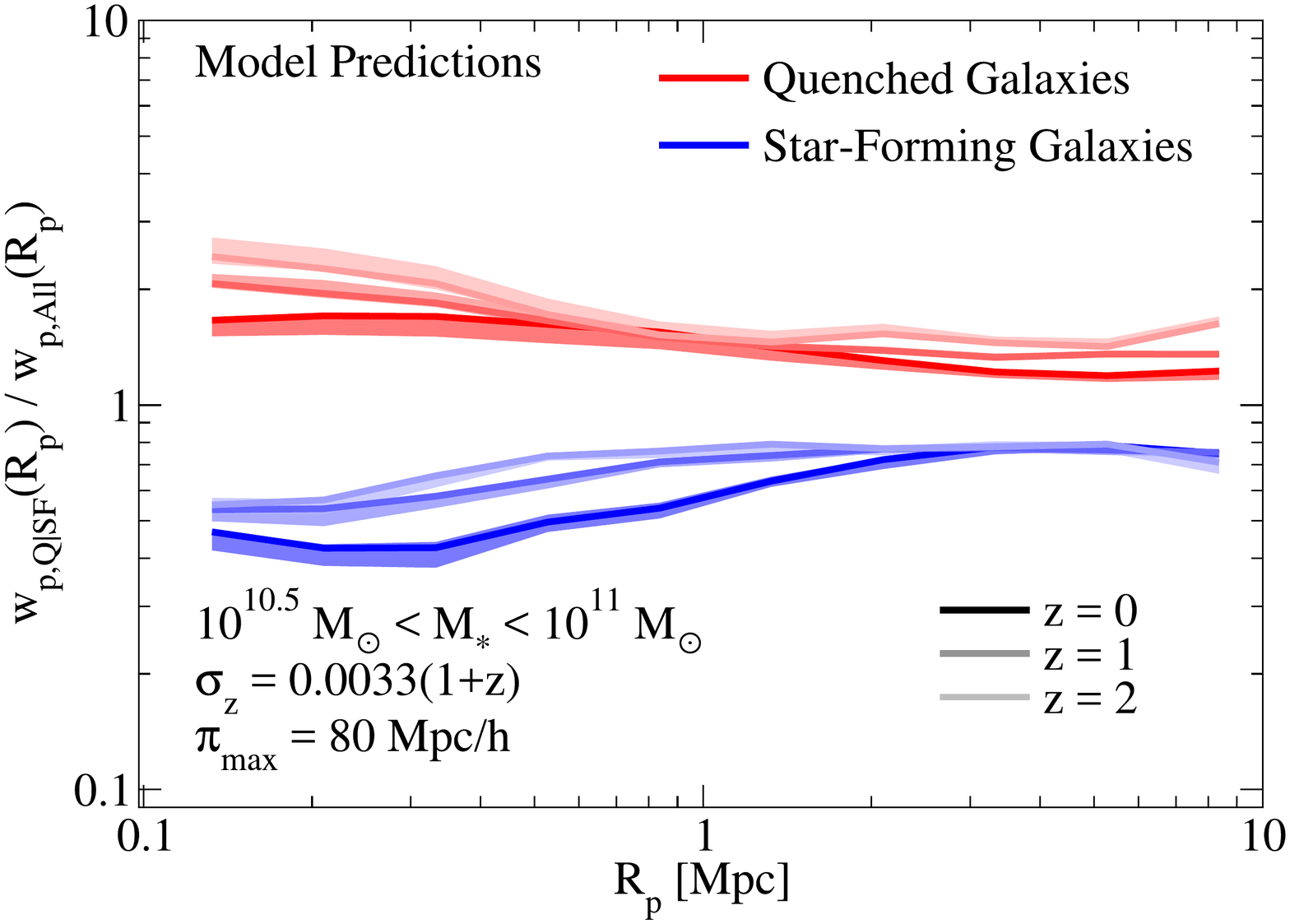}\\[-5ex]
\caption{\textbf{Top-left} panel: predictions for galaxy autocorrelation functions at $z>0$, assuming line-of-sight integration to $\pi_\mathrm{max} =  80$ Mpc $h^{-1}$ and redshift errors of $\sigma_z/(1+z) = 0.0033$, similar to the PRIMUS survey (see Appendix \ref{a:cf}).  \textbf{Top-right} and \textbf{Bottom-left} panels: same as top-left panel for quenched ($SSFR < 10^{-11}$ yr$^{-1}$) and star-forming galaxies, respectively. \textbf{Bottom-right} panel: the ratio of $w_p(R_p)$ for quenched and star-forming galaxies to $w_p(R_p)$ for all galaxies.  Shaded regions show the \onesigdist{} of the model posterior distribution.  All distances are in comoving units.  Predictions for other mass ranges available \href{http://www.peterbehroozi.com/data}{\textbf{online}}.}
\label{f:clustering_pred}
\end{figure*}

\begin{figure*}
\vspace{-5ex}
\plotgrace{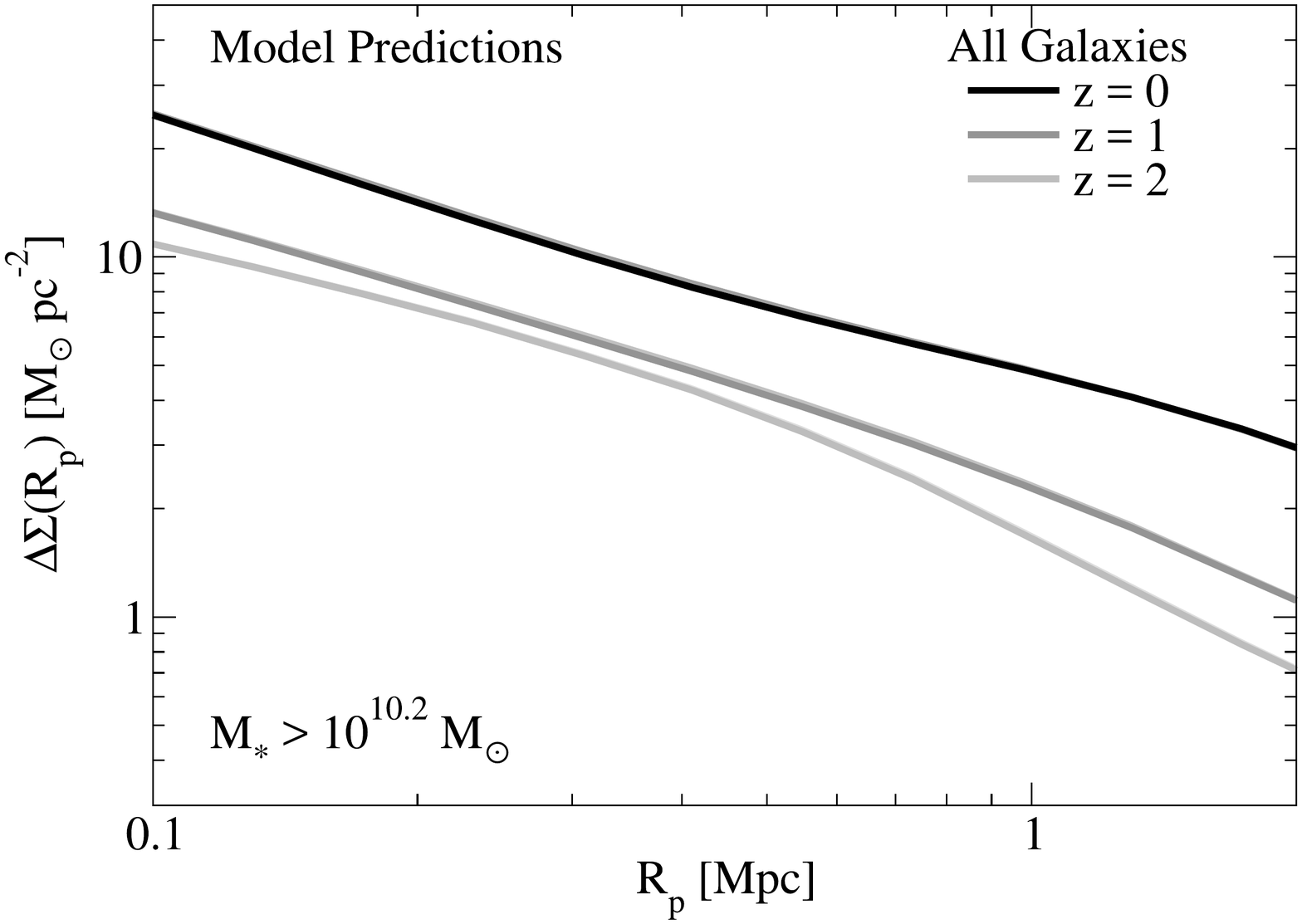}\plotgrace{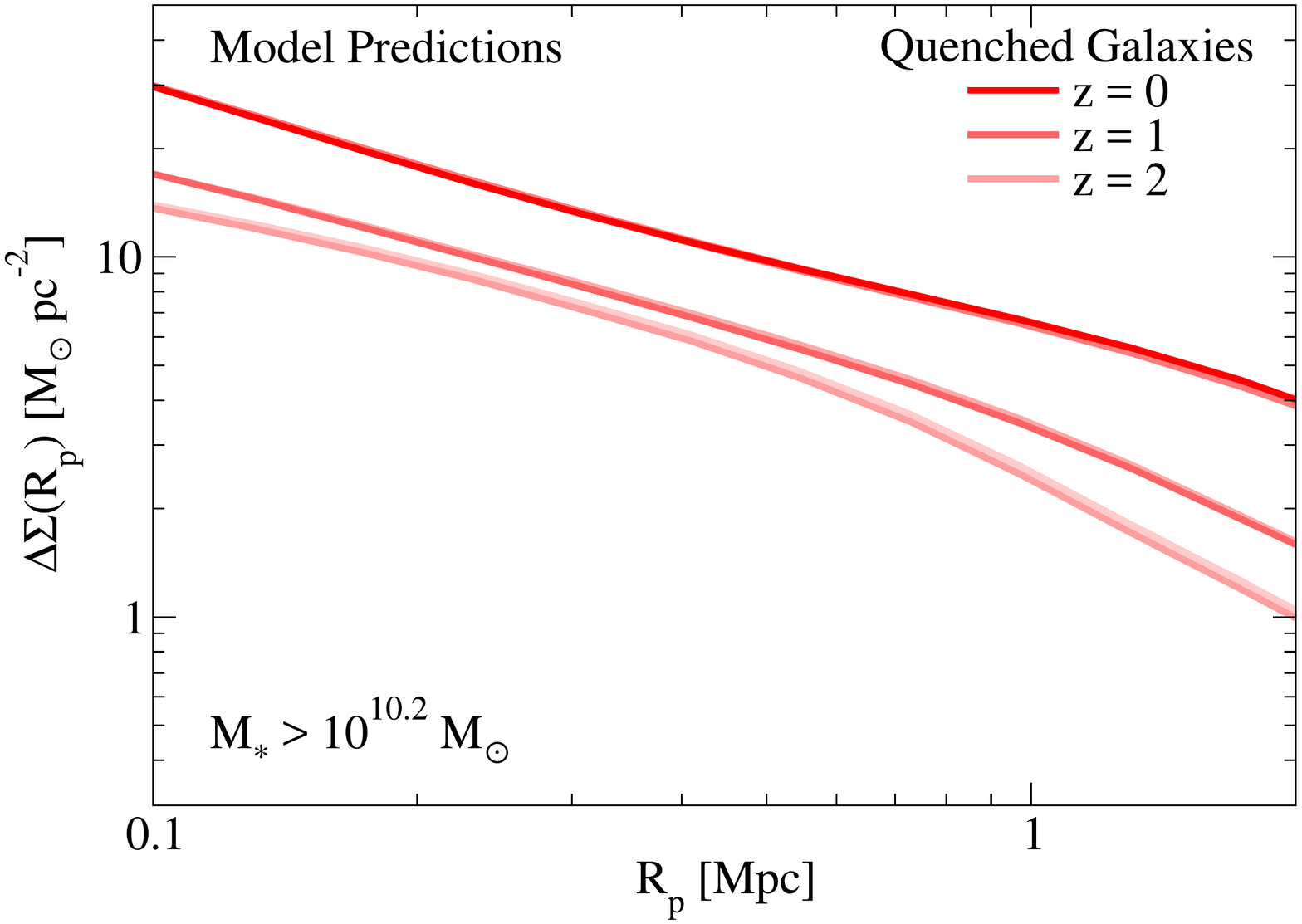}\\[-5ex]
\plotgrace{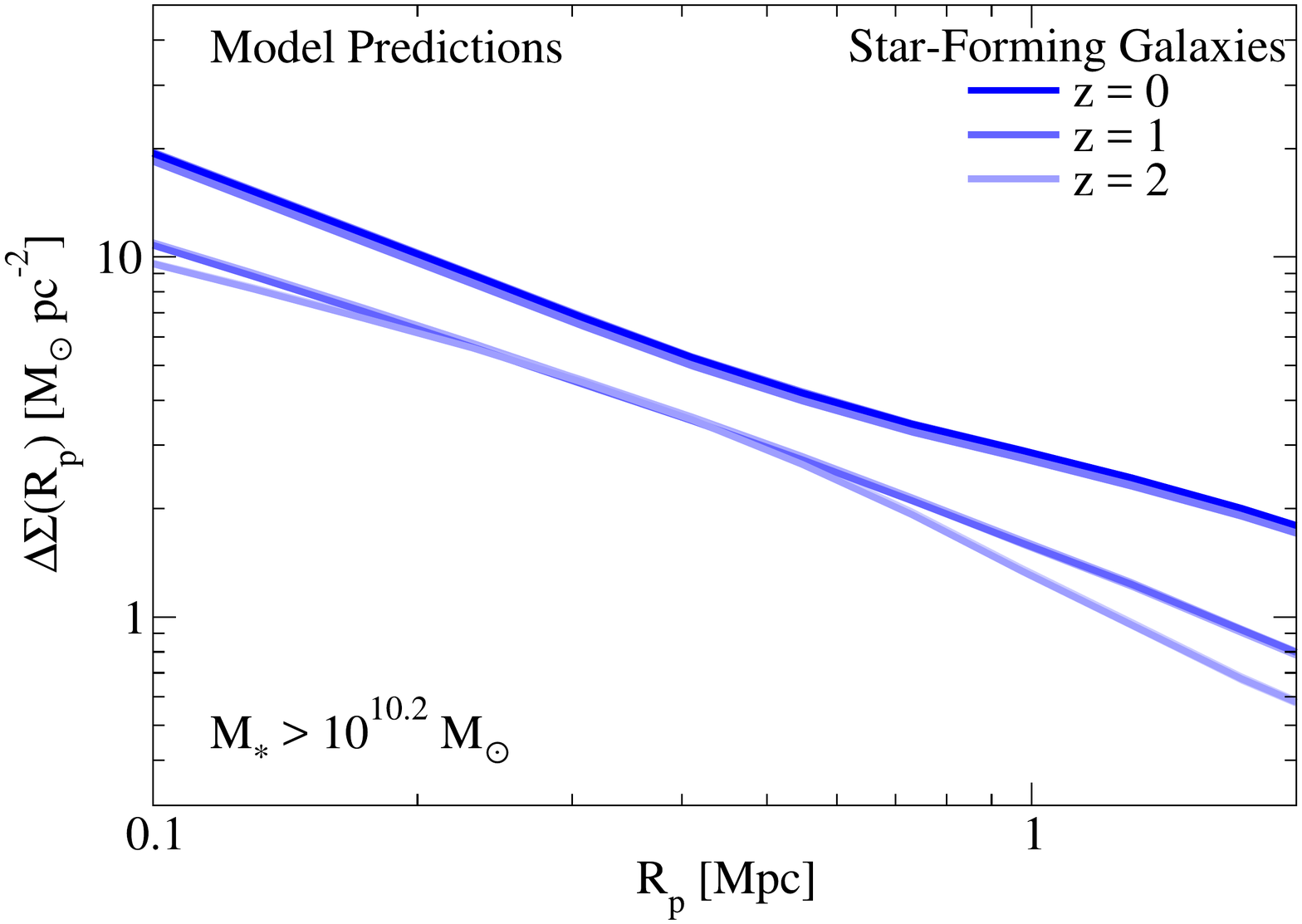}\plotgrace{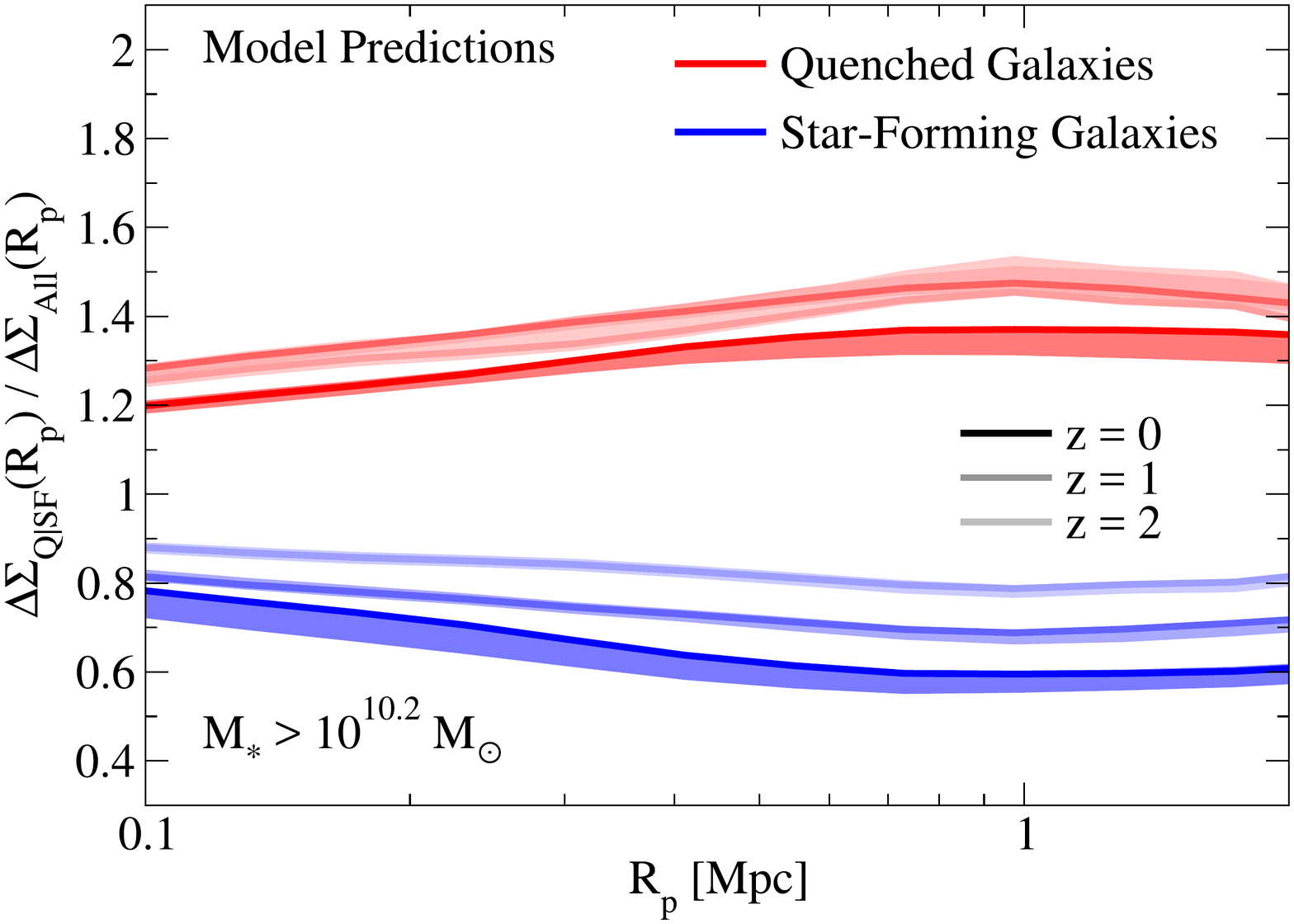}\\[-5ex]
\caption{\textbf{Top-left} panel: predictions for excess surface densities ($\Delta\Sigma$) for all galaxies with $M_\ast > 10^{10.2}\Msun$ as probed by galaxy-galaxy lensing at $z>0$.  \textbf{Top-right} and \textbf{Bottom-left} panels: same as top-left panel for quenched ($SSFR < 10^{-11}$ yr$^{-1}$) and star-forming galaxies, respectively.   \textbf{Bottom-right} panel: the ratio of $\Delta\Sigma$ for quenched and star-forming galaxies to that for all galaxies.   Shaded regions show the \onesigdist{} of the model posterior distribution.  All distances are in comoving units.  Predictions for other mass ranges available \href{http://www.peterbehroozi.com/data}{\textbf{online}}.}
\label{f:wl_pred}
\end{figure*}

\subsection{Fraction of Stellar Mass from In-Situ vs.\ Ex-Situ Growth}

\label{s:in_situ}

The fraction of stellar mass from \textit{ex-situ} vs.\ \textit{in-situ} star-formation increases with increasing halo mass and decreasing redshift (Fig.\ \ref{f:merger_stats}, left panel).  As massive galaxies are mostly quenched (Fig.\ \ref{f:qf_comp}, right panel), their only channel for growth comes via mergers of smaller haloes.  However, because of the shape of the SMHM relation (Fig.\ \ref{f:smhm}), haloes with $M_h < 10^{12}\Msun$ have strongly decreasing stellar fractions towards lower masses.  As a result, mass growth via mergers is only efficient for haloes with $M_h > 10^{12}\Msun$.  This is also reflected in the intrahalo light (IHL) to $M_\ast$ ratio (Fig.\ \ref{f:merger_stats}, right panel).  We find that the IHL / $M_\ast$ ratio reaches $\sim 0.3$ at halo masses of $10^{14}\Msun$ at $z=0$.  This is $\sim 0.5$ dex lower than past results \citep[e.g.][]{Gonzalez07,Moster12,BWC13}, as the SMFs used in this study assign more light to the central galaxy that previously would have been assigned to the IHL \citep{Bernardi13,Kravtsov14}.  Direct comparison with \cite{BWC13} is available \href{http://www.peterbehroozi.com/data}{\textbf{online}}.

\subsection{Predictions for Galaxies' Autocorrelation and Weak Lensing Statistics}
\label{s:corr_pred}

Fig.\ \ref{f:clustering_pred} shows predictions for a hypothetical PRIMUS-like survey extending to $z=2$.  At fixed galaxy mass, the overall clustering signal decreases from $z=0$ to $z=1$ partially due to reduced satellite fractions at higher redshifts.  The clustering signal increases again from $z=1$ to $z=2$ due to the increased rarity (and hence bias) of the host haloes.  As the quenched fraction decreases with increasing redshift, quenched galaxies have increasingly larger offsets relative to the all-galaxy correlation function (Fig.\ \ref{f:clustering_pred}, bottom-right panel).  That is, being quenched at higher redshifts requires increasingly extreme accretion histories, resulting in only the most stripped (and therefore clustered) haloes at high redshift being quenched.  Similarly, star-forming galaxy correlation functions also increase relative to the all-galaxy correlation function with increasing redshift.

We generate weak lensing predictions via projected dark matter surface densities in \textit{Bolshoi-Planck}.  Surface densities are integrated along the full extent (250 Mpc h$^{-1}$) of the $z$-axis in projected radial bins around each halo.  Given the surface density $\Sigma(r_p)$, we compute the excess surface density $\Delta\Sigma$:
\begin{equation}
\Delta\Sigma(r_p) \equiv \frac{\int_0^{r_p} \Sigma(r)\pi r dr}{\pi r_p^2} - \Sigma(r_p).
\end{equation}
Fig.\ \ref{f:wl_pred} shows the resulting predictions; see \cite{Hearin13b} for a discussion of the limitations of this approach.  Similar to autocorrelation functions, the lensing signal decreases from $z=0$ to $z=1$, partially due to lower satellite fractions, and partially due to lower halo concentrations \citep{Diemer13,Diemer15,RP16b}.  The latter is especially evident at halo outskirts.  The same factors continue to affect the lensing signal at higher redshifts; hence, the lensing signal continues to decrease in contrast to the galaxy autocorrelation signal.  Combined with a decreasing number density of lens sources at higher redshifts, this suggests that clustering will offer better halo mass constraints than lensing at high redshifts.  Predictions for clustering at $z>2$ relevant to the \textit{James Webb Space Telescope} are presented in R.\ Endsley, et al. (in prep.).

\subsection{Systematic Uncertainties}

\label{s:syst_uncertainties}

Stellar masses have many systematic uncertainties, including the stellar population synthesis model, dust, metallicities, and star formation histories assumed \citep{Conroy09,Conroy10,Behroozi10}.  Empirical models allow self-consistently treating uncertainties in dust \citep[e.g.,][]{Imara18}, metallicity \citep[e.g.,][]{Lu15b,Lu15c}, and star formation histories \citep[e.g.,][]{Moster12,Moster17,BWC13} when fitting to multi-band luminosity functions, potentially removing many of these systematic uncertainties.  The current model is a first step in this direction, despite the simplicity of the dust model assumed and the lack of non-UV luminosity functions.  A comparison between specific SFRs and stellar mass function evolution reveals strong inconsistencies between observations that would be resolved by $0.3$ dex higher stellar masses (or 0.3 dex lower SFRs) at $z=2$ (Appendix \ref{a:tension}).  Regardless, our inferred ``true'' stellar masses and star formation rates are always within $0.3$ dex of the observed values (e.g., Fig.\ \ref{f:csfr_comp}, left panel).

\subsection{Additional Data Available Online}

\label{s:online_data}

The \href{https://www.peterbehroozi.com/data}{online} data release includes underlying data and documentation for all the figures in this paper, as well as additional mass and redshift ranges where applicable, as well as model posterior uncertainties where possible.  The data release also includes halo and galaxy catalogues for the best-\fit{} model applied to the \textit{Bolshoi-Planck} simulation (both in text format and in an easily-accessible binary format integrated with \textsc{HaloTools}; \citealt{Hearin17}), full star formation and mass assembly histories for haloes at $z=0,1$ and $2$, and mock lightcones corresponding to the five CANDELS fields \citep{Grogin11}.  The data release includes a snapshot of the \textsc{UniverseMachine} code (Appendix \ref{a:code}) that was used for this paper.

\section{Discussion}
\label{s:discussion}

We discuss how results in \S \ref{s:results} impact connections between galaxy and halo assembly (\S \ref{s:gal_halo_assembly}), central (\S \ref{s:central_quenching}) and satellite (\S \ref{s:sat_quenching}) quenching, tracing galaxies across cosmic time (\S \ref{s:tracing}), equilibrium/bathtub models of galaxy formation (\S \ref{s:sharc_bathtub}), ``impossibly early'' galaxies (\S \ref{s:impossible}), uniqueness of the model (\S \ref{s:uniqueness}), and orphan galaxies (\S \ref{s:orphans}); we also compare to previous results (\S \ref{s:smhm_comp}) and discuss evolution in the stellar mass--halo mass relation (\S \ref{s:smhm_evolution}).  We discuss how additional observations and modeling could address current assumptions and uncertainties (\S \ref{s:assumptions}), and finish with future directions for empirical modeling (\S \ref{s:future}).

\subsection{Connections Between Individual Galaxy and Halo Assembly}

\label{s:gal_halo_assembly}

We find that star-forming galaxies reside in significantly more rapidly-accreting haloes than quiescent galaxies (\S \ref{s:corr}).  This is especially clear for satellites, which drive differences in autocorrelation functions between quenched and star-forming galaxies (Fig.\ \ref{f:cf_comp}).  It is also clear for ``backsplash'' galaxies--i.e., those that passed inside a larger host's virial radius before exiting again; these drive the environmental dependence of the quenched fraction for central galaxies (Fig.\ \ref{f:ecq_comp}).

Correlation constraints are much weaker for central galaxies that never interacted with a larger halo.   Indeed, \cite{Tinker11,Tinker16}  argue that \textit{no} correlations exist between low-mass galaxy assembly and quenching except for satellite and backsplash galaxies \citep[as in][]{Peng10,Peng12}.  That said, their main evidence is that star-forming fractions for central galaxies seem not to depend on environmental density except for the highest-density environments.  As shown in Fig.\ \ref{f:ecq_comp}, the model herein has no problem matching this behaviour even with a fairly strong quenching-assembly correlation, similar to the finding in \cite{Wang18}; this is due to the fact that assembly histories do not change significantly for haloes in low- vs.\  median-density environments \citep{Lee16}.  \cite{BehrooziMM} argues that enhanced mass accretion during major mergers does not correlate with galaxy quenching at $z=0$; however, this does not preclude a correlation with smooth accretion.   Determining correlations between smooth matter accretion rates and galaxy formation will require alternate techniques to measure halo mass accretion rates, such as splashback radii \citep[e.g.,][]{More16}.

If quenching \textit{does} correlate with assembly history for isolated centrals, then it becomes very difficult to avoid rejuvenation (\S \ref{s:corr}) in $10^{11}\Msun$ galaxies.  This is because such galaxies' quenched fractions increased slowly over many halo dynamical times---so that changes in assembly history occurred much more rapidly than changes in the quenched fraction.  Thus, the number of such galaxies that quench at any given time due to recently low or negative accretion rates must be approximately balanced by the number that rejuvenate due to recently high accretion rates.  Avoiding this is possible only if central galaxy quenching is not significantly correlated with halo assembly.  As a result, the depth of the green valley in colour space represents another way to test quenching models, as multiple passes through the green valley will lead to a shallower valley than models where galaxies quench only once.

\begin{figure}
\vspace{-5ex}
\plotgrace{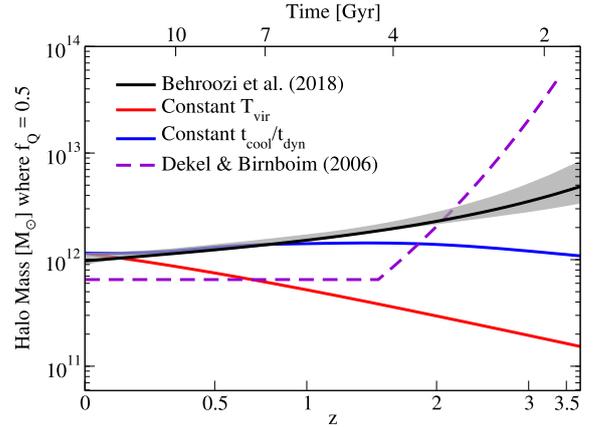}\\[-5ex]
\caption{Empirical halo mass quenching thresholds from this work (\textit{black line}; defined as the halo mass at which $50\%$ of galaxies are quenched) compared to theoretical cooling-based thresholds.  The \textit{red line} shows a model where quenching occurs above a virial temperature threshold of 6.4$\times 10^5$K; the \textit{blue line} shows a model where quenching occurs if the cooling/dynamical time ratio $t_\mathrm{cool}/t_\mathrm{dyn}$ is greater than $0.4$.  The \textit{purple dashed line} shows the model of \protect\cite{Dekel06}.  The \textit{grey shaded region} shows the \onesigdist{} from the posterior distribution for this work.  \textbf{Notes}: see text for definition of $t_\mathrm{cool}$ and $t_\mathrm{dyn}$.  }
\label{f:cooling}
\end{figure}

\subsection{Central Galaxy Quenching}

\label{s:central_quenching}

At fixed halo mass, we find that the galaxy quenched fraction decreases with increasing redshift (\S \ref{s:average_sfrs}).  This is a robust conclusion in other empirical models \citep[e.g.,][]{Moster17}, due to the decreasing galaxy quenched fraction at fixed stellar mass \citep[e.g.,][]{Muzzin13} and the constancy of the stellar mass--halo mass relation from $z=4$ to $z=0$ \citep{Behroozi13}.   Vice versa, as shown in Fig.\ \ref{f:cooling}, the halo mass at which a fixed fraction of galaxies are quenched (e.g., 50\%) increases with increasing redshift.

This latter fact implies that a virial temperature threshold alone is not responsible for quenching.  Virial temperatures increase with redshift at fixed halo mass, so a constant virial temperature quenching threshold would predict that the threshold halo mass for quenching should decrease with increasing redshift (Fig.\ \ref{f:cooling}).  A similar argument applies to thresholds in the ratio of the cooling time to the halo dynamical time (or the free-fall time or the age of the Universe, which are proportional).  These typically give redshift-independent quenched fractions with halo mass \citep[see Fig. 8.6 in][]{MvdBW}.  Here, we adopt a crude cooling time estimate from \cite{MvdBW}:
\begin{equation}
t_\mathrm{cool} \sim 1.5\times10^9\;\mathrm{yr}\; \left(\frac{T_\mathrm{vir}}{10^6\;\mathrm{K}}\right)\left(\frac{10^{-3}\;\mathrm{cm}^{-3}}{\langle n_H \rangle}\right)\left(\frac{10^{-23}\;\mathrm{erg}\;\mathrm{cm}^3\;\mathrm{s}^{-1}}{\Lambda(T_\mathrm{vir})}\right),
\end{equation}
where $n_H$ is the average density of hydrogen atoms in the halo; we take the cooling function $\Lambda(T)$ from \cite{deRijcke13} for a $1/3 Z_\odot$ gas.  For the halo dynamical time, we define as before $t_\mathrm{dyn}\equiv (\frac{4}{3}\pi G\rho_\mathrm{vir})^{-\frac{1}{2}}$.   As shown in Fig.\ \ref{f:cooling}, a model where haloes quench above a constant $t_\mathrm{cool}/t_\mathrm{dyn}$ threshold may be plausible from $z=0$ to $z=1$, but this model is inconsistent with our results at $z>1$.  

\cite{Dekel06} posit that at higher redshifts, cold streams can more effectively penetrate hot haloes ($M_h > 10^{12}\Msun$), allowing for residual star formation.  As shown in Fig.\ \ref{f:cooling}, our best-\fit{} model allows gradually more residual star formation in hot haloes with increasing redshift.  Yet, \cite{Dekel06} predicted a steeper transition for cold streams to exist in hot haloes at redshifts $z>1.5 - 2$ than we find here, suggesting that the quantitative details of quenching are different.  Indeed, this is expected at a basic level because \cite{Dekel06} do not discuss the effect of black holes, which are also expected to play a role in quenching \citep[e.g.,][]{Silk98}. 

Isolated central haloes (as opposed to backsplash haloes) rarely lose matter, and so their quenching in this model is driven by recent mergers and random internal processes.  During a merger, the maximum circular velocity ($\vmax$) will rapidly increase  during first passage, resulting in a burst of star formation; $\vmax$ then rapidly decreases as kinetic energy from the merger dissipates into increased halo velocity dispersion and lower halo concentration \citep{BehrooziMergers}.  In the latter phase, the galaxy will be quenched in our model, resulting in a post-starburst galaxy.  As the quenched fraction decreases toward higher redshifts, only the most extreme merging events will result in quenched centrals, meaning that the fraction of quenched non-satellite galaxies that are post-starburst will increase with redshift.  We hesitate to call this a prediction, since a different equally-reasonable choice of halo assembly proxy  may have different behaviour; instead, it is a testable hypothesis.

\subsection{Satellite Galaxy Quenching}

\label{s:sat_quenching}

For satellites, many quenching mechanisms have been proposed, including ram-pressure/tidal stripping \citep{Gunn72,Byrd90}, strangulation \citep{Larson80}, accretion shocks \citep{Dressler83}, and harassment from other satellites \citep{Farouki81}, among others.  Given the diversity of satellite orbits and infall conditions, it is likely that all of these mechanisms each quench some fraction of satellites.  For example, extremely high fractions of quenched galaxies in cluster centers \citep{Wetzel11} may suggest that galaxies quench rapidly there.  That said, we find that satellite quenching is neither efficient nor necessarily a rapid process for most satellites (\S \ref{s:satellites}), suggesting that accretion shocks may not be dominant.  In addition, harassment from other satellites is problematic because high velocities inside clusters mean that strong interactions are less likely to occur \citep{Binney08}.  This suggests that inefficient ram-pressure/tidal stripping \citep[e.g.,][]{Emerick16} coupled with strangulation is sufficient to explain most satellite quenching \citep[see also][]{Balogh16,Fillingham18}.  In addition, feedback models that launch galaxy gas to significant fractions of the virial radius (leading to efficient stripping) will generically overproduce satellite galaxy quenched fractions.

\subsection{Tracing Galaxies Back in Time}
\label{s:tracing}

\begin{figure}
\vspace{-5ex}
\plotgrace{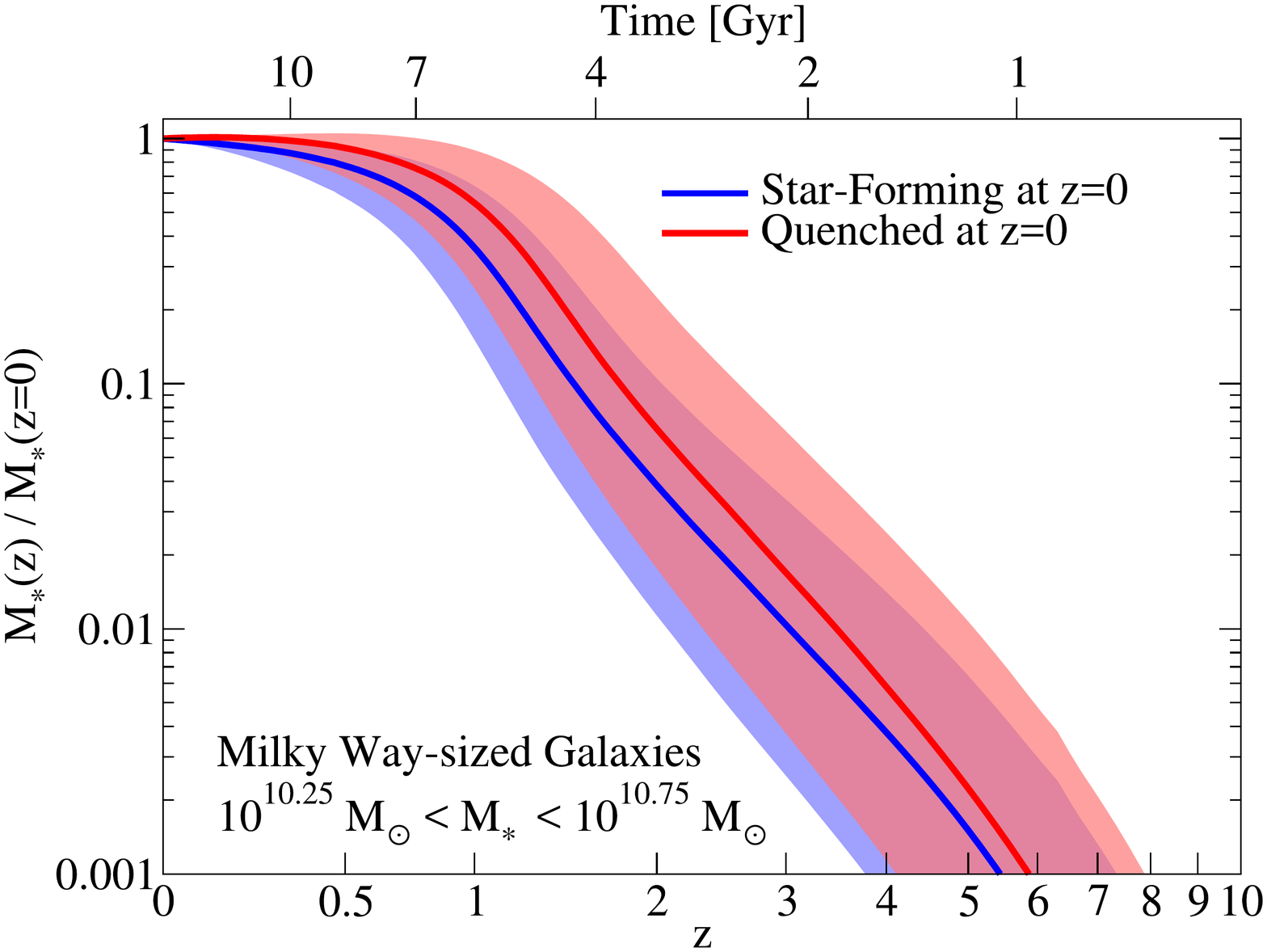}\\[-5ex]
\plotgrace{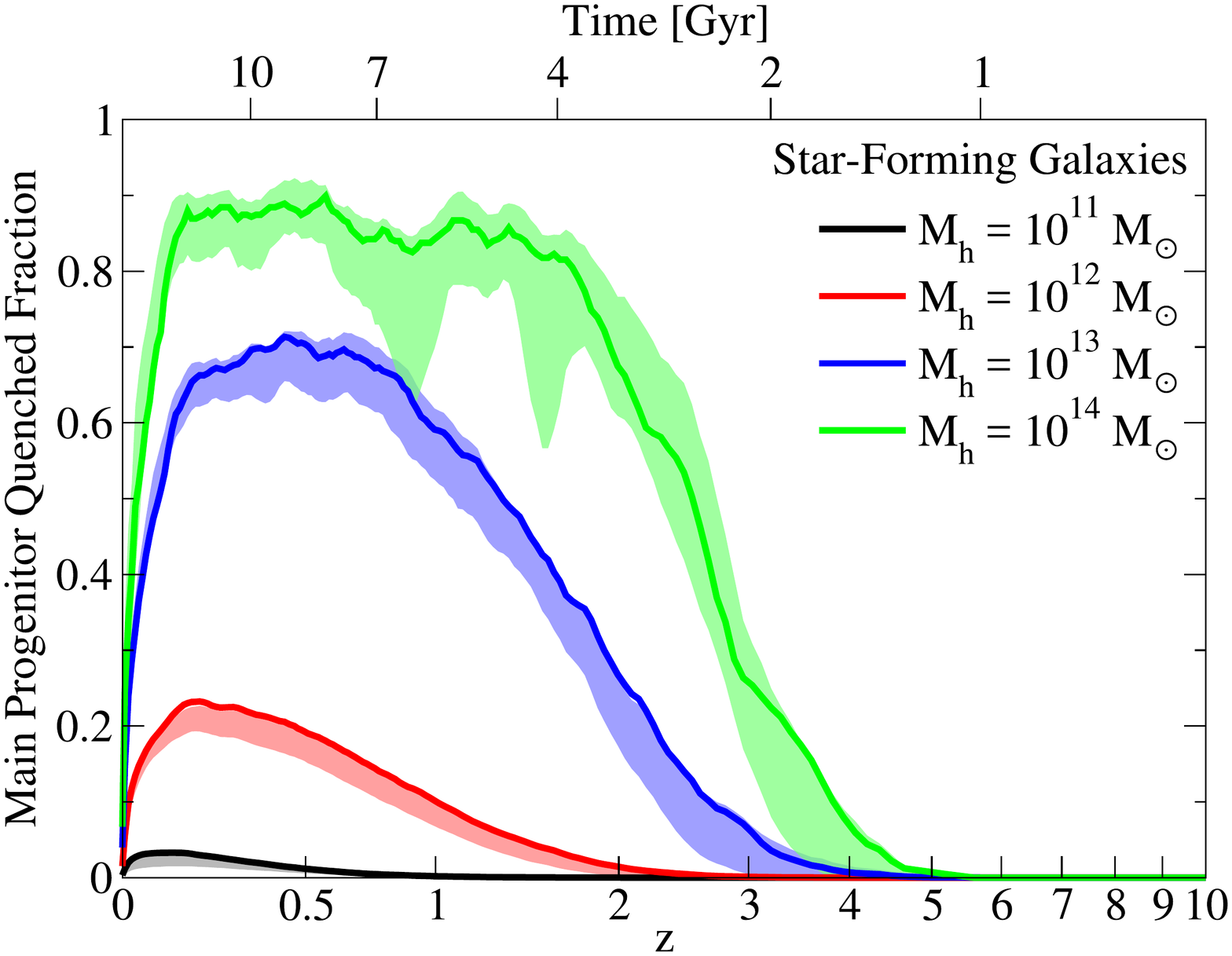}\\[-5ex]
\plotgrace{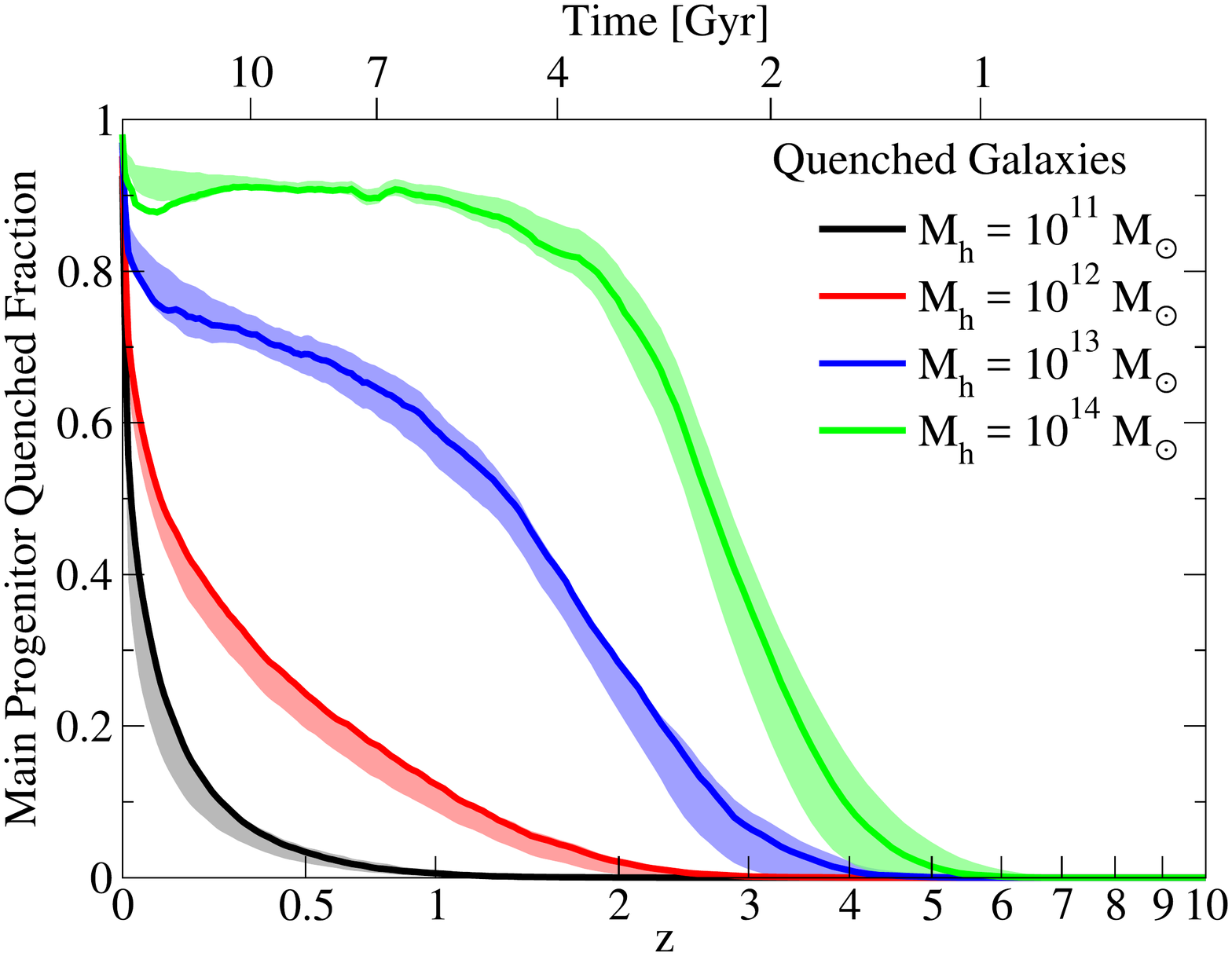}\\[-4ex]
\caption{\textbf{Top} panel: stellar mass histories for star-forming and quenched galaxies at $\sim$Milky-Way masses.  \textbf{Middle} panel: for star-forming galaxies at $z=0$, the fraction of their main progenitors that are quenched as a function of redshift.  By definition, this fraction is 0 at \mbox{$z=0$}.  \textbf{Bottom} panel: same, for quenched galaxies at $z=0$; by definition, the fraction is 1 at $z=0$.  Differences between main progenitor quenched fractions for $z=0$ quenched and star-forming galaxies largely disappear by $z=0.5$.}
\label{f:fq_hist}
\end{figure}

Using cumulative number densities to follow galaxy progenitors \citep[e.g.,][]{Leja13,vanDokkum13,Lin13} has become an increasingly popular approach despite the large scatter in progenitor histories \citep{BehrooziND,Torrey16,Jaacks16,Wellons16}.  Recently, \cite{Clauwens16} noted differences between median progenitor histories for star-forming and quenched galaxies in the EAGLE simulation \citep{Schaye15}.  In our best-\fit{} model, we also find such differences (Fig.\ \ref{f:fq_hist}, top panel), but find as in \cite{Clauwens16} that the scatter in individual progenitor histories dwarfs the median difference at all redshifts.  Joint selection on cumulative number density and SSFR will be explored in future work.  Most of the power in differentiating galaxy properties may only come over galaxies' recent histories; e.g., the difference in progenitor star-forming fractions between quenched and star-forming galaxies largely disappears by $z=0.5$ (Fig.\ \ref{f:fq_hist}, middle and bottom panels).

\subsection{Equilibrium vs.\ Non-Equilibrium Models}

\label{s:sharc_bathtub}

\begin{figure}
\vspace{-5ex}
\plotgrace{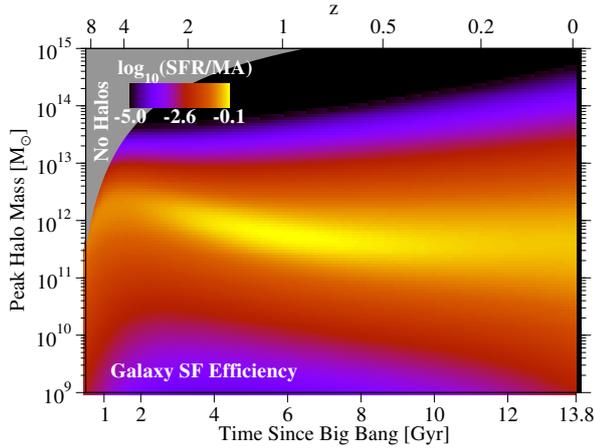}\\[-5ex]
\caption{Average galaxy star formation efficiency, defined as the ratio of average galaxy SFR to the average baryonic mass accretion rate of the host dark matter halo.  This is shown as a function of cosmic time and peak halo mass (at the given cosmic time, as opposed to $z=0$).  The baryonic mass accretion rate is approximated as $f_b \dot{M}_h$, where $f_b=0.16$ is the cosmic baryon fraction.  Compare to analogous figure in \protect\cite{Behroozi13} \href{https://www.peterbehroozi.com/data}{\textbf{online}}.  \textbf{Notes:} relative uncertainties are the same as for average galaxy SFRs, shown in Fig.\ \ref{f:sfr_errors}, left panel.}
\label{f:sfe}
\end{figure}

\begin{figure}
\vspace{-5ex}
\plotgrace{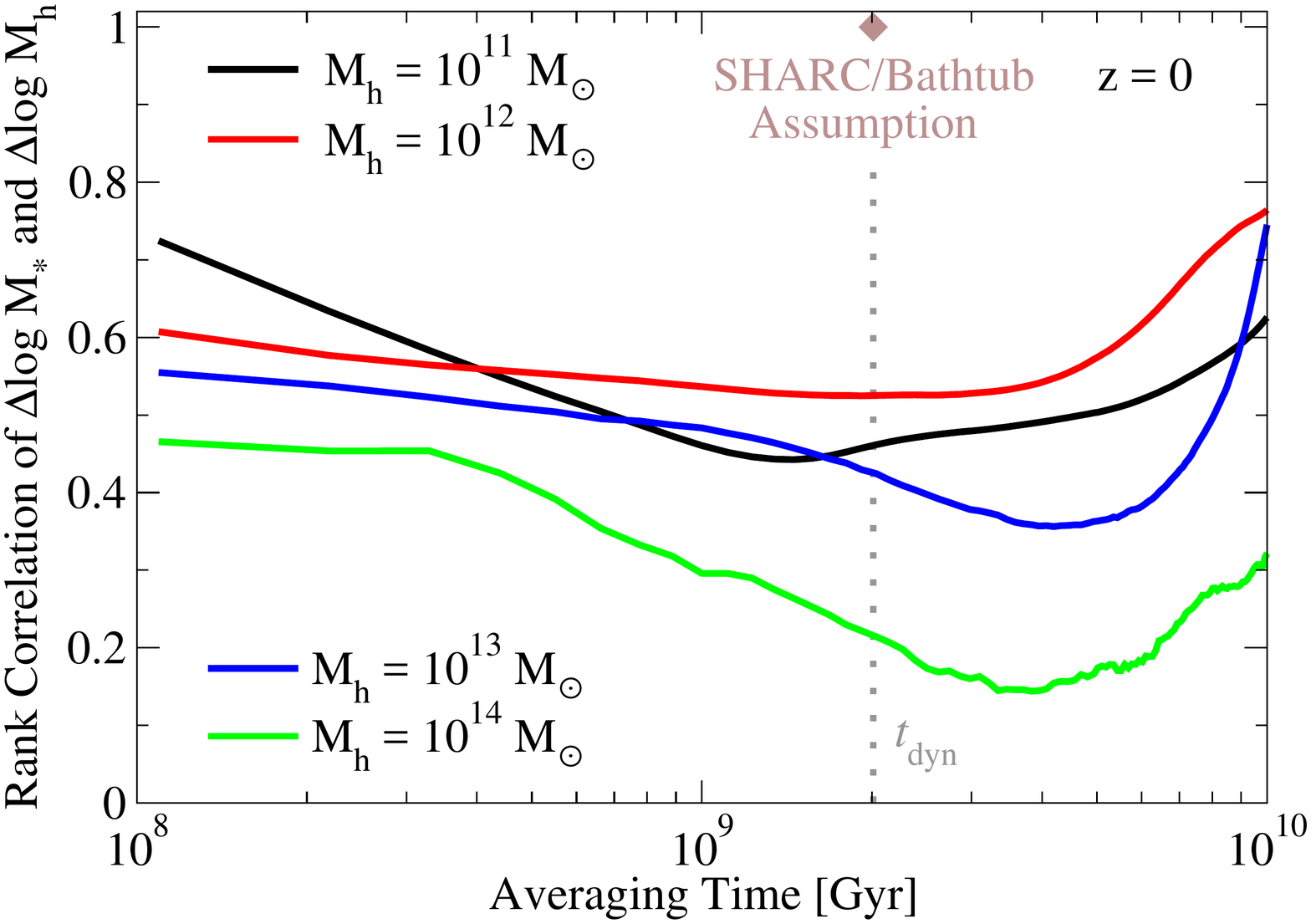}\\[-5ex]
\plotgrace{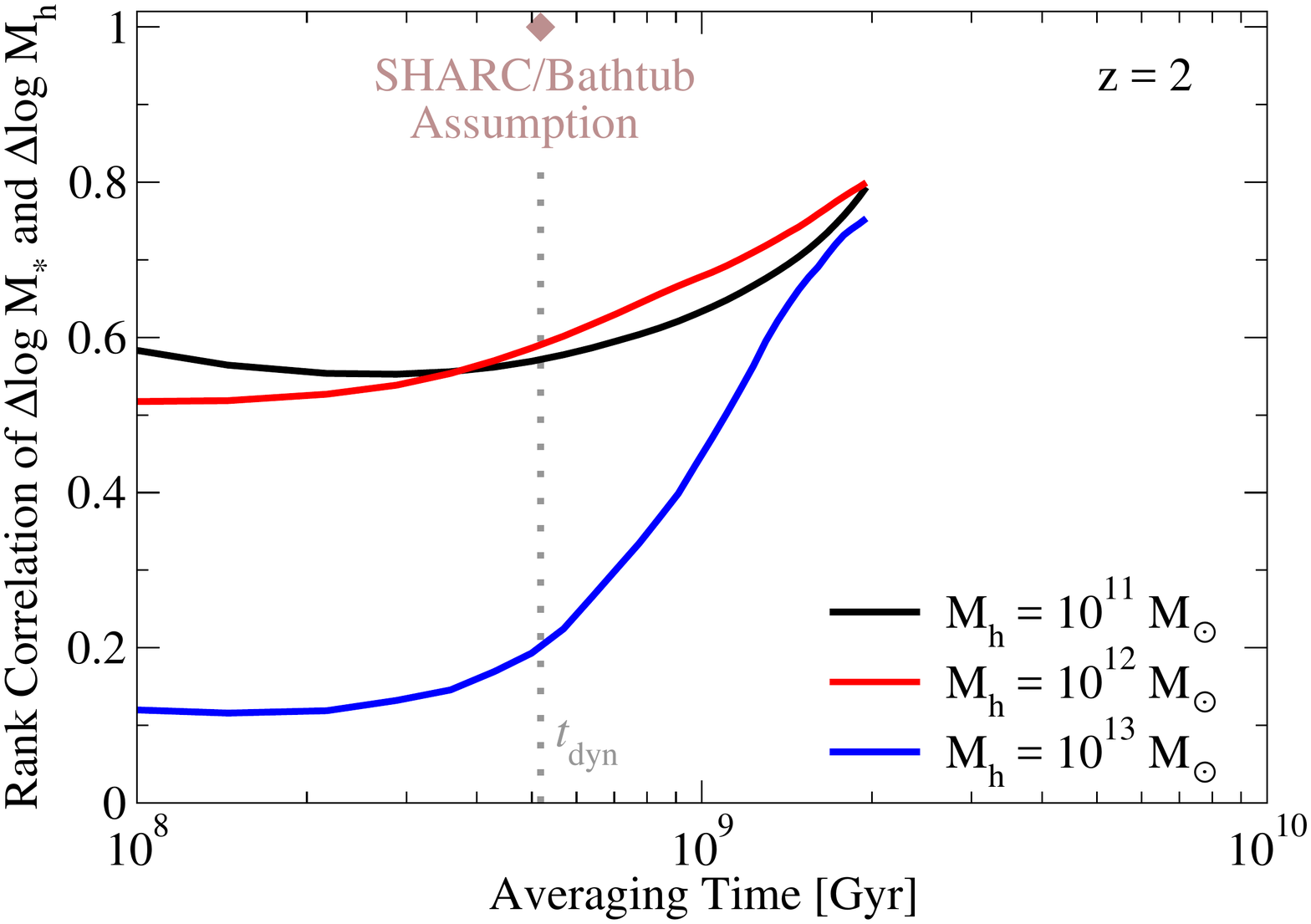}\\[-3ex]
\caption{\textbf{Top} panel: rank correlation of average galaxy specific growth rate ($\langle \frac{d \log{M_\ast}}{dt}\rangle$) to average specific halo mass accretion rate ($\langle \frac{d \log{\mpeak}}{dt}\rangle$) as a function of averaging time at $z=0$ for the best-\fit{} model.  Also shown for comparison is the Stellar-Halo Accretion Rate Coevolution (SHARC) assumption from \protect\cite{RP16}, corresponding to a rank correlation of 1 for averaging over a dynamical time.  \textbf{Bottom} panel: Same, at $z=2$.  In both panels, the \textit{grey dotted line} denotes the dynamical time at the given redshift.}
\label{f:hmsm_correls}
\end{figure}

In equilibrium (a.k.a., ``steady-state'' or ``bathtub'') models of galaxy formation \citep[e.g.,][]{Bouche10,Dave10,Lilly13}, galaxies form stars according to average gas accretion rates scaled by a mass-dependent efficiency.  Observational evidence that galaxies behave in this way \textit{on average} \citep[e.g.][]{Behroozi13} gave rise to empirical models in which \textit{individual} galaxies' SFRs are linearly related to halo gas accretion rates \citep{Becker15,RP16,Sun16,Mitra16,Cohn16,Moster17}.  In principle, the average behaviour could also be reproduced if individual galaxies' positions varied randomly on the SSFR main sequence, with no relation to mass accretion rates.  At $z=0$, two lines of evidence are inconsistent with linear relationships between mass accretion rates and star formation rates; specifically, star-forming satellites' SSFRs are not offset significantly from star-forming centrals' SSFRs \citep{Wetzel11}, and major halo mergers do not result in enhanced star-forming fractions or enhanced SFRs for star-forming galaxies \citep{BehrooziMM}.

The model in this paper reproduces an average ratio between gas accretion and star formation that is nearly constant in time (Fig.\ \ref{f:sfe}), using a strong but imperfect correlation between galaxy assembly and halo assembly ($\sim 0.5 - 0.6$; \S\ref{s:corr}).  If indeed main-sequence SSFRs were perfectly correlated with assembly history at $z\sim 0$, it would be very difficult for satellite fractions to be large enough to explain the autocorrelation function for galaxies; we also find that star-forming central galaxies' SSFRs do not depend much on environment density (Appendix \ref{a:ecq}; see also \citealt{Berti16}), whereas central halo accretion rates are known to do so \citep{Lee16}.

We note that even for models where SFR depends only on halo mass and cosmic time (i.e., not on assembly history), long-term correlations between halo and galaxy growth can result.  Because SFR rises steeply with halo mass for $M_h < 10^{12}\Msun$, a larger growth in halo mass will guarantee a larger change in SFRs and therefore in $M_\ast$.  On the other hand, mergers can reduce correlations between stellar mass growth and halo mass growth due to the shape of the SMHM relation; this is especially true for massive haloes.  The combination of mergers and induced SFR--halo growth correlations results in a very nontrivial shape for how overall galaxy growth---halo growth correlations depend on the averaging timescale (Fig.\ \ref{f:hmsm_correls}). 

\subsection{``Impossibly'' Early Galaxies}

\label{s:impossible}

\cite{Steinhardt16} recently found that $z>4$ galaxy number densities are too large to reconcile with $\Lambda$CDM dark matter halo number densities unless certain conditions are met.  Of these conditions, we find that it is sufficient for stellar fractions in dark matter haloes to increase with increasing redshift (\S \ref{s:smhm_comp}; see also \citealt{Jaacks12,Liu16,BehrooziHighZ,BehrooziSilk16}).  The best-\fit{} model matches both observed stellar mass functions (Fig.\ \ref{f:smf_comp}) and UV luminosity functions (Fig.\ \ref{f:uv_comp}) at $z>4$, using reasonable stellar fractions (Fig.\ \ref{f:smhm}), dust attenuation (Fig.\ \ref{f:irx_comp}), and stellar population synthesis models \citep[FSPS;][]{Conroy09}.  Thus, we do not find the observed $z>4$ number densities to be a cause for concern with $\Lambda$CDM.

\subsection{Uniqueness}

\label{s:uniqueness}

With one-point statistics (SMFs, SFRs, and quenched fractions) alone, there are many mathematically-allowed solutions for galaxies' star formation histories \citep[e.g.,][]{Gladders13,Kelson14,Abramson15,Abramson16}.  This arises because individual galaxies' long-term star-formation histories are not directly measured by such statistics, requiring an additional constraint.  Our method relies on $z<1$ clustering and environmental constraints to anchor the relationship between galaxies and haloes.  As haloes' growth histories are well-measured \citep{Srisawat13}, the uniqueness of the solution is set by the tightness of the galaxy---halo relationship, implying that long-term stellar mass growth histories can be inferred to within $\sim$0.3 dex (Fig.\ \ref{f:indiv_sfr}, top-right panel) if the halo mass is known.  Matching the $z>1$ stellar mass--halo mass relation from independent clustering measurements (\S \ref{s:smhm_comp}) is thus a nontrivial prediction that favours model uniqueness.

That said, constraining recent SFR histories is challenging even if the halo mass is known (Fig.\ \ref{f:indiv_sfr}, top-left panel).  This is especially true at $z<2$, where halo mass has little predictive power for galaxy SFRs for $M_h > 10^{12}\Msun$ galaxies \citep[see also][]{BWC13}.  Knowledge of halo growth rates does appear to help (Figs.\ \ref{f:rank_correl} and \ref{f:hmsm_correls}), but less so for the most massive galaxies.  Physically, this is consistent with more stochastic star formation in massive galaxies at late times (e.g., due to fluctuating activity in the central black hole), but more predictable (i.e., more constrainable) star formation in lower-mass galaxies and at early times.

\begin{figure*}
\vspace{-5ex}
\plotlargegrace{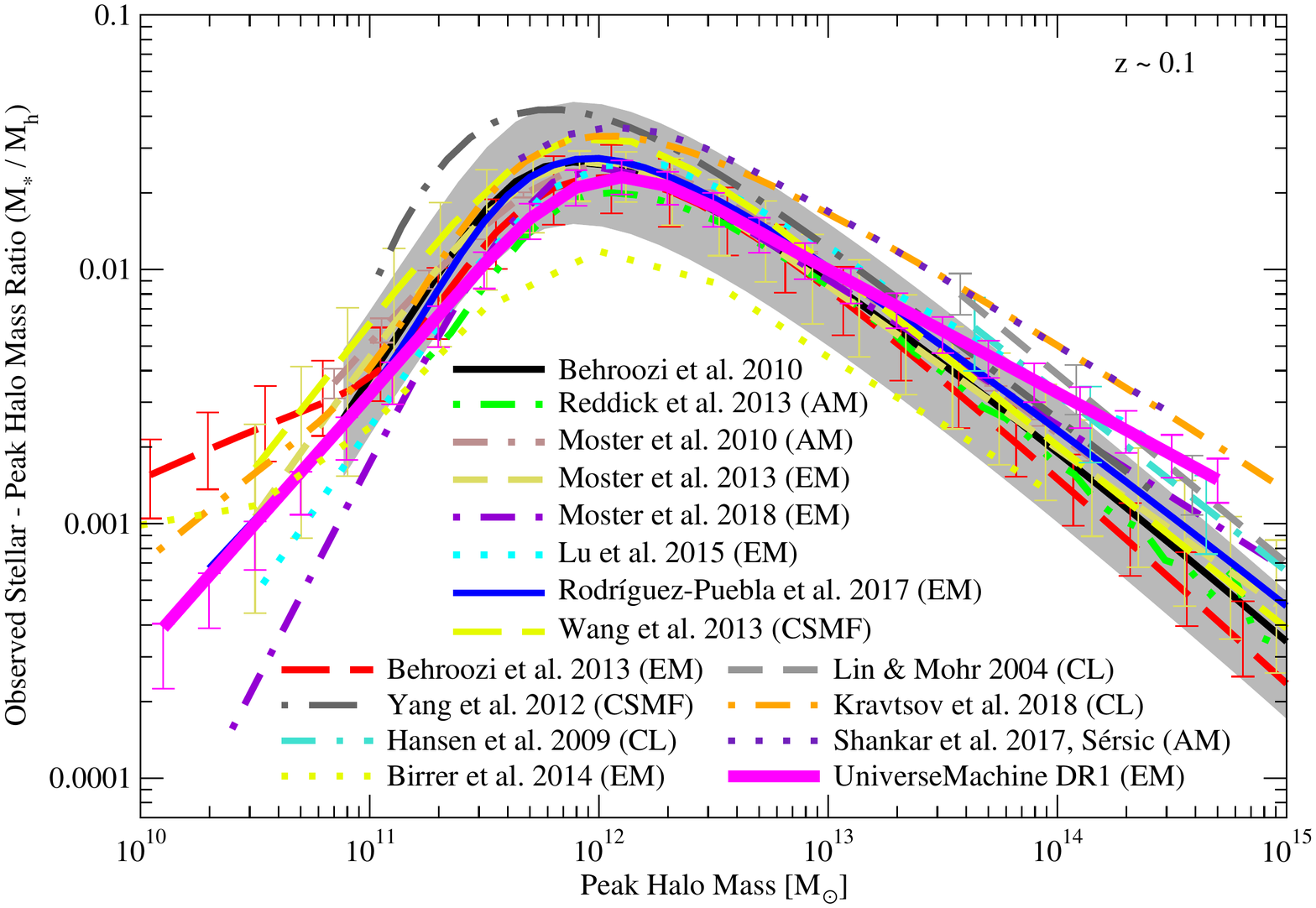}\\[-5ex]
\caption{Median observed stellar mass --- halo mass relation for our best-\fit{} model (labelled as ``UniverseMachine DR1'') compared to previous results at $z=0.1$.  Results compared include those from our previous works \protect\citep{Behroozi10,BWC13}, from empirical modeling \protect\citep[EM;][]{Moster12,Moster17,Birrer14,Lu15,RP17}, from abundance matching \citep[AM;][]{moster-09,Reddick12,Shankar17}, from Conditional Stellar Mass Function (CSMF) modeling \protect\citep{Yang11,Wang12}, and from cluster X-ray mass measurements \protect\citep{LinMohr04,Hansen09,Kravtsov14}.  Grey shaded regions correspond to the \onesigdist{} in \protect\cite{Behroozi10}. The \onesigdist{} of the model posterior distribution is shown by the purple error bars.}
\label{f:comp_z0}
\vspace{-5ex}
\plotgrace{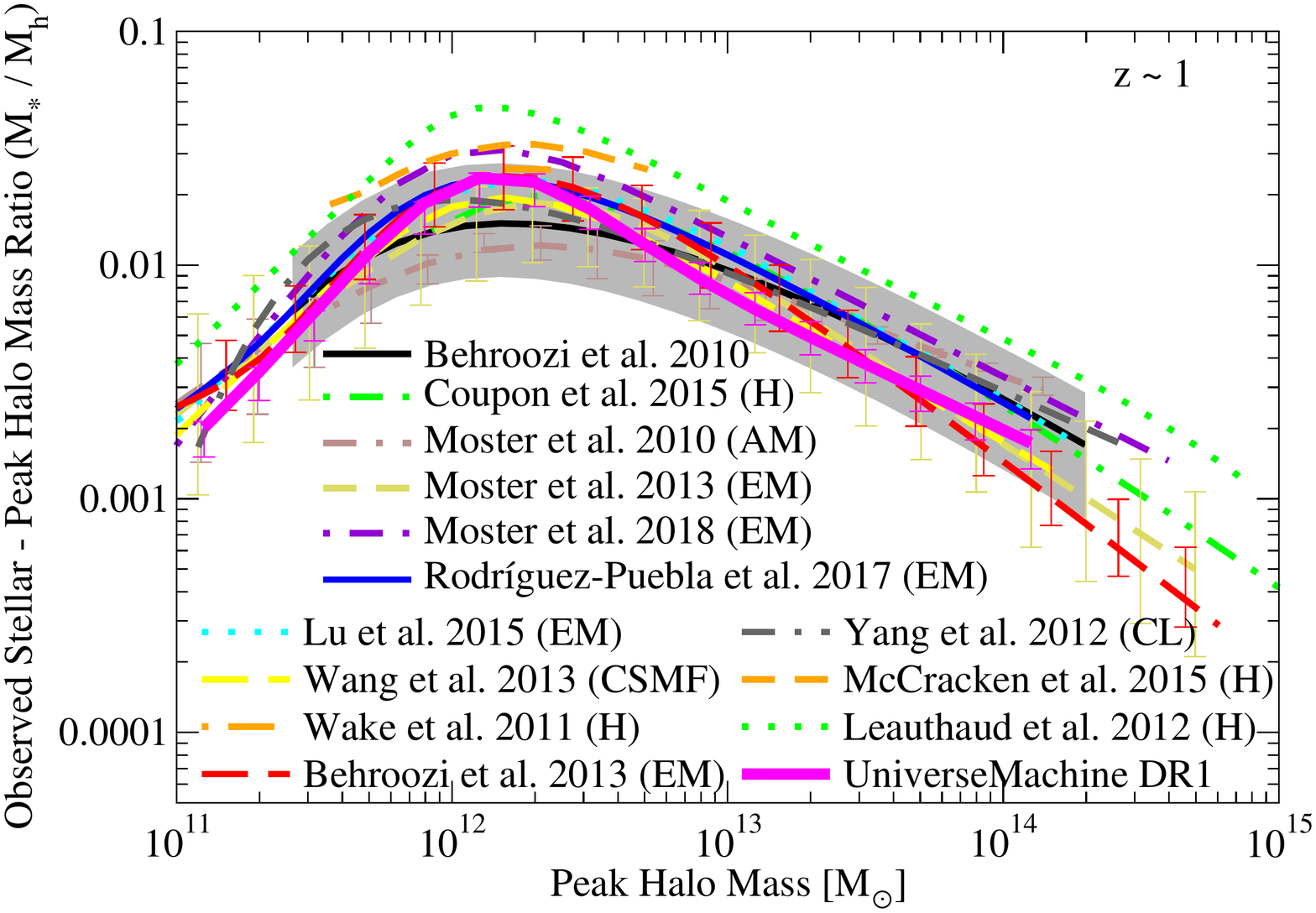}\plotgrace{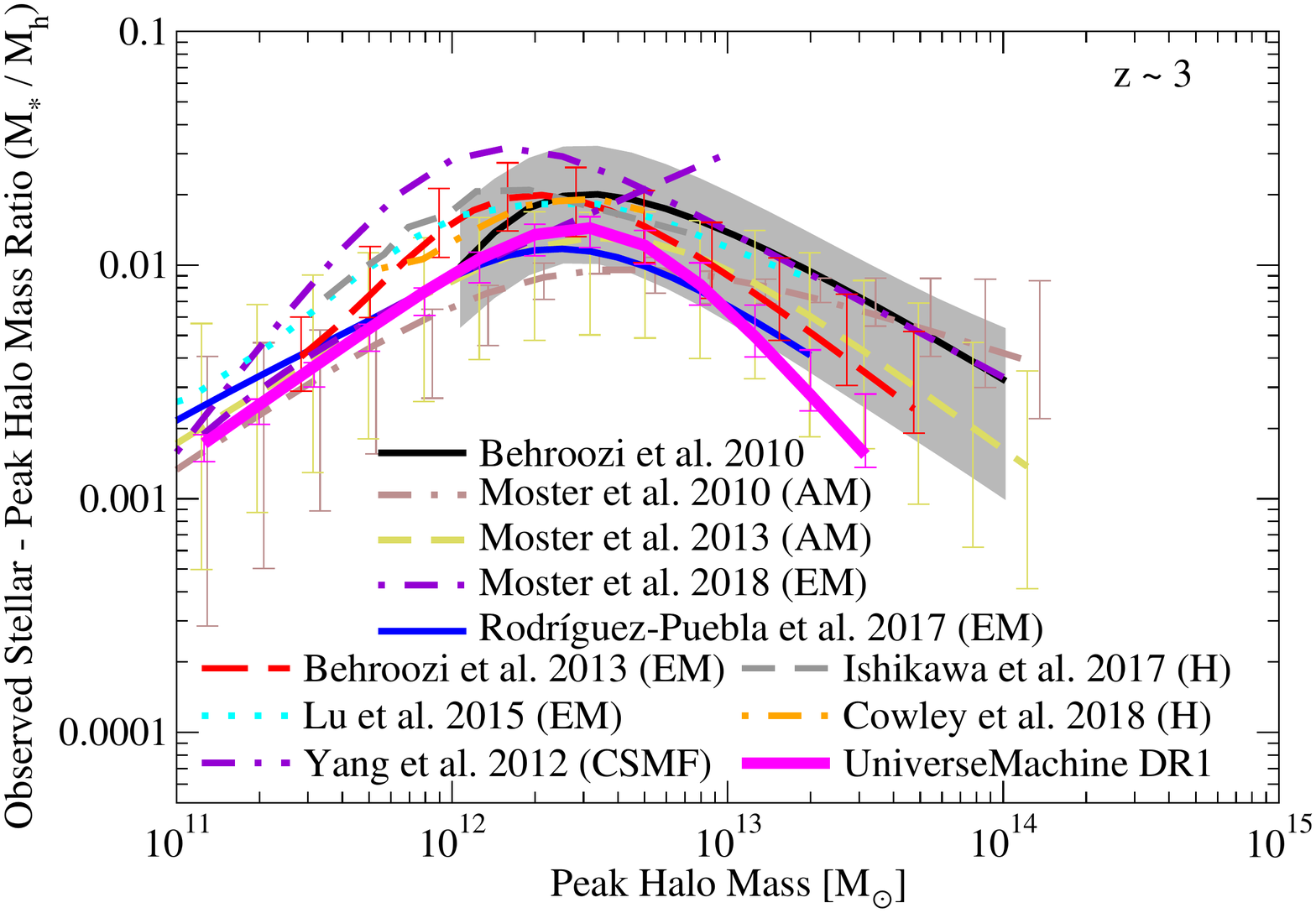}\\[-5ex]
\caption{Median stellar mass --- halo mass relations for our best-\fit{} model compared to previous results at $z\sim 1$ and $z \sim 3$.  Results compared include those from our previous works \protect\citep{Behroozi10,BWC13}, from empirical modeling \protect\citep[EM;][]{Moster12,Moster17,Lu15,RP17}, from abundance matching \citep[AM;][]{moster-09}, from Halo Occupation Distribution modeling \citep[H;][]{Wake11,Leauthaud12,Coupon15,McCracken15,Ishikawa17,Cowley18}, and Conditional Stellar Mass Function modeling \citep[CSMF;][]{Yang11,Wang12}.  \protect\cite{Yang11} reports best fits for two separate stellar mass functions; we show results from SMF2 at $z=3.0$.  Grey shaded regions correspond to the \onesigdist{} in \protect\cite{Behroozi10}.}
\label{f:comp_z1}
\vspace{10ex}
\end{figure*}
\begin{figure*}
\vspace{-5ex}
\plotgrace{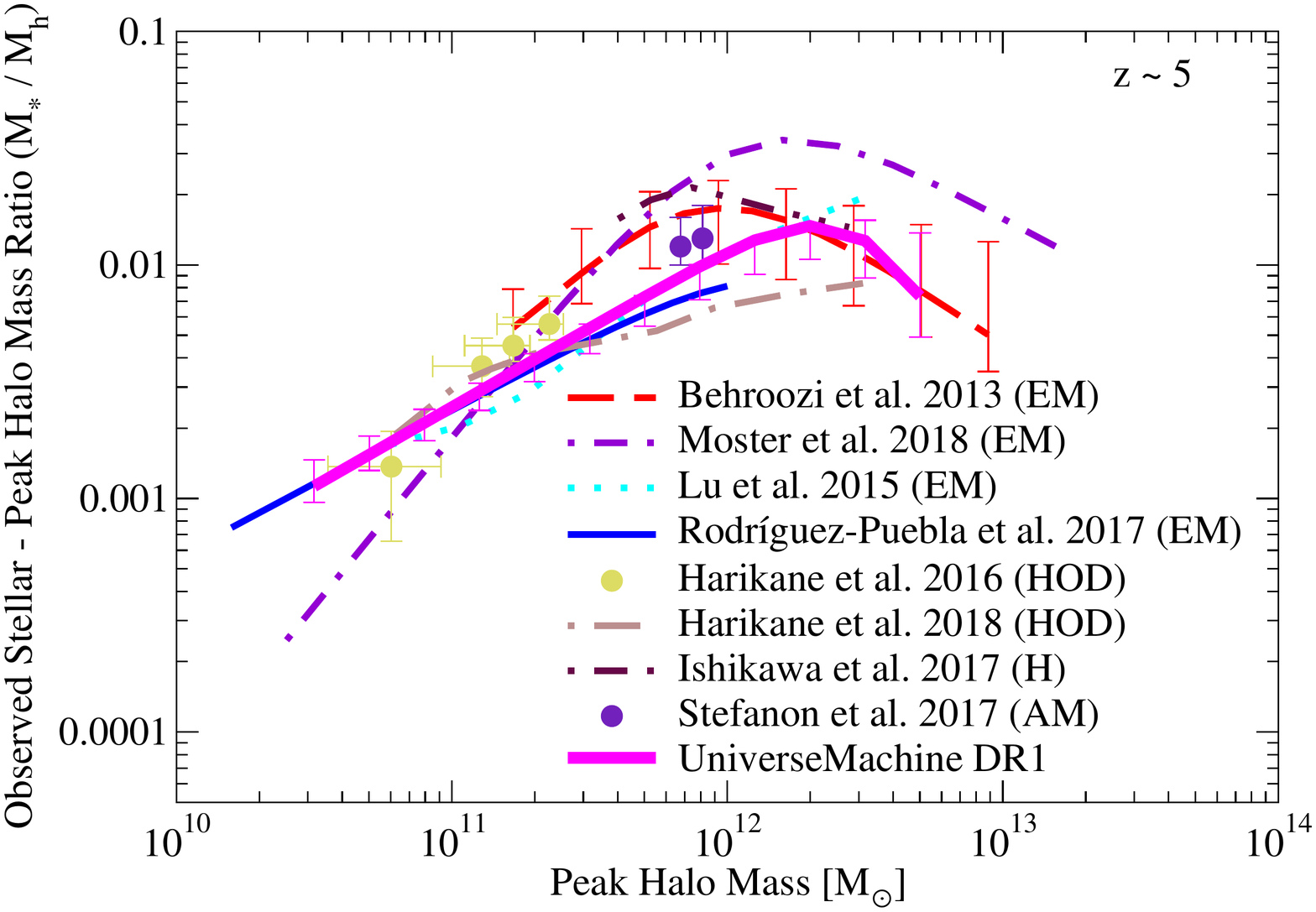}\plotgrace{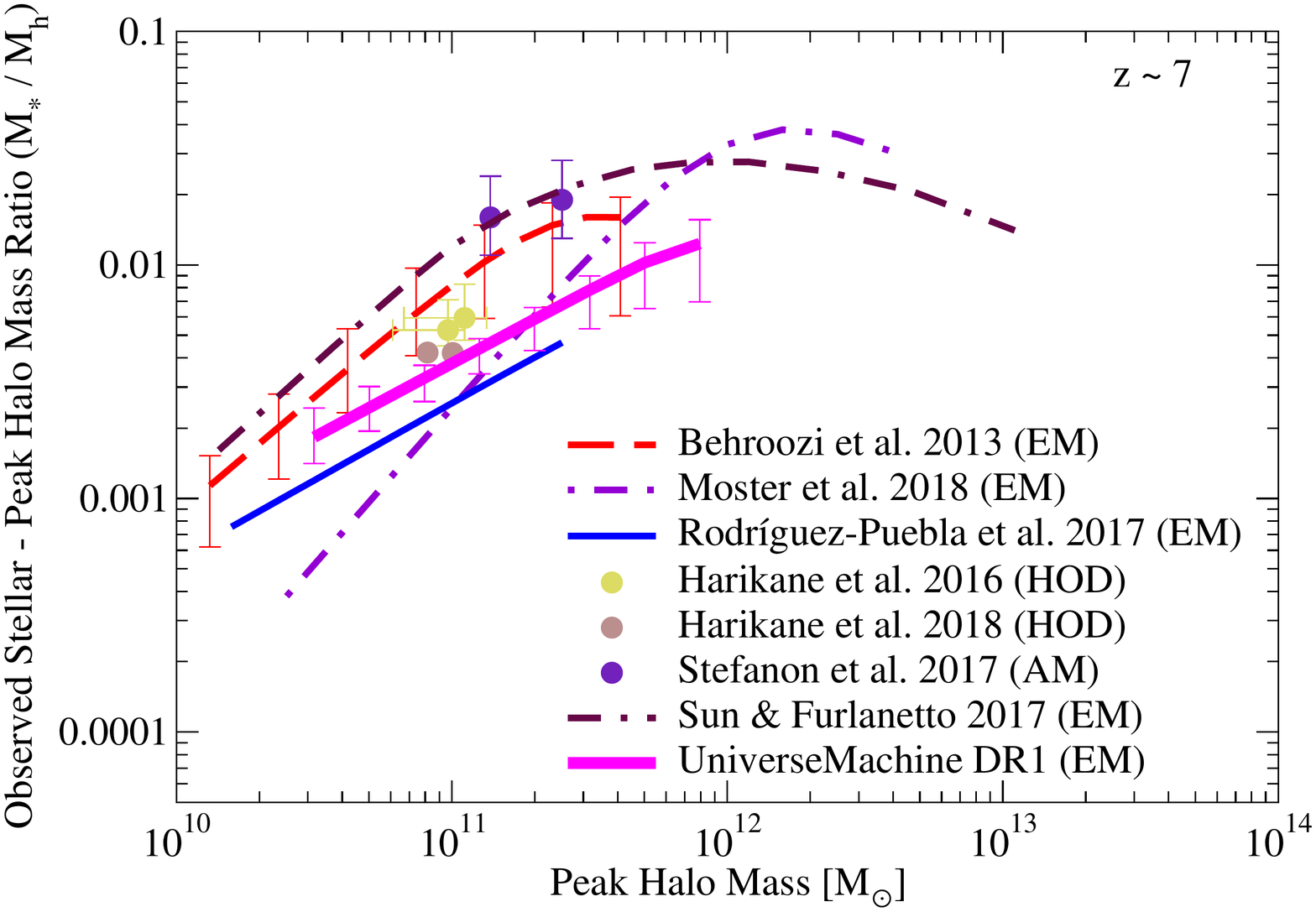}\\[-5ex]
\caption{\textbf{Left}: Median stellar mass --- halo mass relations for our best-\fit{} model compared to previous results at $z\sim 5$ and $z \sim 7$.  Results compared include those from our previous work \protect\citep{BWC13}, from empirical modeling \protect\citep[EM;][]{Lu15,Sun16,RP17,Moster17}, from abundance matching \citep[AM;][]{Stefanon17}, and from Halo Occupation Distribution modeling \citep[HOD;][]{Harikane16,Harikane18,Ishikawa17}.}
\label{f:comp_z4}
\end{figure*}

\begin{figure}
\vspace{-8ex}
\plotgrace{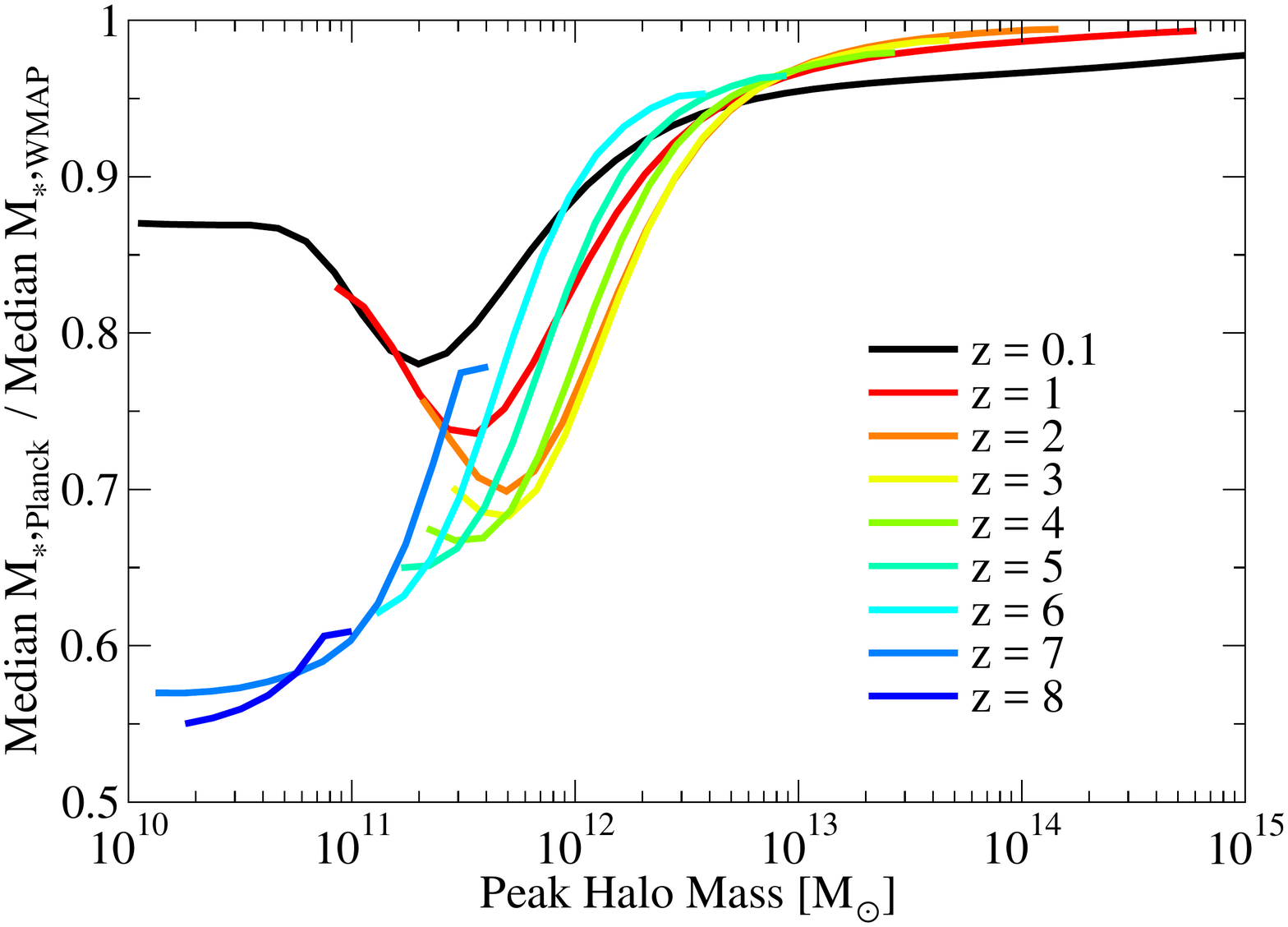}\\[-6ex]
\caption{Change in median stellar mass--halo mass relation between \textit{WMAP} and \textit{Planck} cosmologies for the SMHM relation in \protect\cite{BWC13}.}
\label{f:planck_smhm}
\vspace{-4ex}
\plotgrace{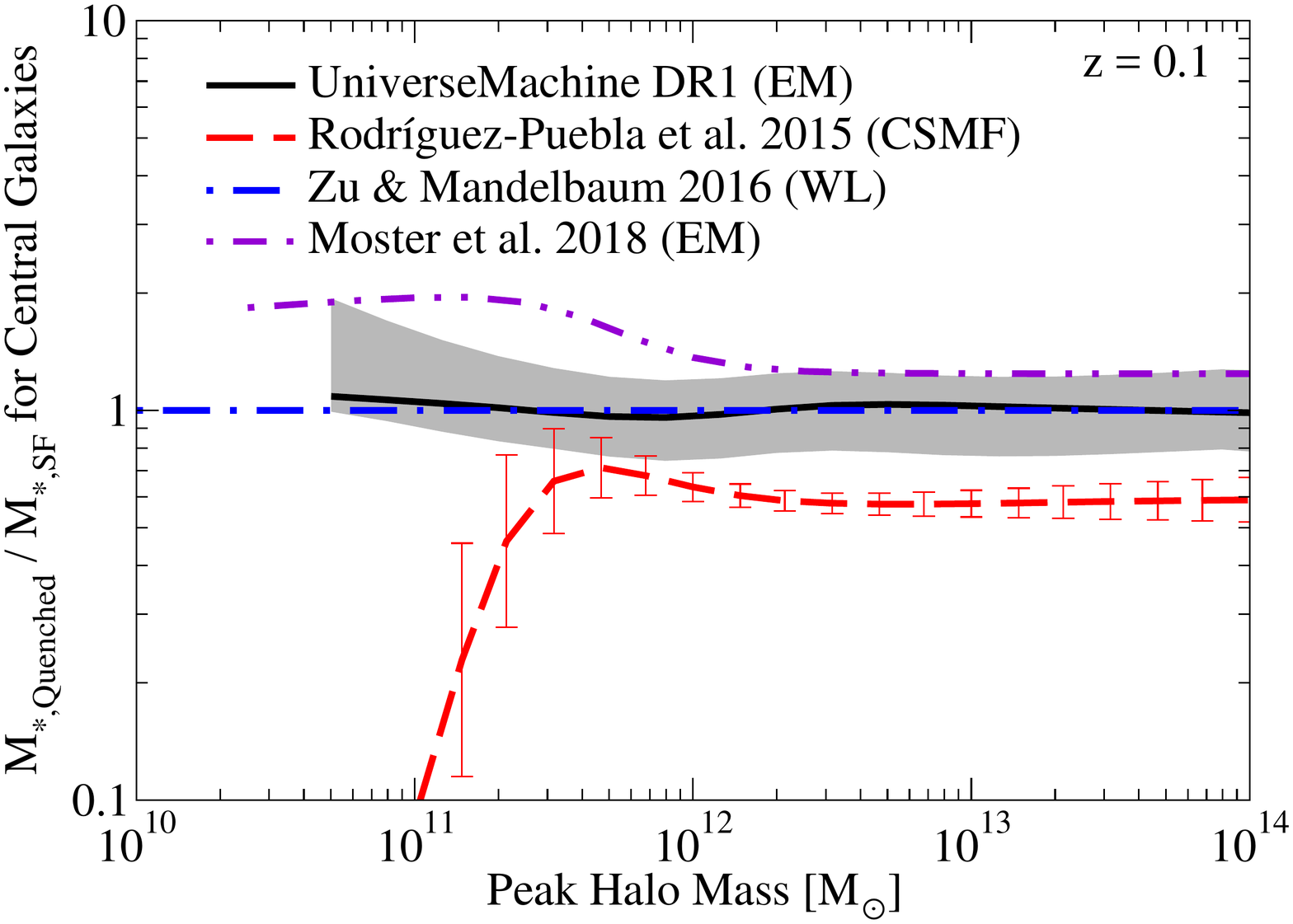}\\[-6ex]
\caption{Ratio of median stellar mass --- halo mass relations for quenched and star-forming central galaxies for our best-\fit{} model compared to previous results at $z=0.1$.  Results compared include those from Conditional Stellar Mass Function modeling \protect\citep[CSMF;][]{RP15}, from lensing \protect\citep[WL;][]{Zu16}, and from empirical modeling \citep[EM;][]{Moster17}.  The \textit{grey shaded region} shows the \onesigdist{} for statistical errors plus an additional 0.1 dex of systematic error in the  recovery of stellar mass for star-forming vs.\ quiescent galaxies (Appendix \ref{a:smf}).}
\label{f:smhm_sf_q_comp}
\end{figure}

\subsection{Orphan Galaxies}
\label{s:orphans}

Orphans are very strongly preferred in our posterior distribution, with satellite fractions roughly $25\%$ larger than would be expected from the N--body simulation alone.  As detailed in Appendix \ref{a:orphans}, this arises because of tension between the need for a large satellite fraction at a given stellar mass (to match the observed autocorrelation for all galaxies) and the need for low satellite star formation rates (to match differences between quenched and star-forming correlation functions).  This tension is not helped by the evolution of the stellar mass--halo mass (SMHM) relation to lower efficiencies at $z=1$ compared to $z=0$, as then satellites have even less stellar mass at infall compared to a non-evolving SMHM relation.  Many potential resolutions are not self-consistent; e.g., using a stellar mass proxy that gives additional stellar mass to satellites \citep[as in][]{Reddick12}.  \cite{Campbell17} suggests that it may be possible to resolve this tension self-consistently through other means, including by allowing satellite galaxies to grow after accretion and to have stellar masses that correlate with halo assembly history. Yet, although our model allows (and includes) both these alternative solutions, orphans are still strongly preferred; \cite{Moster17} reach a similar conclusion.

Observational data could constrain a more complex orphan model than examined here.  In our fiducial model, satellites are retained until they fall below a certain $\frac{\vmax}{\vmp}$ ratio (i.e., $T_\mathrm{merge}$).  That said, the most realistic choice of which satellites to retain could also depend on the orbit, the time since infall, and the simulation's resolution.   Additional data, such as the radial profiles of quenched and star-forming galaxies around groups and clusters \citep[e.g.,][]{Wetzel11}, could then provide observational constraints for these more complex orphan models.

\subsection{Comparison to Previous Results}

\begin{figure*}
\vspace{-5ex}
\plotgrace{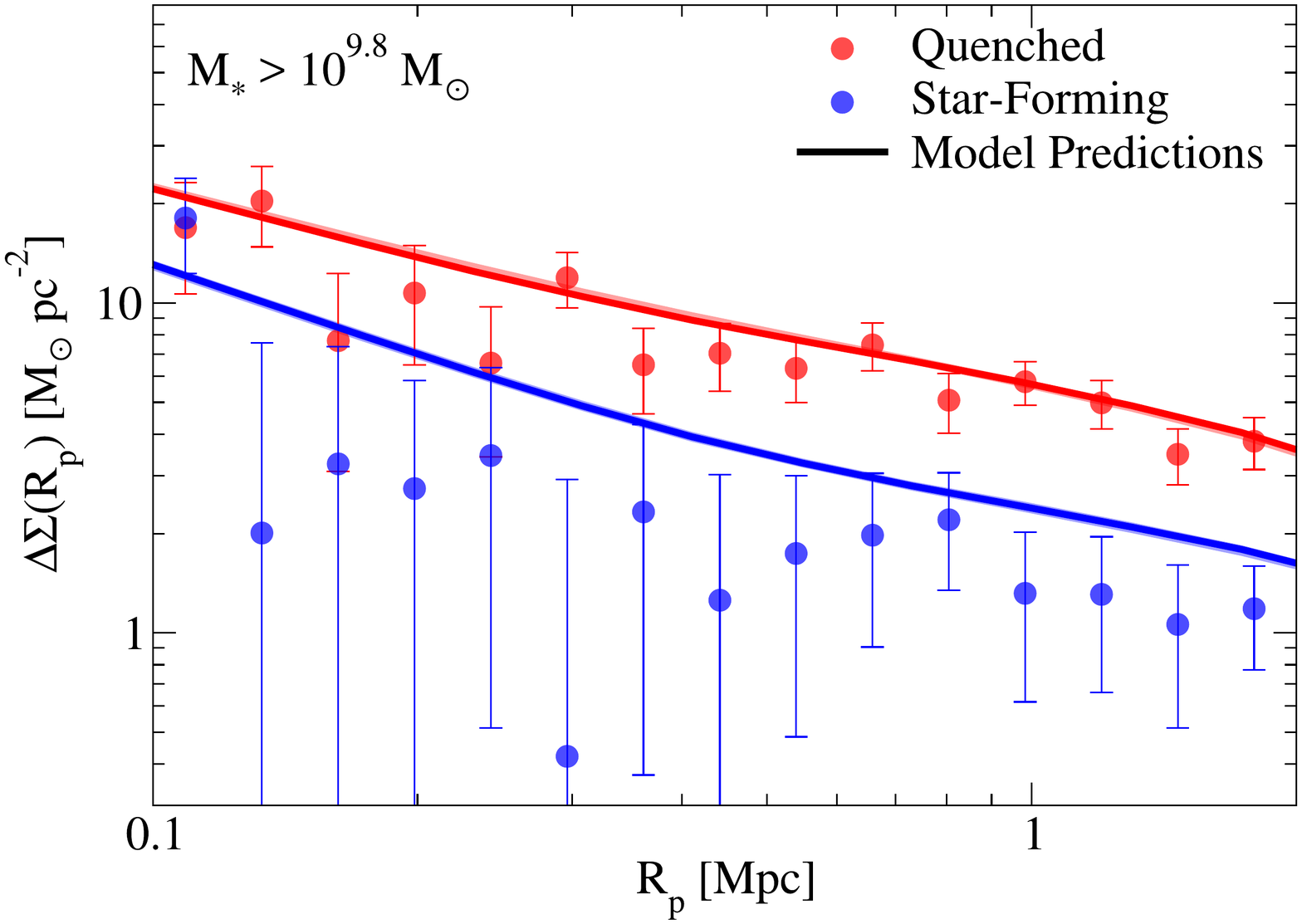}\plotgrace{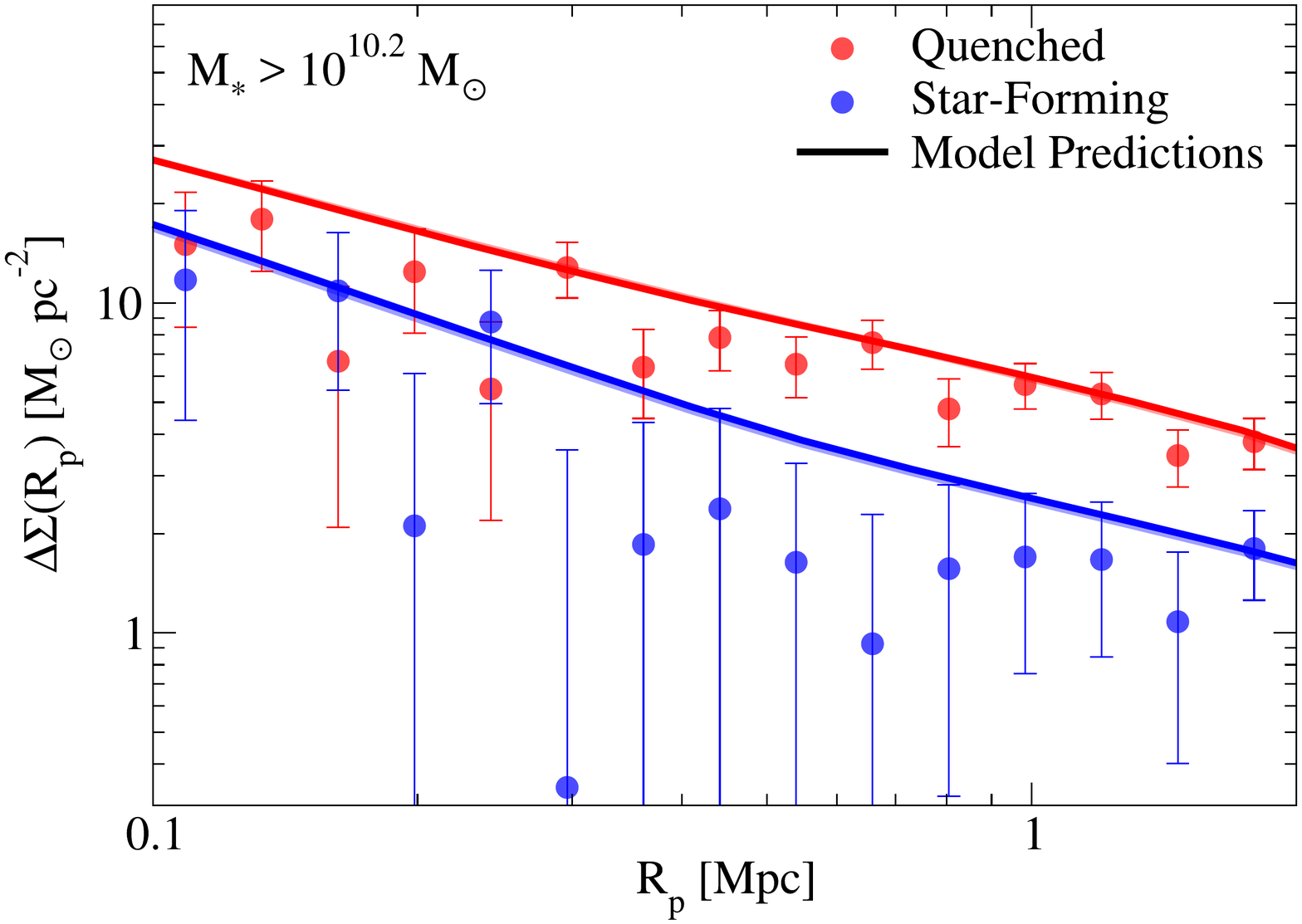}\\[-5ex]
\plotgrace{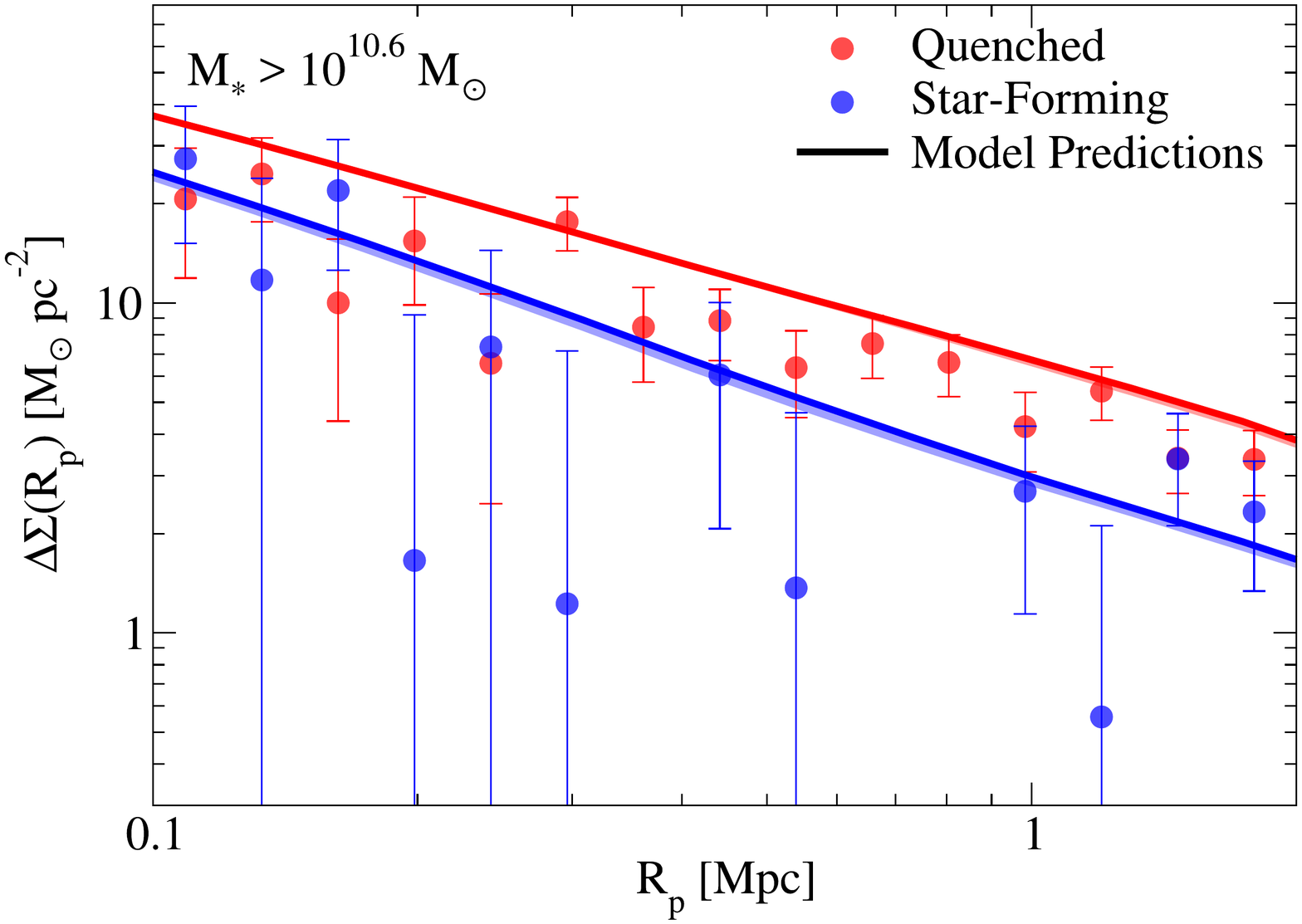}\plotgrace{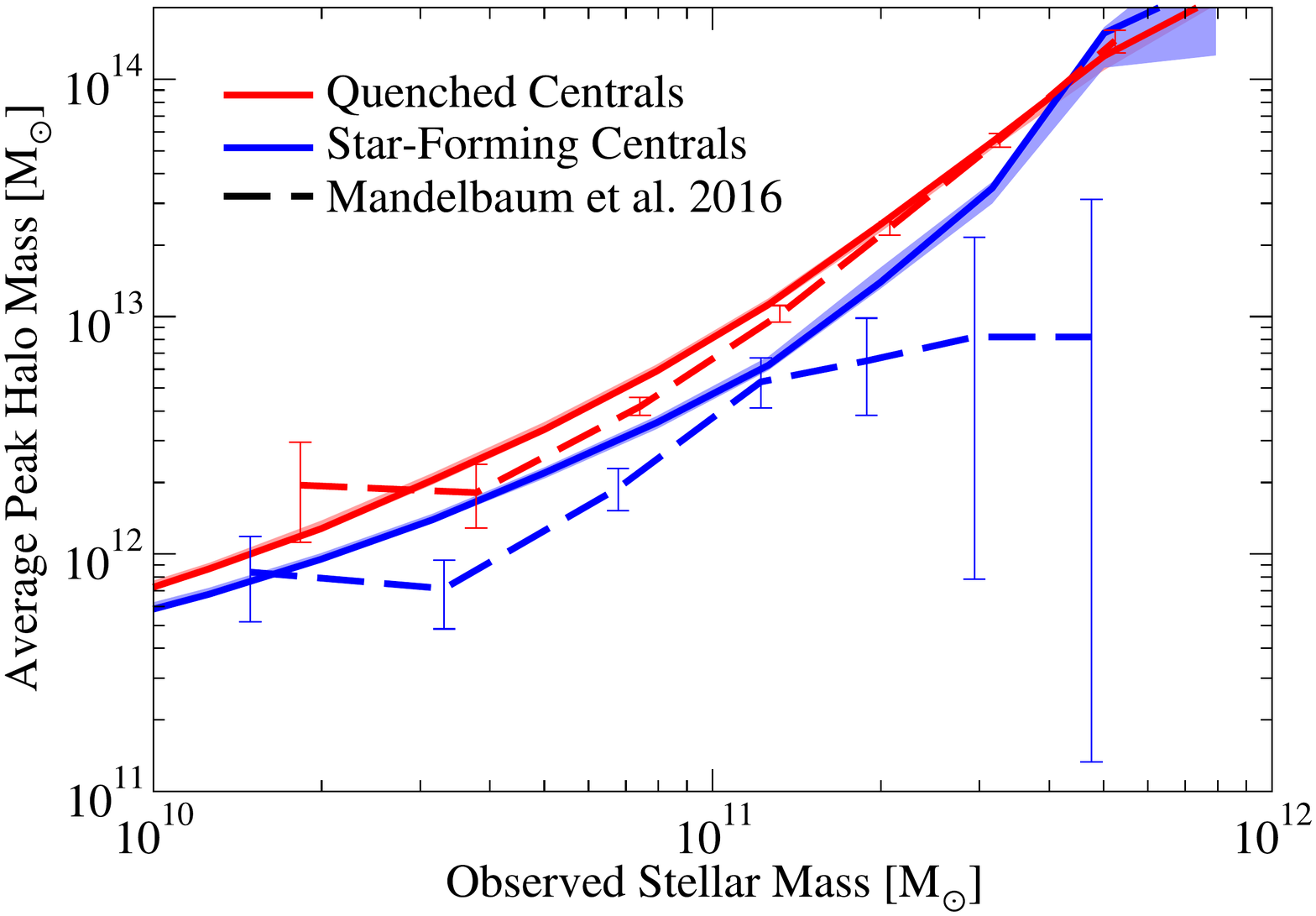}\\[-5ex]
\caption{\textbf{Top} and \textbf{bottom-left} panels: Weak lensing excess surface densities ($\Delta\Sigma$) from the best-\fit{} model at $z=0.1$ compared to observations \citep{Watson15} for three $M_*$ threshold samples. All distances and areas are in comoving units. \textbf{Bottom-right} panel: Comparison of the mean halo mass as a function of stellar mass for star-forming and quenched central galaxies to the results in \protect\cite{Mandelbaum16}.  \textbf{Notes:} consistent with the findings in \protect\cite{Leauthaud17} that lensing and correlation function measurements are difficult to reconcile, our predicted excess surface densities are $10-30$\% higher than the observations.}
\label{f:wl_comp}
\end{figure*}

\label{s:smhm_comp}

We compare the median stellar mass -- halo mass ratios from the model to past results at $z=0.1$ (Fig.\ \ref{f:comp_z0}), $z=1-3$ (Fig.\ \ref{f:comp_z1}), and $z=5-7$ (Fig.\ \ref{f:comp_z4}).  All results have been converted to use the same halo mass definition ($\mvir$; \citealt{mvir_conv}).  Abundance matching results have been further converted to our adopted \textit{Planck} cosmology (\S \ref{s:simulations}); as converting the cosmologies of other techniques requires re-running the original analysis, we have left those as-is.  \textit{Planck}-consistent cosmologies have more haloes at a given mass compared to \textit{WMAP}, especially at higher redshifts \citep{RP16b}.  The corresponding effect on the SMHM relation depends on $\frac{dM_\ast}{dM_h}$, with the result that haloes with $M_h < 10^{12}\Msun$ are more affected than haloes with $M_h > 10^{12}\Msun$ (Fig.\ \ref{f:planck_smhm}).

The comparison agrees extremely well across all redshifts, halo masses, and techniques, with the exception of $M_h > 10^{13}\Msun$ at $z<1$.  As discussed in Appendix \ref{a:smf}, this is due to revised photometry and fitting of massive galaxies in the SDSS, as well as deeper imaging of $z=0-1$ massive galaxies in ULTRAVISTA.  Using the revised stellar mass functions resolves past discrepancies between cluster-based SMHM relations and those derived from empirical models \citep{LinMohr04,Hansen09,Kravtsov14}.  The remaining difference between our results and \cite{Kravtsov14} is due to the latter counting all intracluster light as part of the galaxy.

For low-mass haloes ($M_h < 10^{11}\Msun$), the best-\fit{} model has a weaker upturn in the SMHM ratio than found by \cite{BWC13}; this is because \cite{BWC13} assumed a strong surface-brightness incompleteness correction for faint galaxies that is no longer observationally supported \citep{Williams16}.  This will make it easier to reconcile observed galaxy counts with the HI mass function and observed HI gas fractions in faint galaxies \citep{Popping15}.

The SMHM relation for star-forming vs.\ quiescent galaxies depends on the correlation between galaxy and halo assembly (\S \ref{s:smhm}) and the evolution of the SMHM relation \citep[see also][]{Moster17}.  As shown in Fig.\ \ref{f:smhm_sf_q_comp}, there remain significant differences across studies \citep[see also][]{Wechsler18}.  Our model is flexible in terms of both the SMHM relation evolution and the galaxy--halo assembly correlation, and suggests that the stellar mass--halo mass relation at fixed halo mass is similar for star-forming and quiescent central galaxies, matching the conclusion in \cite{Zu16}.  The results of \cite{Moster17} and \cite{RP15} give opposite conclusions of higher and lower (respectively) median stellar masses for quiescent compared to star-forming galaxies, despite using the same underlying data (correlation functions in the SDSS) to constrain their models.  In part, these divergent conclusions arise because correlation functions and weak lensing measurements are both very sensitive to satellite clustering; hence, small changes to the satellite halo occupation can lead to large changes in the inferred occupation for central galaxies.  Applying a cut to first remove satellites before measuring clustering, environment, or lensing (as in both this study and \citealt{Zu16}) is hence necessary to robustly determine SMHM differences for star-forming and quiescent central galaxies.  As noted in \cite{Zu16} and \cite{Moster17}, having an equivalent median stellar mass at fixed halo mass does \textit{not} imply that the median halo mass at fixed stellar mass will be equal for star-forming and quiescent galaxies.  Because the ratio of star-forming to quiescent galaxies drops rapidly with increasing halo mass, it is much more likely in this case that a given massive star-forming galaxy will be hosted by a lower-mass halo than a massive quiescent galaxy.

We show direct comparisons with lensing ($\Delta\Sigma$) measurements from \cite{Watson15} in Fig.\ \ref{f:wl_comp}.  While the trends are similar, $\Delta\Sigma$ is overestimated by $10-30\%$ for both quenched and star-forming galaxies in the best-\fit{} model.  \cite{Leauthaud17} also finds that matching galaxy autocorrelation functions results in mock catalogues that overpredict $\Delta\Sigma$ measurements by $20-40\%$.  We refer readers to \cite{Leauthaud17} for further discussion of this tension, while noting that correlation functions and weak lensing measurements are affected differently by the satellite occupation distribution adopted.  Fig.\ \ref{f:wl_comp} also shows a comparison with mean halo mass as a function of stellar mass for central galaxies.  \cite{Mandelbaum16} use stellar masses from the NYU-VAGC \citep{Blanton05}; we have corrected these to better correspond to the \cite{Brinchmann04} stellar masses underlying our correlation function measurements using the median offsets between star-forming and quenched galaxies (Fig.\ \ref{f:vagc_offsets}).  We do not attempt to correct the \cite{Mandelbaum16} results for photometry offsets \citep{Bernardi13} as this is less straightforward.  The qualitative effects would be both to increase galaxy masses (for $M_\ast > 10^{11}\Msun$) and to increase inferred host halo masses (via decreasing the scatter in halo mass at fixed galaxy mass; \citealt{Kravtsov14}).

In parallel with empirical models, hydrodynamical simulations and semi-analytic models have shown increased success in matching multiple observations \citep[e.g.,][]{Schaye15,Somerville15,Henriques15,Dubois16,Dolag16,Croton16,Springel18}.  A proper comparison with these results is beyond the scope of this paper; it is hence deferred to future work.

Additional comparisons between figures in this paper and in \cite{BWC13} are available \href{https://www.peterbehroozi.com/data}{\textbf{online}}.

\subsection{Evolution in the Stellar Mass -- Halo Mass Relation at \textit{z}>4}

\label{s:smhm_evolution}

There have been contradictory claims regarding the $z>4$ evolution of the stellar mass--halo mass relation.  Many recent studies, including this one, find either significant evolution \citep{BWC13,BehrooziHighZ,Finkelstein15b,Harikane16,Harikane18,Sun16} or modest evolution \citep{Moster17}.  \cite{Stefanon17} claims to find no evolution, but as is evident from Fig.\ \ref{f:comp_z4}, their findings in fact suggest even stronger evolution than our current results due to the slope of the SMHM relation.  In contrast, \cite{RP17} find very little evolution at $z>4$.  This may be in part due to the fact that their assumed scatter in observed vs.\ true stellar mass grows as $0.1+0.05z$ with no upper bound, so that scatter is $0.5$ dex at $z=8$ and $0.6$ dex at $z=10$, well beyond the $0.3$ dex limit assumed here.  In addition, it may be due to the use of UV--SM relations at high redshifts that could underestimate true stellar masses (Appendix \ref{a:uvsm}).  Despite this, the relative uncertainty in stellar mass estimates at $z>4$ remains large, as shown in Fig.\ 6 of \cite{Moster17}.  Adding to the uncertainty, a non-evolving star formation efficiency (defined as $SFR / \dot{M_h}$) can plausibly fit UV luminosity functions from $z=10$ to $z=0$ \citep{Mason15,Harikane18,Tacchella18}.  As discussed in \cite{Behroozi13}, this is equivalent to a non-evolving SMHM ratio, and so the \cite{Harikane18} star formation efficiency model is in tension with the evolving SMHM ratio found in the same paper via clustering analyses.  While the match to UV luminosity functions is degenerate with the assumed dust evolution, it is clear that a consistent picture of stellar masses, star formation rates, and dust has yet to emerge at $z>4$.  \textit{NIRCam} and \textit{NIRSpec} on the \textit{James Webb Space Telescope} will hence be instrumental in settling this debate \citep[see][for a review]{Kalirai18}.

\subsection{Addressing Assumptions and Uncertainties with Additional Data and Modeling}

\label{s:assumptions}

The empirical model herein has broad classes of assumptions relating to the average connection between galaxy SFR and halo mass, the correlation between individual halo and galaxy growth, and the relevance of dark matter simulations for modeling observed galaxies.  We address each of these in turn.

External validation of predictions including the stellar mass---halo mass relation (\S \ref{s:smhm_comp}) suggests that the model framework is flexible enough to capture how stellar mass and SFR depend on halo mass and redshift.  Additional observations will be helpful especially for $z>4$ galaxies (see \S \ref{s:smhm_evolution} and Appendix \ref{a:uvsm}); the model would also benefit by including more constraints on SFRs in massive clusters at $z\sim1-2$ (see Fig.\ \ref{f:sfr_errors}, left panel).  Nonetheless, the primary sources of uncertainty for most galaxies are systematic ones--e.g., the conversion between luminosity and physical stellar masses and star formation rates (see \S \ref{s:syst_uncertainties} and Appendices \ref{a:data}, \ref{a:uvsm}, and \ref{a:imf}).  Addressing these uncertainties with more observations may be difficult due to their dependence on rare stellar populations \citep[e.g.,][]{Conroy09} and/or uncertain dust geometry distributions \citep[e.g.,][]{Narayanan18}.  Forward modeling directly (and only) to luminosities and colours may help, approaching the method in \cite{Popp15}.  For massive galaxies, there are additional systematics with measuring luminosity profiles at different redshifts (Appendix \ref{a:smf}).  These may be improved with surveys that are both wide enough to find many massive galaxies and deep enough to accurately measure their profiles, such as the Hyper Suprime-Cam Subaru Strategic Program survey (HSC-SSP; \citealt{Aihara18}).

Empirically connecting individual galaxy and halo growth is a new field, so the observables that best constrain this connection are not yet known.  Appendix \ref{a:alternatives} discusses several observables that rule out alternative models.   Clustering gives powerful constraints on satellite behaviour, so it will be important to continue extending the redshift and mass range of clustering measurements.  Despite the observational difficulty, clustering for quiescent galaxies is especially important, as the largest constraining power comes from the contrast between quenched and star-forming galaxy clustering.  For isolated central galaxies (i.e., those that have never crossed the virial radius of a larger halo), the importance of the host halo's assembly history remains debated \citep{Tinker16}.  Connecting observed splashback radii \citep[e.g., as claimed in][]{More16} with dark matter accretion rates is a promising path forward for such galaxies, and is planned for future work.

With dark matter simulations, the most critical issues remain subhalo finding \citep{Onions12,BehrooziNotts}, satellite disruption (see \S \ref{s:orphans} and Appendix \ref{a:orphans}) and the effects of baryons \citep[see, e.g.,][for recent work]{Nadler17,Chua17}.  Central halo finding is a more minor issue, especially when (as in this paper) $\vmax$ is used as the mass proxy \citep{Knebe11,Knebe13}.  Subhalo finding is more than a definitional issue, as different halo finders introduce different orbit dependencies in recovered halo properties \citep{BehrooziNotts}; temporal halo finders (including HBT/HBT+; \citealt{Han12,Han18}) are most immune to this effect.  As noted in \S \ref{s:orphans} observed satellite properties in group catalogues at different redshifts may represent a promising way to calibrate empirical orphan models; this to some extent will include the effect of baryons.  Remaining effects of baryons on halo potential wells (e.g., making them less triaxial; \citealt{Abadi10}) are more difficult to observe, and may require higher-order statistics such as three-point functions to capture.

\subsection{Future Directions in Empirical Modeling}
\label{s:future}

With empirical modeling, physical constraints from different observables can be combined even when the best underlying parameter space is not known.  The same general framework may be applied to many other problems, even outside of galaxy formation.  Whereas this paper constrained galaxy SFR as a function of relevant parameters (halo $\vmax$, accretion rate, and redshift), one may use the same technique to constrain how any observable $X$ depends on arbitrary parameters $Y_1,\ldots,Y_n$.  As of this writing, there are ongoing empirical efforts to constrain how galaxy size, morphology, metallicity, gas content, black hole mass, gamma-ray burst rate, supernova rate, and dust relate to underlying properties of the host dark matter halo, its assembly history, and the resulting galaxy's assembly history.  These applications promise a wealth of new physical insight about our Universe in the years to come \citep{BehrooziDecadal}.

\section{Conclusions}
\label{s:conclusions}
Our model (\S \ref{s:methodology}) flexibly parametrizes the correlation between galaxy assembly and halo assembly, and is able to self-consistently match a broad array of observational data (\S \ref{s:results}, \ref{s:stochasticity}, \ref{s:smhm_comp}).  The following results are \textbf{robust to the modeling uncertainties}.\\

\noindent{}For the stellar mass--halo mass relation:
\begin{itemize}
\item Consistent with past results, haloes near $10^{12}\Msun$ are most efficient (20-40\%) at turning gas into stars at all redshifts (\S \ref{s:smhm}).
\item While the stellar mass--halo mass relation does not evolve significantly from $z=0$ to $z\sim 5$, significant evolution does occur at higher redshifts (\S \ref{s:smhm}) with present data (\S \ref{s:smhm_evolution}).
\item At $z=0$, massive quiescent galaxies reside in higher-mass haloes than massive star-forming galaxies (\S \ref{s:smhm_comp}).  Due to scatter in the stellar mass--halo mass relationship, it is also true that quiescent galaxies are more massive at fixed halo mass than star-forming galaxies for $M_h > 10^{13}\Msun$ at $z=0$ (\S \ref{s:smhm}).
\end{itemize}

\noindent{}For quenching and galaxy--halo assembly correlations:
\begin{itemize}
\item Quenching is highly correlated with halo mass, where a difference of $\lesssim 1.5$ dex in host halo mass separates largely star-forming populations from largely quenched ones (\S \ref{s:average_sfrs}).
\item At $z<1$, satellite quenching becomes more important, so quenching happens over a broader range of halo masses (\S \ref{s:average_sfrs}).
\item The correlation between galaxy and halo assembly is strong, but not perfect (correlation coefficient $\sim 0.5-0.6$; \S \ref{s:corr}).  High-redshift observations of correlation functions and weak lensing will test the galaxy--halo assembly correlation strength at $z>1$; we also make predictions for these measurements (\S \ref{s:corr_pred}).
\item Average quenched fractions robustly decrease with increasing redshift at fixed halo mass (\S \ref{s:average_sfrs}), suggesting that cooling times are not solely responsible for quenching (\S \ref{s:central_quenching}).
\item Except for cluster cores, where satellites quench very quickly, the fraction of quenched satellites minus the fraction of quenched centrals never exceeds $35\%$ at fixed stellar mass (\S \ref{s:satellites}).
\item Satellites quench faster the more massive they are (\S \ref{s:satellites}).
\end{itemize}

\noindent{}In addition:
\begin{itemize}
\item Most galaxy mass formed in-situ for $M_h \le 10^{12}\Msun$ haloes (\S \ref{s:in_situ}).  Low-mass galaxies ($M_\ast < 10^9\Msun$) are predicted to have significant ($\gtrsim 1$ dex) scatter in their intrahalo light (\S \ref{s:in_situ}).
\item \textit{Planck} cosmologies have more low-mass haloes at high redshifts, lowering the inferred stellar mass--halo mass relation by up to 0.3 dex compared to \textit{WMAP} cosmologies (\S \ref{s:smhm_comp}).
\end{itemize}

The following results, while robust given the observational constraints, \textbf{depend on the modeling assumptions}:
\begin{itemize}
\item Satellite galaxies have very broad ($3-5$ Gyr) quenching delay-time distributions after infall (\S \ref{s:satellites}).  
\item If the correlation between galaxy and halo assembly history is also strong for central galaxies that have not interacted with larger haloes, then past rejuvenation (quenching followed by renewed star formation) is common for $M_\ast\sim 10^{11}\Msun$ galaxies and $M_h \sim 10^{12.5}\Msun$ haloes at $z=0$ (\S \ref{s:corr}).
\item Quenched galaxies at $z=0$ have significantly different average star formation histories than star-forming ones out to $\sim 3$ Gyr, which can aid in tracing average galaxy populations through cosmic time (\S \ref{s:stochasticity}).  However, individual galaxies have significant scatter in star formation histories, making the comparison across redshifts less well constrained (\S \ref{s:stochasticity}).
\end{itemize}
All code, mock catalogues, and data products (\S \ref{s:online_data}) are available \href{https://www.peterbehroozi.com/data}{\textbf{online}}.

\section*{Acknowledgements}

We thank Matt Becker, Andreas Berlind, Frank van den Bosch, Rychard Bouwens, Kevin Bundy, Blakesley Burkhart, Joanne Cohn, Alison Coil, Ryan Endsley, Sandy Faber, Steve Finkelstein, Marijn Franx, Alexie Leauthaud, Susan Kassin, Rob Kennicutt, Andrey Kravtsov, Mariska Kriek, Danilo Marchesini, Surhud More, Ben Moster, Adam Muzzin, Camilla Pacifici, Joel Primack, Eliot Quataert, Aldo Rodr\'iguez-Puebla, Joop Schaye, Uro\v{s} Seljak, Rachel Somerville, Mimi Song, Dan Stark, Jeremy Tinker, Ryan Trainor, Ramin Skibba, Doug Watson, Sarah Wellons, Andrew Wetzel, Martin White, Simon White, Andrew Zentner, and Zheng Zheng for very stimulating discussions.  We also thank the referee, Darren Croton, for several very helpful comments that improved the clarity of this paper.

We thank Mariangela Bernardi, Jarle Brinchmann, Alison Coil, Danilo Marchesini, John Moustakas, Adam Muzzin, Camilla Pacifici, Aldo Rodr\'iguez-Puebla, Samir Salim, Brett Salmon, Mimi Song, Adam Tomczak, and Kate Whitaker for providing their data in electronic form.  Data compilations for other studies used in this paper were made much more accurate and efficient by the online \textsc{WebPlotDigitizer} code.\footnote{\url{https://automeris.io/WebPlotDigitizer/}}  This research has made extensive use of the arXiv and NASA's Astrophysics Data System.

PB was partially supported by a Giacconi Fellowship from the Space Telescope Science Institute.  PB was also partially supported through program number HST-HF2-51353.001-A, provided by NASA through a Hubble Fellowship grant from the Space Telescope Science Institute, which is operated by the Association of Universities for Research in Astronomy, Incorporated, under NASA contract NAS5-26555.

PB, RHW, and APH thank the Aspen Center for Physics and the NSF (Grant \#1066293) for hospitality during the development of this paper.  This research was also supported in part by the National Science Foundation (NSF PHY11-25915), through a grant to KITP during the ``Cold Universe'' and the ``Galaxy-Halo Connection Across Cosmic Time'' programs.  This research was also supported by the Munich Institute for Astro- and Particle Physics (MIAPP) of the DFG cluster of excellence ``Origin and Structure of the Universe.''

This research used the Edison and Cori supercomputers of the National Energy Research Scientific Computing Center, a DOE Office of Science User Facility supported by the Office of Science of the U.S. Department of Energy under Contract No. DE-AC02-05CH11231.  An allocation of computer time from the UA Research Computing High Performance Computing (HPC) at the University of Arizona is gratefully acknowledged.  This work performed (in part) at SLAC under DOE Contract DE-AC02-76SF00515.  The \textit{Bolshoi-Planck} simulation was performed by Anatoly Klypin within the Bolshoi project of the University of California High-Performance AstroComputing Center (UC-HiPACC; PI Joel Primack).  Resources supporting this work were provided by the NASA High-End Computing (HEC) Program through the NASA Advanced Supercomputing (NAS) Division at Ames Research Center.  The \textit{MultiDark-Planck2} simulation was performed by Gustavo Yepes on the SuperMUC supercomputer at LRZ (Leibniz-Rechenzentrum) using time granted by PRACE, project number 012060963 (PI Stefan Gottloeber).

Funding for SDSS-III\footnote{\url{http://www.sdss3.org/}} has been provided by the Alfred P. Sloan Foundation, the Participating Institutions, the National Science Foundation, and the U.S. Department of Energy Office of Science.

SDSS-III is managed by the Astrophysical Research Consortium for the Participating Institutions of the SDSS-III Collaboration including the University of Arizona, the Brazilian Participation Group, Brookhaven National Laboratory, Carnegie Mellon University, University of Florida, the French Participation Group, the German Participation Group, Harvard University, the Instituto de Astrofisica de Canarias, the Michigan State/Notre Dame/JINA Participation Group, Johns Hopkins University, Lawrence Berkeley National Laboratory, Max Planck Institute for Astrophysics, Max Planck Institute for Extraterrestrial Physics, New Mexico State University, New York University, Ohio State University, Pennsylvania State University, University of Portsmouth, Princeton University, the Spanish Participation Group, University of Tokyo, University of Utah, Vanderbilt University, University of Virginia, University of Washington, and Yale University.

GAMA\footnote{\url{http://www.gama-survey.org/}} is a joint European-Australasian project based around a spectroscopic campaign using the Anglo-Australian Telescope. The GAMA input catalogue is based on data taken from the Sloan Digital Sky Survey and the UKIRT Infrared Deep Sky Survey. Complementary imaging of the GAMA regions is being obtained by a number of independent survey programmes including GALEX MIS, VST KiDS, VISTA VIKING, WISE, Herschel-ATLAS, GMRT and ASKAP providing UV to radio coverage. GAMA is funded by the STFC (UK), the ARC (Australia), the AAO, and the participating institutions.

{\footnotesize
\bibliography{master_bib}
}

\appendix
\section{Alternate Parameterizations of Halo Assembly History}
\label{a:alternatives}

The following alternate conditional abundance matching (CAM) parameters were tried and rejected, as detailed below.

\subsection{Classical Age-Matching ($z_\mathrm{starve}$)}

CAM using the $z_\mathrm{starve}$ parameter is known to reproduce galaxy clustering, weak lensing, and radial quenching profiles around groups and clusters \citep{Hearin13,Hearin14,Watson15}.  While $z_\mathrm{starve}$ is a complicated function, for most haloes it is equivalent to rank ordering on $\frac{c_\mathrm{acc}}{a_\mathrm{acc}}$, where $c_\mathrm{acc}$, and $a_\mathrm{acc}$ are the concentration and scale factor (respectively) at accretion (for satellites) or at the present day (for centrals).

There are several minor issues with $z_\mathrm{starve}$ that, while all fixable, would have made the parameter even less transparently physical than it already is.  These include:\\[-4ex]
\begin{enumerate}
\item Special treatment for satellites (in the form of $c_\mathrm{acc}$ and $a_\mathrm{acc}$) leading to unphysical discontinuities for backsplash haloes (i.e., haloes that enter and leave the virial radius of a larger halo).
\item No orbit-dependent SFRs for satellites.
\item Increased concentrations during the early stages of major mergers would result in \textit{decreased} SFRs, instead of the expected boost prior to the merger.  Afterwards, decreased concentrations would result in \textit{increased} SFRs, instead of the expected post-starburst phase.
\end{enumerate}

\subsection{Mass Accretion History}

This has been used by many other studies \citep{Becker15,RP16,Cohn16,Moster17}.  Satellites are the clearest issue: while the fraction of satellite haloes increases towards lower halo masses \citep{Rockstar,RP16b}, the fraction of quenched galaxies decreases \citep{Salim07,Geha12}.  Hence, many satellite galaxies are still forming stars even though their host (sub)haloes are losing mass.  While it's possible to rank-order satellites on their mass-loss rate, measuring subhalo masses is notoriously difficult \citep{Onions12}, and is thus extremely sensitive to the halo finder.  Alternate solutions include measuring the time since $M_\mathrm{peak}$ was reached \citep{Moster17}, although this erases any differences between different orbits after infall.

With mass accretion rates, an open question is which mass definition best correlates with galaxy assembly (e.g., a specified spherical overdensity, the mass within the splashback region, etc.).  A related issue is that pseudo-evolution (i.e., ``mass accretion'' due to a changing overdensity definition) dominates the mass accretion rate at $z=0$ \citep{Diemer13}, so that for most central haloes, the mass accretion rate is really measuring the halo concentration (i.e., it becomes equivalent to using $z_\mathrm{starve}$).  As shown in \cite{Wetzel15}, this may not be a problem after all, as this ``mass accretion rate'' is a reasonable estimate of the gas that actually makes it to the halo center---also suggesting that $z_\mathrm{starve}$ works in part due to its correlation with mass assembly.

A more serious issue is that current studies correlating halo mass growth with galaxy SFRs do not distinguish between smooth mass accretion and mergers.  \cite{BehrooziMM} showed that field galaxies in haloes undergoing major mergers (observationally identifiable as close galaxy pairs) have nearly identical quenched fractions as more typical field galaxies; mock observations applied to mock catalogues with mass accretion --  SFR correlations were significantly discrepant from the real observations.  This may be fixable by correlating only smooth mass accretion rates with galaxy SFRs (i.e., discounting the effects of major and minor mergers), but doing so would require higher resolution simulations than are currently practical to use.

\subsection{Tidal Forces}

Tidal forces are robustly calculable for satellites and would also introduce clearly orbit-dependent effects in their star formation histories.  Yet, connecting high tidal forces with quenching would require that voids (i.e., regions with low tidal forces) should primarily contain star-forming galaxies, which has less observational support.  We tested this in the Sloan Digital Sky Survey (SDSS), selecting extremely isolated $L^*$ galaxies (i.e., a stellar mass-complete sample with $10^{10} < \mstar / \Msun < 10^{10.5}$; no larger neighbour within 4 Mpc projected or 2000 km/s redshift distance) and comparing them to ``normal'' $L^*$ field galaxies (i.e., same stellar mass cut; no larger neighbour within 500 kpc projected or 1000 km/s redshift distance).  Using mock catalogues from \cite{BehrooziMM}, we verified that the majority of extremely isolated galaxies had tidal forces in the 15$^\mathrm{th}$ percentile or lower, and that 90\% had tidal forces in the 50$^\mathrm{th}$ percentile or lower, as compared to the normal field galaxies.  However, in the SDSS, quenched fractions were indistinguishable between the extremely isolated galaxies and the normal field galaxies (see also \citealt{Croton08}).

Separately, \cite{Lee16} has examined central halo assembly as a function of environment.  While many halo properties (concentration, spin, mass accretion history) are strongly affected in dense environments with large tidal forces, halo properties in extremely underdense environments (with very low tidal forces) are almost indistinguishable from halo properties in median-density environments.  Hence, weak tidal forces do not appear to correlate with internal halo (and presumably galaxy) properties to the same extent as strong ones do.

\subsection{Simpler $\vmax$ Prescriptions}

We considered many simpler $\vmax$ prescriptions, including $\frac{v_\mathrm{max}(z_\mathrm{now})}{v_\mathrm{max}(z_\mathrm{dyn})}$.  However, this prescription required \textit{continuing} mass loss for quenching.  Hence, reinfalling satellites were not quenched at high enough rates, regardless of the number of orbits they had made.  This resulted in not enough separation between correlation functions for quenched and star-forming galaxies.  We also considered $\frac{v_\mathrm{max}(z_\mathrm{now})}{\vmp}$ (and its close relative $\frac{v_\mathrm{max}(z_\mathrm{now})}{v_\mathrm{peak}}$).  This dramatically improved clustering, but then caused issues for central galaxies: most centrals have $v_\mathrm{max}(z_\mathrm{now})= \vmp$, as they are continuously growing in mass.  Similarly, $\frac{v_\mathrm{max}(z_\mathrm{now})}{v_\mathrm{peak}}$ was strongly peaked---so that small errors in recovering $\vmax$ (e.g., from the halo finder) would result in huge variations in halo rank order.

\section{Orphan Galaxies}
\label{a:orphans}

\subsection{The Need for Orphans}

\begin{figure}
\vspace{-5ex}
\plotgrace{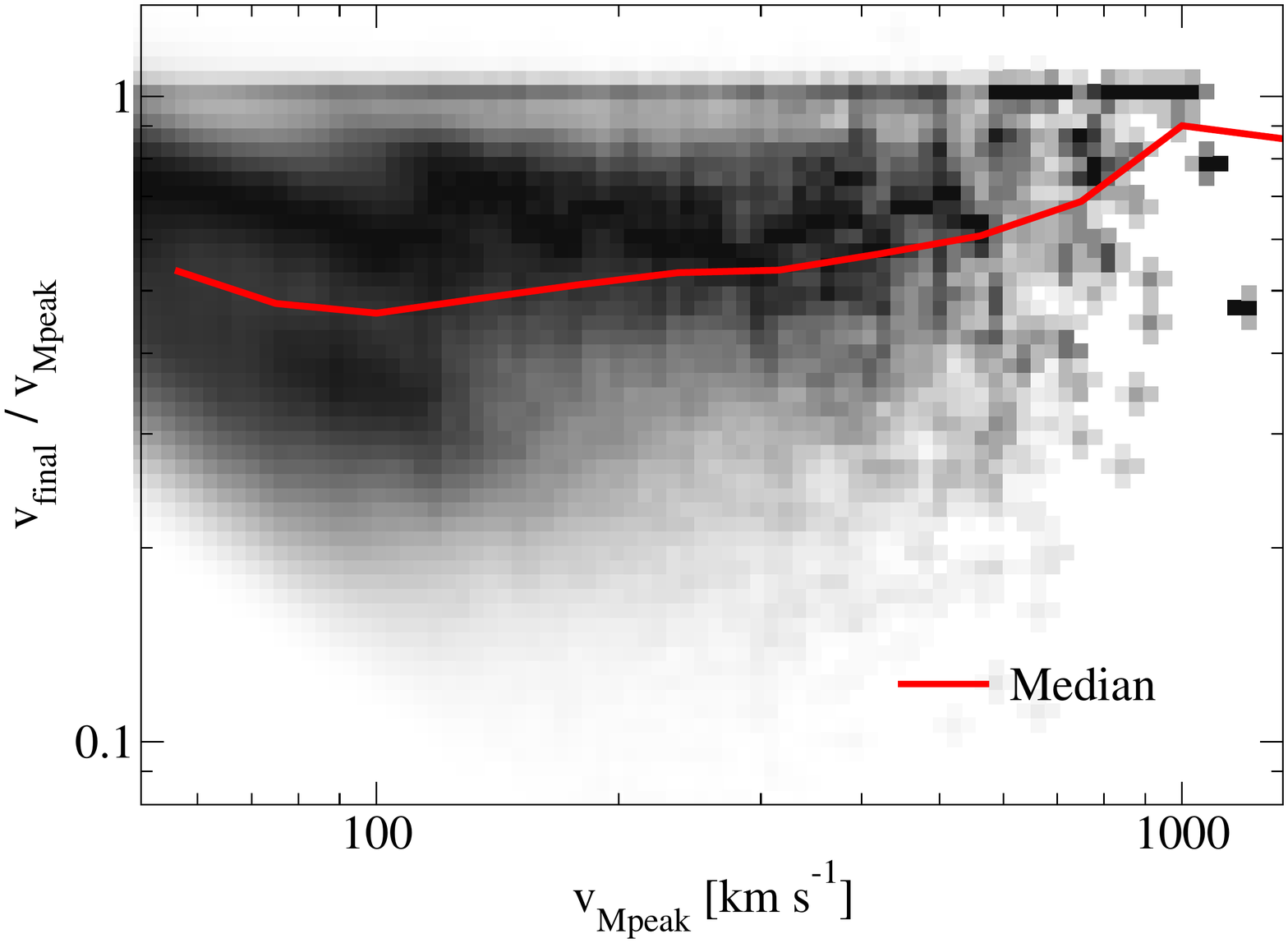}\\[-5ex]
\caption{Conditional distribution of the ratio between the final detectable $\vmax$ and $\vmp$ for subhaloes that disappeared between $z=1$ and $z=0$ in the \textit{Bolshoi-Planck} simulation. A quarter of massive subhaloes ($\vmp>200$ km s$^{-1}$) had more than 70\% of their initial $\vmax$ at the time they disappeared.  The nominal completeness limit for \textit{Bolshoi-Planck} is $\vmax = $50 km s$^{-1}$.}
\label{f:orphans}
\end{figure}

``Orphan'' galaxies (i.e., galaxies whose host subhaloes are no longer detectable by a halo finder) have not been required by most previous empirical models \citep{Reddick12} to match clustering constraints.  This is because satellite fractions can be boosted to the required level either by increasing the stellar mass at fixed satellite mass (or $\vmax$) \textit{or} by increasing the number of satellites at a given mass---i.e., by adding orphans.  Hence, if orphans are not included, one is forced to conclude that satellite haloes have larger stellar masses than central haloes at fixed $\mpeak$ at $z=0$ \citep{RP12,Reddick12,Watson13}.

However, we contrast this conclusion with the known evolution of satellite galaxies.  Satellites are more quenched than field galaxies \citep[e.g.,][]{Wetzel11}, and hence formed their stars earlier (at $z=1-2$) than central galaxies of the same $\mpeak$.  At $z=1-2$, the stellar mass--halo mass ratio was likely lower (and certainly no higher) than the ratio at $z=0$ \citep{Moster12,BWC13}.  Hence, satellites started out with lower stellar masses, could not grow as efficiently (due to being more quenched), and lost more stellar mass from passive stellar evolution (due to older average stellar ages) compared to central galaxies of the same $\mpeak$.  We thus are forced to conclude that adding orphans is a more self-consistent way to reproduce observed galaxy clustering constraints (see also \citealt{Campbell17}).

A separate reason to include orphans is shown in Fig.\ \ref{f:orphans}.  The extent to which satellites survive is dependent on the simulation \citep{Klypin99}, the halo finder \citep{Onions12}, and the baryonic physics included \citep{Zolotov12}.  The point at which haloes disappear in our simulation (\textit{Bolshoi-Planck}) may have no relationship with when galaxies tidally disrupt.  For example, considering massive haloes only ($\vmp > 200$ km s$^{-1}$), a quarter of those that disappeared had $\vmax > 0.7 \vmp$---i.e., $>16\%$ of their infall mass remaining \citep{Jiang16}.  While it is plausible that a satellite galaxy could tidally disrupt with this much stripping, it's also plausible that satellite galaxies persist for much longer.  Hence, not including orphans imposes a strong, arbitrary prior on satellite evolution, yielding artificially tight constraints on quenching timescales and star formation histories.

\subsection{Tracking Method}

When a satellite halo becomes undetectable, we identify the parent halo (i.e., a larger halo containing the satellite within its virial radius) to which it is most bound, and follow the satellite's evolution according to a softened gravity law:
\begin{equation}
\dot{\mathbf{v}} = -\frac{GM(<r)}{(r+0.1 R_\mathrm{vir})^2}\hat{\mathbf{r}},
\end{equation}
where $\mathbf{v}$ is the satellite's peculiar velocity, $R_\mathrm{vir}$ is the parent's virial radius, and $M(<r)$ is the mass of the parent halo enclosed within the satellite--parent distance ($r$)---assumed to follow a \cite{NFW97} profile:
\begin{equation}
M(<r) = 4\pi \rho_0 r_s^3 \left[\ln\left(1+\frac{r}{r_s}\right) - \frac{r}{r+r_s}\right].
\end{equation}
The force law is softened as hard collisions are impossible with spatially-extended satellites; as with past approaches, we use leapfrog integration with 10 sub-timesteps per simulation timestep \citep{BehrooziUnbound,BehrooziTree}.  Each sub-timestep is then on the order of $1\%$ of a dynamical time.

To track the satellite's mass loss, we adopt the orbit-averaged mass-loss prescription of \cite{Jiang16}, with a small modification.  Both synthetic \citep{Knebe11} and cosmological \citep[Fig.\ 1 of][]{BehrooziMergers} tests show that subhaloes lose almost no mass on infall, but instead lose mass (and $\vmax$) steadily after passing pericenter. We hence take:
\begin{eqnarray}
\dot{m}_\mathrm{infalling} & = & 0\\
\dot{m}_\mathrm{outgoing} & = & -1.18\frac{m}{t_\mathrm{dyn}}\left(\frac{m}{M}\right)^{0.07},
\end{eqnarray}
where $m$ is the satellite's mass, $M$ is the parent halo's mass, $\rho_\mathrm{vir}$ is the virial overdensity from \cite{mvir_conv}, and we have adjusted the leading constant for $\dot{m}_\mathrm{outgoing}$ from \cite{Jiang16} for scatter, our different definition of $t_\mathrm{dyn}$ ($(\frac{4}{3}\pi G\rho_\mathrm{vir})^{-\frac{1}{2}}$), and our assumption that all the mass loss occurs over half the orbit.

To determine $\vmax$, we recast the fitting formula from \cite{Jiang16}:
\begin{equation}
\frac{d \log \vmax}{d\log m} = 0.3 - 0.4\frac{m}{m+m_i},
\end{equation}
where $m_i$ is the satellite's mass at first infall.

As discussed in \S \ref{s:mergers}, satellites are no longer tracked once the ratio $\frac{\vmax}{\vmp}$ falls below $T_\mathrm{merge}$.  We also remove a very small fraction ($<1\%$) of satellites that are highly unbound (kinetic $> 2\times$ potential energy), as those would otherwise travel unphysical distances and require much greater communications overhead between processors.

\section{Observational Data Compilation, Calibrations, Exclusions, and Uncertainties}

\label{a:data}

\subsection{Overview}

\label{a:sys_overview}

\begin{table}
\caption{Adopted Modeling Assumptions for Stellar Mass Functions}
\begin{tabular}{rl}
\hline
Assumption & Publication\\
\hline
Initial Mass Function & \cite{Chabrier03}\\
Stellar Population Synthesis & \cite{bc-03}\\
Dust & \cite{calzetti-00}\\ 
\hline
\end{tabular}
\label{t:smf_assumptions}
\end{table}

Here, we discuss the stellar mass functions (SMFs), star formation rates (SFRs), quenched fractions(QFs), correlation functions (CFs), and higher-order galaxy statistics used as model constraints.  Modeling assumptions are inherent to galaxy stellar masses and SFRs (see \citealt{Conroy13b}, \citealt{Madau14}, and \citealt{Mobasher15} for reviews).  In this paper, we standardize assumptions for SMFs (Table \ref{t:smf_assumptions}) because discontinuities in modeling assumptions manifest as unphysical discontinuities in galaxy evolution.  Notably, assumptions in Table \ref{t:smf_assumptions} were the only ones for which data were consistently available across the entire redshift range considered.

For SFRs, standardizing on the IMF appears important, but using the same SPS and dust model does not guarantee self-consistency between SMFs and SFRs \citep[and references therein]{Madau14,Leja15,Tomczak15}.  Yet, the growth of SMFs appears reasonable within the current \textit{range} of assumptions used to calculate SFRs and SSFRs \citep{Moster12,BWC13}.   Hence, as in \cite{BWC13}, we conducted a review of cosmic and specific SFR constraints published since 2007 \citep[see also][]{Madau14,Speagle14}.  Earlier constraints were excluded due to incorrect dust assumptions strongly affecting derived $z>2$ SFRs (see discussion in \citealt{BWC13}).

We measure CFs and higher-order statistics directly from the Sloan Digital Sky Survey (SDSS) instead of using past results.  This ensures that observed data and models are treated in exactly the same way, that corrections for stellar mass modeling assumptions can be applied self-consistently, and that covariance matrices can be computed using the same code.

All data in this compilation are available in a convenient standard format \href{http://www.peterbehroozi.com/data.html}{\textbf{online}}.

\begin{figure}
\vspace{-5ex}
\plotgrace{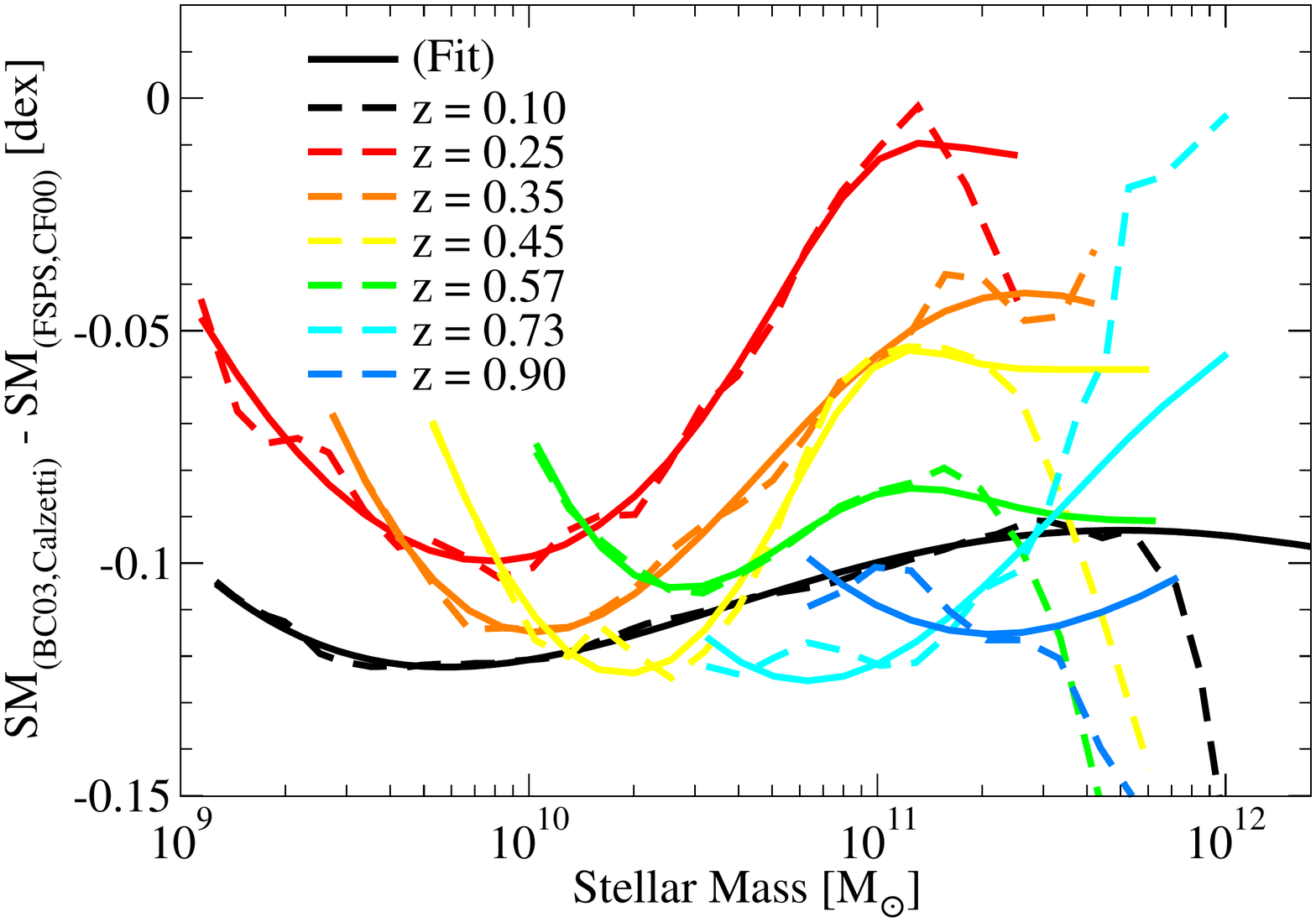}\\[-5ex]
\plotgrace{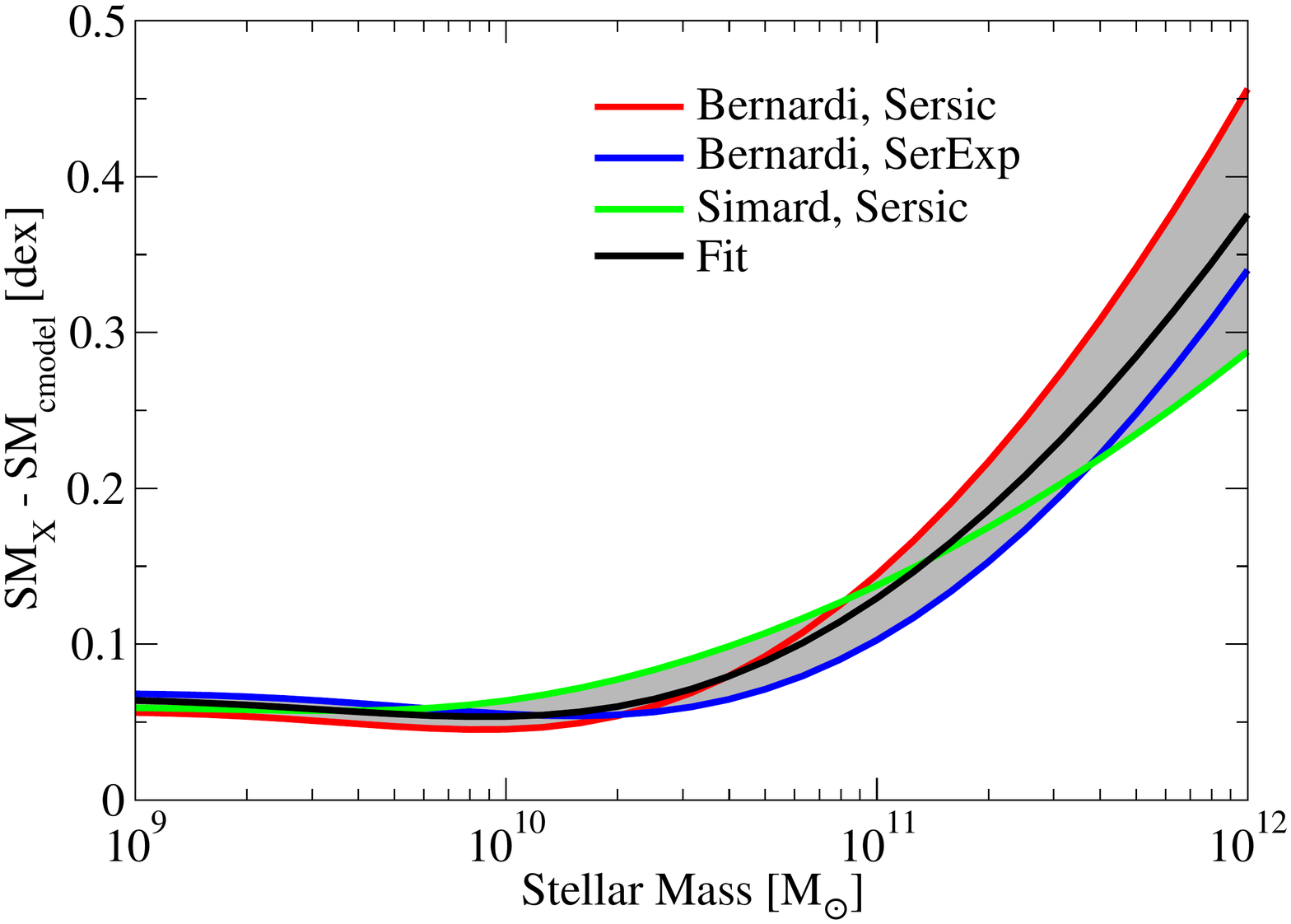}\\[-5ex]
\plotgrace{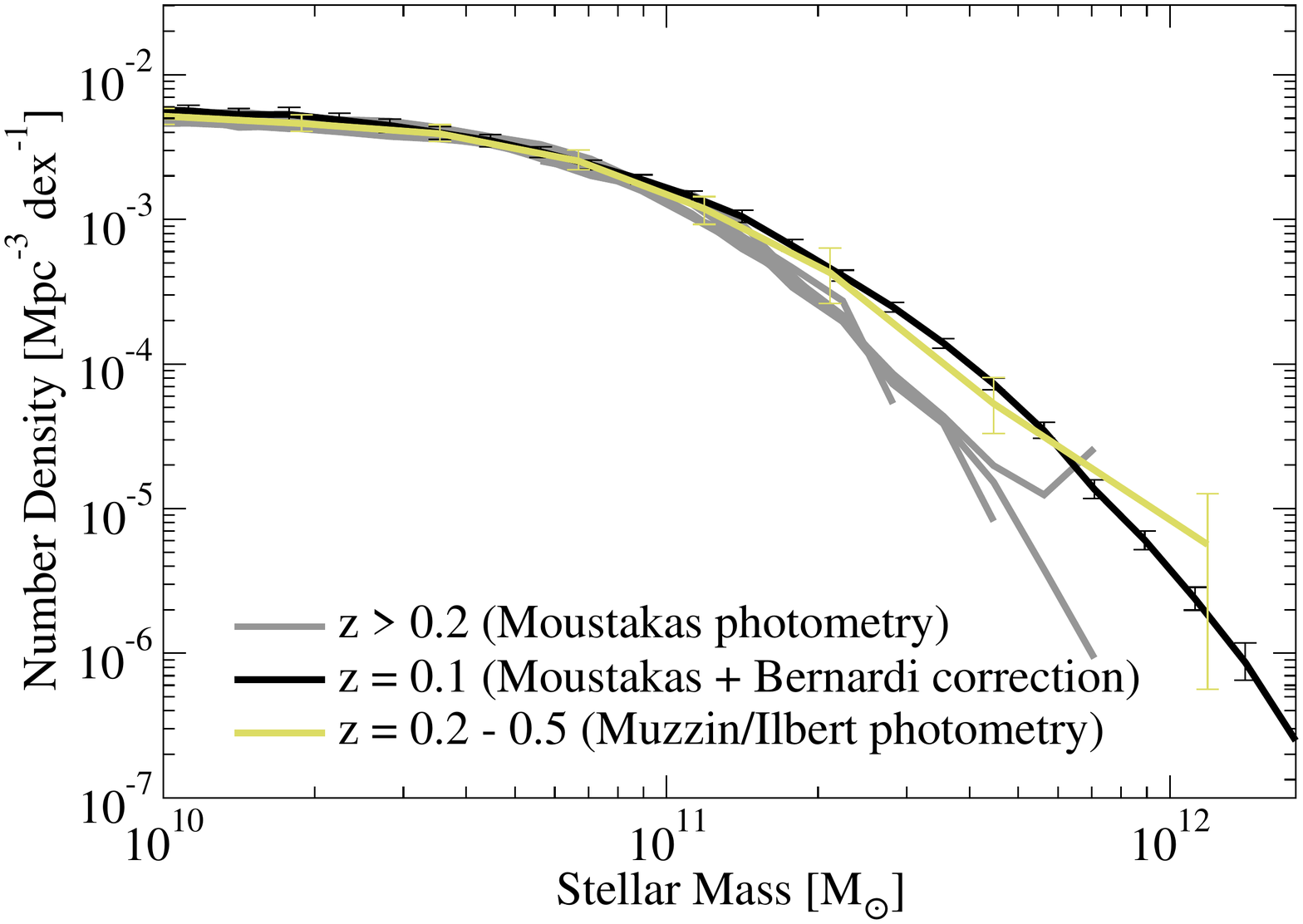}\\[-5ex]
\caption{\textbf{Top} panel: stellar mass differences between model assumptions in \protect\cite{Moustakas13} (i.e., the FSPS SPS model and Calzetti dust law) and our adopted assumptions (Table \protect\ref{t:smf_assumptions}).  \textbf{Middle} panel: range of corrections to SDSS stellar masses using \texttt{cmodel} photometry at $z\sim 0.1$.  \textit{Black} line corresponds to our adopted correction; the \textit{grey shaded region} corresponds to the increase in the error budget from photometry uncertainties.  \textbf{Bottom} panel: effect of photometry choices on SMFs for massive galaxies at $z<1$. }
\label{f:mous_corr}
\end{figure}

\begin{figure}
\vspace{-5ex}
\plotgrace{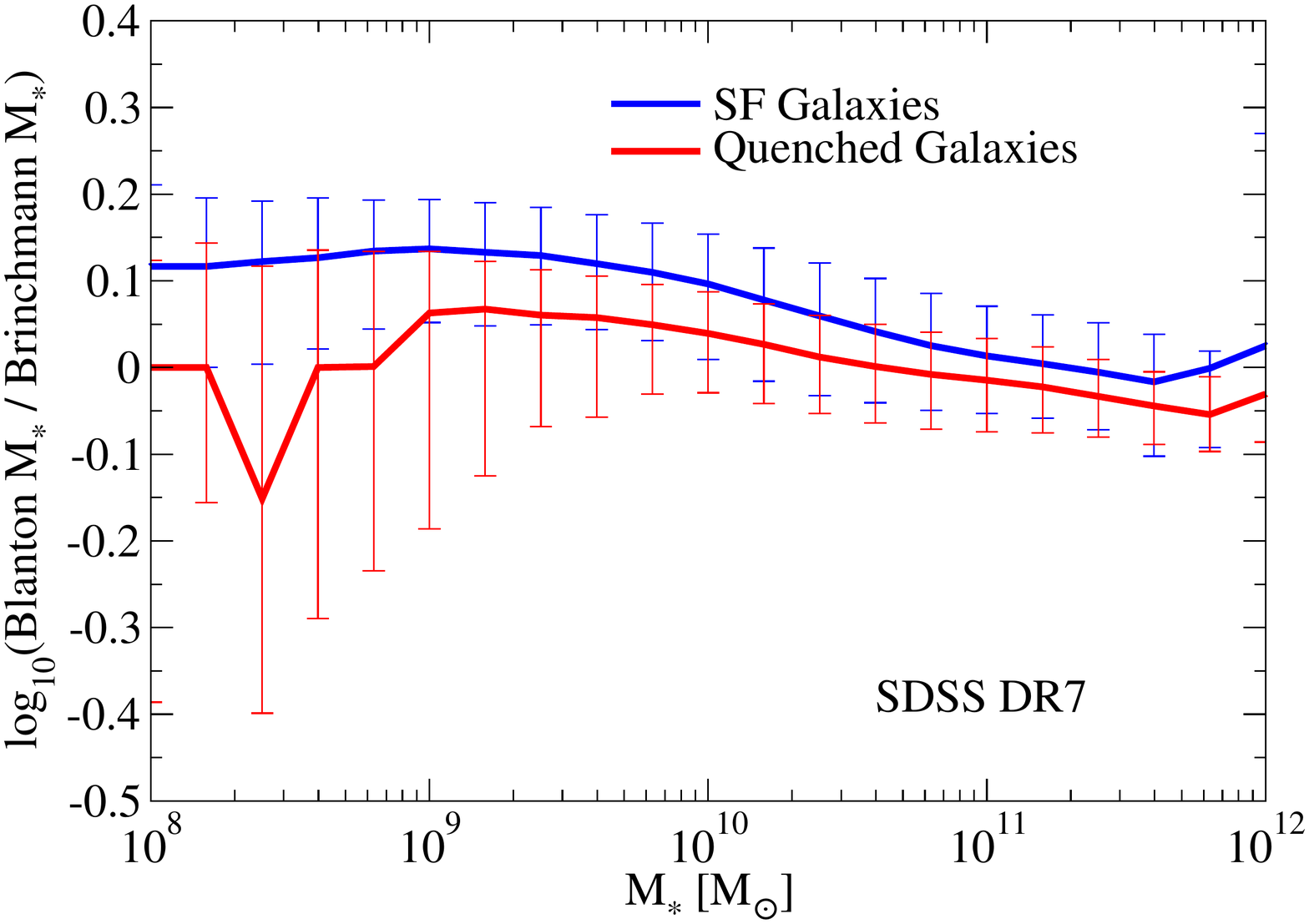}\\[-5ex]
\caption{Median offsets between stellar masses in the \protect\cite{Brinchmann04} and \protect\cite{Blanton05} catalogues, split for star-forming and quenched galaxies.  Error bars correspond to 68\% of the scatter between the two catalogues.}
\label{f:vagc_offsets}
\end{figure}

\subsection{Stellar Mass Functions}

\label{a:smf}
\begin{table}
\caption{Observational Constraints on the Stellar Mass Function}
\begin{tabular}{rcccc}
\hline
Publication & Redshifts & colours & Area (deg$^2$) & Notes\\
\hline
\cite{Baldry12} & 0.002-0.06 & \textit{ugriz} & 143  & N\\
\cite{Moustakas13} &  0.05 - 1 & UV-MIR & 9  & SDP\\
\cite{Tomczak13} & 0.2 - 3 & UV-K$_\mathrm{s}$ & 0.08 & N\\
\cite{Ilbert13} & 0.2 - 4 & UV-K$_\mathrm{s}$ & 1.5  & NM\\
\cite{Muzzin13} & 0.2 - 4 & UV-K$_\mathrm{s}$ & 1.5  & INM\\
\cite{Song15} & 4 - 8 & UV-MIR & 0.08  & NUF\\
\hline
\end{tabular}
\parbox{1.03\columnwidth}{\textbf{Notes.} Letters correspond to conversions applied to the published results: I (Initial Mass Function), S (Stellar Population Synthesis model), D (Dust model), P (Photometry), U ($M_\ast - $UV relation).  The letter ``N'' indicates that the constraint was not previously used in \cite{BWC13}.  The letter ``M'' indicates that results from \cite{Ilbert13} and \cite{Muzzin13} were merged into a single set of constraints to avoid double-counting the same data sources.  ``F'' indicates that the original data were re-fit using different assumptions (see Appendix \ref{a:uvsm}).  Local results ($z<0.2$) in the \cite{Moustakas13} mass functions are taken from the SDSS and cover an area of 2505 deg$^2$.}
\label{t:smf}
\end{table}

Table \ref{t:smf} lists our adopted SMF constraints.  All of the SMFs in this study used the model assumptions in Table \ref{t:smf_assumptions} except for those from PRIMUS \citep{Moustakas13}, which used FSPS \citep{Conroy09} as the SPS model and CF00 \citep{cf-00} as the dust model.  Alternate PRIMUS mass functions using the assumptions [FSPS, Calzetti] and [BC03, CF00] were available; however, the combination [BC03, Calzetti] was not available.  As a compromise, we computed the stellar mass offsets at fixed cumulative number density between the PRIMUS default mass function ([FSPS, CF00]) and the [FSPS, Calzetti] mass function, and used smooth fits to these offsets to shift the masses in the [BC03, CF00] mass function to approximate a [BC03, Calzetti] mass function.  On average, the offsets were $-0.05$ to $-0.1$ dex when converting from the CF00 dust model to the Calzetti dust model, as shown in Fig.\ \ref{f:mous_corr}, top panel.

The \texttt{cmodel} magnitudes used for the $z\sim 0.1$ SMF in \cite{Moustakas13} underestimate galaxies' true luminosities due to the SDSS DR7 sky-subtraction algorithm and the profile-fitting routines used \citep{Simard11,Bernardi13,Bernardi16b,Bernardi16,DSouza15}.  \cite{Kravtsov14} argues that using the \cite{Moustakas13} SMFs overestimates both the impact of AGN feedback and the scatter in the halo mass---stellar mass relation for massive galaxies; the latter effect also reduces clustering for massive galaxies \citep{Bernardi16}.  We compute the stellar mass offset at constant number density between the \texttt{cmodel} SMF and the average of the SMFs for the three fitting methods in \cite{Bernardi13} (i.e., \citealt{Simard11} S\'ersic, \citealt{Bernardi13} S\'ersic, and \citealt{Bernardi13} S\'erExp) and apply this as a correction (Fig.\ \ref{f:mous_corr}, middle panel) to the \cite{Moustakas13} $z\sim 0.1$ SMF.  We also expand the error budget to encompass the minimum and maximum correction among the three methods.  The resulting $z\sim0.1$ SMF is shown in Fig.\ \ref{f:mous_corr}, bottom panel.  It is then apparent that the $z>0.2$ SMFs in \cite{Moustakas13} suffer from similar photometry problems---i.e., it is unphysical for the number density of $\mstar > 10^{11.5}\Msun$ galaxies to be non-evolving from $z=1$ to $z=0.2$ but then to double from $z=0.2$ to $z=0.1$ (Fig.\ \ref{f:mous_corr}, bottom panel).  By comparison, the SMFs from \cite{Ilbert13} and \cite{Muzzin13} are measured from much deeper photometry and appear more consistent with the corrected $z\sim 0.1$ SMF (Fig.\ \ref{f:mous_corr}, bottom panel).  Hence, we truncate the $z>0.2$  constraints from \cite{Moustakas13} to galaxies with $\mstar < 10^{11}\Msun$ to avoid inconsistencies in the photometry of massive galaxies.

We do not apply photometry corrections to the \cite{Baldry12} SMF, as they infer stellar masses from S\'ersic magnitudes (see \citealt{Hill11} for a comparison to SDSS \texttt{cmodel} magnitudes).  However, the GAMA source data overlaps entirely with the SDSS for bright galaxies over their redshift range ($0<z<0.06$), so we exclude data points from their SMF with $M_* > 10^9\Msun$ (i.e., the lower limit of the \citealt{Moustakas13} $z\sim 0.1$ SMF).  In contrast to \cite{BWC13}, we do not apply a surface brightness incompleteness correction, as recent evidence suggests it is not necessary \citep{Williams16}.

Both the \cite{Ilbert13} and \cite{Muzzin13} SMFs are measured from near-identical data sources (i.e., same fields, same filters, very similar depths) with near-identical stellar mass modeling assumptions.  Encouragingly, both studies' measurements are largely within each others' error bars, but there are occasions (especially $2 < z < 2.5$) where this is not the case.  To avoid double-counting the same data, we merge the two sets of SMFs into a single set by taking the geometric average between the two studies SMFs, after dividing the \cite{Muzzin13} stellar masses by a factor of 1.07 to convert from their \cite{Kroupa01} IMF to our adopted \cite{Chabrier03} IMF.  Where necessary, we increase the error budget to accommodate the 68\% confidence regions of both studies' results.

Stellar masses at $z>3.5$ suffer large systematic uncertainties \citep{Stefanon15,Stefanon17,RP17,Moster17}, with literature estimates of mass functions and UV--SM relations differing by up to 0.5 dex \citep{Duncan14,Song15}.  We investigate in Appendix \ref{a:uvsm}, finding that uncertainties in star formation history (SFH) priors dominate other sources of error.  Combining the SED data in \cite{Song15} with constraints on SFHs from the redshift evolution of UV luminosity functions \citep{Papovich11} and constraints on galaxy dust content \citep{Bouwens16}, we derive improved estimates of UV--stellar mass relations as a function of redshift.  Briefly, reproducing the UV--stellar mass relations in \cite{Song15} requires unphysically short star formation histories; more realistic extended histories result in $\sim 0.3$ dex higher median stellar masses at fixed luminosity, even after accounting for nebular emission lines.  

Finally, we note that the choice of SPS and dust models results in a bias that is generically different for star-forming and quenched galaxies.  For example, Fig.\ \ref{f:vagc_offsets} shows differences between the \cite{Brinchmann04} and \cite{Blanton05} stellar mass estimates.  The offsets between the two are consistently higher for star-forming galaxies compared to quenched ones.  As the offsets with true galaxy stellar masses are unknown, caution must be taken when comparing properties for quenched and star-forming galaxies that have a strong dependence on stellar mass (e.g., host halo mass).

\subsection{Cosmic and Specific SFRs}
\label{a:csfr_ssfr}

\begin{table}
\caption{Observational Constraints on the Cosmic Star Formation Rate}
\label{t:csfr}
\begin{tabular}{rcccc}
\hline
Publication & Redshifts & Type & Area (deg$^2$) & Notes\\
\hline
\cite{Robotham11} & 0-0.1 & UV & 833  & I\\
\cite{Salim07} & 0-0.2 & UV & 741  & \\
\cite{Gunawardhana13} & 0-0.35 & H$\alpha$ & 144  & IN\\
\cite{Ly10} & 0.8 & H$\alpha$ & 0.8  & I\\
\cite{Zheng07} & 0.2-1 & UV/IR & 0.46  & \\
\cite{Rujopakarn10} & 0-1.2 & FIR & 0.4-9  & I\\
\cite{Drake15} & 0.6-1.5 & [OII] & 0.63  & N\\
\cite{Shim09}  & 0.7-1.9 & H$\alpha$ & 0.03  & I\\
\cite{Sobral14} & 0.4-2.3 & H$\alpha$ & 0.02-1.7  & N\\
\cite{Magnelli11} & 1.3-2.3 & IR & 0.08  & I\\
\cite{Karim11} & 0.2-3 & Radio & 2  &\\
\cite{Santini09} & 0.3-2.5 & IR & 0.04 & IN\\
\cite{Ly12} & 1-3 & UV & 0.24  & I\\
\cite{Kajisawa10} & 0.5-3.5 & UV/IR & 0.03  & I\\
\cite{Schreiber15} & 0-4 & FIR & 1.75 & IN\\
\cite{PlanckSFR} & 0-4 & FIR & 2240 & IN\\
\cite{Dunne09} & 0-4 & Radio & 0.8  & I\\
\cite{Cucciati11} & 0-5 & UV & 0.6  & I\\
\cite{LeBorgne09} & 0-5 & IR-mm & varies & I\\
\cite{vdBurg10} & 3-5 & UV & 4  & I\\
\cite{Yoshida06} & 4-5 & UV & 0.24  & I\\
\cite{Finkelstein15} & 3.5-8.5 & UV & 0.084  & IN\\
\cite{Kistler13} & 4-10.5 & GRB & varies & IRN\\
\hline
\end{tabular}
\parbox{1.03\columnwidth}{\textbf{Notes.} Letters correspond to conversions applied to the published results: I (Initial Mass Function), R (CSFR normalization for GRBs).   The letter ``N'' indicates that the constraint was not previously used in \cite{BWC13}.    The technique of \cite{LeBorgne09} (parametric derivation of the cosmic SFH from counts of IR-sub mm sources) uses multiple surveys with different areas.  \cite{Kistler13} used GRB detections from the \textit{Swift} satellite, which has a FOV of $\sim 3000$ deg$^2$ (fully coded) and $\sim$10000 deg$^2$ (partially coded).}
\end{table}

\begin{table}
\caption{Observational Constraints on Average SSFRs}
\label{t:ssfr}
\begin{tabular}{rcccc}
\hline
Publication & Redshifts & Type & Area (deg$^2$) & Notes\\
\hline
\cite{Salim07} & 0-0.2 & UV & 741  & \\
\cite{Bauer13} & 0-0.35 & H$\alpha$ & 144 & N\\
\cite{Whitaker14} & 0-2.5 & UV/IR & 0.25  & N\\
\cite{Zwart14} & 0-3 & Radio & 1 & IN\\
\cite{Karim11} & 0.2-3 & Radio & 2  &\\
\cite{Kajisawa10} & 0.5-3.5 & UV/IR & 0.03  & I\\
\cite{Schreiber15} & 0-4 & FIR & 1.75 & IN\\
\cite{Tomczak15} & 0.5-4 & UV/IR & 0.08 & N\\
\cite{Salmon15} & 3.5-6.5 & SED & 0.05 & IN\\
\cite{Smit14} & 6.6-7 & SED &  0.02 & IN\\
\cite{Labbe12} & 7.5-8.5 & UV/IR & 0.040 & I\\
\cite{McLure11} & 6-8.7 & UV & 0.0125  & I\\
\hline
\end{tabular}
\parbox{1.03\columnwidth}{\textbf{Notes.} Letters correspond to conversions applied to the published results: I (Initial Mass Function).   The letter ``N'' indicates that the constraint was not previously used in \cite{BWC13}.  While \cite{Bauer13} did not provide average SSFR constraints directly, they released a public catalogue of SSFRs \citep[included in][]{Liske15}, from which we computed averages.  We obtained average SSFRs from \cite{Schreiber15} by multiplying average SSFRs for star-forming galaxies by $(1-f_q$), where $f_q$ was the reported quenched fraction.  \cite{Zwart14} assumed a \cite{Chabrier03} IMF for stellar masses but a \cite{Salpeter55} IMF for SFRs.}
\end{table}

Tables \ref{t:csfr} and \ref{t:ssfr} list our adopted CSFR and SSFR constraints, respectively.  As for stellar mass functions, we standardize on the \cite{Chabrier03} IMF for specific (SSFRs) and cosmic (CSFRs) star formation rates.  For SSFRs, we divide stellar masses derived from \cite{Kroupa01} and \cite{Salpeter55} IMFs by factors of 1.07 and 1.7, respectively, to convert to a \cite{Chabrier03} IMF.  For both SSFRs and CSFRs, we divide SFRs by factors of 1.06 and 1.58, respectively, for the same conversions \citep[from][]{Salim07}.

As noted in Appendix \ref{a:sys_overview}, we do not enforce further assumptions in Table \ref{t:smf_assumptions}, as those do not guarantee self-consistency between SMFs and SFRs. For $z<4$ constraints, we excluded studies using fewer than 50 galaxies per redshift bin.  We also excluded SSFR studies that did not use mass-selected samples (e.g., main-sequence-only studies), as incorporating dozens of individual studies' selections would have dramatically increased the complexity of the modeling code.  Notably, this requirement excluded using SSFRs from H$\alpha$-selected samples.  Studies that did not provide enough details to model their results---including not providing the assumed IMF, the redshift range, or (in the case of SSFRs) a stellar mass range---were also excluded.  Notably, the redshift range requirement led to a preference for studies using photometric or spectroscopic redshifts, as opposed to, e.g., dropout selection techniques.  Finally, we excluded studies whose results were superseded by a later work.

Emission-line (EL)-derived SFRs (e.g., from H$\alpha$) deserve special consideration.  \cite{Gunawardhana13} argue that spectroscopic surveys allow better estimation of contamination and extinction, but suffer from bivariate selection (i.e., preselecting galaxies by their optical magnitude before determining EL flux) so that they require incompleteness corrections at low EL luminosities.  They also demonstrate that H$\alpha$ narrowband surveys at $z<0.1$ have significant scatter ($\sim 1$ dex) due to cosmic variance; \cite{Drake13} mentions further evidence of significant ($\sim 0.5$ dex) cosmic variance even for $\sim 1$ deg$^2$ narrowband surveys to $z=0.4$.  Hence, we exclude uncorrected spectroscopic EL and $z<0.5$ narrowband EL measurements from this compilation.  We note that some studies using ELs assume constant extinction \citep[e.g.][]{Sobral14}.  As EL extinction correlates with stellar mass \citep[e.g.,][]{Hopkins01,An14}, assuming constant extinction will underestimate SFRs for massive galaxies and overestimate them for faint ones.  That said, we have tested the net effect on the CSFR from mass-dependent extinction according to \cite{An14} and found a difference of $\sim0.1$ dex compared to the constant extinction assumption in \cite{Sobral14}, which is less than other major sources of systematic error at that redshift \citep[see also][]{Gunawardhana13}.  We note that the assumption in \cite{Drake15} of a constant [OIII]/H$\beta$ ratio is incorrect \citep{Dickey16} and significantly underestimates the CSFR; hence, we exclude [OIII]-derived CSFRs from that study.

For $z>4$ constraints, it was no longer possible to require mass-selected samples for SSFRs.  However, observed quenched fractions decline from $z=0$ to $z\sim 3-4$ \citep{Muzzin13} and theoretical expectations \citep{Jaacks12b,BehrooziHighZ} suggest that the majority of $z>4$ galaxies should be star-forming.  However, we excluded SSFR studies that did not account for nebular emission lines, which would otherwise bias stellar masses high by $\sim$0.3 dex \citep{Stark13}.

At $z>4$, the majority of CSFR constraints are based on UV-derived SFRs.  Because of steep reported faint-end slopes in the UVLF \citep{Bouwens15}, estimating the total CSFR is subject to large uncertainties, and most studies choose to integrate to a specified lower luminosity limit.  In this work, we adopt the lower luminosity limit of \cite{Finkelstein15} of $M_{UV}=-17$ and convert other UV-based studies to this threshold using their provided Schechter-function fits for the UVLF.  Low number statistics at $z\sim9$ and $z\sim10$ \citep{Bouwens15} pose a special challenge.  \cite{Oesch14} report an exceptionally low CSFR compared to extrapolations of lower-redshift trends, as well as expectations from SSFRs at lower redshifts \citep{BehrooziHighZ,Finkelstein15}.  At least in part, this was because \cite{Oesch14} assumed that luminosity bins with no detected galaxies did not contribute to the CSFR; a more physically reasonable prior (e.g., a Schechter function) substantially revises their estimate for the CSFR (S.\ Finkelstein et al., in prep.).  Because of the resulting bias, we do not use CSFR constraints from UV studies at $z>8$, instead opting to compare to the luminosity functions directly.

We note that estimates of the CSFR using long gamma ray bursts (GRBs) are still maturing.  GRBs can probe star formation in faint galaxies, but the evolution of the GRB--CSFR normalization is still uncertain, especially at $z>3$ \citep{Kistler13}.  Moreover, several authors have recently argued for a redshift-dependent GRB luminosity function \citep[e.g.,][]{Yu15GRB,Petrosian15,Pescalli15}, also up to $z\sim 3$.  Here, we include the CSFR estimates in \cite{Kistler13} (renormalized from the \citealt{Hopkins06} CSFR fit to the \citealt{BWC13} CSFR fit), but inflate the lower error estimates to include possible continued evolution in the GRB luminosity function according to \cite{Pescalli15}.

As in \cite{BWC13}, we note that many studies' reported error bars are smaller than the inter-publication variance.  We adopt the same error bars (i.e., the inter-publication scatter) of 0.28 dex for SSFRs and redshift-dependent errors for CSFRs (i.e., 0.13 dex for $z<0.9$, 0.17 dex for $0.9 < z < 1.5$, 0.19 dex for $1.5 < z < 3$, and 0.27 dex for $z>3$) as calculated in \cite{BWC13}.

\begin{figure}
\vspace{-5ex}
\plotgrace{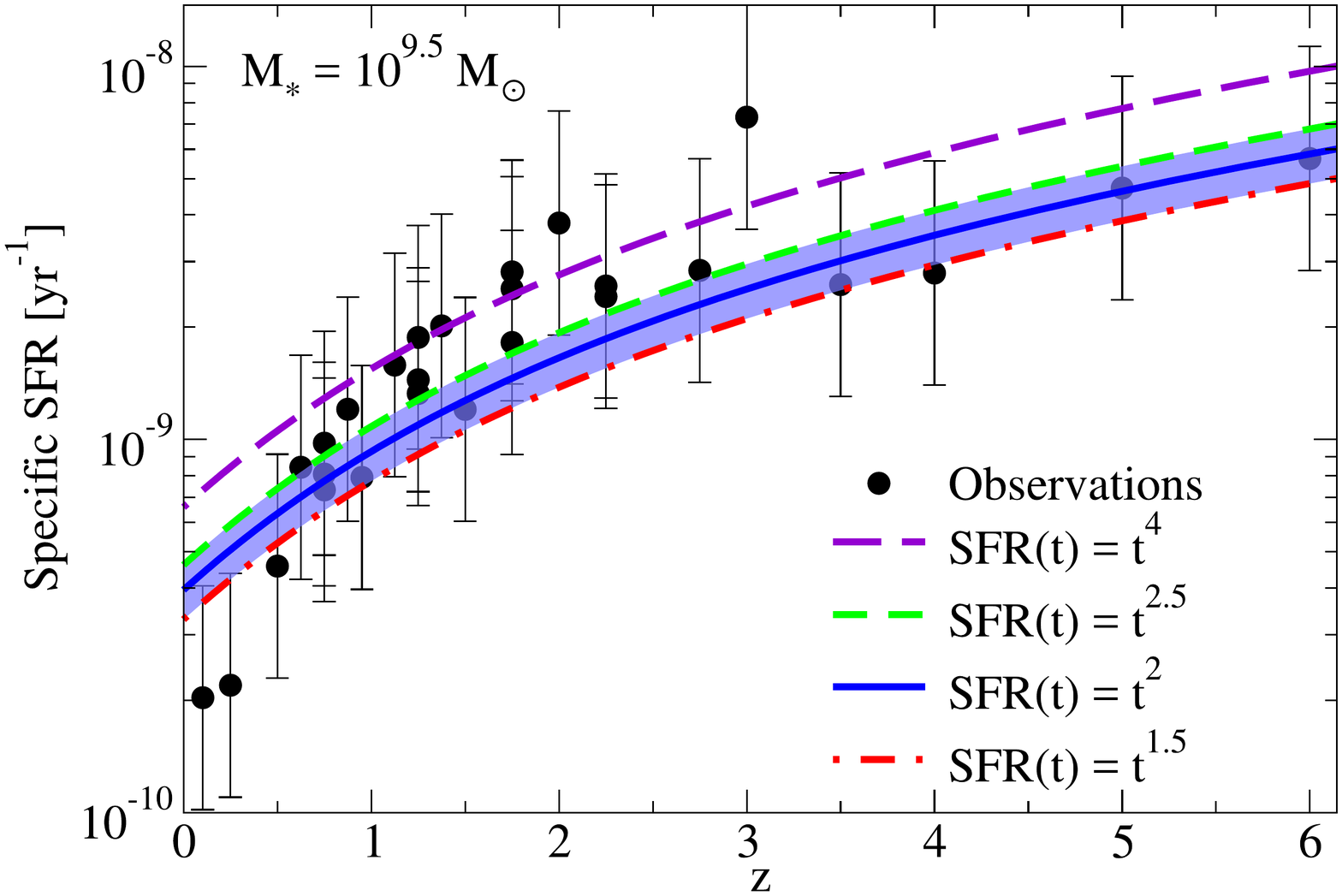}\\[-5ex]
\plotgrace{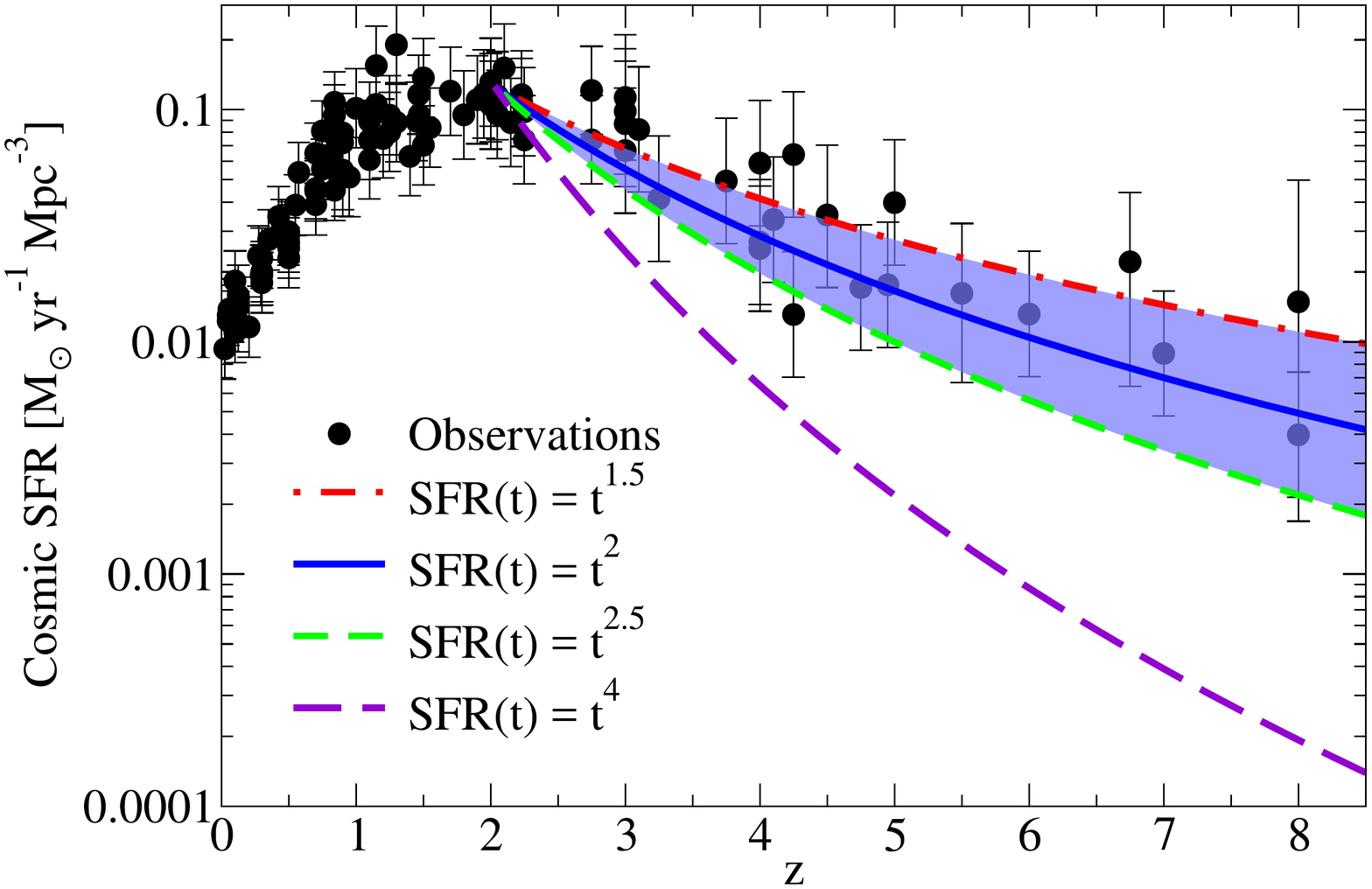}\\[-5ex]
\caption{Illustrations of tension between observed SSFRs and the evolution of SMFs and CSFRs.  \textbf{Top} panel shows observed average SSFRs (Appendix \ref{a:csfr_ssfr}) for galaxies with $M_\ast \sim 10^{9.5}\Msun$.  Overlaid are predictions for average SSFRs (including effects of stellar mass loss) for different power-law star formation histories.  The \textit{blue shaded} region shows a very generous interpretation of current constraints on galaxy star formation histories at $z\sim2$ \citep{Papovich11,BWC13}.  Nonetheless, most observations are $\sim 0.3$ dex higher.  As discussed in the text, this would imply a typical $t^4-t^5$ star formation history, which would have drastic consequences for, e.g., the evolution of SMFs and CSFRs (\textbf{bottom} panel).}
\label{f:ssfr_tension}
\end{figure}

\subsection{Tension Between SMFs and SSFRs}

\label{a:tension}

Tensions between integrated SFRs and the evolution of SMFs have been repeatedly noted, especially at $z\sim2$ \citep{Hopkins06,Wilkins08,Leja15,Yu16}.  Given current systematic uncertainties on stellar masses and SFRs ($\sim 0.3$ dex), it is possible to find self-consistent solutions that approximately match all data sets \citep[e.g.,][]{Moster12,BWC13}.  The source of these tensions has remained unclear, but major factors include a lack of calibration data for SFR indicators (e.g., radio and 24$\mu$m) at higher redshifts, as well as the fact that star formation history priors for stellar mass fits are not fully self-consistent with SMF evolution.  Examples of this at $z>4$ are discussed in Appendix \ref{a:uvsm}.  Here, we show explicitly how this affects $z\sim2$ SSFRs.

Low-mass galaxies at $z\sim 2$ are still rapidly increasing in stellar mass, with the evolution of stellar mass and luminosity functions giving typical star formation histories of $t^{1.7}$ \citep{Papovich11} to $t^2$ \citep{BWC13}.  For a given power-law star formation history, $SFR(t) \propto t^n$, the ratio of the SFR to total stellar mass formed is:
\begin{equation}
SSFR_\mathrm{total}(t) = \frac{n+1}{t}.
\end{equation}
Returned stellar mass ($< 30\%$) will increase this; a slightly smaller effect is averaging log-normal scatter in observed SFRs at fixed SM ($\sim 0.3$ dex scatter results in a mean SSFR that is $27\%$ higher than the median).  Combined, these corrections mean that $SSFR_\mathrm{obs}(t) < 1.8 \frac{n+1}{t}$.  These observed SSFR predictions are overplotted on observations of $M_\ast \sim 10^{9.5}\Msun$ galaxies (Appendix \ref{a:csfr_ssfr}).  Even for an overly generous range of possible star formation histories ($t^{1.5}$ to $t^{2.5}$), most observed SSFRs at $z=2$ are $\sim 0.2-0.3$ dex above the reasonable range (Fig.\ \ref{f:ssfr_tension}, top panel).

Because SSFRs are proportional to the index $n+1$, the observations at face value would imply $t^4-t^5$ star formation histories for typical low-mass galaxies at $z=2$.  This steep history is inconsistent with both the evolution in SMFs and with CSFRs  (Fig.\ \ref{f:ssfr_tension}, bottom panel).  Hence, we adopt a parameter ($\kappa$, Eq.\ \ref{e:kappa}) to allow for discrepancies between SFRs and SMs that peaks at $z\sim2$ and smoothly falls to nearly zero by $z=0$ and $z=4$.

\subsection{UV Luminosity Functions}

\label{a:uvlfs}

We adopt the UV luminosity functions of \cite{Finkelstein15} from $z=4$ to $z=8$, and the \cite{Bouwens15b} luminosity functions for $z\sim9$ and $z\sim10$.  As both studies use similar fields, the \cite{Finkelstein15} results are preferable at low redshifts due to the fact that they use photo-$z$ selection instead of LBG selection, and are hence easier to model.  In addition, they are consistently selected with the galaxies that constrain our UV--stellar mass relations (Appendix \ref{a:uvsm}).  On the other hand, the \cite{Finkelstein15} data do not extend to $z>8$, and so we adopt the \cite{Bouwens15b} constraints at higher redshifts, approximating the LBG cuts as window functions ($z=8.4-9.5$ and $z=9.5-11$, at $z\sim 9$ and $z\sim 10$, respectively).

\subsection{Quenched Fractions}

\label{a:qf}

\begin{figure}
\vspace{-5ex}
\plotgrace{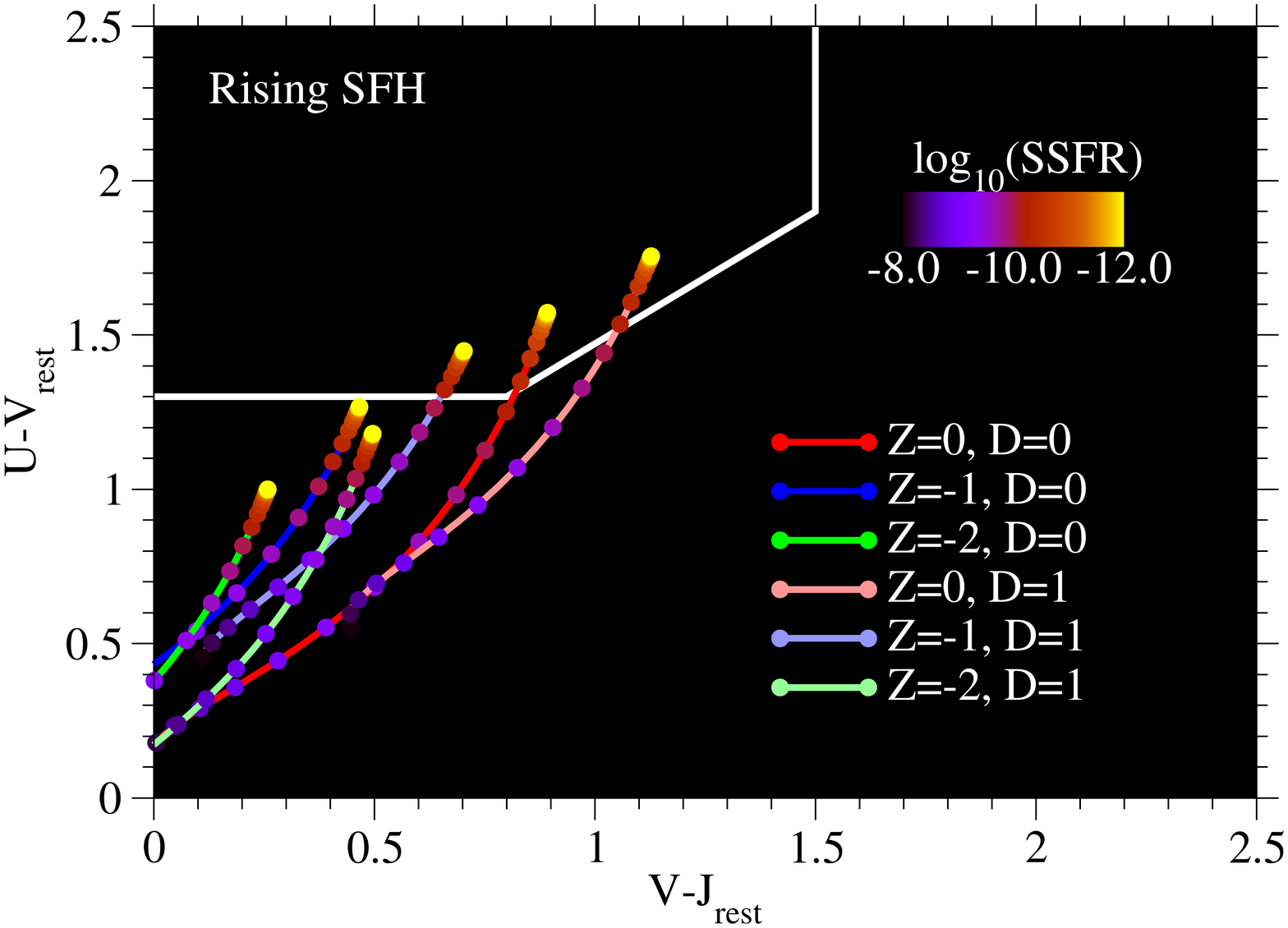}\\[-5ex]
\plotgrace{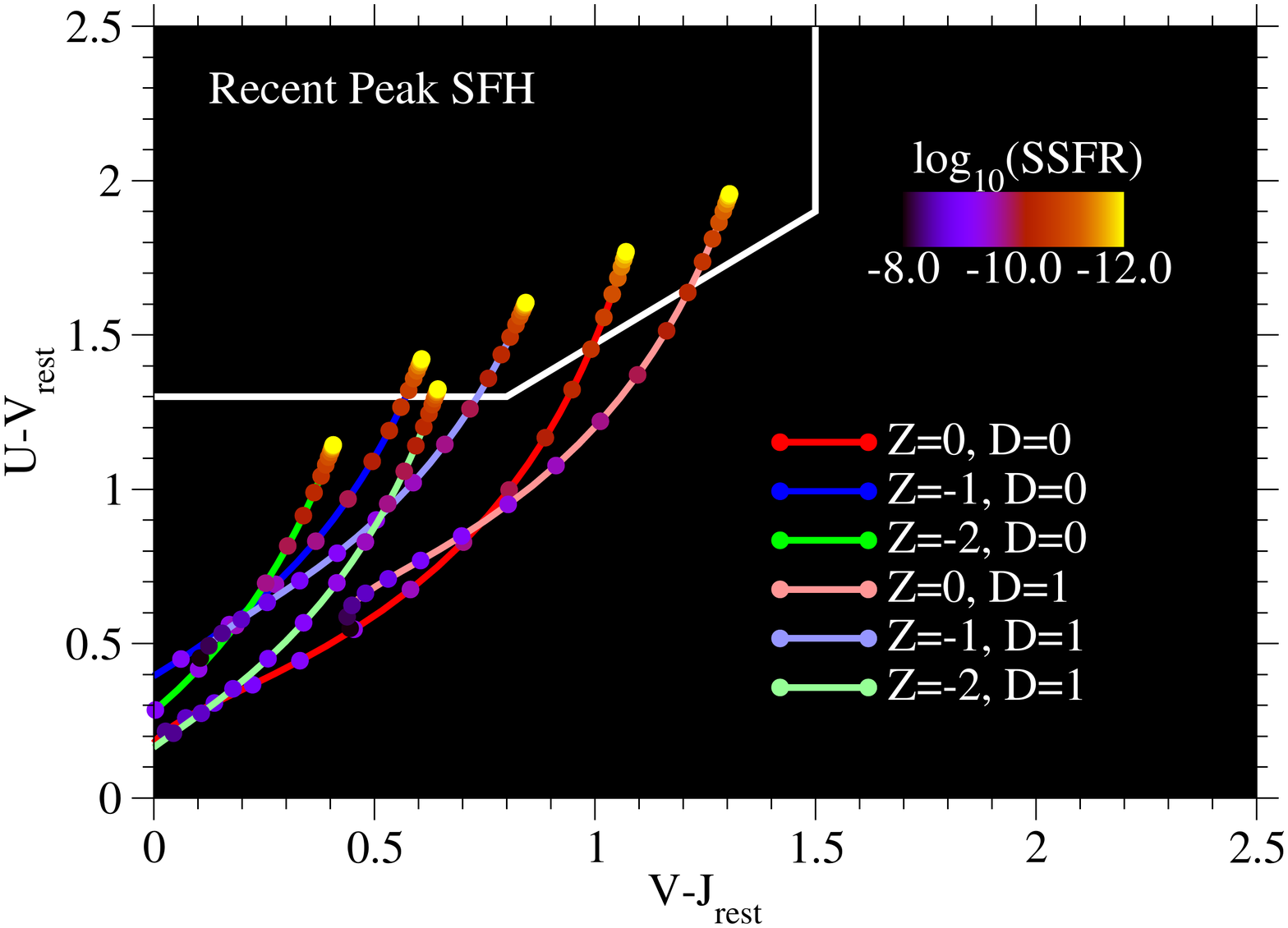}\\[-5ex]
\plotgrace{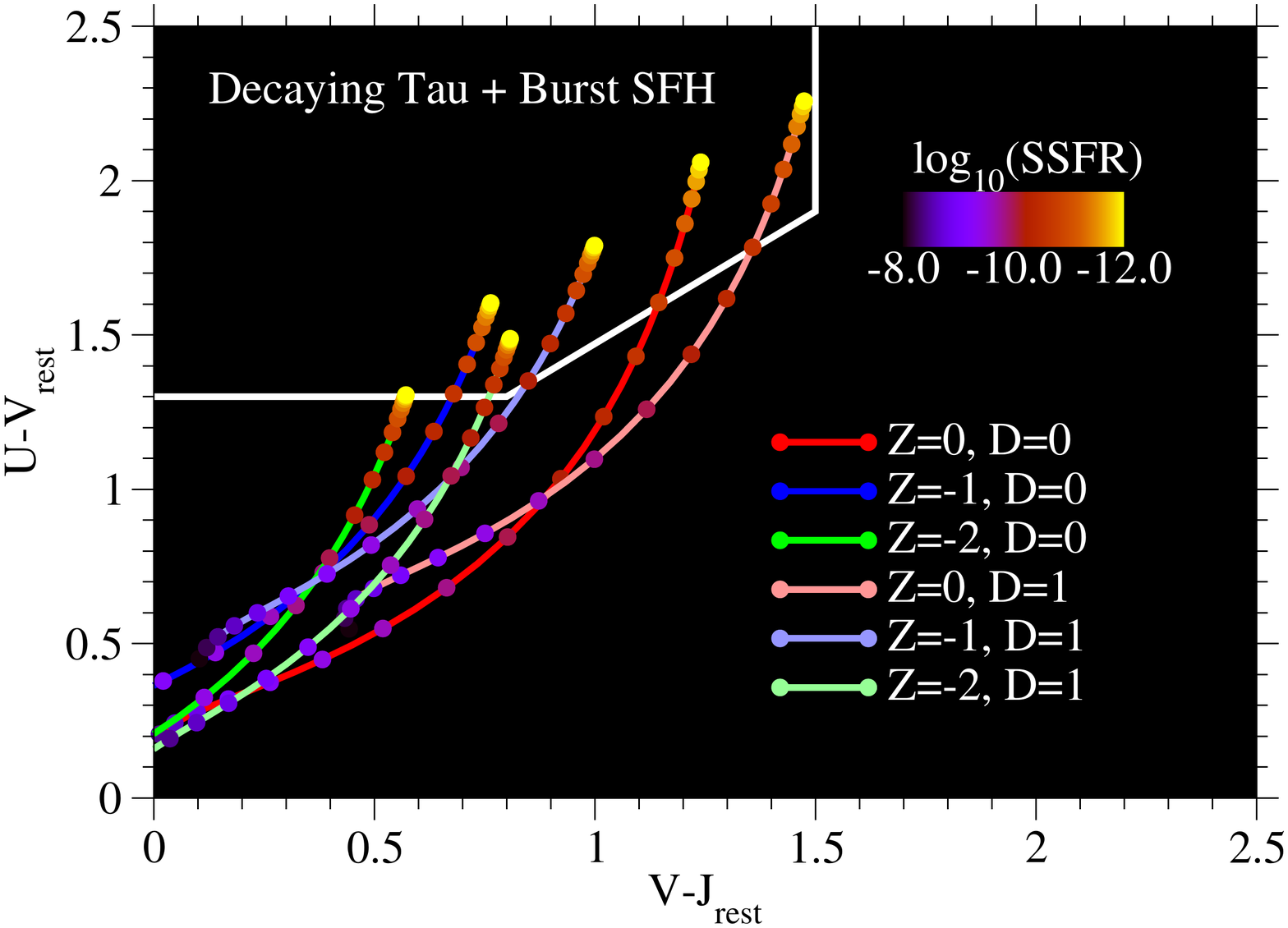}\\[-5ex]
\caption{Galaxy position in the UVJ quenching diagram as a function of recent SSFR; different panels show the effects of different past star-formation histories (see text).  Different line colours correspond to different stellar metallicities (0.01 solar to solar); different line saturations correspond to different dust prescriptions (D=0 implies no dust; D=1 implies \protect\citealt{cf-00} dust).  With some exceptions for low-metallicity / low-dust galaxies, the UVJ diagram separates quenched from star-forming galaxies at similar SSFRs.}
\label{f:uvj_syst}
\end{figure}

Constraints on galaxy quenched fractions as a function of stellar mass (QFs) are taken from \cite{Bauer13}, \cite{Moustakas13}, \cite{Muzzin13}.  Each paper has a different definition of ``quenched,'' so we calculate QFs in three different ways to compare as directly as possible to the observations.  We computed the quenched fraction at $z\sim 0$ from the raw \cite{Bauer13} data, defining ``quenched'' as an observed SSFR below $10^{-11}$ yr$^{-1}$.  

When comparing to \cite{Moustakas13} QFs, we use their adopted quenched/star-forming cut on the observed SSFR:
\begin{equation}
\log_{10}(SSFR_q \cdot \mathrm{yr}) = -10.49 - 0.35 \log_{10}\left(\frac{\mstar}{10^{10}\Msun}\right) + 1.07 (z-0.1).\label{e:ssfr_q}
\end{equation}
We excluded \cite{Moustakas13} data at $z\sim 0$, because the SSFR cut for massive $10^{12}\Msun$ galaxies ($\sim 10^{-11.2}$) was lower than could be reliably measured, resulting in a spuriously low quenched fraction for such galaxies.

When comparing to QFs in \cite{Muzzin13}, we use their adopted cut in UVJ colour space \citep[see also][]{Labbe05,Williams09}, specifically:
\begin{eqnarray}
U-V > 1.3,\, V-J<1.5\, && \textrm{[all redshifts]} \label{e:uvj_boundaries} \\
U-V > (V -J)\times 0.88+0.69 && \textrm{[0.0}<z<\textrm{1.0]}\\
U-V > (V -J)\times 0.88+0.59 && \textrm{[1.0}<z<\textrm{4.0]}.
\end{eqnarray}
\cite{Muzzin13} also explored the systematic errors arising from the UVJ cut location, and found that it may lead to $\sim 10\%$ differences in the resulting quenched fraction; we adopt $\pm 5\%$ systematic errors here.

When comparing to colour-based cuts, a caveat is that the choice of metallicity and dust can strongly affect galaxy luminosities.  Since metallicity and dust are not treated self-consistently for the modeled galaxies, it is important to check whether incorrect metallicity / dust could significantly affect resulting quenched fractions.  We generated three sets of star formation histories (rising, recent peak, and decaying tau) as a function of scale factor ($a$) according to the following equations for $0 < a < 1$:
\begin{eqnarray}
SFH(a) \propto a^2 && \textrm{[rising]} \\
SFH(a) \propto a^2 \exp(-4a) && \textrm{[recent peak]} \\
SFH(a) \propto a^3 \exp(-10a) && \textrm{[decaying tau]}.
\end{eqnarray}
Then, for each initial history above, we generated 41 new star formation histories by substituting the last 300 Myr of the initial history with a constant SSFR between $10^{-12}$ yr$^{-1}$ and $10^{-8}$ yr$^{-1}$.   We generated colours for each of these histories using FSPS with three different metallicity assumptions (solar, 0.1 solar, 0.01 solar) and two different dust assumptions (no dust, \citealt{cf-00} dust).  This test allows us to see how well the UVJ diagram captures a recent change in the SSFR given a wide range of assumptions about the older stellar population history, the metallicity, and the dust.

Several trends are apparent in the results (Fig.\ \ref{f:uvj_syst}).  Galaxies with older pre-existing stellar populations (e.g., the decaying tau SFHs) require less recent star formation to exit the UVJ quenched box, so position in the UVJ diagram does not translate uniquely to SSFR.  However, for a fixed star formation history, galaxies with solar metallicities and dusty low-metallicity galaxies (0.1 solar) all enter the quenched box in the UVJ diagram at similar SSFRs.  As these galaxies make up the majority of quenched galaxies in the Universe \citep{Muzzin13}, this would suggest that modest inaccuracies in assumptions about metallicity and dust will not affect the resulting quenched fraction as long as the underlying star formation history is correct.  That said, low-metallicity, low-dust galaxies with rising star formation histories could lie below traditional quenched boxes (Fig.\ \ref{f:uvj_syst}, upper panel); and similarly, extremely dusty galaxies with very old populations (Fig.\ \ref{f:uvj_syst}, lower panel) could lie to the right of traditional quenched boxes.  This suggests that the present vertical and horizontal boundaries used (Eq.\ \ref{e:uvj_boundaries}) are artificial and might be better removed or significantly relaxed as, e.g.,  $U-V > 0.8$ and $V-J < 1.7$.

\subsection{Galaxy Correlation Functions and Covariance Matrices}
\label{a:cf}

\begin{figure}
\vspace{-5ex}
\plotgrace{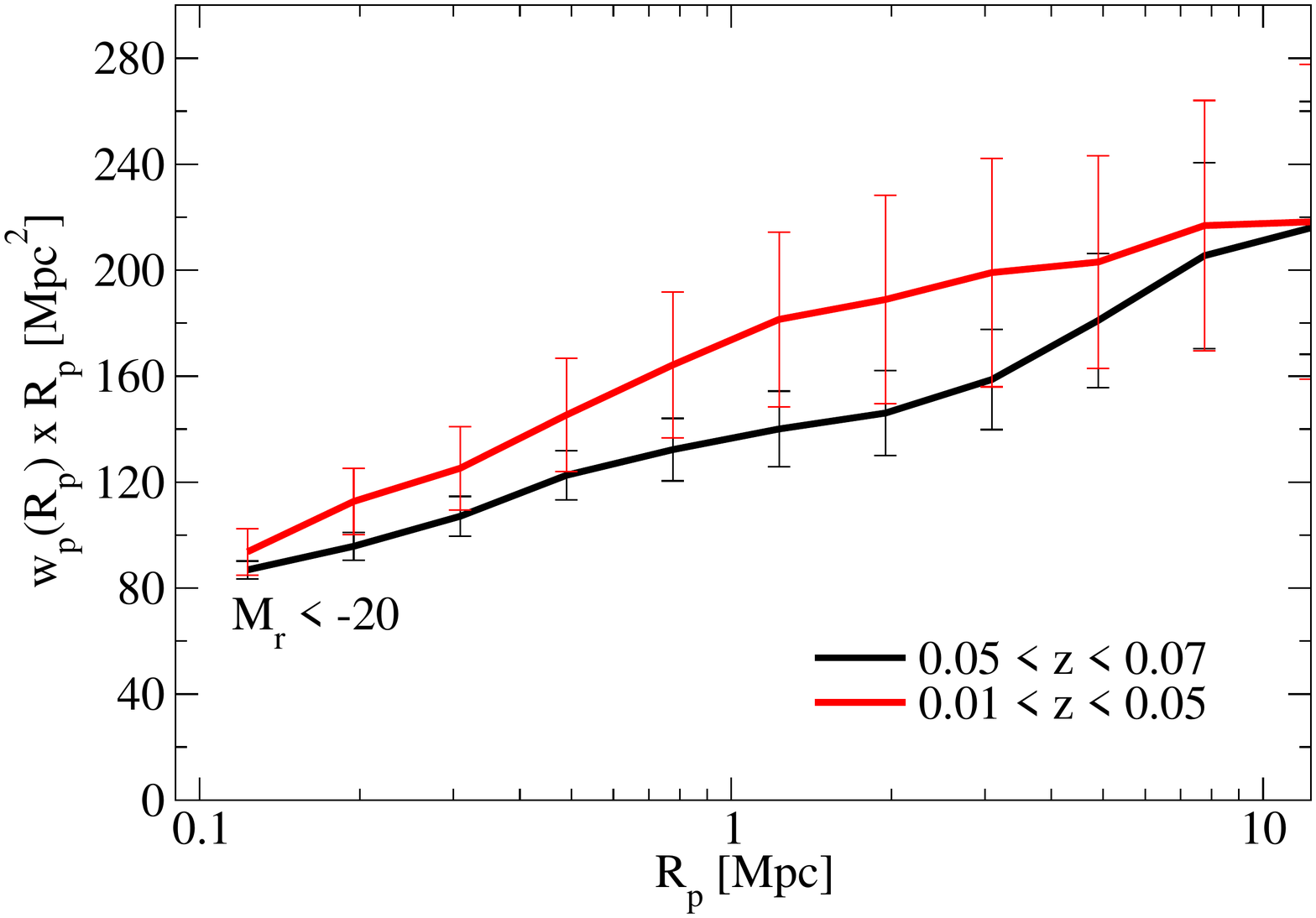}\\[-5ex]
\caption{Excess clustering in the SDSS at $z<0.05$ for an autocorrelated luminosity threshold sample ($M_r < -20$) as compared to higher redshifts.  Errors shown here are from 16 jackknife bins.}
\label{f:cf_anomaly}
\end{figure}

\begin{table}
\caption{Measured Correlation Functions from the SDSS}
\begin{tabular}{ccccccc}
\hline
$\mstar^\mathrm{min}$ & $\mstar^\mathrm{max}$ & $z_\mathrm{min}$ & $z_\mathrm{max}$ & $V/$Gpc$^3$ & N & $f_\mathrm{SF}$\\
\hline\\[-2ex]
$10^{10.3}\Msun$ & $10^{10.5}\Msun$ & 0.045 & 0.069 & 0.012 & 10081 & 0.51\\
$10^{10.5}\Msun$ & $10^{11.0}\Msun$ & 0.045 & 0.070 & 0.013 & 16953 & 0.40\\
$10^{11.0}\Msun$ & $10^{13.0}\Msun$ & 0.090 & 0.121 & 0.050 & 16912 & 0.29\\
\hline
\end{tabular}
\label{t:cfs}
\end{table}

Galaxy correlation functions (CFs) for all, quenched, and star-forming galaxies are derived from the SDSS DR7 \citep{Abazajian09}.  Stellar masses and star formation rates were taken from the public catalogues of \cite{Kauffmann03} and \cite{Brinchmann04}, respectively, as updated for DR7.  Stellar masses were translated to match the adjusted \cite{Moustakas13} SMF (Appendix \ref{a:smf}) by abundance matching at fixed cumulative number density to the SDSS DR7 SMF in \cite{BehrooziMM}; the resulting offsets were $<$0.03 dex for all stellar mass bin edges up to $10^{11}\Msun$, consistent with findings in \cite{DSouza15}.  Although both \cite{Brinchmann04} and \cite{Bauer13} SFRs are H$\alpha$-based, we found modest discrepancies.  To match the quenched fraction inferred from \cite{Bauer13} (using a cut of SSFR$<10^{-11}$ yr$^{-1}$), we found that we had to apply a quenched threshold of SSFR$<10^{-11.3}$ yr$^{-1}$ in the data from \cite{Brinchmann04}.

We first generate volume-limited catalogues in each mass bin according to the cut in \cite{BehrooziMM}:
\begin{equation}
-0.25 - 1.9\log_{10}\left(\frac{\mstar}{\Msun}\right) + 5 \log_{10} \left(\frac{D_L(z)}{10\mathrm{pc}}\right) < 17.77. \label{e:lum_cut}
\end{equation}
We then calculate the two-point 3D CF $\xi(r_p, \pi)$ out to $\pi = \pm 20$ Mpc (comoving) in bins of $1$ Mpc in $\pi$ using the \cite{Landy93} estimator ($\xi = (DD-2DR+RR)/RR$) with 10$^6$ randoms distributed with constant number density over the same angular (6261.75 deg$^2$) and redshift footprint.  For cross-correlations, we use the modified estimator $\xi = (D_1 D_2 - D_1 R_2 - D_2 R_1 + R_1 R_2)/R_1 R_2$, which reduces to the \cite{Landy93} estimator for autocorrelations. 

We then integrate $\xi$ along the line-of-sight direction ($\pi$) to obtain $w_p(r_p)$.  The limited line-of-sight integration is necessary because several CFs are calculated over narrow redshift slices ($<100$ Mpc), so longer integrations in $\pi$ would not be possible for the majority of the galaxies.  No correction is required, however, as we perform the identical integration on the generated mock galaxy catalogues when comparing to data.  As is typically done \citep{Skibba14} for galaxies at redshift-space positions $\mathbf{x}_1$ and $\mathbf{x}_2$, we define:
\begin{eqnarray}
\mathbf{s} = \mathbf{x}_1 - \mathbf{x}_2, & \quad & \mathbf{l} = 0.5( \mathbf{x}_1 + \mathbf{x}_2)\\
\mathbf{\pi} \equiv (\mathbf{s}\cdot\mathbf{l}) / |\mathbf{l}|, & \quad & r_p \equiv \sqrt{s^2 - \pi^2}.
\end{eqnarray}
We correct for fiber collisions by weighting galaxy counts.  Briefly, some areas of the SDSS were observed multiple times, with the result that for every observed pair within the fiber collision radius (55"), 2.08 galaxies have an unobserved pair \citep{Patton13}.  For each such galaxy pair, we therefore add each of their individual contributions to the $DD$ counts 1.04 extra times to account for collided galaxies not present in the spectroscopic sample.  At large separations ($>100$ kpc for $\mstar < 10^{11}$ galaxies and $>200$ kpc for $\mstar > 10^{11}\Msun$ galaxies), the change in $w_p(r_p)$ is $<5\%$.  We do not attempt to calculate $w_p(r_p)$ at closer separations, partially as much more sophisticated algorithms are necessary to robustly recover $w_p$ \citep[e.g.,][]{Guo15}.

Table \ref{t:cfs} lists the redshift and stellar mass bins we use for measuring autocorrelation functions.  We additionally measure the cross-correlation between galaxies with $M_\ast > 10^{11}\Msun$ and galaxies with $10^{10.3}\Msun < M_\ast < 10^{10.5}\Msun$, in the same redshift bin as for the $10^{10.3}\Msun < M_\ast < 10^{10.5}\Msun$ autocorrelation function.    Redshift bin constraints arise from volume completeness requirements, SDSS Great Wall avoidance ($0.07 < z < 0.09$; \citealt{Gott05}), and a $1-2\sigma$ clustering excess at $z<0.045$.  The clustering excess is shown in Fig.\ \ref{f:cf_anomaly}; the excess is robust to choice of luminosity or stellar mass bin, choice of stellar masses (e.g., from \citealt{Blanton05}), method of distributing randoms (e.g., randomizing sky position of galaxies instead of assuming constant number density, correcting for redshift-space distortions in volume density), local flow corrections \citep[e.g.,][]{Tonry00}, method of counting randoms (e.g., direct pair counts or Monte Carlo integration), and method for computing $\mathbf{\pi}$ or $r_p$ (e.g., sky angle times mean distance).

Covariance matrices are crucial when comparing modeled to observed CFs, as CF bins are often extremely covariant.  We compute covariance matrices for the observations by generating mock lightcones from the MDPL2 simulation at the mean redshifts of the samples in Table \ref{t:cfs}; notably, this simulation did not include orphan galaxies.  Stellar masses were assigned by abundance-matching the $z\sim 0.1$ SMF from Appendix \ref{a:smf} to haloes on $\vmp$ with $0.2$ dex of log-normal scatter (following \citealt{Reddick12}).  Quenching was correlated with $\Delta\vmax$ (Eq.\ \ref{e:dvmax}) with $r_c=0.7$ (\S \ref{s:fform}), with the fraction of galaxies varying as a function of $\vmp$ according to Appendix \ref{a:fq}.  For each sample in Table \ref{t:cfs}, 5000 mock lightcones were generated with the same footprint as the SDSS DR7, and auto- and cross CFs for all, quenched, and star-forming galaxies were calculated using the identical code used to calculate observed auto- and cross CFs.  Matrices for the simultaneous covariance of all, quenched, and star-forming galaxies were computed, with $N_\mathrm{bins} = 24-39$ for each sample in Table \ref{t:cfs}.  For model covariance matrices, the observed volume is much smaller than the volume of \textit{Bolshoi-Planck} for the two lower-mass bins in Table \ref{t:cfs}, so we assume that the model covariance matrices are subdominant to the observed ones.  For the highest-mass bin, we assume that the model covariance matrix is equivalent to the observational covariance matrix, as the observational and simulated volumes are nearly equal.

Computing covariance matrices requires a volume $\sim 10 \times V_\mathrm{obs}\times N_\mathrm{bins}$ for the resulting error estimates to be accurate at the $\sim 10\%$ level \citep{Dodelson13}.  Our largest available high-resolution simulation (MDPL2) has a volume of  3.22 Gpc$^3$.  This volume is appropriate for the two lower-mass samples in Table \ref{t:cfs}, but likely is oversampled for the highest-mass bin.  To investigate this, we performed a Principal Component (PC) Analysis on all covariance matrices to obtain PCs ($\mathbf{PC}_1$\ldots$\mathbf{PC}_n$) and their corresponding standard deviations ($\sigma_1\ldots\sigma_n$).  For the highest-mass sample, only 7 PCs have standard deviations of $>10\%$; the volume of MDPL2 suggests that we should trust the largest $\sim 3.22$ Gpc$^3 / (10 V_\mathrm{obs}) = 6$ PCs.  However, we note that systematic errors inherent in calculating CFs (including errors in modeling fiber collisions, edge effects, systematic variations in photometry across the sky, different typical colours of central vs.\ satellite galaxies leading to different systematic stellar mass offsets) set a maximum achievable accuracy of no better than 10\%.  Hence, for \textit{all} samples, we take the effective error on the $i$th PC to be $\sigma_{\mathrm{eff},i} \equiv \max(10\%, \sigma_i)$.  The resulting contribution ($\Delta\chi^2$) to the total error calculation from a given correlation function then becomes:
\begin{equation}
\Delta\chi^2 = \sum_{i=1}^n \left(\frac{\mathbf{PC}_i\cdot(\mathbf{w}_{p,\mathrm{obs}} - \mathbf{w}_{p,\mathrm{model}})}{\sigma_{\mathrm{eff},i}}\right)^2.
\end{equation}

Results for the measured auto- and cross correlation functions are shown in Fig.\ \ref{f:cf_comp}.

\begin{figure}
\vspace{-5ex}
\plotgrace{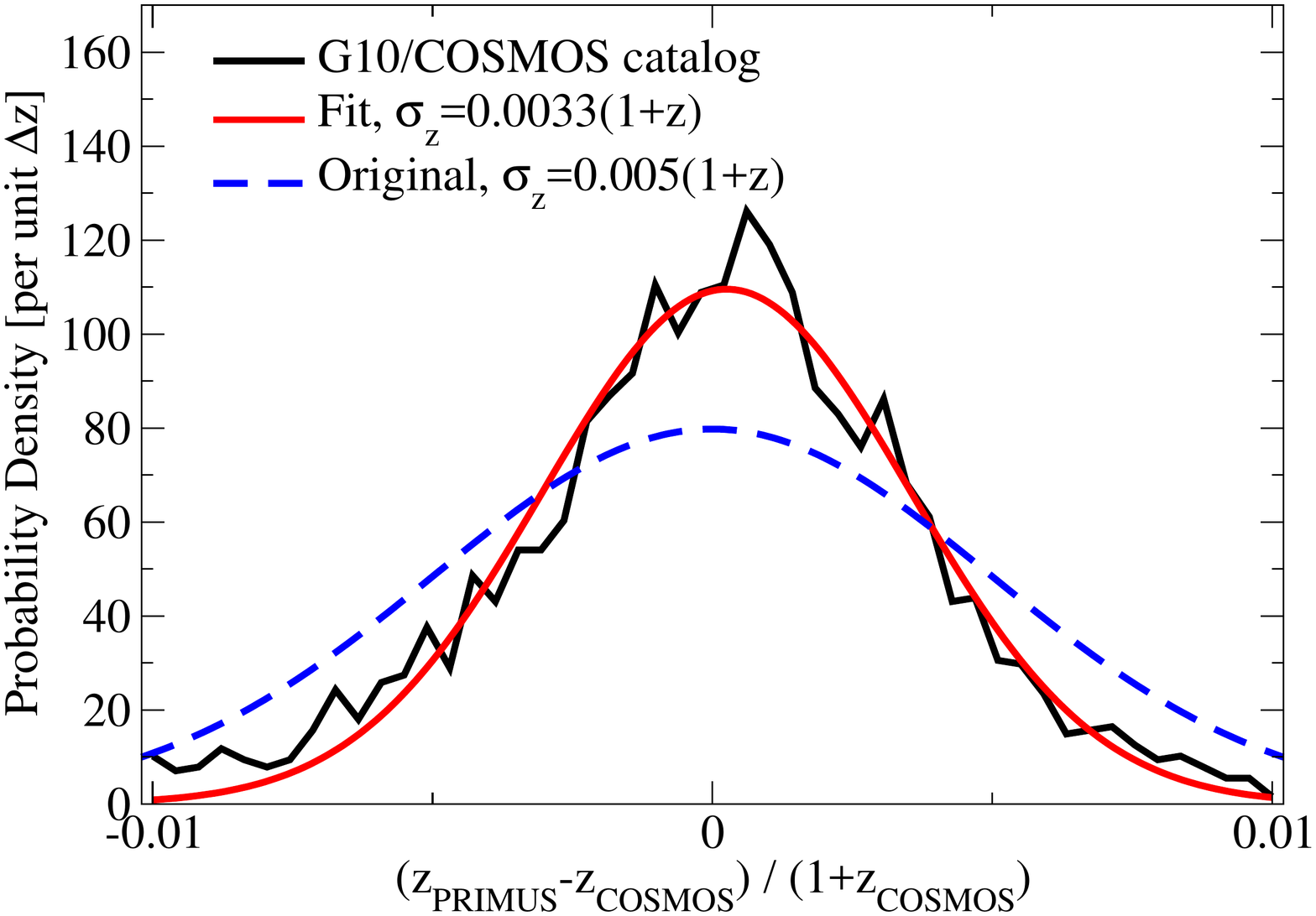}\\[-5ex]
\caption{Comparison between redshifts in the G10/COSMOS catalogue \protect\citep{Davies15} and the PRIMUS DR1 catalogue \protect\citep{Coil11}.  This test suggests that the original redshift errors reported in \protect\cite{Coil11} overestimate the true redshift errors in PRIMUS.}
\label{f:primus_errs}
\end{figure}

At $z>0$, we adopt the PRIMUS correlation functions from \cite{Coil17}.  Of the reported correlation functions, only one mass bin ($10^{10.5}\Msun < M_\ast < 10^{11}\Msun$ at $0.2 < z < 0.7$) is complete and has both star forming and quenched samples available.  Covariance matrices for this data were generated in the same way as for the SDSS (i.e., from MDPL2, using a lightcone with a footprint matching the observed angular area).  Tests on these mock lightcones revealed that it was not possible to match the observed PRIMUS clustering amplitudes with reported redshift errors \citep{Coil11} of $\sigma_z = 0.005(1+z)$.  Cross-comparing redshifts between the PRIMUS DR1 catalogues and the G10/COSMOS catalogues \citep{Davies15} revealed much lower typical errors of $\sigma_z = 0.0033(1+z)$, as shown in Fig.\ \ref{f:primus_errs}.  We therefore adopt this lower estimate of the redshift errors when comparing to the \cite{Coil17} data.

\subsection{Environmental Dependence of Central Galaxy Quenching}
\label{a:ecq}
\begin{figure}
\vspace{-5ex}
\plotgrace{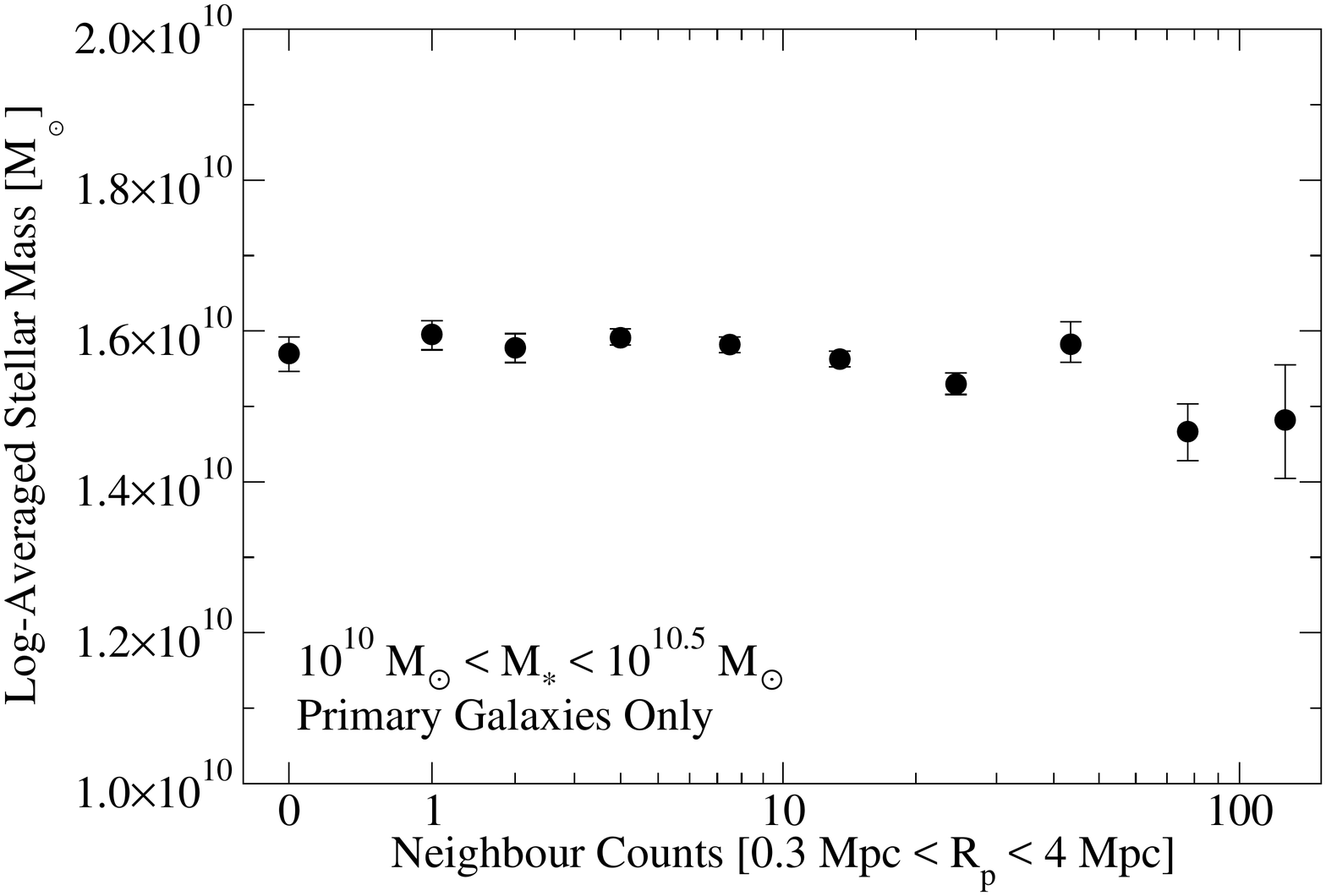}\\[-5ex]
\plotgrace{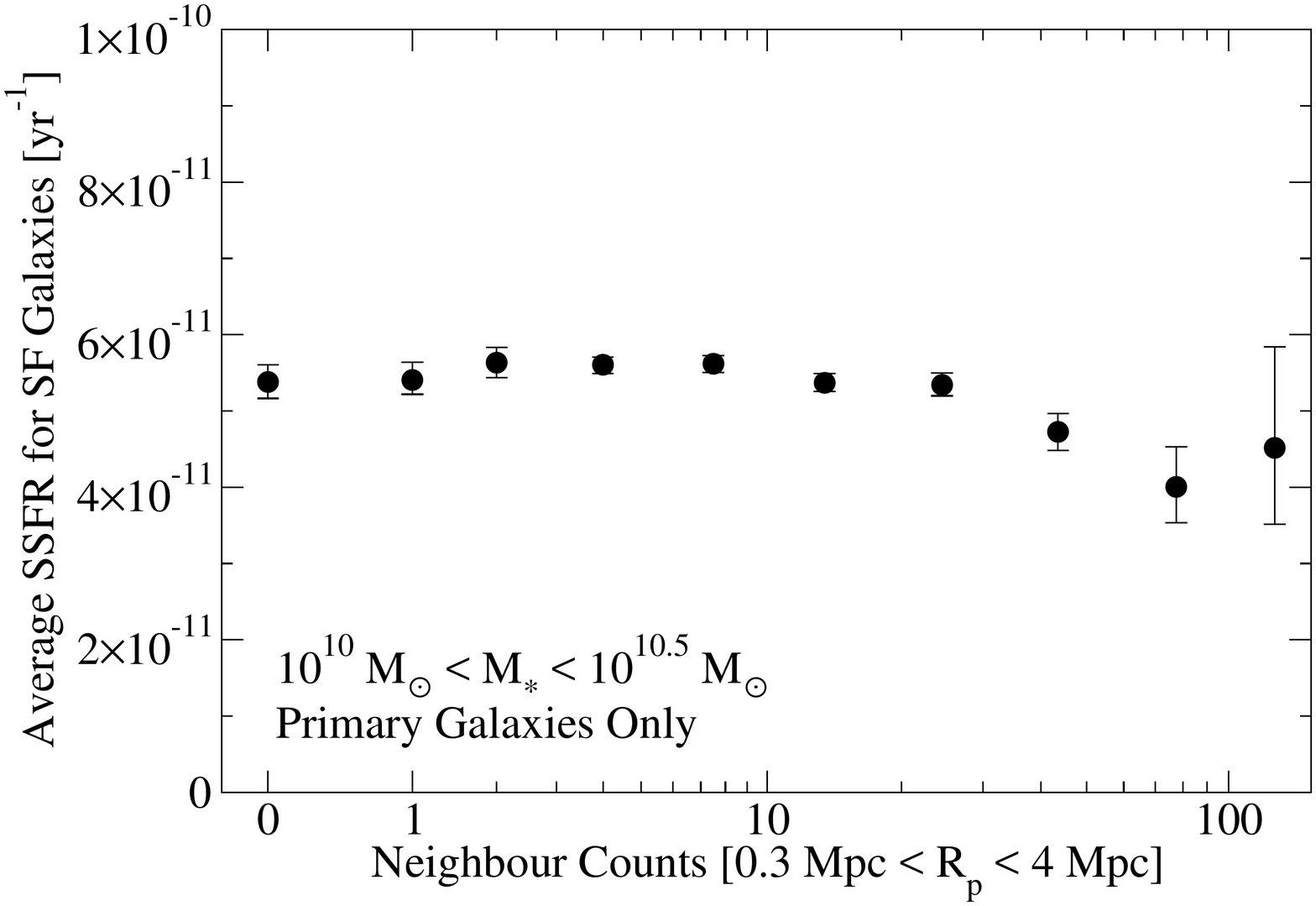}\\[-5ex]
\caption{\textbf{Top} panel: average stellar mass of primary galaxies as a function of neighbour count, showing that the chosen environment proxy does not bias the stellar mass distribution.  \textbf{Bottom} panel: average specific SFR for star-forming primary galaxies as a function of environment.}
\label{f:c_quenching}
\end{figure}

Even though satellites make up a small fraction of galaxies ($\lesssim 30\%$ for $\mstar > 10^{10}\Msun$; \citealt{Reddick12}), correlation functions (CFs) weight galaxies by density.  This results in satellites dominating the CF signal to large distances; well into the traditional 2-halo regime, there are significant contributions from satellite-central pairs and satellite-satellite pairs with different host haloes.  As a result, CFs of star-forming and quenched galaxies provide an excellent constraint on satellite quenching, but are less useful for constraining central galaxy quenching.  Lensing of star-forming and quenched galaxies has a similar problem, as increasing the satellite fraction will boost the shear signal significantly.  Hence, higher-order statistics that explicitly exclude satellites are required to test quenching models for central galaxies.

The most well-known such statistic is two-halo conformity \citep{Kauffmann13}---i.e., the correlation between SSFRs for central galaxies separated by up to 4 Mpc (projected).  Two-halo conformity plausibly arises because of correlations between galaxy assembly history and environment \citep{Hearin14,Hearin15}.\footnote{The alternate physical model of preheating by nearby radio-loud AGN suggested by \cite{Kauffmann15} has largely indistinguishable effects.  \cite{Kauffmann15} notes that central galaxies in groups and clusters are much more likely to be radio-loud than field galaxies of the same stellar mass.  Yet, tidal forces at a given distance are greater from groups and clusters than from haloes of field galaxies, so the radio-loudness of (and preheating from) nearby galaxies is in fact strongly correlated with the tidal field.  As tidal fields significantly influence halo assembly \citep{Hahn09,BehrooziMergers,Hearin15}, it then becomes challenging to discriminate which is the underlying cause for spatially correlated quenching of central galaxies.}  As galaxies share similar environments on large scales (evidenced by $\sigma_8 \sim 1$), central galaxy quenching is spatially correlated.

Two-halo conformity thus measures a very indirect correlation (galaxy 1's SSFR $\leftarrow$ environment $\rightarrow$ galaxy 2's SSFR), and this has resulted in significant debate about its existence \citep[e.g.,][]{Sin17,Treyer18}.  Here, we achieve much better signal/noise for the same data by measuring the direct dependence of the quenched fraction of central galaxies on environment.  

To this end, we select $10^{10}\Msun < \mstar < 10^{10.5}$ galaxies from the SDSS DR7 over the redshift range $0.01 < z < 0.057$, using the same stellar masses and star formation rates discussed in Appendix \ref{a:cf}.  To select primarily central galaxies ($\equiv$ primary galaxies), we exclude galaxies with larger (in stellar mass) neighbours within 500 kpc in projected distance and 1000 km s$^{-1}$ in redshift; we also excluded those near bright cluster galaxies lacking spectra as discussed in \cite{BehrooziMM}.  Tests on mock catalogues suggest that this selection has a purity of 97.4\% and a completeness of 77\% for central galaxies \citep{BehrooziMM}.  We define environment as the number of neighbouring galaxies within 0.3 to 4 Mpc (i.e., in an annulus excluding satellites) in projected distance, 1000 km s$^{-1}$ in redshift, and with stellar masses between $0.3 - 1$ times the mass of the central galaxy.  Without this cut on stellar mass, larger galaxies would preferentially have many more neighbours due to their higher bias.  As detailed in \cite{BehrooziMM}, we weight galaxies by the inverse observable volume of their dimmest potential neighbours using Eq.\ \ref{e:lum_cut}, and we exclude galaxies whose dimmest potential neighbours are not observable.  

Results are shown in Fig.\ \ref{f:ecq_comp}.  As a basic test of the selection technique, we also show that our environmental measure does not correlate with the stellar mass of the central galaxy (Fig.\ \ref{f:c_quenching}, top panel).  We note in passing that SSFRs of star-forming galaxies change relatively little as a function of environment (Fig.\ \ref{f:c_quenching}), suggesting (as in \citealt{Wetzel13b} and \citealt{Berti16}) a rapid transition from being star-forming to being quiescent.

\section{Deriving UV--Stellar Mass Relations at \textit{z}>4}

\label{a:uvsm}

\begin{figure}
\vspace{-5ex}
\plotgrace{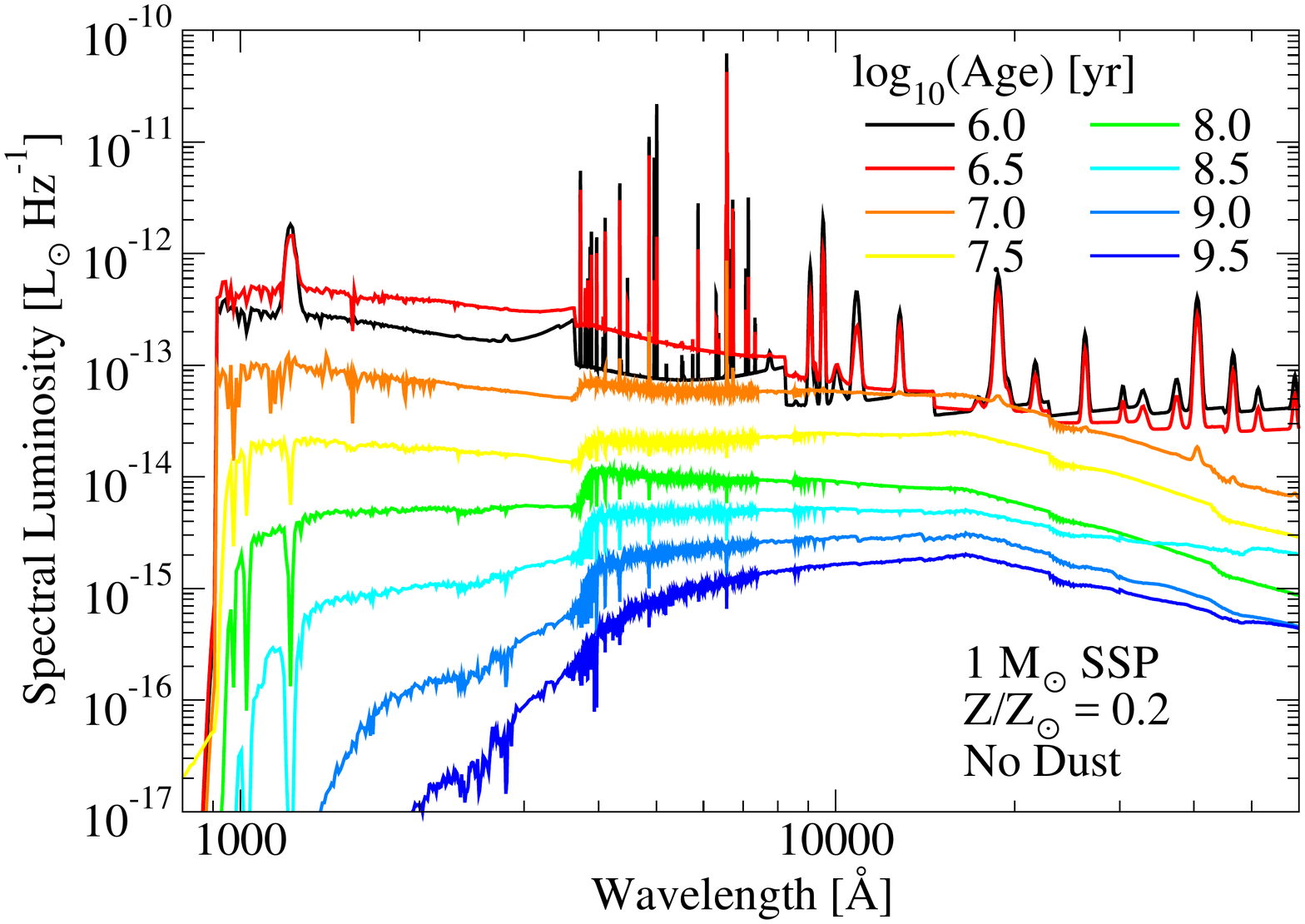}\\[-5ex]
\plotgrace{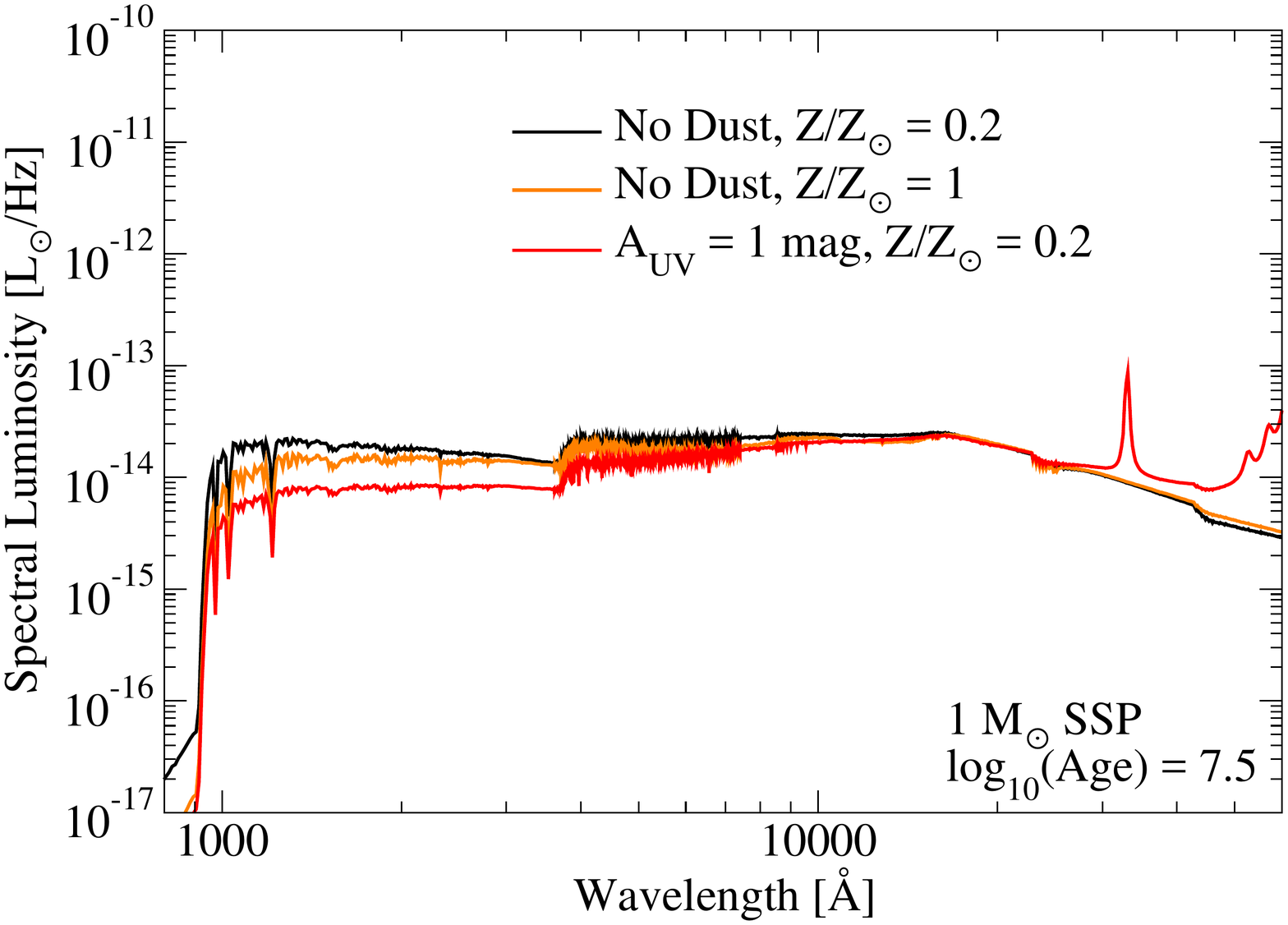}\\[-3ex]
\caption{\textbf{Top}: Spectral luminosity (i.e., the spectral energy distribution) for a 1$\Msun$ single stellar population (SSP), with nebular emission, no dust, and 0.2 solar metallicity.  Population age results in significant mass/light ratio variations.  \textbf{Bottom}: Spectral luminosity for a 1$\Msun$ SSP, varying dust and metallicity.  While mass/light ratio variations are significantly less, broadband spectral features sensitive to age (e.g., UV slopes) also depend on dust and metallicity.  All spectra from FSPS \citep{Conroy09}, with a \protect\cite{Chabrier03} IMF.}
\label{f:spectra}
\end{figure}

\begin{figure}
\vspace{-5ex}
\plotgrace{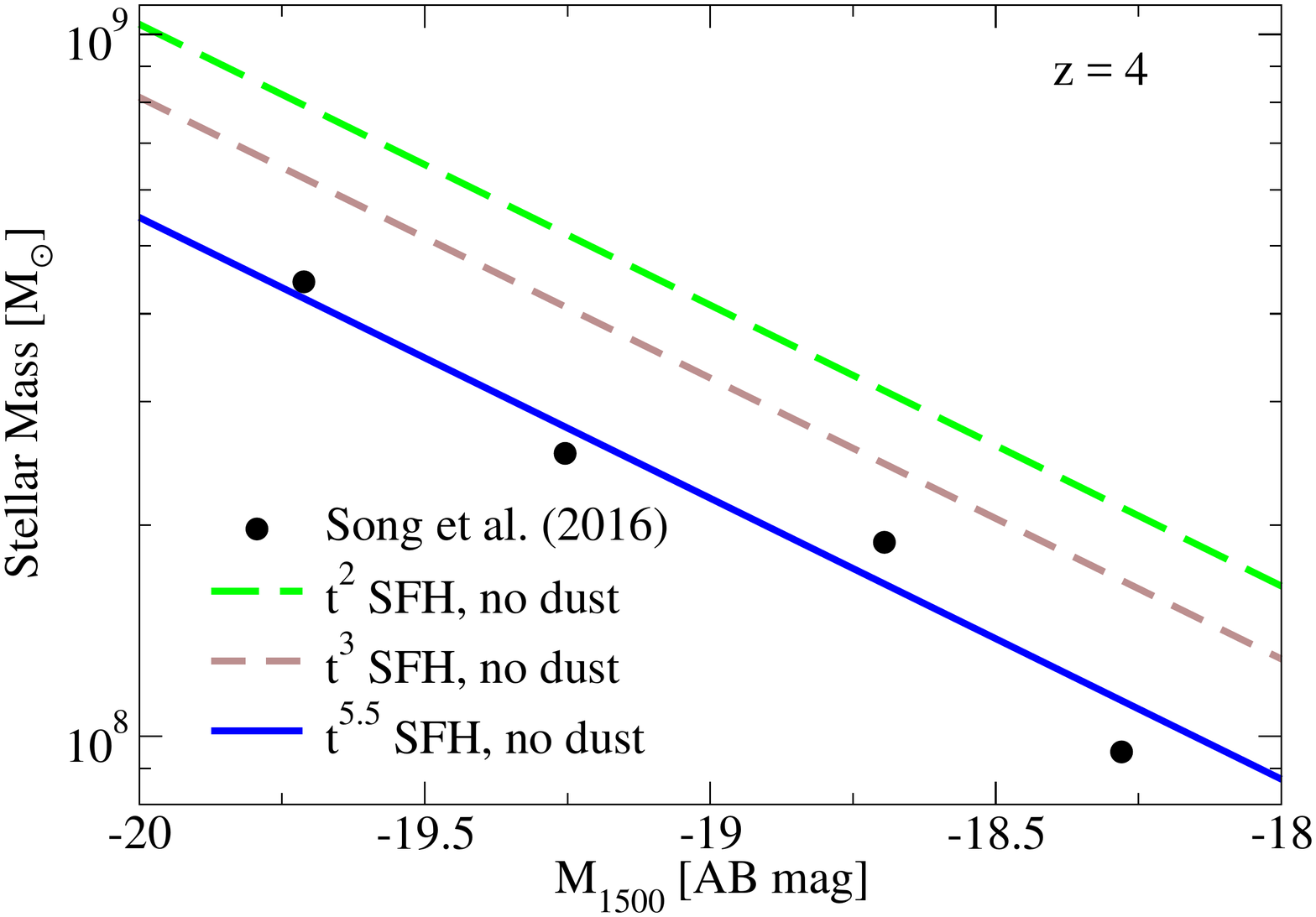}\\[-5ex]
\plotgrace{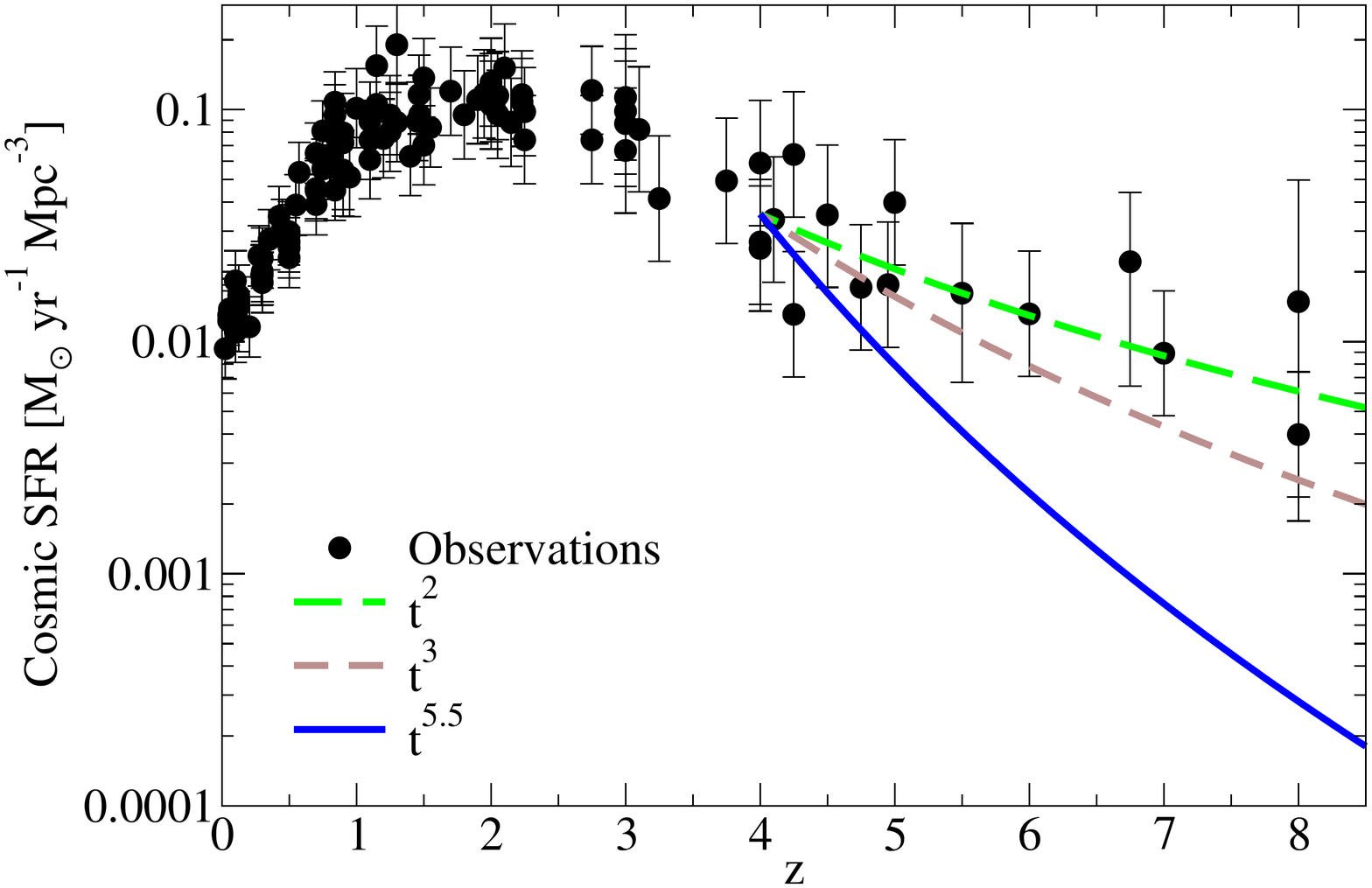}\\[-3ex]
\caption{\textbf{Top}: The \protect\cite{Song15} UV--stellar mass relation at $z=4$ (from median stacks), compared with expected UV--stellar mass relations for low-metallicity ($Z/Z_\odot = 0.03$) SFHs with no dust from FSPS \citep{Conroy09}.  At low luminosities, the \protect\cite{Song15} relation is best fit by a very steep $t^{5.5}$ SFH---requiring very short timescales for star formation.   Adding dust or metallicity would require even steeper SFHs to compensate; using \protect\cite{bc-03} instead of FSPS has minimal effect ($<$0.06 mag).   \textbf{Bottom}: growth rate of the cosmic star formation rate for several different power-law SFHs.  If galaxies typically had $t^{5.5}$ or similar SFHs at $z=4$, observed CSFRs would be very different at high redshifts.  More reasonable typical SFHs of $t^2$ \citep{Papovich11,BWC13,Salmon15} would result in 0.3 dex larger inferred masses at a given UV luminosity (see top panel).  \textbf{Notes}: \protect\cite{Song15} results are converted to a \protect\cite{Chabrier03} IMF to match analysis here.}
\label{f:uvsm_sfh}
\end{figure}

\begin{table*}
\caption{Stellar Mass Fitting Models.}
\begin{tabular}{lcccccc}
\hline
Legend (Fig.\ \ref{f:uv_sm_params}) & SFH & $\alpha$-prior & Burst & Dust & Dust Prior & $Z$ prior\\
\hline
\multicolumn{7}{l}{\textit{Pure Power-law SFHs}}\\
$0<\alpha<4$; ALMA dust$^*$ & $t^\alpha$ & $0<\alpha<4$ & no & 1c & \cite{Bouwens16} & \cite{Maiolino08}\\
$0<\alpha<4$; no dust prior & $t^\alpha$ & $0<\alpha<4$ & no & 1c & none & \cite{Maiolino08}\\
$0<\alpha<4$; ALMA dust; Free Z & $t^\alpha$ & $0<\alpha<4$ & no & 1c & \cite{Bouwens16} & \cite{Maiolino08} + arb.\ offset\\
$0<\alpha<10$; no dust priors & $t^\alpha$ & $0<\alpha<10$ & no & 1c & none & \cite{Maiolino08}\\
\hline
\multicolumn{7}{l}{\textit{Power-law + 20 Myr Burst SFHs}}\\
$0<\alpha<10$; no dust prior & $t^\alpha$ + burst & $0<\alpha<10$ & 20 Myr & 1c & none & \cite{Maiolino08}\\
$0<\alpha<4$; 1c ALMA dust & $t^\alpha$ + burst & $0<\alpha<4$ & 20 Myr & 1c & \cite{Bouwens16} & \cite{Maiolino08}\\
$0<\alpha<4$; 2c ALMA dust & $t^\alpha$ + burst & $0<\alpha<4$ & 20 Myr & 2c & \cite{Bouwens16} & \cite{Maiolino08}\\
\hline
\multicolumn{7}{l}{\textit{Power-law + Arbitrary-length Burst SFHs}}\\
$0<\alpha<10$; no dust prior & $t^\alpha$ + burst & $0<\alpha<4$ & Arb. & 1c & none & \cite{Maiolino08}\\
$0<\alpha<4$; 2c ALMA dust & $t^\alpha$ + burst & $0<\alpha<4$ & Arb. & 2c & \cite{Bouwens16} & \cite{Maiolino08}\\
\hline
\end{tabular}
\parbox{2.15\columnwidth}{\textbf{Notes.} 1c = 1-component dust (i.e., fixed optical depth for full SFH), 2c = 2-component SFH (i.e., different optical depth for younger stars, similar to \citealt{cf-00}).  ``Bursts'' are treated as periods of constant star formation at the end of the SFH.  $^*$ denotes the fiducial model.}
\label{t:sfh_params}
\end{table*}

\begin{figure*}
\vspace{-8ex}
\plotgrace{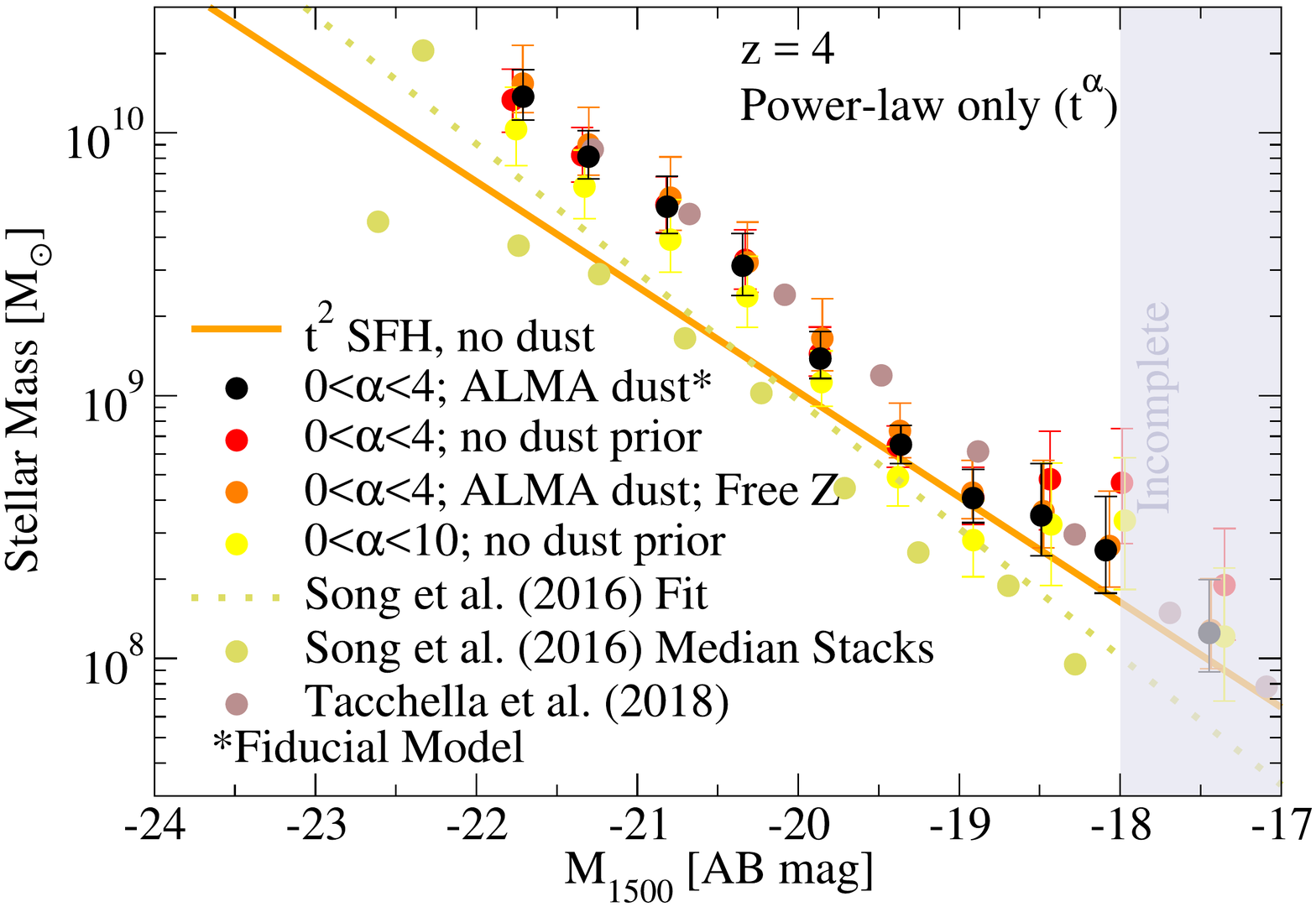}\plotgrace{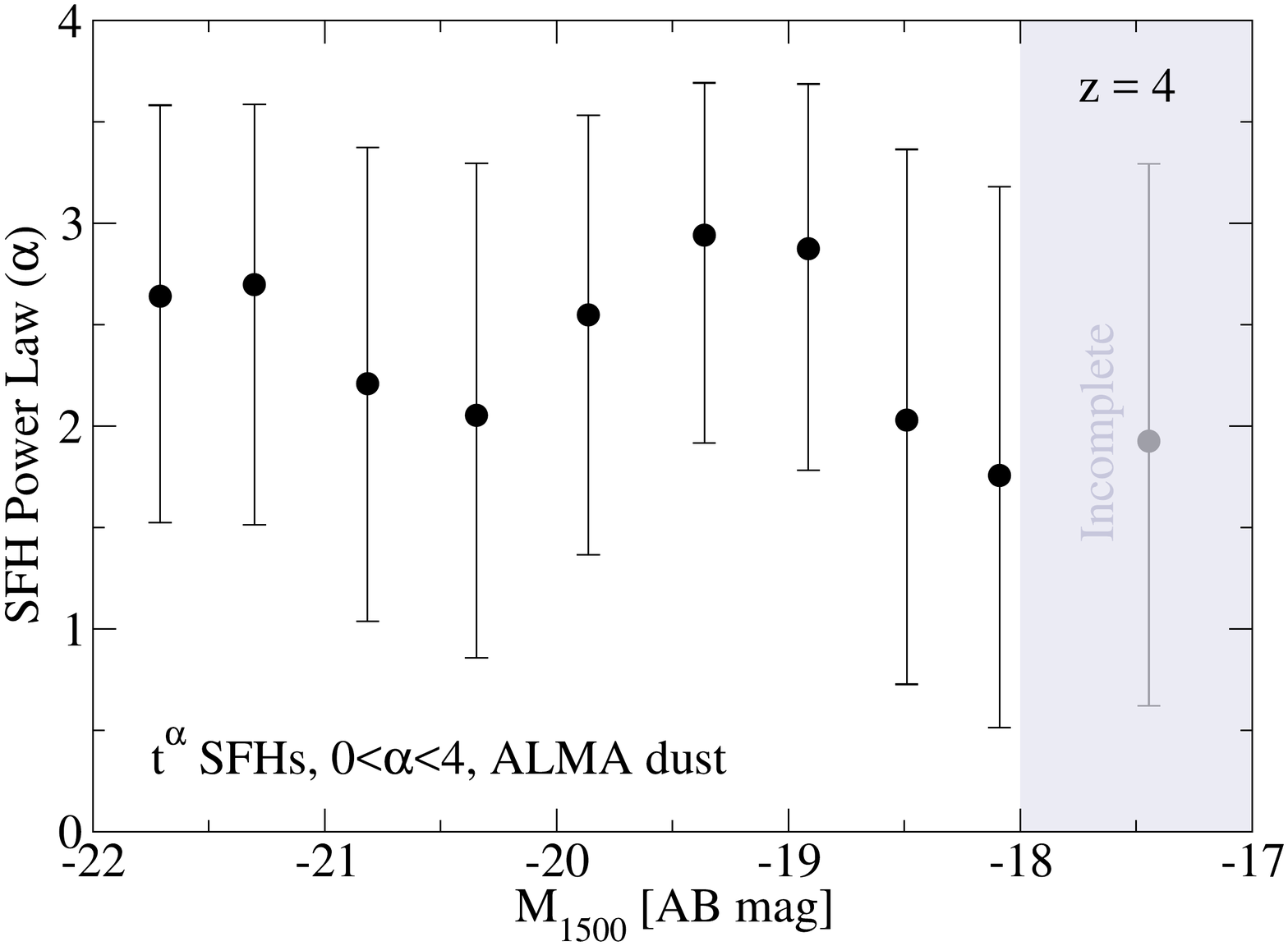}\\[-6ex]
\plotgrace{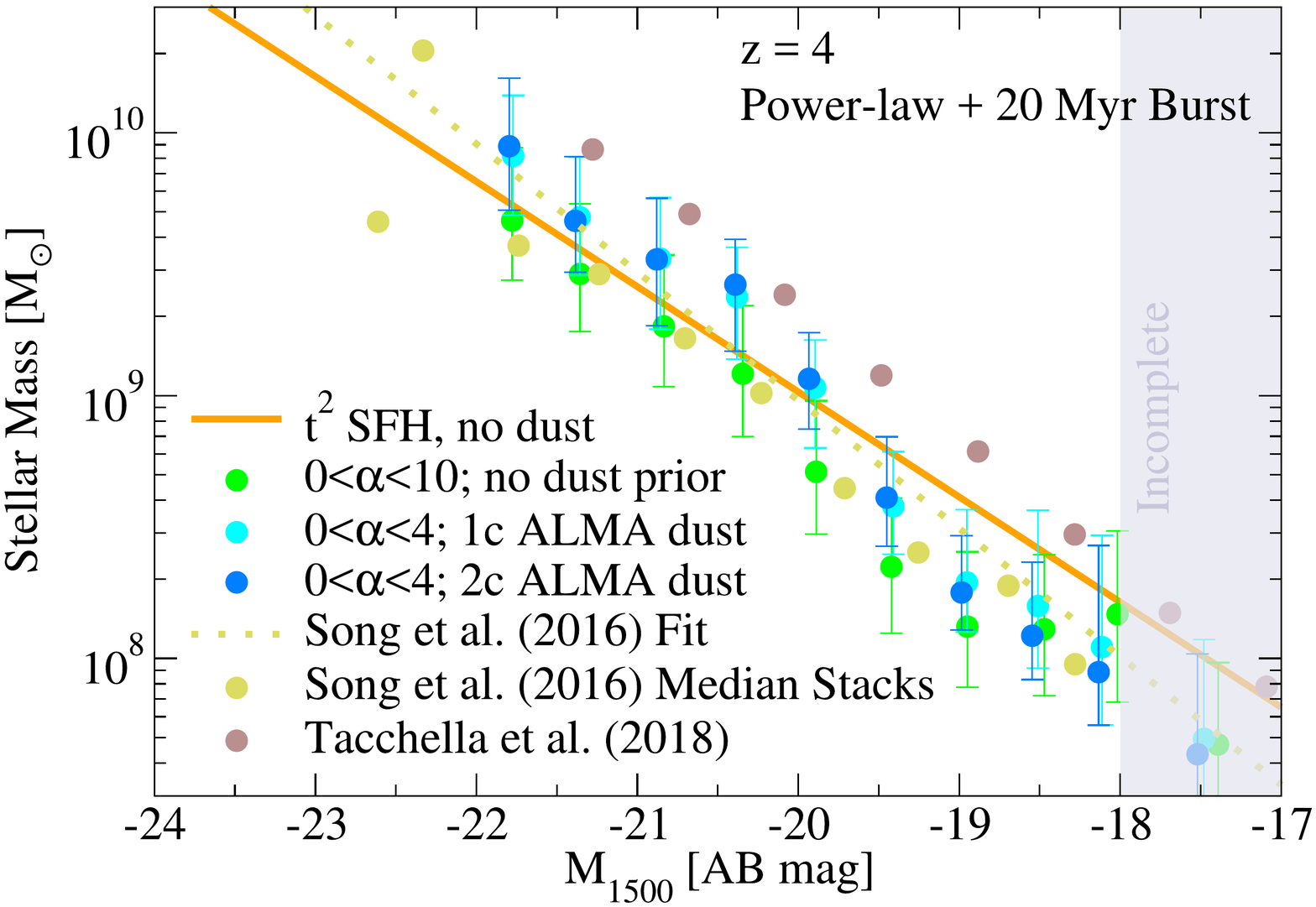}\plotgrace{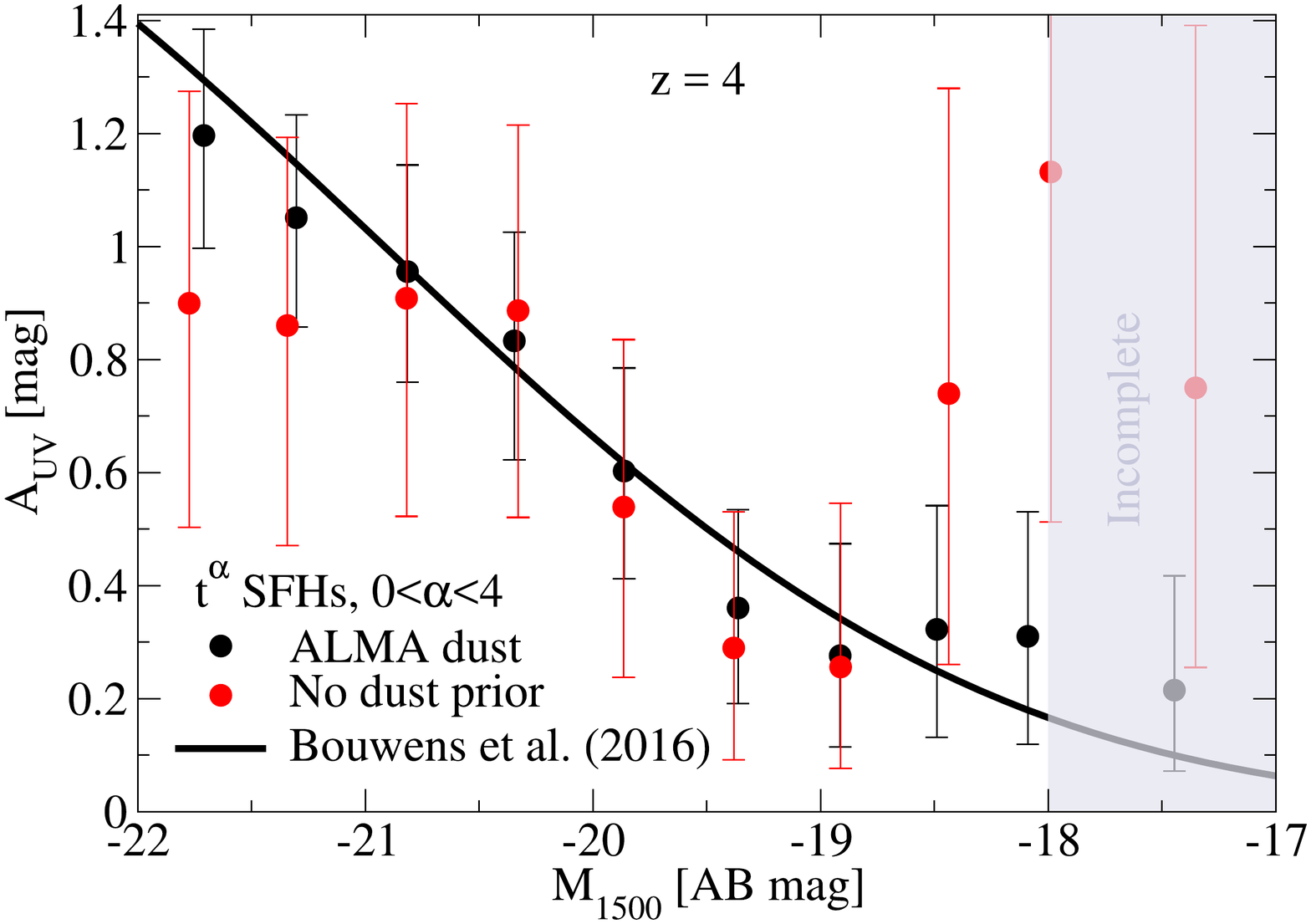}\\[-6ex]
\plotgrace{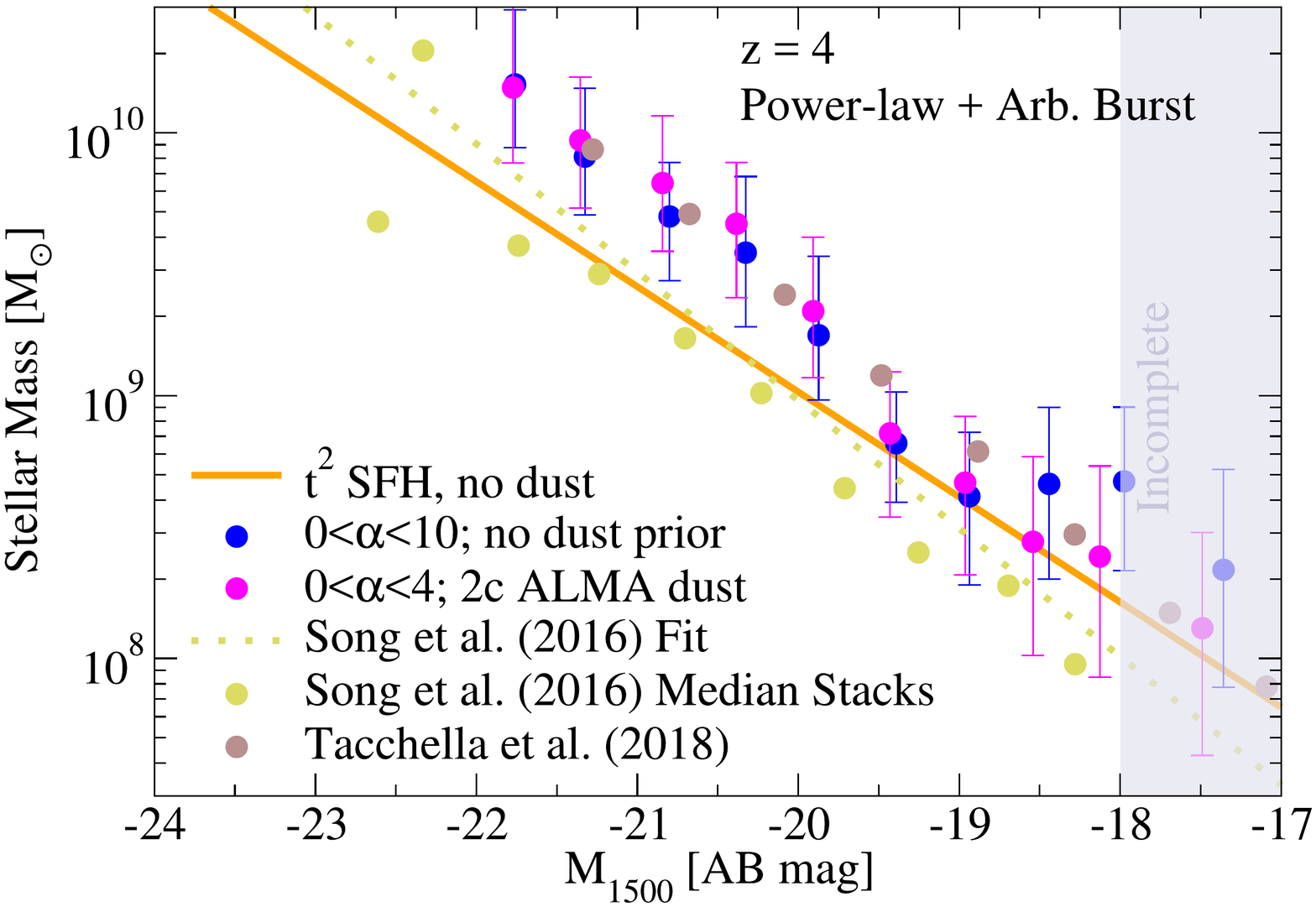}\plotgrace{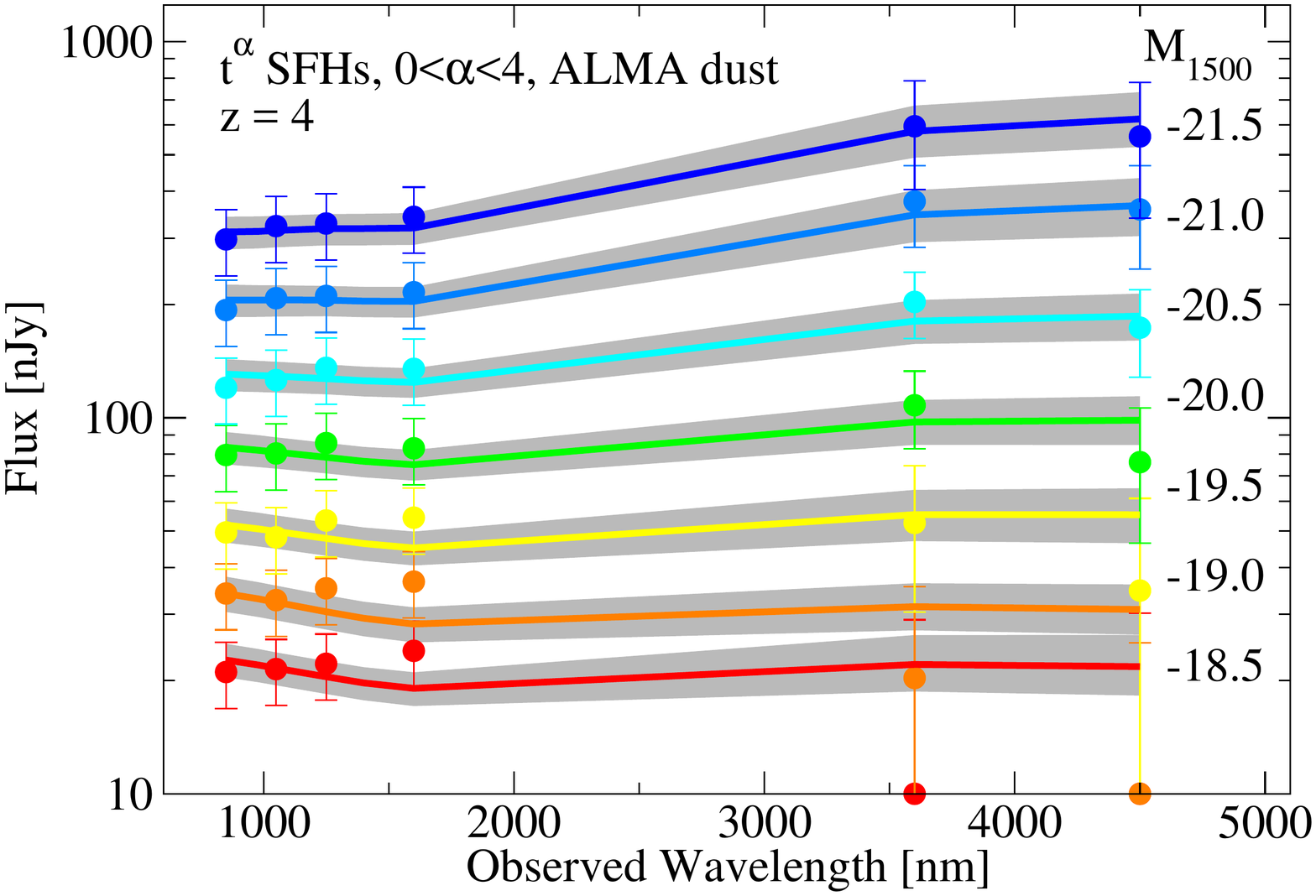}\\[-5ex]
\caption{\textbf{Left} panels:  Effects of priors and parameterizations (see Table \protect{\ref{t:sfh_params}} for full descriptions) on inferred mass-to-light ratios at $z=4$, based on fits to \protect\cite{Song15} SED stacks.  Coloured points with error bars show the median and $68^\mathrm{th}$ percentile range of the posterior distributions for each prior/parameterization set.  Coloured points without error bars show comparisons to median relations from \protect\cite{Song15} and \protect\cite{Tacchella18}.  Using pure power-law SFHs (\textbf{top-left}) or power-law SFH plus arbitrary-length bursts (\textbf{bottom-left}) gives higher mass-to-light ratios than using power-law SFHs with fixed 20 Myr bursts (\textbf{middle-left} panel).   \textbf{Top-right} panel: inferred power-law index for galaxy SFHs.  While constant SFHs are modestly disfavoured, the index is otherwise poorly constrained due to degeneracies, and it is hence heavily influenced by the prior.  \textbf{Middle-right} panel: inferred total attenuation ($A_{UV}$) at $z=4$ with and without dust priors from \protect\cite{Bouwens16}.  Inferred $A_{UV}$ values without dust priors are consistent with the \protect\cite{Bouwens16} results except at low luminosities, where dust content is poorly constrained by the optical/near-infrared SEDs.   \textbf{Bottom-right} panel: posterior distributions (\textit{grey bands}; \textit{coloured lines} are best-\fit{} results) for the broadband SED stacks from \protect\cite{Song15} (\textit{coloured points}).  Red UV slopes at faint luminosities are in modest tension with the low \textit{Spitzer} flux and low dust emission inferred in \protect\cite{Bouwens16}.  \textit{Light steel-blue shaded regions} denote the luminosity completeness thresholds in \protect\cite{Song15}.    \textbf{Notes}: \protect\cite{Song15}  and \protect\cite{Tacchella18} results are converted to a \protect\cite{Chabrier03} IMF to match analysis here.}
\label{f:uv_sm_params}
\end{figure*}

\begin{figure*}
\vspace{-8ex}
\plotgrace{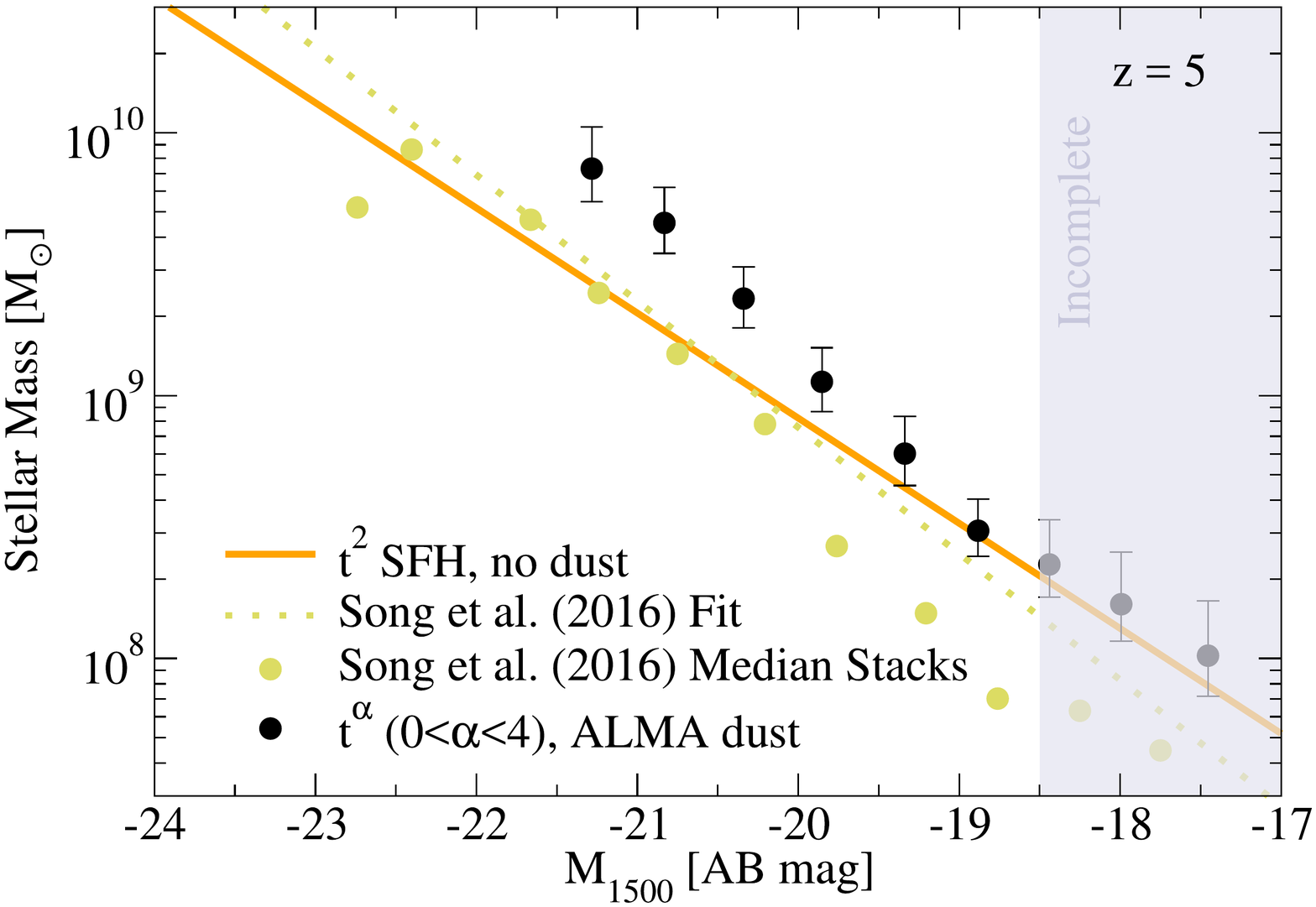}\plotgrace{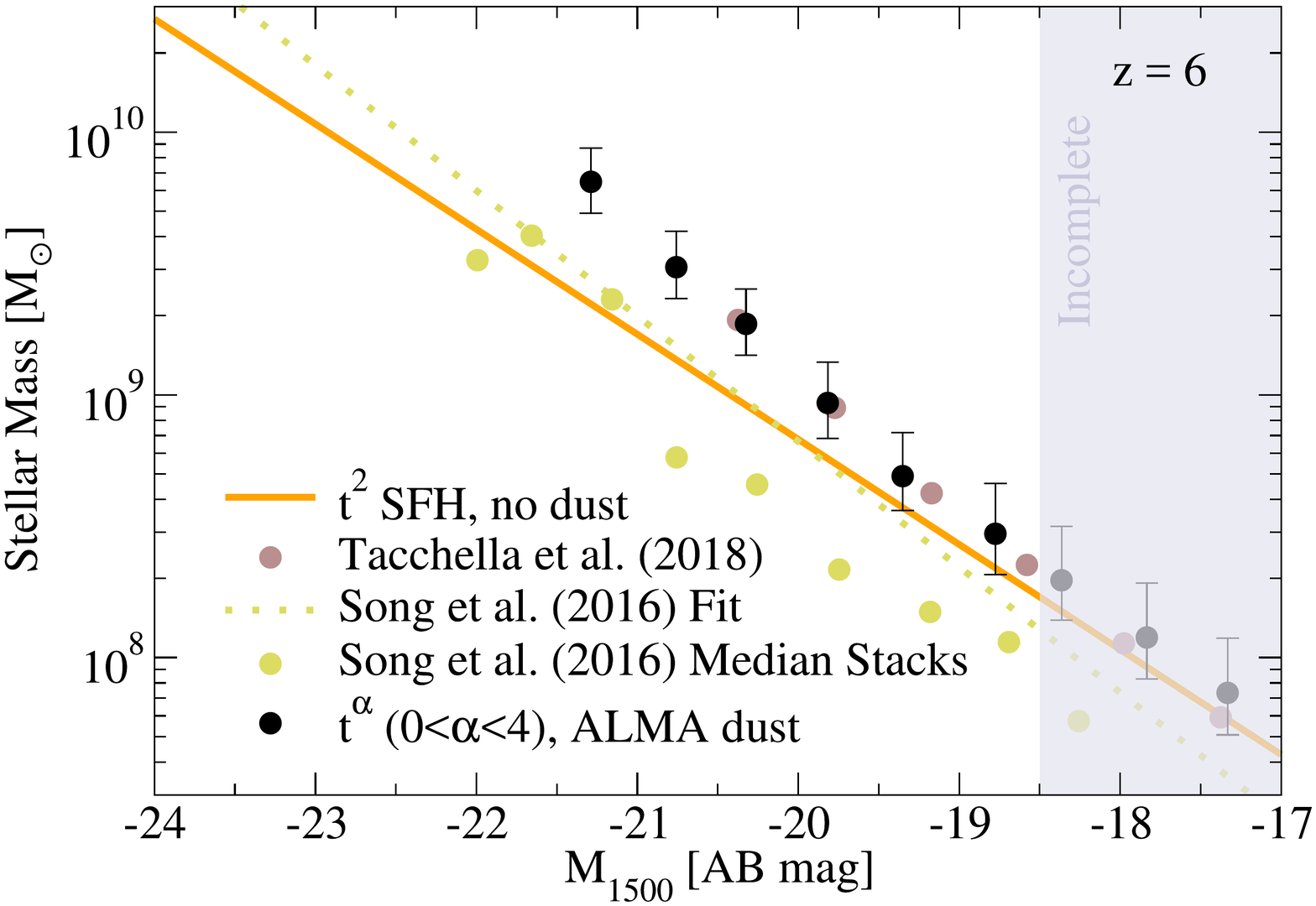}\\[-6ex]
\plotgrace{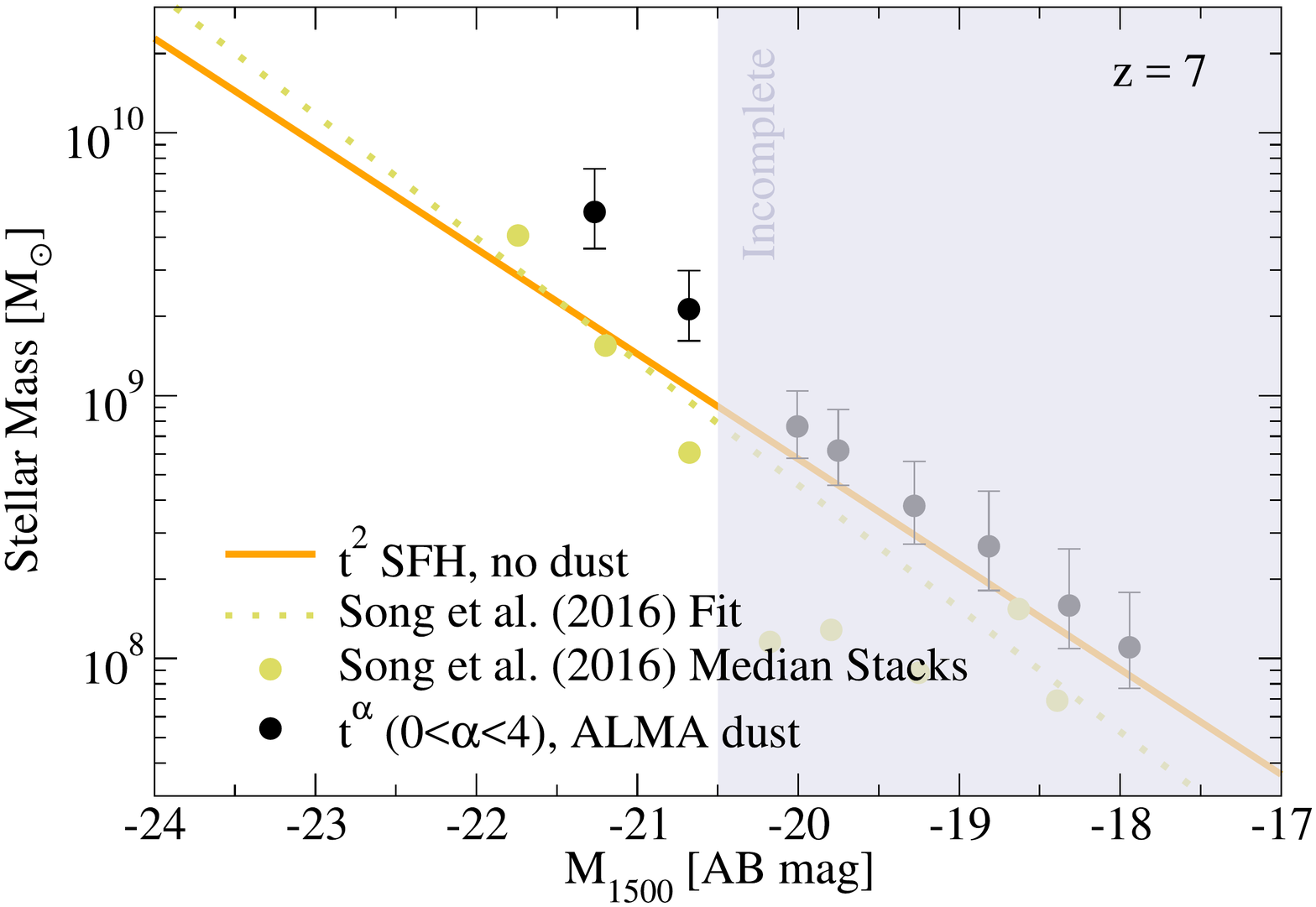}\plotgrace{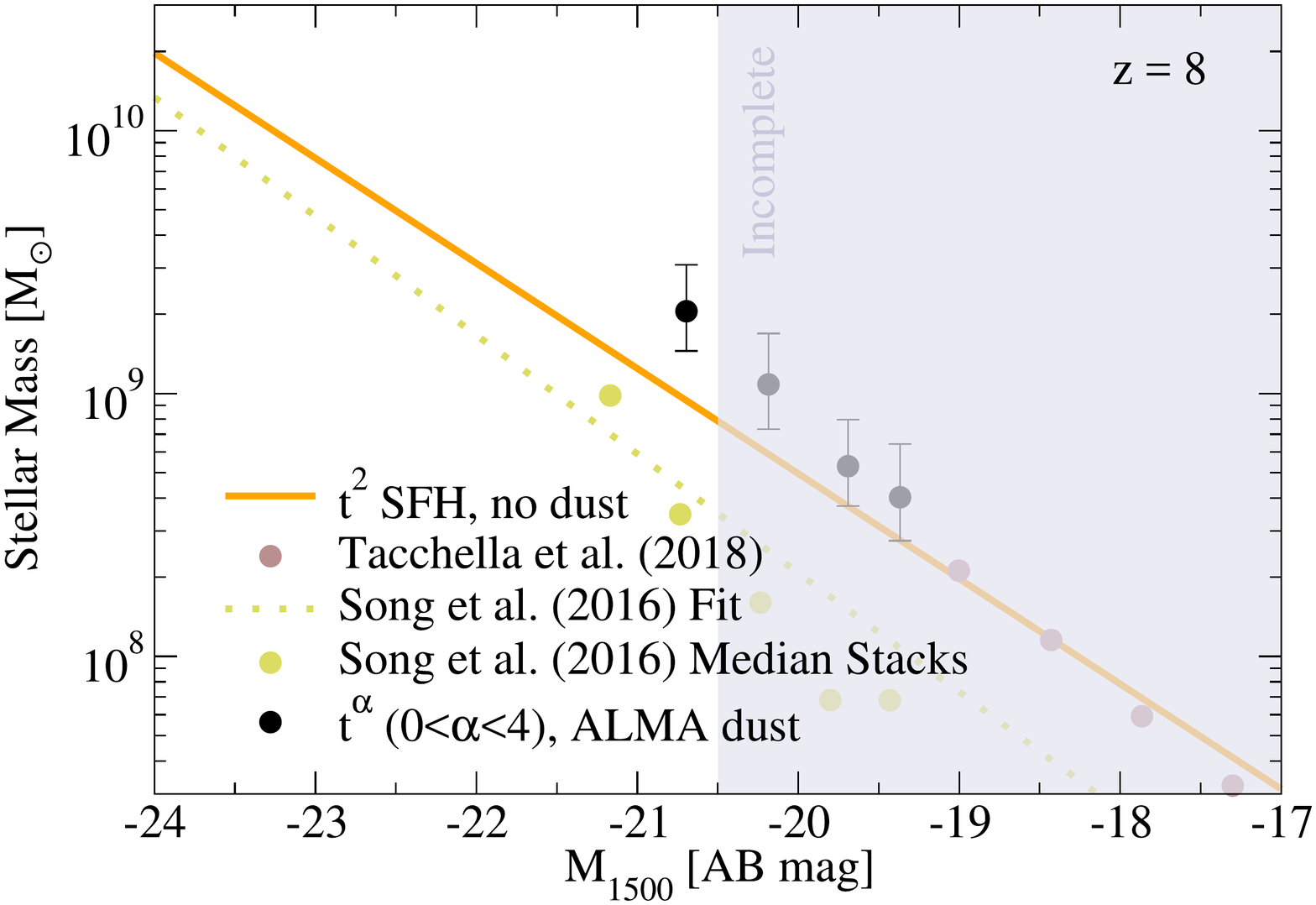}\\[-5ex]
\caption{UV--stellar mass relations for $z=5$ to $z=8$ obtained from \textsc{SEDition} and the \protect\cite{Song15} SED stacks using fiducial priors ($t^{\alpha}$ SFHs with $0<\alpha<4$, dust priors from \citealt{Bouwens16}, and metallicity histories consistent with \citealt{Maiolino08}). Black points with error bars show the median and $68^\mathrm{th}$ percentile range of the posterior distributions for the fiducial prior/parameterization set. Luminosity bins with fewer than 10 ($z=5$) or 5 ($z=6-8$) galaxies in \protect\cite{Song15} were not fit.   coloured points without error bars show comparisons to median relations from \protect\cite{Song15} and \protect\cite{Tacchella18}.    \textit{Light steel-blue shaded regions} denote the luminosity completeness thresholds in \protect\cite{Song15}.    \textbf{Notes}: \protect\cite{Song15} and \protect\cite{Tacchella18} results are converted to a \protect\cite{Chabrier03} IMF to match analysis here.  \protect\cite{Tacchella18} relations at $z=8$ are fully within the incomplete region.}
\label{f:uvsm_stacks}
\vspace{-6ex}
\plotgrace{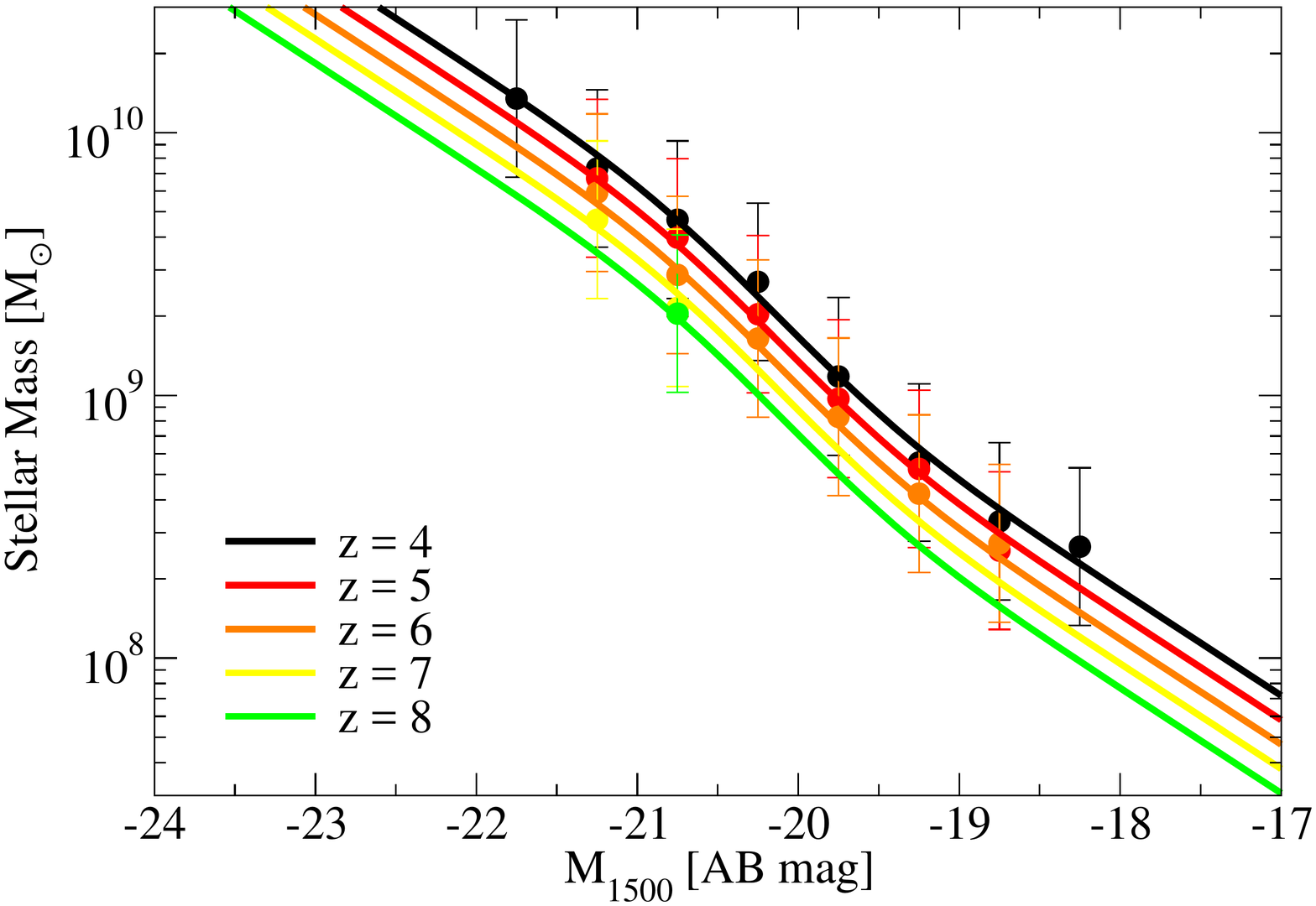} \plotgrace{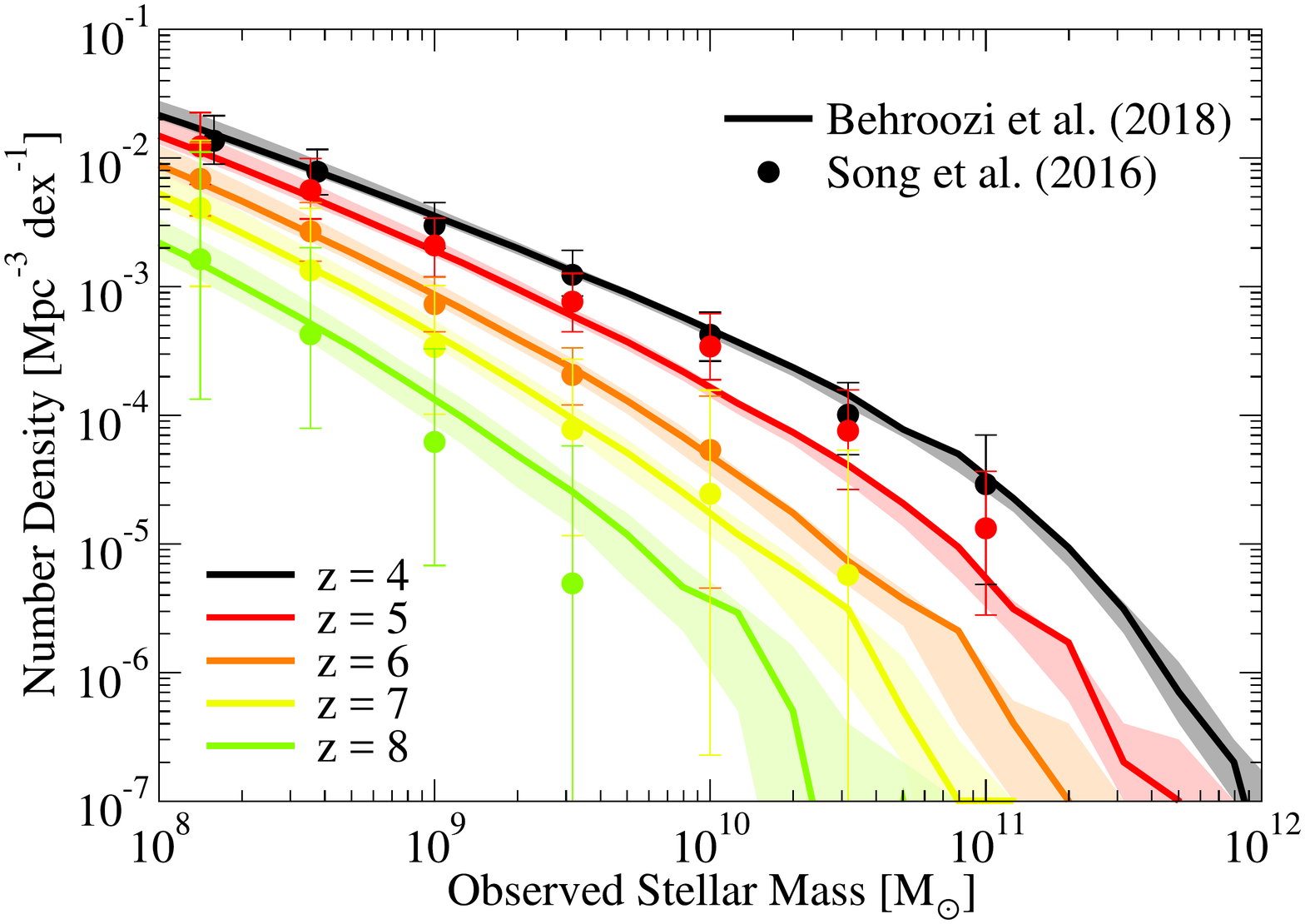}\\[-4ex]
\caption{\textbf{Left} panel: Fits (Eq.\ \ref{e:uvsm_fit}) to median UV--stellar mass relations for $z=4$ to $z=8$ obtained from \textsc{SEDition}.  \textbf{Right} panel: $z=4-8$ stellar mass functions, derived from the \textsc{UniverseMachine}.  \textit{Lines} are the best-\fit{} model, and \textit{shaded regions} show the \onesigdist{} of the posterior distribution.  These principally depend on the UVLFs from \protect\cite{Finkelstein15} and the UV--SM relations in the left panel (no $z\ge 4$ SMFs were used as input constraints).  These are not significantly different from \protect\cite{Song15} except at $z=8$.  Data for both panels is available \href{http://www.peterbehroozi.com/data/}{\textbf{online}}.}
\label{f:uvsm_fit}
\end{figure*}

Converting light to stellar mass requires assumptions for photometry, redshift determinations, stellar population synthesis, star formation histories (SFHs), metallicities, dust amounts/sizes/geometries, and initial mass functions, introducing 0.3-0.5 dex systematic uncertainties \citep[see][for details]{Conroy10,Behroozi10}.  Determining stellar masses from broadband spectral energy distributions (SEDs) is especially difficult, as the main age-sensitive features are degenerate with dust, metallicity, and nebular emission.  For example, while the UV slope depends on age (Fig.\ \ref{f:spectra}, top panel), it also depends on metallicity and dust (Fig.\ \ref{f:spectra}, bottom panel).  The strength of the Balmer/4000\AA{} breaks also depend on age, but a large broadband ``break'' can arise both from an old, dust-free population and from a dusty population of young (<10 Myr) stars with significant nebular emission.  Due to these degeneracies, inferred mass-to-light ratios are very sensitive to priors on age, dust, and metallicity.

Flat priors, as used by many fitting codes, can lead to unphysical results.  For example, the median UV--SM relation for $z\sim4$ galaxies from \cite{Song15} requires very low mass-to-light ratios, typical of recent burst or steeply rising star formation histories.  Comparing the \cite{Song15} relation to those for power-law SFHs, we find that $t^{5.5}$ or steeper SFHs are required to achieve similar mass-to-light ratios (Fig.\ \ref{f:uvsm_sfh}), even with no dust and low metallicities.  However, if typical galaxies had such steep SFHs, very different UVLF and CSFR evolution would result (e.g., Fig.\ \ref{f:uvsm_sfh}, bottom panel; compare with Fig.\ \ref{f:ssfr_tension}).  If SFHs that matched the CSFR evolution ($t^2$) or UVLF/SMF evolution at $z>4$ ($t^{1.4}-t^{2}$; \citealt{Papovich11,BWC13,Salmon15}) were used instead, the inferred stellar masses would be 0.3 dex higher at fixed UV luminosity (Fig.\ \ref{f:uvsm_sfh}, upper panel).  While binning on UV luminosity does select galaxies with younger stellar populations, even very steep $t^3$ SFHs would still result in 0.2 dex larger stellar masses.

The SED data available in \cite{Song15} are nonetheless among the best currently published, as they include deep \textit{Spitzer} observations that probe rest-frame optical colours.  Hence, we re-fit the median SED stacks in \cite{Song15} with empirical priors for dust, metallicity, and age to obtain more self-consistent stellar masses.  To this end, we develop a new Bayesian SED-fitting code (\textsc{SEDition}, the Spectral Energy Distribution Inference Tool for Infrared/Optical Normalization).  Briefly, the code relies on FSPS to generate broadband luminosities as a function of stellar age, metallicity, and dust (including nebular continuum and emission lines).  These luminosities are combined with an affine MCMC algorithm \citep{Goodman10,emcee} to explore SFH, dust, and metallicity parameter space.    The code's two main features are the flexible inclusion of empirical priors on galaxy properties as well as the wide variety of built-in parameterizations for galaxy SFHs, dust histories, and metallicity histories.

The fiducial model used herein assumes a pure power-law SFH ($t^{\alpha}$, with $0<\alpha<4$), uniform optical depth (i.e., 1-component dust) with a \cite{calzetti-01} attenuation law and total $A_{UV}$ priors from \cite{Bouwens16}.  For the latter, we fit $A_{UV}(M_{1500})$ under the assumption of evolving dust temperatures \citep[e.g.,][]{Bethermin15}:
\begin{equation}
A_{UV}(M_{1500}) =  1+\mathrm{erf}\left(\frac{M_{1500}+20.93}{-3}\right).
\end{equation}
For metallicity histories, the fiducial model adopts the gas-phase metallicities of \cite{Maiolino08} for star formation.  Specifically, the metallicity of new stars is assumed to be the same as the observed gas-phase metallicity at the time of formation, which depends on redshift and the total remaining mass of previously-formed stars; we use a smooth fit to the \cite{Maiolino08} relation for $\log_{10}(Z(M_\ast, z)) \equiv 12 +  \log_{10}(O/H)$:
\begin{eqnarray}
\log_{10}(M_0(z)) &=& 11.22+0.47z \label{e:maiolino1}\\
K_0(z) & = & 9.07-0.07z\\
\log_{10}(Z(M_\ast, z)) & = & K_0(z) - 0.086\left[\log_{10}\left(\frac{M_\ast}{M_0(z)}\right)\right]^2. \label{e:maiolino3}
\end{eqnarray}
As in \S \ref{s:sm_uv}, we set a lower metallicity floor of $\log_{10}(Z/Z_\odot)=-1.5$ to avoid unphysically low metallicities at high redshifts and low stellar masses.  Besides this fiducial model, we test many other possible assumptions for SFHs, dust, and metallicities (Table \ref{t:sfh_params}).

For full details on the methods and data sources used for the SED stacks, see \cite{Song15}.  Briefly, SEDs were median-stacked for CANDELS galaxies in redshift bins from $z=4$ to $z=8$ ($\Delta z = 1$), and in rest-frame UV magnitude bins of 0.5 mag.  Available data included nonuniform coverage of $B_{435}$, $V_{606}$, $i_{775}$, $I_{814}$, $z_{850}$, $Y_{098}$, $Y_{105}$, $J_{125}$, $JH_{140}$, and $H_{160}$ bands \citep[see Table 1 of][for details]{Finkelstein15}, as well as $\textit{Spitzer}$ $3.6\mu$m and $4.5\mu$m bands.  Following \cite{Song15}, we exclude bands containing Ly$\alpha$ and bands for which $<50\%$ of galaxies had a detection.  Further, we exclude bands containing the Lyman break, and we use a dust law that does not include the 2175 \AA{} UV bump \citep[i.e.,][]{calzetti-01}; such features are expected to be smoothed because the galaxy SEDs were not $k$-corrected to the same redshift prior to stacking (M.\ Song, priv.\ comm.).  Finally, following \cite{Finkelstein15}, we assumed minimum 20\% photometric uncertainties, due to a combination of Poisson statistics, redshift errors, fitting errors, and stacking uncertainties.

Table \ref{t:sfh_params} lists the different parameterizations and priors tried for fitting the \cite{Song15} SED stacks.  Most assumptions gave significantly higher mass-to-light ratios than \cite{Song15}, and were consistent with $t^2$ dust-free mass-to-light ratios for low-luminosity galaxies (Fig.\ \ref{f:uv_sm_params}, left panels).  The largest impact came from changing the SFH parameterization.  Between pure power-law SFHs, power laws with 20 Myr constant-SFR bursts, and power laws with arbitrary-length bursts, the SFHs with 20-Myr bursts had significantly lower mass-to-light ratios than the other two parameterizations (Fig.\ \ref{f:uv_sm_params}, left panels).  Similarly, allowing very steep power-law SFHs ($t^{\alpha}$ with $0<\alpha<10$) resulted in lower mass-to-light ratios than our fiducial prior ($0<\alpha<4$).  Indeed, while constant SFHs were modestly disfavoured, the SEDs gave very weak constraints on the steepness of the SFH (Fig.\ \ref{f:uv_sm_params}, upper-right panel).  Hence, SFH parameter spaces biased toward younger ages gave lower mass-to-light ratios; as the \cite{Song15} SFH parameter space included SFHs as short as 1 Myr, this could be one explanation for the lower mass-to-light ratios therein.  Consistency with external SFH estimates from UVLF and SMF evolution was therefore primary motivation for our choice of fiducial SFH prior.

Changing the dust parameterization from 1-component (i.e., identical $A_{UV}$ for all stars) to 2-component (i.e., different $A_{UV}$ for recent stellar populations vs.\ older stellar populations as in \citealt{cf-00}) made little difference, as did allowing metallicity to vary (Fig.\ \ref{f:uv_sm_params}, left panels).  Including dust priors from ALMA measurements \citep{Bouwens16} made little difference for bright galaxies, because they were consistent with SED-derived dust estimates (Fig.\ \ref{f:uv_sm_params}, middle-right panel).  However, the ALMA priors reduced mass-to-light ratios for low-luminosity galaxies (Fig.\ \ref{f:uv_sm_params}, middle-right panel), possibly because of corresponding uncertainties in stacking \textit{Spitzer} flux (Dan Stark, priv.\ comm.).  As shown in the SED posterior distributions (Fig.\ \ref{f:uv_sm_params}, bottom-right panel), low-luminosity galaxies had somewhat red median UV slopes ($\beta < 2$) and faint \textit{Spitzer} fluxes, in modest tension with the lack of ALMA dust emission.  Given that other studies have found blue median UV slopes in this luminosity range \citep[e.g.,][]{Bouwens14}, we expect that including ALMA priors makes the mass estimates somewhat more realistic (i.e., less dust and lower mass-to-light ratios).

To summarize, our fiducial priors yield results consistent with external constraints on UVLF and SMF evolution, galaxy gas-phase metallicities, and ALMA dust measurements, but significantly different assumptions can still fit the stacked SEDs.  We thus adopt the median of the posterior distribution from our fiducial priors (Fig.\ \ref{f:uv_sm_params}, upper-left panel, and Fig.\ \ref{f:uvsm_stacks}), but we increase the errors to 0.3 dex for all bins as an estimate of the systematic errors inherent in our modeling assumptions.  Additionally, we exclude luminosity bins with fewer than 10 galaxies ($z=4-5$) or 5 galaxies ($z=6-8$) to reduce the influence of Poisson variations, and we adopt the same UV luminosity completeness cuts as in \cite{Song15}.  The resulting data points (Fig.\ \ref{f:uvsm_fit}) are available in electronic form \href{http://www.peterbehroozi.com/data/}{online}.

In the past, median UV--SM relations have been fit with single power-laws \citep[e.g.,][]{Gonzalez10,Stark13,Duncan14,Salmon15,Song15}.  At $z>4$, brighter galaxies also have more dust attenuation on average \citep[e.g.][]{Bouwens14}, so all fits to date have resulted in steeper slopes than -0.4 (i.e., that expected for constant mass-to-light ratios).  However, faint galaxies at $z>4$ have not to date shown significant dust attenuation \citep{Bouwens14,Bouwens16}, so it is unphysical to expect a steep power-law to continue fainter than, e.g., $M_{UV}=-19$.  We find that the following equation fits all our data well (Fig.\ \ref{f:uvsm_fit}) for $z=4$ to $z=8$:
\begin{eqnarray}
\log_{10}(M_\ast(M_{1500})) & = & -0.4(M_{1500} - 4.04 + 0.232z) \nonumber \\
&& + 0.189\mathrm{erf}(-20.11-M_{1500}). \label{e:uvsm_fit}
\end{eqnarray}

As shown in Table \ref{t:obs_summary}, we also include the \cite{Bouwens16} estimates of dust attenuation as a direct model constraint.  We allow for offsets between galaxies' observed and true stellar masses (\S \ref{s:systematics}), and so a highly-offset ``true'' stellar mass for galaxies could remain consistent with UV luminosity functions and the ``observed'' relation between UV luminosity and stellar mass, provided there was a large amount of dust attenuation.  Hence, the estimated constraint from  Fig.\ 14 of \cite{Bouwens16} on infrared excess as a function of observed UV luminosity places strong limits on this degeneracy.  We assume errors in the estimate that are the greater of $\pm 0.5$mag or the difference between the two provided dust temperature models.

In the main body of the paper, we forward-model the evolution of galaxies in dark matter haloes, determining which models can match both the observed UVLFs and median UV--SM relations.  No $z>4$ SMFs are used as input constraints (see \S \ref{s:obs}), so the posterior distributions for SMFs at $z>4$ are a nontrivial result.  As shown in Fig.\ \ref{f:uvsm_fit}, we find SMFs largely in agreement with those from \cite{Song15} from $z=4$ to $z=7$---our higher normalization for the median UV--SM relation is largely balanced by fewer galaxies skewing to very large stellar masses.  At $z=8$, however, we find a significantly higher SMF; the resulting SMF evolution from $z=7$ to $z=8$ is more in line with UVLF evolution over the same redshift range \citep{Finkelstein15} as compared to \cite{Song15}.

\section{Verifying Functional Forms}

\label{a:fforms}

\subsection{Quenched Fraction}

\begin{figure}
\vspace{-5ex}
\plotgrace{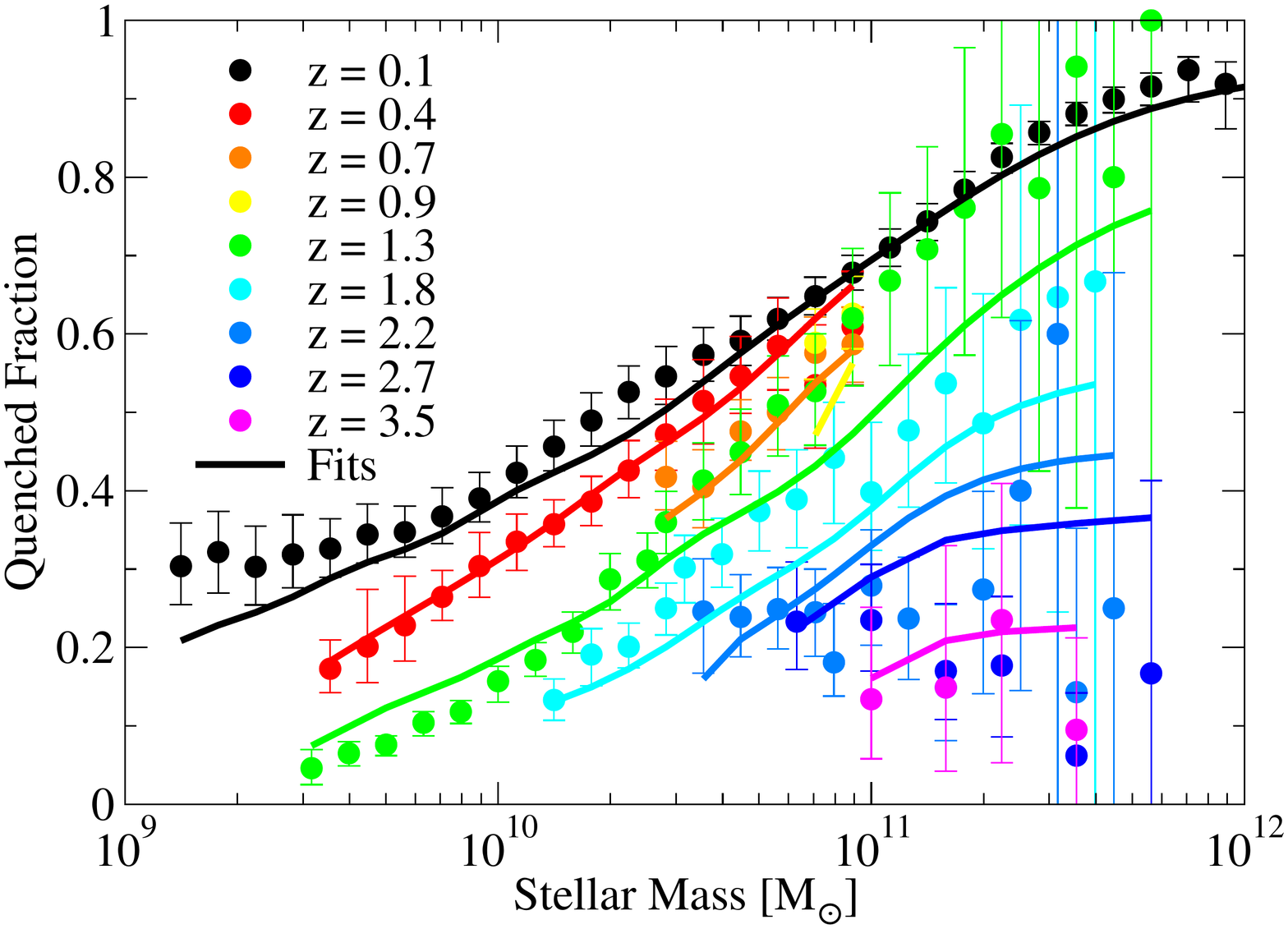}\\[-5ex]
\plotgrace{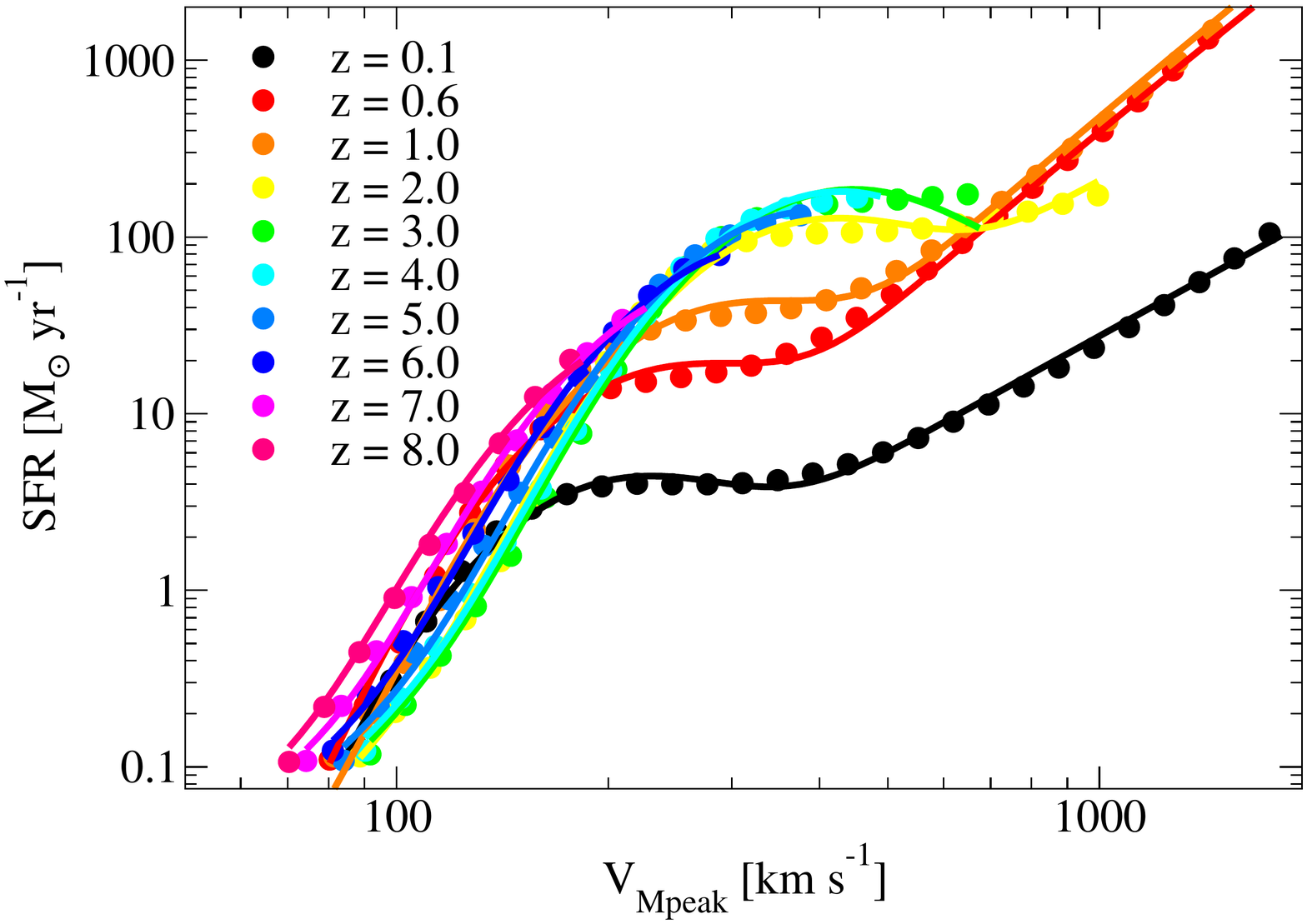}\\[-5ex]
\caption{\textbf{Top} panel: test that the parametrization for $\fq(\vmp,z)$ in Eqs.\ \ref{e:fq_1}--\ref{e:fq_3} (\textit{lines}) is flexible enough to match the observed data (\textit{points}) in Appendix \ref{a:qf} (see Appendix \ref{a:fq}).  \textbf{Bottom} panel: test that the parametrization for $\sfrsf$ in Eqs.\ \ref{e:v}--\ref{e:delta} (\textit{lines}) is flexible enough to match constraints from \protect\cite{BWC13} (\textit{points}), as discussed in Appendix \ref{a:sfr_sf}.  }
\label{f:ff_test}
\end{figure}

\label{a:fq}

The quenched fraction as a function of $\mstar$ is measurable directly (Appendix \ref{a:qf}), but the quenched fraction as a function of $\vmp$ is not.  Yet, assuming that the scatter in $\mstar$ at fixed $\vmp$ is not too correlated with SFR, the two are related by a convolution:
\begin{equation}
\fq(\mstar) = \int_0^\infty \fq(\vmp)P(\mstar|\vmp) d\vmp , \label{e:fq_test}
\end{equation}
where $P(\mstar|\vmp)$ is the probability distribution of $\mstar$ as a function of $\vmp$.  Given that the quenched fraction is bounded between 0 and 1 and that it increases with stellar mass (and therefore increases with $\vmp$), it is natural to try a sigmoid function for $\fq(\vmp)$, as in Eq.\ \ref{e:fq_1}.  We note that there may be an increase in the quenched fraction toward extremely low stellar masses \citep{Wetzel15b}; however, these are well below the halo mass resolution limit of the simulations we use.

We test the parametrization for $\fq(\vmp,z)$ in Eqs.\ \ref{e:fq_1}-\ref{e:fq_3} via forward-modeling through Eq.\ \ref{e:fq_test} and comparing to the observed $\fq(\mstar,z)$ (Fig.\ \ref{f:ff_test}, upper panel).  Briefly, we obtain $P(\mstar|\vmp)$ at the median redshift of each stellar mass bin in \cite{Moustakas13} and \cite{Muzzin13} via abundance matching the corresponding SMFs to haloes rank-ordered by $\vmp$, with 0.2 dex scatter (matching \citealt{Reddick12}).  Any choice of parameters in Eqs.\ \ref{e:fq_2}-\ref{e:fq_3} fully specifies $\fq(\vmp,z)$, from which Eq.\ \ref{e:fq_test} predicts the observed $\fq(\mstar)$.  Using the standard likelihood function ($\exp(-0.5\chi^2)$) and the \textsc{emcee} algorithm to explore parameter space \citep{emcee}, we find that Eqs.\ \ref{e:fq_1}--\ref{e:fq_3} are flexible enough to fit the observed data.

\subsection{Star-formation Rates for Star-forming Galaxies}

\label{a:sfr_sf}

Constraints on $\langle SFR(\mpeak, z)\rangle$ in \cite{BWC13} guide our functional form for $\sfrsf(\vmp, z)$.  The median $\vmp$ as a function of halo mass and redshift in \textit{Bolshoi-Planck} is
\begin{eqnarray}
\vmp(M_h,a) & = & 200 \mathrm{km}\, \mathrm{s}^{-1} \left[\frac{M_h}{M_{200\mathrm{kms}}(a)}\right]^3 \label{e:vmp}\\
M_{200\mathrm{kms}}(a) & = & \frac{1.64 \times 10^{12} \Msun}{\left(\frac{a}{0.378}\right)^{-0.142} + \left(\frac{a}{0.378}\right)^{-1.79}}.
\end{eqnarray}
Then, $\langle SFR \rangle$ and $\sfrsf$ are related as
\begin{equation}
\sfrsf(\vmp,z) \approx C_\sigma \frac{\langle SFR(\mpeak(\vmp,a),z)\rangle}{1-\fq(\vmp, z)},
\end{equation}
where $C_\sigma$ is a constant depending on the scatter in SFR at fixed $\vmp$; the equation is not exact due to very modest scatter ($<0.1$ dex) in $\vmp$ at fixed $\mpeak$.  For $\fq$, we use the best fit from Appendix \ref{a:fq}; the resulting estimate of $\sfrsf$ is shown as filled circles in Fig.\ \ref{f:ff_test}, bottom panel.  As in Appendix \ref{a:fq}, we use the \textsc{emcee} algorithm to explore parameter space, finding again that Eqs.\ \ref{e:v}--\ref{e:delta} are flexible enough to fit constraints in \cite{BWC13}.  Notably, simpler parameterizations (including double power-laws) would not be sufficient, due to an extra bump in efficiency near the transition between low-$\vmp$ and high-$\vmp$ power laws.

\section{Code Implementation, Parallelization, and Scaling}

\label{a:code}

\begin{figure}
\vspace{-5ex}
\plotgrace{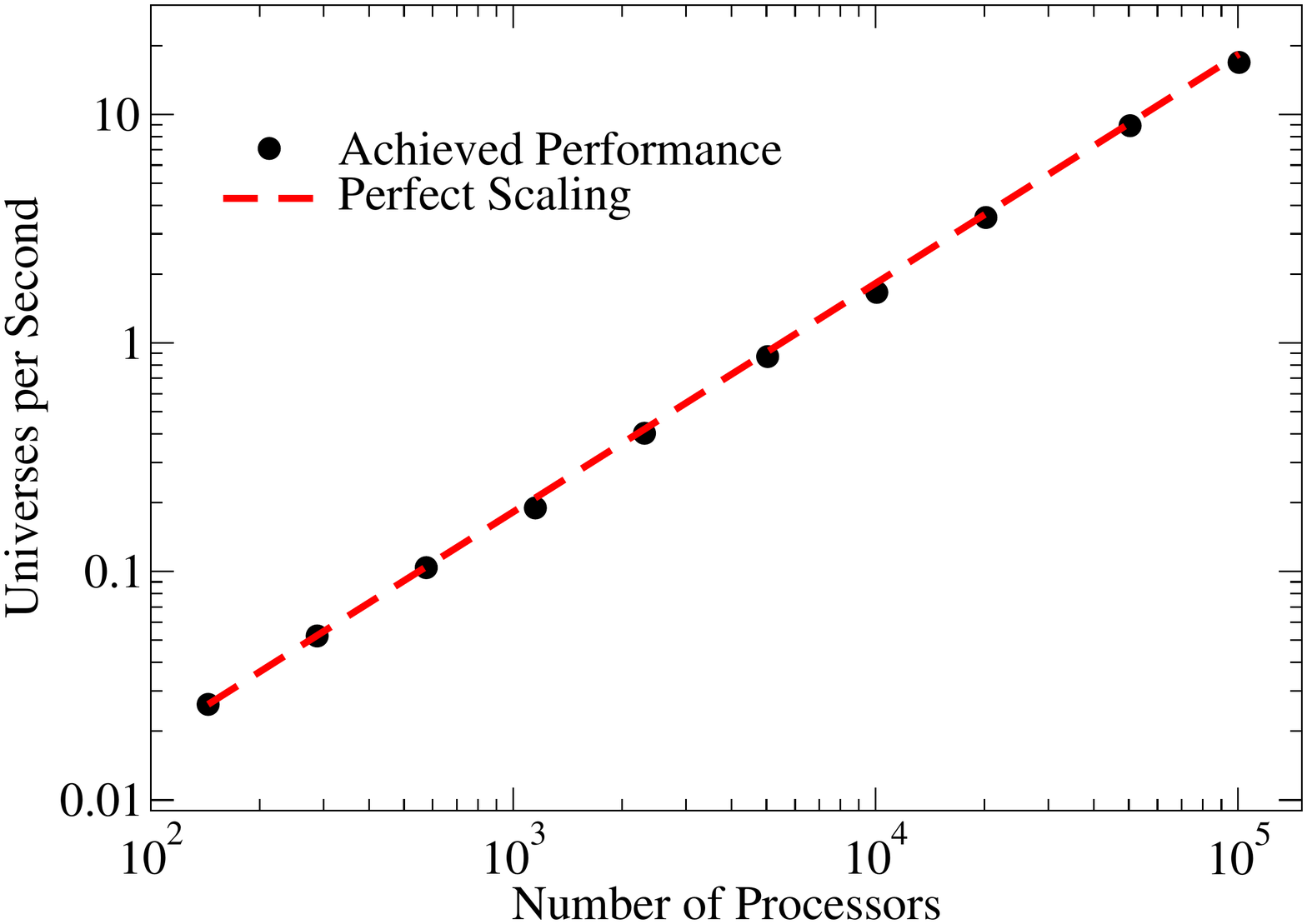}\\[-5ex]
\caption{Strong scaling performance of the \textsc{UniverseMachine} code on Edison.  Near-linear speedup with the number of processors is achieved.}
\label{f:perf}
\vspace{-5ex}
\plotgrace{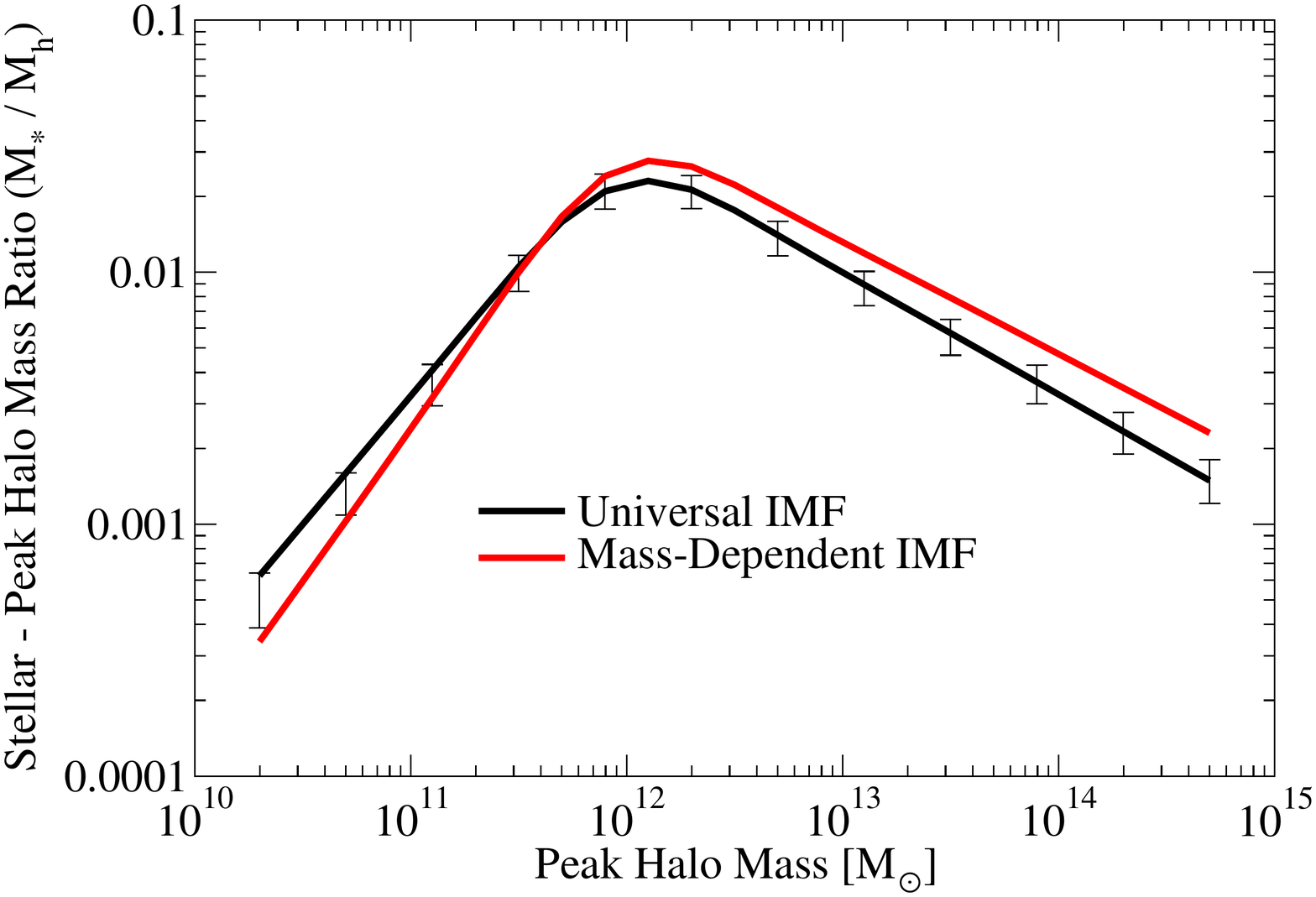}\\[-5ex]
\renewcommand\thefigure{G1} 
\caption{Approximate effect of a mass-dependent IMF (Eqs.\ \ref{e:sigma_mstar}--\ref{e:imf_relation}) on the stellar mass--halo mass relation at $z=0$.  The error bars on the non-varying IMF model show 68\% confidence intervals for other uncertainties in determining stellar masses.}
\renewcommand\thefigure{\thesection\arabic{figure}} 
\label{f:imf}
\end{figure}

The \textsc{EmCee} MCMC algorithm \citep{emcee} is naturally parallelizable; we use a custom implementation in C.  In this algorithm, multiple walkers (100--1000) simultaneously traverse parameter space, leapfrogging over each other to select new points to explore.  No communication is required between walkers, with the result that each walker can be run on a different set of processors (or even on a different supercomputer) with near-perfect scaling (Fig.\ \ref{f:perf}).  

Each walker accepts a point in parameter space, populates haloes from the \textit{Bolshoi-Planck} simulation with galaxies according to \S \ref{s:methodology}, and generates mock galaxy catalogues (i.e., containing locations, velocities, stellar masses, and SFRs).  Since the halo catalogue for \textit{Bolshoi-Planck} is $\sim$300 GB even in binary format (i.e., more than most single machines' memory capacity), each walker divides the computations over 144 processors.  Each processor handles a rectangular subvolume, pre-computed such that each processor generates galaxies for the same number of haloes.  Processors work on each subvolume independently, processing haloes in chronological order.  When a given processor finishes computation for haloes at a given timestep for its subvolume, it notifies neighbouring processors so that they may compute statistics (e.g., correlation functions) that require information outside their own subvolumes.  To reduce sampling noise in the $\chi^2$ surface, simulated data is analytically convolved with observational errors (i.e., stellar mass, SFR, redshift, and photometric errors) for all observables.  For two-point and higher-order correlations, all galaxies are weighted according to their net probability of being included in an observational bin.  While this increases computational time per model evaluation, it also allows the use of gradient-based algorithms for finding the best fit as well as making an initial Fisher matrix-based guess for the initial walker distribution.

To reduce communication, \textit{average} observables (e.g., the \textit{average} SSFR for all galaxies, as opposed to the \textit{median} SSFR) are used for constraints whenever possible.  To reduce strain on the cluster I/O system when loading haloes, a Bittorrent-like method is employed, whereby walkers that have already loaded the data share it with others; thus, the data is loaded only once, and the achieved data throughput increases exponentially in time.  At the time Fig.\ \ref{f:perf} was generated, each walker used 1.5 cpu-hours (36 wall-clock seconds; measured using Ivy-Bridge 2.4GHz processors on Edison) per point in parameter space to process the 2.3 billion haloes in \textit{Bolshoi-Planck}.  Recent improvements have lowered this to $0.8$ cpu-hours per point in parameter space.  Source code and documentation are available at \url{https://bitbucket.org/pbehroozi/universemachine}.

\section{Non-Universal Initial Mass Functions}

\label{a:imf}

Many studies have found that the Initial Mass Function (IMF) becomes increasingly bottom-heavy at larger stellar velocity dispersions, although the exact amount and redshift dependence is debated \citep{Conroy12,Conroy13,Geha13,MartinNavarro15,LaBarbera15,Posacki15}.  \cite{vanDokkum17} note that correcting for the radial gradient of the IMF significantly reduces these discrepancies.  We can thus make a crude estimate of the effect of a mass-dependent IMF, starting with a stellar mass--$\sigma$ relation \citep{Zahid16}:
\begin{align}
\sigma(M_\ast) = 118.3 \textrm{km s}^{-1} \left(\frac{M_\ast}{1.8\times 10^{10}\Msun}\right)^{\beta} \label{e:sigma_mstar}\\
\beta = \begin{cases}
0.403 & \textrm{if } M_\ast < 1.8\times 10^{10}\Msun\\
0.250 & \textrm{if } M_\ast \ge 1.8\times 10^{10}\Msun
\end{cases}
\end{align}
where $\beta$ for $M_\ast \ge 1.8\times 10^{10}\Msun$ has been reduced by a factor 1.17; this accounts for our corrections to SDSS luminosities for massive galaxies (Appendix \ref{a:smf}; Fig.\ \ref{f:mous_corr}, middle panel).  This is then combined with an IMF--$\sigma$ relation \citep{Posacki15}:
 \begin{equation}
 \log_{10}(\alpha_\mathrm{P15}) = 0.38 \log_{10}\left(\frac{\sigma}{200 \textrm{km s}^{-1}}\right) + 0.17,
 \end{equation}
where $\alpha_\mathrm{P15}$ is the mass-to-light ratio compared to that for a \cite{Chabrier03} IMF.  We reduce the \cite{Posacki15} $\alpha_\mathrm{P15}$ according to the aperture correction in \cite{vanDokkum17} as
\begin{equation}
\alpha_\mathrm{vD17} = 1 + 0.75(\alpha_\mathrm{P15} - 1). \label{e:imf_relation}
\end{equation}
Perhaps coincidentally, Eq.\ \ref{e:imf_relation} matches the IMF offset for ultrafaint dwarfs in \cite{Geha13}.  Fig.\ \ref{f:imf} shows the resulting effect on the stellar mass--halo mass relation.  This would suggest that IMF uncertainties remain substantial--as large as other modeling uncertainties combined.  On the other hand, we note that none of the current observations used (Appendix \ref{a:data}) are sensitive to the low-mass stars that dominate IMF uncertainties.  As a result, as improved IMF determinations become available, the star formation rates derived here can be multiplied by a factor $\alpha_\mathrm{IMF}(M_\ast, z)$ to rescale the results appropriately.
\section{Best-Fitting Model Parameters}

\label{a:bestfit}

The resulting best-\fit{} and 68\% confidence intervals for the posterior parameter distribution follow:\\

\noindent\textbf{Median Star Formation Rates:}
\begin{align}
\allowdisplaybreaks
\rlap{Characteristic $\vmp$ [km s$^{-1}$]:}\phantom{100000000}\nonumber\\
\log_{10}(V) = & \, 2.151^{+0.039}_{-0.074} + (-1.658^{+0.892}_{-0.201})(1-a) \nonumber\\
& \,+ (1.680^{+0.149}_{-0.680})\ln(1+z) + (-0.233^{+0.119}_{-0.027})z\nonumber \displaybreak[0]\\
\rlap{Characteristic SFR [$\Msun$ yr$^{-1}$]:}\phantom{100000000}\nonumber\\
\log_{10}(\epsilon) = & \, 0.109^{+0.204}_{-0.249} + (-3.441^{+3.137}_{-1.035})(1-a) \nonumber\\
& \, + (5.079^{+0.749}_{-2.780})\ln(1+z) + (-0.781^{+0.592}_{-0.131})z\displaybreak[0]\nonumber\\
\rlap{Faint-end slope of SFR--$\vmp$ relation:}\phantom{100000000}\nonumber\\
\alpha = & \, -5.598^{+0.211}_{-0.995} + (-20.731^{+12.145}_{-1.253})(1-a) \nonumber\\
& \, + (13.455^{+1.211}_{-7.936})\ln(1+z) + (-1.321^{+0.957}_{-0.240})z\nonumber\displaybreak[0]\\
\rlap{Massive-end slope of SFR--$\vmp$ relation:}\phantom{100000000}\nonumber\\
\beta = & \, -1.911^{+0.804}_{-1.334} + (0.395^{+5.739}_{-3.341})(1-a) \nonumber\\
& \, + (0.747^{+0.801}_{-1.015})z\nonumber\displaybreak[0]\\
\rlap{Strength of Gaussian SFR efficiency boost:}\phantom{100000000}\nonumber\\
\log_{10}(\gamma) = & \, -1.699^{+0.343}_{-0.214} + (4.206^{+0.947}_{-1.012})(1-a) \nonumber\\
& \, + (-0.809^{+0.366}_{-0.302})z\nonumber\displaybreak[0]\\
\rlap{Width of Gaussian SFR efficiency boost:}\phantom{100000000}\nonumber\\
\delta = & \, 0.055^{+0.026}_{-0.010}\nonumber\displaybreak[0]
\end{align}
\noindent\textbf{Quenched Fractions:}
\begin{align}
\allowdisplaybreaks
\rlap{Minimum quenched fraction:}\phantom{100000000}\nonumber\\
Q_\mathrm{min} = & \, \max(0, -1.944^{+1.769}_{-1.669} + (-2.419^{+3.874}_{-2.698})(1-a)) \displaybreak[0]\nonumber\\
\rlap{Characteristic $\vmp$ for quenching [km s$^{-1}$]:}\phantom{100000000}\nonumber\\
\log_{10}(V_Q) = & \, 2.248^{+0.011}_{-0.026} + (-0.018^{+0.152}_{-0.081})(1-a) \nonumber\\
& \, + (0.124^{+0.026}_{-0.050})z\nonumber\displaybreak[0]\\
\rlap{Characteristic $\vmp$ width for quenching [dex]:}\phantom{100000000}\nonumber\\
\sigma_{VQ} = & \, 0.227^{+0.062}_{-0.038} + (0.037^{+0.337}_{-0.543})(1-a) \nonumber\\
& \, + (-0.107^{+0.299}_{-0.178})z\nonumber\displaybreak[0]
\end{align}
\textbf{Galaxy-Halo Assembly Correlations:}
\begin{align}
\allowdisplaybreaks
\rlap{SFR scatter for star-forming galaxies:}\phantom{100000000}\nonumber\\
\sigma_\mathrm{SF} =  & \, \min(-4.361^{+4.227}_{-3.673} + (26.926^{+16.474}_{-21.836})(1-a), 0.3)\;\mathrm{dex} \displaybreak[0]\nonumber\\
\rlap{Minimum correlation for halo--galaxy assembly:}\phantom{100000000}\nonumber\\
r_\mathrm{min} = & \, 0.140^{+0.147}_{-0.230}\nonumber\displaybreak[0]\\
\rlap{Characteristic $\vmp$ for assembly correlations [km s$^{-1}$]:}\phantom{100000000}\nonumber\\
\log_{10}(V_R) = & \, 2.125^{+1.338}_{-1.374} + (-2.976^{+3.149}_{-1.323})(1-a) \nonumber\displaybreak[0]\\
\rlap{Characteristic width of assembly correlation change [dex]:}\phantom{100000000}\nonumber\\
r_\mathrm{width} = & \, 6.989^{+1.861}_{-4.256}\nonumber\displaybreak[0]\\
\rlap{Fraction of short- (vs.\ long-) timescale random variations in SFR:}\phantom{100000000}\nonumber\\
f_\mathrm{short} = & \, 0.465^{+0.398}_{-0.276}\nonumber\displaybreak[0]
\end{align}
\textbf{Galaxy Mergers:}
\begin{align}
\allowdisplaybreaks
\rlap{Sat.\ merger threshold ($\vmax/\vmp$), 300 km s$^{-1}$ host haloes:}\phantom{100000000}\nonumber\\
T_\mathrm{orphan,300} = & \, 0.544^{+0.067}_{-0.043}\phantom{100000000000000000000000000000}\nonumber\displaybreak[0]\\
\rlap{Sat.\ merger threshold ($\vmax/\vmp$), 1000 km s$^{-1}$ host haloes:}\phantom{100000000}\nonumber\\
T_\mathrm{orphan,1000} = & \, 0.466^{+0.033}_{-0.048}\phantom{10000000000000000000000000000}\nonumber\displaybreak[0]\\
\rlap{Central galaxy merger threshold ($R$/$R_\mathrm{vir,host}$):}\phantom{100000000}\nonumber\\
f_\mathrm{merge} = & \, 0.400^{+0.117}_{-0.050}\nonumber\displaybreak[0]
\end{align}
\textbf{Dust:}
\begin{align}
\allowdisplaybreaks
\rlap{Characteristic UV luminosity [AB mag] for dust to become important:}\phantom{100000000}\nonumber\\
M_\mathrm{dust} = & \, -20.594^{+0.191}_{-0.506} + (-0.054^{+0.170}_{-0.103})(\max(4,z)-4) \nonumber\displaybreak[0]\\
\rlap{UV luminosity change [mag] for increases in dust attenuation:}\phantom{100000000}\nonumber\\
\alpha_\mathrm{dust} = & \, 0.559^{+0.102}_{-0.016}\nonumber\displaybreak[0]
\end{align}
\textbf{Systematics:}
\begin{align}
\allowdisplaybreaks
\rlap{Offset between observed and true stellar masses [dex]:}\phantom{100000000}\nonumber\\
\mu = & \, 0.041^{+0.053}_{-0.077} + (-0.044^{+0.060}_{-0.058})(1-a)\phantom{10000000} \nonumber\displaybreak[0]\\
\rlap{Offset between observed and true SFRs at $z=2$ [dex]:}\phantom{100000000}\nonumber\\
\kappa = & \, 0.314^{+0.028}_{-0.029}  \nonumber\displaybreak[0]\\
\rlap{Random error in recovering observed stellar masses:}\phantom{100000000}\nonumber\\
\sigma_\mathrm{SM} = & \, \min(0.070 + (0.071^{+0.009}_{-0.012})z, 0.3)\; \mathrm{dex} \nonumber\displaybreak[0]
\end{align}

\begin{figure*}
\vspace{-15ex}
\plotlargegrace{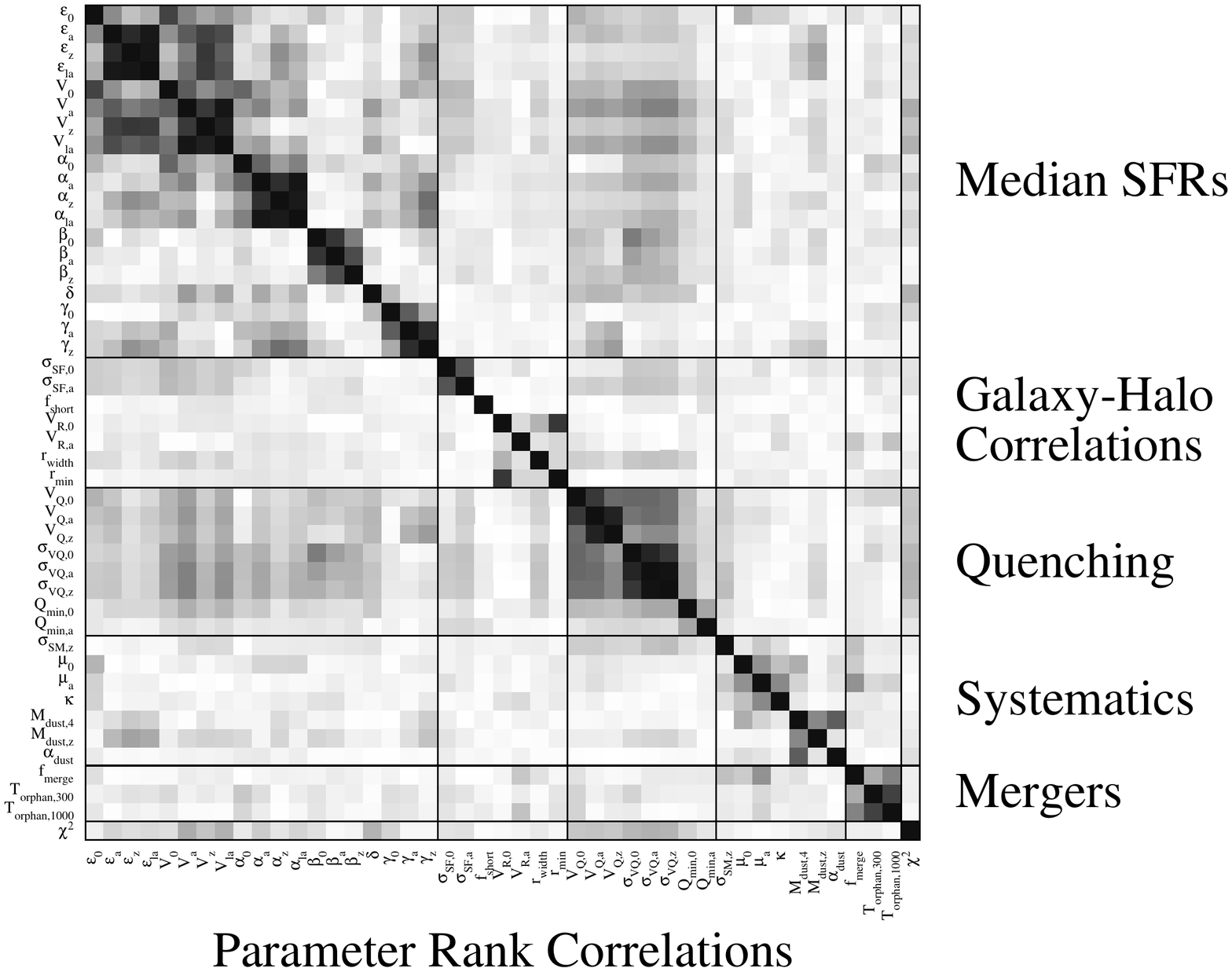}\\[-5ex]
\renewcommand\thefigure{I1} 
\caption{Rank correlations between parameters in the model posterior distribution.  Darker shades indicate higher absolute values of rank correlation coefficients (both positive and negative).  An interactive version of this figure that includes easy access to the underlying joint parameter distributions is available \href{http://www.peterbehroozi.com/data}{online}.}
\renewcommand\thefigure{\thesection\arabic{figure}} 
\label{f:parameter_corrs}
\end{figure*}

\section{Parameter Correlations}

\label{a:correlations}

Rank correlations between parameters in the model posterior distribution are shown in Fig.\ \ref{f:parameter_corrs}.  Correlations between parameters that specify the redshift scaling are common and expected.  For example, $\ln(1+z)$ scales similarly at $z<0.25$ to $1-a$ and to $z$, resulting in strong natural degeneracies between $\epsilon_\mathrm{a}$, $\epsilon_\mathrm{la}$, and $\epsilon_\mathrm{z}$.  Note that this does \textit{not} imply that an equally good fit could be obtained by removing two of these parameters---only that they are not orthogonal.  In fact, the \textsc{UniverseMachine} internally uses linear combinations of parameters to increase orthogonality, reducing burn-in time.  See Appendix \ref{a:fforms} for an explanation of the chosen parameterization; the redshift scalings were chosen due to ease of interpretation and implementation, rather than orthogonality.

Several correlations are physically intuitive.  Increasing both median SFRs \textit{and} quenched fractions appropriately will leave predicted SMFs unchanged, leading to natural degeneracy between the parameter groups.  Systematics parameters are degenerate with both median SFRs and quenched fractions for the same reason.  Within the systematics parameters, $M_\mathrm{dust}$ correlates with $\mu_0$, suggesting that the UV luminosity function can be left unchanged by increasing both dust and the total stellar mass produced.  Luminosity functions in redder bands are hence critical to break this type of degeneracy.  Since the total intrahalo light is not being used as a constraint, the distance for satellites to merge into the intrahalo light is modestly degenerate with the threshold for keeping satellites as orphan galaxies.

Two non-correlations are also worth noting.  Within median SFRs, the scaling for low-mass galaxies ($\alpha$) is independent of the scaling for high-mass galaxies ($\beta$).  This is expected as the double power-law functional form for $SFR_\mathrm{SF}$ effectively isolates the two parameters, and predictions for the constraining data sets (total SMFs for $\alpha$ and high-mass SSFRs for $\beta$) can be adjusted relatively independently.  In addition, parameters for the strength of the galaxy-halo assembly connection are relatively uncorrelated with median SFRs, as the former have little effect on the total stellar mass produced (which constrains median SFRs), and median SFRs have little effect on the separation in autocorrelation functions for quenched and star-forming galaxies (which constrains the galaxy-halo assembly correlation).

\section{Fits to Stellar Mass--Halo Mass Relations}

\label{a:smhm_fits}

To fit median stellar mass--halo mass relationships, we use:
\begin{align}
\allowdisplaybreaks
\log_{10}\left(\frac{M_\ast}{M_1}\right) = & \, \epsilon - \log_{10}\left(10^{-\alpha x} + 10^{-\beta x}\right) + \gamma\exp\left[-0.5\left(\frac{x}{\delta}\right)^2\right]\\
x \equiv & \log_{10} \left(\frac{M_\mathrm{peak}}{M_1}\right)
\end{align}
This is a double power-law plus a Gaussian, similar to Eq.\ \ref{e:sfr}.  The adopted redshift scaling is:
\begin{align}
\log_{10}\left(\frac{M_1}{\Msun}\right) = & \, M_0 + M_a(a-1) - M_\mathrm{lna}\ln(a)+M_z z\\
\epsilon = &\, \epsilon_0 + \epsilon_a(a-1) - \epsilon_\mathrm{lna}\ln(a)+\epsilon_z z\\
\alpha = &\, \alpha_0 + \alpha_a(a-1) - \alpha_\mathrm{lna}\ln(a)+\alpha_z z\\
\beta = & \, \beta_0 + \beta_a(a-1) + \beta_z z\\
\delta = & \, \delta_0\\
\log_{10}(\gamma) = &\, \gamma_0 + \gamma_a(a-1) + \gamma_z z
\end{align}
These formulae are fit to the mock catalogues' actual median SM--HM relations at 22 redshifts ($z=0.1$ and $z=0-10$ in steps of $\Delta z = 0.5$) in bins of width $0.2$ dex for $10^{10.5}\Msun < M_\mathrm{peak} < 10^{15}\Msun$.  Best-\fit{} parameters and uncertainties are shown in Table \ref{t:smhm_fits} for many different selection cuts.  Fits were performed with a tolerance of 0.03 dex; $\chi^2$ values of $>200$ in Table \ref{t:smhm_fits} indicate that the fit often exceeded this tolerance.  These cases only happen for quenched galaxies, when low-number statistics prevent good fits at high redshifts; nonetheless, if such fits are used, then comparison with the raw data accompanying this paper is necessary to ensure that the fits are not being used outside their well-fitting range.

The data in Table \ref{t:smhm_fits} and \textsc{Python} code to evaluate the fits at arbitrary redshifts are both available \href{http://www.peterbehroozi.com/data}{online}.

\begin{landscape}
\begin{table}
\caption{Fit Parameters for Stellar Mass--Halo Mass Relations}\label{t:smhm_fits}
\small
\begin{tabular}{cccccccccccccc}
\hline
SM & Q/SF & Cen/Sat & IHL & $\epsilon_0$ & $\epsilon_a$ & $\epsilon_\mathrm{lna}$ & $\epsilon_z$ & $M_0$ & $M_a$ & $M_\mathrm{lna}$ & $M_z$ & $\alpha_0$ & $\alpha_a$ \\ 
 & & & & $\alpha_\mathrm{lna}$ & $\alpha_z$ & $\beta_0$ & $\beta_a$ & $\beta_z$ & $\delta_0$ & $\gamma_0$ & $\gamma_a$ & $\gamma_z$ & $\chi^2$\\ 
\hline
Obs. & All & All & Excl. & $-1.435^{+0.023}_{-0.075}$& $1.831^{-0.066}_{-2.818}$& $1.368^{+0.089}_{-2.557}$& $-0.217^{+0.466}_{-0.048}$& $12.035^{+0.008}_{-0.100}$& $4.556^{+0.076}_{-2.453}$& $4.417^{+0.237}_{-2.255}$& $-0.731^{+0.464}_{-0.064}$& $1.963^{+0.163}_{-0.010}$& $-2.316^{+0.855}_{-1.361}$\\
 & & & & $-1.732^{+0.519}_{-1.271}$& $0.178^{+0.196}_{-0.084}$& $0.482^{+0.053}_{-0.002}$& $-0.841^{+0.709}_{+0.188}$& $-0.471^{+0.288}_{+0.049}$& $0.411^{+0.060}_{-0.087}$& $-1.034^{+0.264}_{-0.177}$& $-3.100^{+2.496}_{-0.363}$& $-1.055^{+0.973}_{-0.127}$ & 157\\[2ex]
Obs. & All & Cen. & Excl. & $-1.435^{+0.015}_{-0.076}$& $1.813^{+0.044}_{-2.839}$& $1.353^{+0.242}_{-2.606}$& $-0.214^{+0.481}_{-0.075}$& $12.081^{+0.008}_{-0.080}$& $4.696^{+0.036}_{-2.398}$& $4.485^{+0.159}_{-2.237}$& $-0.740^{+0.446}_{-0.057}$& $1.957^{+0.137}_{-0.002}$& $-2.650^{+0.977}_{-1.406}$\\
 & & & & $-1.953^{+0.632}_{-1.310}$& $0.204^{+0.186}_{-0.116}$& $0.474^{+0.057}_{+0.000}$& $-0.903^{+0.746}_{+0.210}$& $-0.492^{+0.303}_{+0.062}$& $0.386^{+0.047}_{-0.068}$& $-1.065^{+0.266}_{-0.229}$& $-3.243^{+2.577}_{-0.346}$& $-1.107^{+1.010}_{-0.119}$ & 156\\[2ex]
Obs. & All & Sat. & Excl. & $-1.449^{+0.102}_{-0.046}$& $-1.256^{+1.289}_{-0.377}$& $-1.031^{+1.076}_{-0.583}$& $0.108^{+0.156}_{-0.178}$& $11.896^{+0.033}_{-0.069}$& $3.284^{+0.427}_{-1.425}$& $3.413^{+0.513}_{-1.276}$& $-0.580^{+0.271}_{-0.098}$& $1.949^{+0.198}_{-0.022}$& $-4.096^{+1.604}_{+0.015}$\\
 & & & & $-3.226^{+1.204}_{-0.109}$& $0.401^{+0.043}_{-0.200}$& $0.477^{+0.016}_{-0.053}$& $0.046^{+0.150}_{-0.308}$& $-0.214^{+0.111}_{-0.157}$& $0.357^{+5.809}_{-0.060}$& $-0.755^{+0.027}_{-2.410}$& $0.461^{-0.084}_{-4.909}$& $0.025^{+0.062}_{-1.674}$ & 170\\[2ex]
Obs. & Q & All & Excl. & $-1.471^{+0.027}_{-0.133}$& $-1.952^{+1.689}_{-0.153}$& $-2.508^{+1.920}_{-0.266}$& $0.499^{+0.094}_{-0.454}$& $12.021^{+0.062}_{-0.072}$& $3.368^{+0.129}_{-2.588}$& $3.615^{+0.164}_{-2.666}$& $-0.645^{+0.634}_{-0.018}$& $1.851^{+0.081}_{-0.211}$& $-4.244^{+4.170}_{+0.745}$\\
 & & & & $-4.402^{+4.680}_{+1.056}$& $0.803^{-0.258}_{-1.395}$& $0.505^{+0.058}_{-0.009}$& $-0.125^{+0.172}_{-0.159}$& $-0.094^{+0.097}_{-0.112}$& $0.461^{+0.069}_{-0.029}$& $-0.858^{+0.313}_{-0.099}$& $-0.933^{+0.492}_{-0.836}$& $-0.098^{+0.069}_{-0.198}$ & 184\\[2ex]
Obs. & Q & Cen. & Excl. & $-1.480^{+0.028}_{-0.081}$& $-0.831^{+1.297}_{-0.731}$& $-1.351^{+1.433}_{-0.868}$& $0.321^{+0.225}_{-0.315}$& $12.069^{+0.032}_{-0.047}$& $2.646^{+0.466}_{-1.486}$& $2.710^{+0.599}_{-1.550}$& $-0.431^{+0.347}_{-0.159}$& $1.899^{+0.082}_{-0.090}$& $-2.901^{+2.210}_{-0.442}$\\
 & & & & $-2.413^{+2.492}_{-0.279}$& $0.332^{+0.035}_{-0.613}$& $0.502^{+0.047}_{-0.014}$& $-0.315^{+0.234}_{-0.132}$& $-0.218^{+0.134}_{-0.069}$& $0.397^{+0.067}_{-0.015}$& $-0.867^{+0.197}_{-0.110}$& $-1.146^{+0.432}_{-0.973}$& $-0.294^{+0.179}_{-0.349}$ & 202$\dag$\\[2ex]
Obs. & SF & All & Excl. & $-1.441^{+0.037}_{-0.088}$& $1.697^{-0.186}_{-3.365}$& $1.326^{+0.041}_{-3.146}$& $-0.221^{+0.576}_{-0.052}$& $12.054^{+0.015}_{-0.095}$& $4.554^{-0.052}_{-2.725}$& $4.484^{+0.089}_{-2.536}$& $-0.750^{+0.522}_{-0.033}$& $1.976^{+0.146}_{-0.039}$& $-2.123^{+0.639}_{-1.519}$\\
 & & & & $-1.617^{+0.449}_{-1.328}$& $0.162^{+0.207}_{-0.091}$& $0.465^{+0.049}_{-0.023}$& $-1.071^{+1.051}_{+0.098}$& $-0.659^{+0.460}_{+0.000}$& $0.436^{+0.091}_{-0.108}$& $-1.016^{+0.240}_{-0.525}$& $-2.862^{+1.942}_{-0.897}$& $-0.941^{+0.800}_{-0.217}$ & 176\\[2ex]
Obs. & SF & Cen. & Excl. & $-1.426^{+0.011}_{-0.116}$& $1.588^{+0.064}_{-3.211}$& $1.237^{+0.298}_{-3.066}$& $-0.210^{+0.539}_{-0.080}$& $12.071^{+0.025}_{-0.075}$& $4.633^{-0.048}_{-2.488}$& $4.527^{+0.050}_{-2.286}$& $-0.757^{+0.483}_{-0.030}$& $1.985^{+0.120}_{-0.053}$& $-2.492^{+0.820}_{-1.406}$\\
 & & & & $-1.860^{+0.609}_{-1.222}$& $0.188^{+0.196}_{-0.108}$& $0.448^{+0.067}_{-0.005}$& $-1.121^{+1.064}_{+0.111}$& $-0.665^{+0.442}_{+0.017}$& $0.407^{+0.113}_{-0.075}$& $-1.149^{+0.396}_{-0.295}$& $-3.221^{+2.266}_{-0.120}$& $-1.022^{+0.883}_{-0.038}$ & 178\\[2ex]
\hline
True & All & All & Excl. & $-1.430^{+0.024}_{-0.143}$& $1.796^{-0.285}_{-2.992}$& $1.360^{-0.070}_{-2.830}$& $-0.216^{+0.495}_{-0.040}$& $12.040^{+0.004}_{-0.102}$& $4.675^{-0.003}_{-2.512}$& $4.513^{+0.154}_{-2.281}$& $-0.744^{+0.446}_{-0.054}$& $1.973^{+0.158}_{-0.016}$& $-2.353^{+0.748}_{-1.315}$\\
 & & & & $-1.783^{+0.466}_{-1.294}$& $0.186^{+0.191}_{-0.080}$& $0.473^{+0.065}_{+0.005}$& $-0.884^{+0.748}_{+0.249}$& $-0.486^{+0.300}_{+0.071}$& $0.407^{+0.079}_{-0.074}$& $-1.088^{+0.336}_{-0.170}$& $-3.241^{+2.649}_{+0.085}$& $-1.079^{+1.002}_{-0.053}$ & 115\\[2ex]
True & All & All & Incl. & $-1.466^{+0.008}_{-0.177}$& $1.852^{+0.041}_{-2.940}$& $1.439^{+0.137}_{-2.853}$& $-0.227^{+0.493}_{-0.059}$& $12.013^{+0.015}_{-0.106}$& $4.597^{+0.084}_{-2.550}$& $4.470^{+0.257}_{-2.416}$& $-0.737^{+0.488}_{-0.069}$& $1.965^{+0.146}_{-0.053}$& $-2.137^{+0.763}_{-1.400}$\\
 & & & & $-1.607^{+0.572}_{-1.291}$& $0.161^{+0.215}_{-0.090}$& $0.564^{+0.091}_{+0.013}$& $-0.835^{+0.680}_{+0.115}$& $-0.478^{+0.293}_{+0.041}$& $0.411^{+0.082}_{-0.062}$& $-0.937^{+0.355}_{-0.077}$& $-2.810^{+2.417}_{+0.459}$& $-0.983^{+0.911}_{+0.135}$ & 130\\[2ex]
True & All & Cen. & Excl. & $-1.431^{+0.024}_{-0.131}$& $1.757^{-0.146}_{-2.914}$& $1.350^{-0.021}_{-2.787}$& $-0.218^{+0.476}_{-0.049}$& $12.074^{+0.011}_{-0.083}$& $4.600^{+0.100}_{-2.297}$& $4.423^{+0.236}_{-2.179}$& $-0.732^{+0.429}_{-0.060}$& $1.974^{+0.136}_{-0.020}$& $-2.468^{+0.614}_{-1.448}$\\
 & & & & $-1.816^{+0.422}_{-1.388}$& $0.182^{+0.205}_{-0.077}$& $0.470^{+0.061}_{+0.004}$& $-0.875^{+0.744}_{+0.198}$& $-0.487^{+0.307}_{+0.066}$& $0.382^{+0.050}_{-0.067}$& $-1.160^{+0.363}_{-0.121}$& $-3.634^{+2.907}_{+0.154}$& $-1.219^{+1.132}_{+0.029}$ & 126\\[2ex]
True & All & Cen. & Incl. & $-1.462^{+0.006}_{-0.180}$& $1.882^{-0.033}_{-2.995}$& $1.446^{+0.117}_{-2.765}$& $-0.224^{+0.480}_{-0.055}$& $12.055^{+0.003}_{-0.101}$& $4.667^{+0.101}_{-2.515}$& $4.471^{+0.254}_{-2.356}$& $-0.735^{+0.458}_{-0.067}$& $1.956^{+0.134}_{-0.037}$& $-2.570^{+0.941}_{-1.044}$\\
 & & & & $-1.904^{+0.678}_{-1.088}$& $0.198^{+0.169}_{-0.119}$& $0.558^{+0.093}_{+0.017}$& $-0.840^{+0.686}_{+0.128}$& $-0.472^{+0.287}_{+0.036}$& $0.393^{+0.046}_{-0.066}$& $-1.004^{+0.388}_{-0.016}$& $-2.983^{+2.514}_{+0.380}$& $-0.996^{+0.916}_{+0.079}$ & 142\\[2ex]
True & All & Sat. & Excl. & $-1.432^{+0.083}_{-0.069}$& $-1.231^{+1.233}_{-0.415}$& $-0.999^{+0.969}_{-0.596}$& $0.100^{+0.127}_{-0.154}$& $11.889^{+0.041}_{-0.049}$& $3.236^{+0.131}_{-1.531}$& $3.378^{+0.264}_{-1.395}$& $-0.577^{+0.295}_{-0.050}$& $1.959^{+0.172}_{-0.022}$& $-4.033^{+1.786}_{-0.005}$\\
 & & & & $-3.175^{+1.391}_{-0.197}$& $0.390^{+0.056}_{-0.229}$& $0.464^{+0.019}_{-0.041}$& $0.130^{+0.248}_{-0.233}$& $-0.153^{+0.135}_{-0.133}$& $0.319^{+3.829}_{-0.060}$& $-0.812^{+0.079}_{-1.933}$& $0.522^{-0.047}_{-3.413}$& $0.064^{+0.031}_{-0.882}$ & 127\\[2ex]
True & Q & All & Excl. & $-1.491^{+0.045}_{-0.145}$& $-2.313^{+1.835}_{-0.044}$& $-2.778^{+2.055}_{-0.370}$& $0.492^{+0.192}_{-0.438}$& $12.005^{+0.088}_{-0.057}$& $3.294^{+0.350}_{-2.069}$& $3.669^{+0.257}_{-2.309}$& $-0.683^{+0.603}_{-0.027}$& $1.852^{+0.070}_{-0.207}$& $-3.922^{+3.359}_{+0.432}$\\
 & & & & $-4.052^{+3.886}_{+0.636}$& $0.692^{-0.200}_{-1.087}$& $0.511^{+0.044}_{-0.024}$& $-0.028^{+0.092}_{-0.252}$& $-0.041^{+0.055}_{-0.160}$& $0.506^{+0.018}_{-0.076}$& $-0.858^{+0.276}_{-0.118}$& $-0.902^{+0.499}_{-0.897}$& $-0.041^{+0.033}_{-0.269}$ & 133\\[2ex]
True & Q & Cen. & Excl. & $-1.462^{+0.022}_{-0.136}$& $-0.732^{+0.798}_{-1.226}$& $-1.273^{+0.970}_{-1.471}$& $0.302^{+0.296}_{-0.266}$& $12.072^{+0.042}_{-0.036}$& $3.581^{+0.000}_{-2.035}$& $3.665^{+0.109}_{-1.998}$& $-0.634^{+0.425}_{-0.056}$& $1.928^{+0.053}_{-0.126}$& $-3.472^{+1.880}_{-0.756}$\\
 & & & & $-3.119^{+2.116}_{-0.431}$& $0.507^{+0.009}_{-0.522}$& $0.488^{+0.047}_{-0.010}$& $-0.419^{+0.349}_{+0.016}$& $-0.256^{+0.226}_{-0.023}$& $0.406^{+0.083}_{-0.027}$& $-0.980^{+0.273}_{-0.090}$& $-1.443^{+0.695}_{-0.702}$& $-0.335^{+0.252}_{-0.231}$ & 153\\[2ex]
True & SF & All & Excl. & $-1.494^{+0.083}_{-0.105}$& $1.569^{-0.202}_{-3.177}$& $1.293^{-0.010}_{-2.898}$& $-0.215^{+0.514}_{-0.027}$& $12.059^{-0.000}_{-0.106}$& $4.645^{-0.160}_{-2.402}$& $4.544^{+0.000}_{-2.244}$& $-0.757^{+0.446}_{-0.032}$& $1.905^{+0.216}_{+0.037}$& $-2.555^{+0.815}_{-1.204}$\\
 & & & & $-1.875^{+0.607}_{-1.180}$& $0.197^{+0.187}_{-0.110}$& $0.509^{+0.032}_{-0.053}$& $-0.889^{+1.073}_{+0.347}$& $-0.538^{+0.441}_{+0.103}$& $0.460^{+0.069}_{-0.125}$& $-0.807^{+0.093}_{-0.623}$& $-1.859^{+1.187}_{-0.659}$& $-0.637^{+0.526}_{-0.185}$ & 98\\[2ex]
True & SF & Cen. & Excl. & $-1.459^{+0.040}_{-0.155}$& $1.515^{-0.058}_{-3.018}$& $1.249^{+0.104}_{-2.703}$& $-0.214^{+0.488}_{-0.048}$& $12.060^{+0.026}_{-0.074}$& $4.609^{-0.072}_{-2.215}$& $4.525^{+0.017}_{-2.098}$& $-0.756^{+0.423}_{-0.032}$& $1.972^{+0.142}_{-0.030}$& $-2.523^{+0.728}_{-1.469}$\\
 & & & & $-1.868^{+0.591}_{-1.382}$& $0.188^{+0.219}_{-0.097}$& $0.488^{+0.059}_{-0.044}$& $-0.965^{+1.083}_{+0.306}$& $-0.569^{+0.450}_{+0.106}$& $0.391^{+0.331}_{-0.059}$& $-0.958^{+0.260}_{-0.440}$& $-2.230^{+1.598}_{-0.282}$& $-0.706^{+0.588}_{-0.165}$ & 98\\[2ex]
\hline
\end{tabular}
\normalsize
\\
\parbox{0.85\columnwidth}{\textbf{Notes.} \textsc{Python} code to generate SMHM relations for all these fits at arbitrary redshifts is included in the \textsc{UniverseMachine} distribution \href{http://www.peterbehroozi.com/data}{\textbf{online}}.  Large numbers show the parameters for the best-\fit{} model, and uncertainties show the 68\% confidence interval for the model posterior distribution.  $\dag$Fits with $\chi^2>200$ indicate features in the underlying raw data that are not well-captured.  Direct comparison with the raw data included in the data release should be done to verify that the fit is not being used inappropriately.}
\end{table}
\end{landscape}
\clearpage

\end{document}